\documentclass[preprint,authoryear,10pt,3p]{elsarticleFTP}

\usepackage{amssymb}
\usepackage{tikz}
\usepackage{graphicx}
\usepackage{caption}
\usepackage{mathtools}
\usepackage{amsmath}
\usepackage{multirow}
\usepackage{colortbl}
\usepackage{color}
\usepackage{tabulary}
\usepackage{bm}
\usepackage{amsmath}
\usepackage{eqparbox}
\usepackage{tabularx}
\usepackage{tabulary}
\usepackage[utf8]{inputenc}
\usepackage{booktabs}
\usepackage{kpfonts}
\usepackage{rotating}
\usepackage{enumitem}
\usepackage{hyperref}
\usepackage{setspace}
\usepackage{epigraph}
\usepackage{longtable}
\usepackage{pdflscape}
\usepackage{dirtytalk}

\journal{International Journal of Production Research}

\begin{document}

\begin{frontmatter}
\title{Operations \& Supply Chain Management: Principles and Practice}

\author[label1]{Fotios~Petropoulos\corref{cor1}}
\cortext[cor1]{Corresponding author: f.petropoulos@bath.ac.uk; fotios@bath.edu}
\author[label2]{Henk~Akkermans}
\author[label3]{O.~Zeynep~Aksin}
\author[label206]{Imran~Ali}
\author[label4]{Mohamed~Zied~Babai}
\author[label5]{Ana~Barbosa-Povoa}
\author[label4]{Olga~Battaïa}
\author[label7]{Maria~Besiou}
\author[label8]{Nils~Boysen}
\author[label1]{Stephen~Brammer}
\author[label1]{Alistair~Brandon-Jones}
\author[label103]{Dirk~Briskorn}
\author[label10]{Tyson~R.~Browning}
\author[label11]{Paul~Buijs}
\author[label12]{Piera~Centobelli}
\author[label202]{Andrea~Chiarini}
\author[label13]{Paul~Cousins}
\author[label14]{Elizabeth~A.~Cudney}
\author[label204]{Andrew~Davies}
\author[label104]{Steven~J.~Day}
\author[label15]{René~de~Koster}
\author[label16]{Rommert~Dekker}
\author[label205]{Juliano~Denicol}
\author[label17]{Mélanie~Despeisse}
\author[label18]{Stephen~M.~Disney}
\author[label19]{Alexandre~Dolgui}
\author[label20]{Linh~Duong}
\author[label207]{Malek~El-Qallali}
\author[label21]{Behnam~Fahimnia}
\author[label101]{Fatemeh~Fakhredin}
\author[label22]{Stanley~B.~Gershwin}
\author[label23]{Salar~Ghamat}
\author[label1]{Vaggelis~Giannikas}
\author[label25]{Christoph~H.~Glock}
\author[label26]{Janet~Godsell}
\author[label27]{Kannan~Govindan}
\author[label29]{Claire~Hannibal}
\author[label30]{Anders~Haug}
\author[label31]{Tomislav~Hernaus}
\author[label32]{Juliana~Hsuan}
\author[label33]{Dmitry~Ivanov}
\author[label7]{Marianne~Jahre}
\author[label17]{Björn~Johansson}
\author[label210]{Madan~Shankar~Kalidoss}
\author[label36]{Argyris~Kanellopoulos}
\author[label27]{Devika~Kannan}
\author[label3]{Elif~Karul}
\author[label39]{Konstantinos~V.~Katsikopoulos}
\author[label1]{Ayse~Begüm~Kilic-Ararat}
\author[label102]{Rainer~Kolisch}
\author[label41]{Maximilian~Koppenberg}
\author[label42]{Maneesh~Kumar}
\author[label201]{Yong-Hong~Kuo}
\author[label43]{Andrew~Kusiak}
\author[label1]{Michael~A.~Lewis}
\author[label46]{Stanley~Frederick~W.~T.~Lim}
\author[label47]{Veronique~Limère}
\author[label26]{Jiyin~Liu}
\author[label1]{Omid~Maghazei}
\author[label31]{Matija~Marić}
\author[label101]{Joern~Meissner}
\author[label41]{Miranda~Meuwissen}
\author[label51]{Pietro~Micheli}
\author[label52]{Samudaya~Nanayakkara}
\author[label53]{Bengü~Nur~Özdemir}
\author[label54]{Thanos~Papadopoulos}
\author[label1]{Stephen~Pavelin}
\author[label52]{Srinath~Perera}
\author[label20]{Wendy~Phillips}
\author[label106]{Dennis~Prak}
\author[label57]{Hubert~Pun}
\author[label105]{Sharfah~Ahmad~Qazi}
\author[label58]{Usha~Ramanathan}
\author[label59]{Gerald~Reiner}
\author[label60]{Ewout~Reitsma}
\author[label1]{Jens~K.~Roehrich}
\author[label62]{Nada~R.~Sanders}
\author[label63]{Joseph~Sarkis}
\author[label64]{Nico~André~Schmid}
\author[label1]{Christoph~G.~Schmidt}
\author[label65]{Andreas~Schroeder}
\author[label65]{Kostas~Selviaridis}
\author[label105]{Stefan~Seuring}
\author[label67]{Chuan~Shi}
\author[label68]{Byung-Gak~Son}
\author[label65]{Martin~Spring}
\author[label70]{Brian~Squire}
\author[label2]{Wendy~van~der~Valk}
\author[label11]{Dirk~Pieter~van~Donk}
\author[label203]{Geert-Jan~van~Houtum}
\author[label72]{Miriam~Wilhelm}
\author[label73]{Finn~Wynstra}
\author[label25]{Ting~Zheng}

\address[label1]{School of Management, University of Bath, Bath, UK}
\address[label2]{Department of Information Systems \& Operations Management, Tilburg School of Economics and Management, Tilburg University, Netherlands}
\address[label3]{College of Administrative Sciences and Economics, Koc University, Turkey}
\address[label206]{School of Business and Law, Central Queensland University, Australia}
\address[label4]{Kedge Business School, France}
\address[label5]{Center for Management Studies, Instituto Superior Técnico, University of Lisbon, Portugal}
\address[label7]{Kühne Logistics University, Hamburg, Germany}
\address[label8]{Friedrich-Schiller-Universität Jena, Lehrstuhl für Operations Management, Germany}
\address[label103]{Bergische Universität Wuppertal, Germany}
\address[label10]{Neeley School of Business, Texas Christian University, Fort Worth, Texas, USA}
\address[label11]{Department of Operations, University of Groningen, Netherlands}
\address[label12]{Department of Industrial Engineering, University of Naples “Federico II”, Italy}
\address[label202]{Department of Management, University of Verona, Italy}
\address[label13]{Management School, University of Liverpool, UK}
\address[label14]{John E. Simon School of Business, Maryville University of Saint Louis, Missouri, USA}
\address[label204]{Science Policy Research Unit, University of Sussex Business School, Brighton, UK}
\address[label104]{School of Management, Zhejiang University, China}
\address[label15]{Rotterdam School of Management, Erasmus University, The Netherlands}
\address[label16]{Erasmus School of Economics, Erasmus University, Netherlands}
\address[label205]{Bartlett School of Sustainable Construction, The Bartlett Faculty of the Built Environment, University College London, UK}
\address[label17]{Department of Industrial and Material Sciences, Chalmers University of Technology, Sweden}
\address[label18]{Center for Simulation, Analytics, and Modelling, University of Exeter Business School, UK}
\address[label19]{IMT Atlantique, LS2N-CNRS, Nantes, France}
\address[label20]{College of Business and Law, University of the West of England, UK}
\address[label207]{Rabdan Academy, UAE}
\address[label21]{Institute of Transport and Logistics Studies, The University of Sydney Business School, Australia}
\address[label101]{Kühne Logistics University, Germany}
\address[label22]{Department of Mechanical Engineering, Massachusetts Institute of Technology, Massachusetts, USA}
\address[label23]{Lazaridis School of Business and Economics, Wilfrid Laurier University, Canada}
\address[label25]{Technical University of Darmstadt, Germany}
\address[label26]{Loughborough Business School, Loughborough University, UK}
\address[label27]{Adelaide Business School, University of Adelaide, Australia}
\address[label29]{Robert Gordon University, UK}
\address[label30]{Department of Business and Sustainability, University of Southern Denmark, Denmark}
\address[label31]{Faculty of Economics and Business, University of Zagreb, Croatia}
\address[label32]{Department of Operations Management, Copenhagen Business School, Denmark}
\address[label33]{Berlin School of Economics and Law, Department of Business Administration, Germany}
\address[label210]{Department of Technology and Innovation, University of Southern Denmark, Denmark}
\address[label36]{Operations Research \& Logistics Group, Wageningen University and Research, Netherlands}
\address[label39]{Department of Decision Analytics and Risk, University of Southampton Business School, UK}
\address[label102]{TUM School of Management, Technical University of Munich, Germany}
\address[label41]{Wageningen University and Research, Netherlands}
\address[label42]{Cardiff Business School, Cardiff University, UK}
\address[label201]{Department of Data and Systems Engineering, The University of Hong Kong, Hong Kong}
\address[label43]{Department of Industrial and Systems Engineering, University of Iowa, Iowa, USA}
\address[label46]{Eli Broad College of Business, Michigan State University, Michigan, USA}
\address[label47]{Department of Business Informatics and Operations Management, Ghent University, Belgium}
\address[label51]{Warwick Business School, University of Warwick, UK}
\address[label52]{Centre for Smart Modern Construction, School of Engineering, Design and Built Environment, Western Sydney University, Australia}
\address[label53]{University of Wisconsin-Whitewater, Wisconsin, USA}
\address[label54]{Department of Analytics, Operations and Systems, Kent Business School, University of Kent, UK}
\address[label106]{Department of Industrial Engineering and Business Information Systems, University of Twente, Netherlands}
\address[label57]{Ivey Business School, Western University, Canada}
\address[label105]{Faculty of Economics and Management, University of Kassel, Germany}
\address[label58]{Department of Management, College of Business Administration, University of Sharjah, United Arab Emirates}
\address[label59]{Vienna University of Economics and Business, Austria}
\address[label60]{Department of Supply Chain and Operations Management, School of Engineering, Jönköping University, Sweden}
\address[label62]{Northeastern University, Massachusetts, USA}
\address[label63]{Business School, Worcester Polytechnic Institute, Massachusetts, USA}
\address[label64]{IESEG School of Management, France}
\address[label65]{Lancaster University Management School, Lancaster University, UK}
\address[label67]{School of Data Science, The Chinese University of Hong Kong, Shenzhen, China}
\address[label68]{Bayes Business School, City University London, UK}
\address[label70]{University of Bristol Business School, University of Bristol, UK}
\address[label203]{School of Industrial Engineering, Eindhoven University of Technology, The Netherlands}
\address[label72]{Department of Information Systems and Operations Management, Vienna University of Economics and Business, Austria}
\address[label73]{Department of Technology and Operations Management, Rotterdam School of Management, Erasmus University, Netherlands}

%%%%%%%%%%%%%%%%%%%%%%%%%%%%%%%%%%%%%%%%%%%%%%%%%%%%%%%%%%%%%%%%%%%%%%%%%%%%%%%%%%%%%%%%%%%%%%%%%%%%%%%%%%%%%%%%%

\end{frontmatter}

%\epigraph{A quote here.}{\cite{Ackoff1956-qh}}

\section*{Abstract}
\label{sec:Abstract}
Operations and Supply Chain Management (OSCM) has continually evolved, incorporating a broad array of strategies, frameworks, and technologies to address complex challenges across industries. This encyclopedic article provides a comprehensive overview of contemporary strategies, tools, methods, principles, and best practices that define the field’s cutting-edge advancements. It also explores the diverse environments where OSCM principles have been effectively implemented. The article is meant to be read in a nonlinear fashion. It should be used as a point of reference or first-port-of-call for a diverse pool of readers: academics, researchers, students, and practitioners.

\noindent \textbf{Keywords:} review; encyclopedia; theory; practice; principles; strategy.

%\clearpage
\setcounter{tocdepth}{3}
\small
\tableofcontents
\normalsize

\clearpage

\section[Introduction (Joseph Sarkis)]{Introduction\protect\footnote{This subsection was written by Joseph Sarkis.}}
\label{sec:introduction}
Thousands of years of civilisation has seen modern humans evolve from hunter-gatherers to agrarian societies to city-states and nations. Operations management (OM) and supply chain management (SCM) principles have been ingrained in each of these stages. OM and SCM principles and practices existed even before scientific explanations and theories were established to help us understand and improve them. Practices included managing production, transportation, storage, and distribution -- in many cases as part of historical documents.

Here is an example. A popular western Judeo-Christian bible story -- also appearing in the Torah and the Qur’an -- is of Joseph, son of Jacob. In the book of Genesis (41-47). The story focuses on Joseph’s management of grains during his time in Egypt. He was appointed the senior administrator in what is likely the Egyptian department of agriculture and commerce.  Joseph planned (forecasted) for seven years of surplus and seven years of famine -- trying level supply and demand during peak and trough times. This was an empire-wide effort to ensure efficient grain production (harvesting), delivery, storage, production, financial transactions, and meticulous recordkeeping of thousands of farms and schedules. Inventory management was necessary to provide food and prevent spoilage. Human resources management was needed throughout the process. Joseph didn’t necessarily have textbooks with formalised and scientific tools and techniques (as presented in this compendium paper) -- but the concepts and practices did exist before we gave them their names.

This story -- whether fact or fictional -- exemplifies a period where centralised and mass production concepts were practised.  It also exemplifies a long history of operations and supply chain management principles.  These principles were important enough to appear in religious and historical documents and literature.

Sometimes we lose perspective and believe that pre-modern practices were not as complex as today. Clearly planning for and feeding a whole empire over decades, even in agrarian societies, showed that significant complexities existed in early civilisations. OM and SCM have been ingrained in society for thousands of years and the practices have been intertwined with civilisation and modernity progression.

Many operations and supply chain textbooks provide a review of the historical nature of operations and supply chain management \citep[e.g.,][]{Stevenson2020-yl_JS,Gupta2014-jx_JS}. In fact, many introductory OM and SCM textbooks include the historical evolution of operations and supply chain principles in their first two chapters.

General business history books also have similar outlines. Pre-industrial management principles from the Middle East, Asia, and Africa and the role of the Church and Feudalism are all part of this business history \citep[e.g.,][]{Wilson2021-gx_JS}.

In the book, \textit{A World Lit Only by Fir}e \citep{Manchester1993-mn_JS}, business and trade helped further the growth of the Renaissance and the University system.  Where languages, law, and business became central to a time period where global supply chains started to function.

Modern business history -- after the industrial revolution commences -- has management advances in fixating on productivity and efficiency improvements.  OM and SCM textbooks include economic principal milestones such as Adam Smith’s division of labour allowing greater specialisation with resultant work efficiency; steam power and the industrial revolution (technology); and parts and materials standardisation (processes and designs) including the cotton gin and military weaponry. These principles and tools contributed to productivity.  Then scientific principles emerged (Frederick Winslow Taylor) -- and the initial stages of just-in-time and supply chain (vertical) integration (Henry Ford).  These were the formalisation stages.

During the past century science, mathematics, economics, and technology were interspersed.  Society knew how to build stable bridges even before theoretical science and mathematics emerged to support and improve practice.  To help develop innovations and replicate success theory and practice supported each other. Eventually, modern operations research tools and computerisation resulted in optimisation and greater automated control within OM.

Prior to the 1970’s most operations were considered as shop floor concerns, and sometimes included inter-organisational management of multi-echelon inventory.  Operations was primarily focused on short-term and tactical planning and management concerns -- usually focusing on efficiencies and cost management.  Eventually, operations became strategic \citep{Skinner1969-qq_MALMA}. As a strategy, operations can help organisations differentiate themselves and build competitive advantage.  This strategic perspective grew during the 1970’s and 1980’s as various global competitiveness issues arose -- moving away from mass production to other principles and more nuanced operations perspectives.

Many of us lived and remember a time when operations management was positioned as management sciences and operations research.  But operations and its strategic content is more complex than specific mathematical principles can capture -- wicked problems could not be solved just by single mathematical models and tools \citep{Churchman1967-wa_JS}. Thus, more qualitative management and business perspectives were introduced.  For example, Total Quality Management -- which included analytical models -- also introduced managerial, culture, and qualitative principles such as \textit{continuous improvement}.  This period also portended World Class Manufacturing \citep{Flynn1999-sm_JS} and Operational Excellence \citep{Oakland2020-vc_JS}.

In the later 1980's the term `supply chain management' was introduced.  The adoption and diffusion of the concept started to take hold in the 1990’s with initial programs and textbooks on supply chain management -- an outgrowth of purchasing, logistics, transportation, operations, and marketing -- began to enter the academic lexicon \citep{Mentzer2001-qp_WPLD}.

During these periods -- as exemplified by a historical perspective on sustainability \citep{Sarkis2018-lj_JS} and sustainable development goals \citep[SDGs;][]{Sarkis2022-ox_JS} within the \textit{International Journal of Production Research} -- the focus was mainly on economic and business optimisation.  Supply chain management was mostly focused on the traditional topics, many of which are covered in this paper.

As the strategic focus expanded and business-societal interactions became more pronounced, the OM-SCM research evolved.  Broader and non-traditional manufacturing production concerns further emerged.  Issues such as humanitarian operations, triple-bottom-line sustainability, and healthcare operations entered the research literature during the early 2000's into the 2020's. Highly integrated, smarter, multiple stakeholder technologies also transpired.  Multiple stakeholder perspectives and engaged transdisciplinary involvement in research opened new doors for OM and SCM scholars to step through, into fields that that traditionalists may not have envisioned such as of medical, biological conservation, philosophy ethics, and feminist theory fields.

External pressures and interventions meant that OM and SCM needed greater resilience, decarbonisation, equity, and circularity -- at global and local levels. OM and SCM research catalysts were no longer intra-organisational worries, but societal shifts and concerns. In many cases we went from addressing current problems to prospective and real problems.

Traditional optimisation, operational efficiencies, and productivity remain concerns; but strategic and external environmental and institutional issues are significantly influencing OM and SCM research and application. Oftentimes scholars are a step ahead of practice -- preemptively providing prospective OM and SCM solutions for industry. Exploring solutions that can benefit society, an important dimension of responsible research.

In this scholarly compendium we examine current and emergent OM and SCM topics.  OM topics are initially introduced.  Perspectives on twenty OM themes are covered by leading global scholars. SCM topics are then overviewed by another set of leading scholars.  Eleven SCM sections provide a comprehensive view with state-of-the-art insights presented. Broader society, environment, global economics, and institutional concerns span both OM-SCM. These OM-SCM spanning themes are more prevalent in practice and in theory.

The earlier sections of this paper focus on conceptual and academic perspectives. But, as the saying goes, the ``proof of the pudding is in the eating'' -- the last major section considers various applications from traditional manufacturing and production applications, to agriculture, healthcare, and humanitarian operations applications.  The broad variety of applications represent how the various theories and conceptualisations can help us understand, manage, and support diverse contexts beyond the shopfloor.  These broad applications are evidence that OM and SCM are maturing and more societally important. OM-SCM has become part of the popular discourse -- especially after the interconnected global web of material, information, financial, and capital flows were disrupted during the COVID crisis.

This compendium covers a wide range of topics with scholars from throughout the world. Many are leading scholars who have lived through some of the most significant changes in our field occurring over the past five decades. Some are junior scholars whose perspectives are insightful and needed to guide us into the future. Each topic (section and sub-section) provides what the author of that section believes are important topical elements and with thoughts for future directions.  Many also provide some recommendations on further reading that they feel can help both novice and advanced researchers and practitioners.

We hope that you enjoy this broad compendium of the OM-SCM field.  We believe it can be a valuable reference resource but realise it is a snapshot in time of where we have been, where we are, and the road ahead.

\clearpage

\section{Operations Management}
\label{sec:Operations_Management}

\subsection[Operations strategy (Michael A. Lewis \& Malek El-Qallali)]{Operations strategy\protect\footnote{This subsection was written by Michael A. Lewis and Malek El-Qallali.}}
\label{sec:Operations_strategy}

The study of Operations Strategy (OS) began with \citeauthor{Skinner1969-qq_MALMA}'s (\citeyear{Skinner1969-qq_MALMA}) seminal argument that corporate strategy should be aligned with business unit strategies and, in turn, functional (e.g., operations) strategies \citep{Ward2000-ks_MALMA}. This process should ensure that all levels of strategy are coherent and support strategic objectives. Today, there are many definitions of OS but in their best-selling textbook \cite{Slack2024-lz_MALMA} describe it as the ``total pattern of decisions that shape the long-term capabilities of any type of operation''. This captures the systemic character of OS, combining structure and function, or more pragmatically, the `what' (how big, where, what suppliers, etc.) and `how' of operations capabilities being reconciled with (market) requirements. The specific mechanisms for this reconciliation give rise to alternative OS models.

The first and arguably most influential model is that of `cumulative’' capability building, an approach strongly linked to lean/quality concepts (discussed at length in other sections in this paper). Although exact trajectories (e.g., quality first, then reliability, flexibility, and cost efficiency, etc.) remain the subject of debate \citep{Sarmiento2018-he_MALMA,Singh2015-ty_MALMA,Narasimhan2013-dn_MALMA}, the sequence of building capabilities does matter \citep{Rosenzweig2004-yx_MALMA}. The second model is the idea of `trade-offs', which observes that, in simple terms, if one thing increases, another must decrease \citep{Boyer2002-ie_MALMA}. For instance, a warehouse can carry a few large or many small items. Of course, the concept of a trade-off suggests strategic choices made with complete knowledge of the dis/function of each setup. Interestingly, most of the evidence of firm performance used to underpin the cumulative capabilities logic came from measured improvement within individual plants over time (to lead their competitors in almost every dimension of performance). In contrast, trade-off data came from comparisons across plants at a given time. This may suggest that the two approaches are not in conflict \citep{Schmenner1998-lz_MALMA}. The third model is to be focused; an OS concentrating on a limited set of tasks will likely generate more productive outcomes. Beyond \citeauthor{Skinner1974-ik_MALMA}'s (\citeyear{Skinner1974-ik_MALMA}) original focused factory, there are also well-known examples of focus in healthcare systems around the world, including the Shouldice Hernia Hospital in Canada (taught in virtually every business school), the Aravind Eye Clinic in India and Coxa Hospital for Joint Replacement in Finland. OS research generally \citep{Ketokivi2006-zw_MALMA} supports the benefits of focus \citep{Tsikriktsis2007-jw_MALMA,Huckman2008-qg_MALMA}, and there is a strong logic for the benefits, such as the static economies of scale that appear as either reductions in average cost via amortising of fixed investments or improvements in average quality from learning, depth of know-how, etc. There may also be benefits in `related spillovers'; \cite{Fong-Boh2007-tx_MALMA}, for example, found that software team productivity was more strongly impacted by the group's average experience with related activities than the average experience with the focal activity.

Regardless of the specific approach to OS, measurement and metrics are crucial, providing the foundation for operational control and organisational learning \citep{Sitkin1994-ax_MALMA}. The relationship between strategy and performance measurement is well-documented, with effective measurement systems enhancing strategic alignment at various levels of the organisation \citep{Micheli2010-qb_MALMA}. Interestingly, the idea of programmatic strategic alignment shares similarities with the concept of Hoshin Kanri, the planning process that involves multi-directional communication \citep{Witcher2001-lx_MALMA}.

As well as this emphasis on the process of OS, and although scholars have referred to them in different ways, including `decision areas', `policy areas', `sub-strategies' or `tasks', a great deal of research explores aspects of OS content. From consideration of the (dis)economic scale of capacity and its increments to the timing of any changes \citep{Olhager2001-dy_MALMA} to the set of broad and long-term decisions governing how the operation is improved on a continuing basis \citep[e.g., Six Sigma:][]{Linderman2003-lo_MALMA}, any elements of an operating model that involves more significant decisions (i.e. more existential, capital intensive/complex investments, etc.) with considerable lead-times are of interest to OS researchers. For example, consider the process management triangle \citep{Klassen2007-cn_MALMA}, a heuristic for understanding an implied trade-off relationship between capacity utilisation, variability in terms of input and processing characteristics, and inventory. These interactions suggest generic guidelines for OS decisions. If an operation has high levels of utilisation (a busy factory), it will generate a backlog in the presence of any meaningful variability (such as product variety). A specific capacity strategy for such a situation would be to drive down variability by, for instance, focusing on particular product types, only allowing scheduled demand, etc. Alternatively, if additional capacity can be added, the impact of variability can be significantly reduced, meaning that the system is responsive (low/no backlog) but at the ‘cost’ of low absolute levels of utilisation and high relative costs. Recognition of the balance between costs and flexibility/responsiveness – and echoing the \citep{Schmenner1998-lz_MALMA} argument that trade-off and cumulative capability can be complementary approaches -- just reflecting different aspects of performance improvement over time –- the balance of cost and flexibility is also central to discussions of mass customisation, where firms like BMW have had great success using modular architectures to deliver customised products efficiently \citep{Kortmann2014-di_MALMA}.

Other decisions, such as make-versus-buy (MvB), which sets the boundaries for internal and external resources and capabilities \citep{McIvor2005-iu_MALMA}, have also long been an OS priority. This decision has evolved from its transactional cost economics roots to include OS considerations like protecting and developing resources and managing supply risks. This decision area also highlights how OS has evolved in an era of global business ecosystems \citep[e.g.,][]{Adner2017-pv_MALMA}, the sharing economy \citep{Benjaafar2020-pg_MALMA} and a rapidly evolving set of digital infrastructures \citep{Choi2022-qh_MALMA}. Consequently, the evolution of OS in many industries has shifted away from internal resource management to the dynamic management of external assets. This transformation necessitates re-evaluating traditional operating strategies and MvB's issues. \cite{Maghazei2022_CGS}, for example, explored the acquisition of emerging operational technologies and drones and observed a dynamic interaction between an ecosystem keen to hype a solution and a more tentative market pull from companies. Likewise, \cite{Giustiziero2023-mt_MALMA} noted how many digital firms are both very narrow in scope and very large scale, concluding this reflects different economics of MvB in contemporary digital technology ecosystems. This increasingly underscores the importance of managing relationships with partners, competitors, and customers. OS has always stressed that the health of a firm's supply network influences its performance. Still, this shift to ``hyper specialisation'' requires OS to more pro-actively consider ecosystem health alongside internal capabilities \citep{Iansiti2004-ga_MALMA}. Indeed, many operational technologies are perhaps better characterised by a relational rather than entitative ontology? As \citeauthor{Faraj2022-hv_MALMA} (\citeyear{Faraj2022-hv_MALMA}, p.777) argued when describing how autonomous vehicles rely on GPS, AI, or connection to a charging network, ``[s]ome of these relations may be performed within the firm, but many are performed involving a diverse set of external actors via constant real-time data flows. Thus, no entity within this network of relations is meaningful outside the scope of the relations that constitute it''. This relational framing has potentially profound implications for OS. For example, traditional internal performance metrics must also be complemented by measures of ecosystem health, such as structure, predictability, and vulnerability.

OS's evolution means it is now more global \citep{Gray2014-ly_MALMA} and technologically enabled \citep{Choi2022-qh_MALMA}. There has been a corresponding supply chain turn, and sustainability \citep{Longoni2015-yz_MALMA} is now increasingly (but perhaps insufficiently) a significant concern. There has also been a fuller incorporation of regulatory and other governance concerns \citep{Fan2022-gz_MALMA}. At its core, effective OS is still concerned with the alignment between different levels and functions. Cumulative capabilities, trade-offs, and focus remain critical logics, albeit today, we have an increased understanding of the influence of behavioural factors \citep{Fahimnia2019-sd_KKBF} -- from overconfidence, where individuals believe they know more than they do, to anchoring decisions on otherwise irrelevant past experiences and then making insufficient adjustments to these anchored estimates -- in considering how the OS process works.

\subsection[Capacity management (Joern Meissner \& Fatemeh Fakhredin)]{Capacity management\protect\footnote{This subsection was written by Joern Meissner and Fatemeh Fakhredin.}}
\label{sec:Capacity_management}
Capacity management is a crucial aspect of strategic planning in businesses, focusing on optimizing resource use to achieve maximum efficiency and productivity. \cite{Slack2010-jk_MDBJ} emphasise the importance of aligning capacity with demand to ensure operational efficiency. In essence, capacity management ensures that an organisation has the right amount and mixture of resources at the right time to meet current and future demands. Done correctly, it optimally allocates resources to meet customer demand both efficiently and effectively \citep{Chopra2021-lr_JGSD}. It involves a deep understanding of the organisation's capabilities, allowing for a proactive approach in aligning those capabilities with business needs. \cite{hopp2012_JM} describe capacity management as balancing supply and demand, a process that includes identifying the resources needed, measuring current capacity, and making necessary adjustments to meet customer needs without overextending resources.

Research shows that effective capacity management not only reduces costs but also enhances supply chain performance by improving lead times and service levels \citep{Axsater2015-vc_ZBDP}. In a dynamic market, the ability to respond quickly to fluctuations in demand is paramount. Thus, a well-executed capacity management strategy provides businesses with the agility needed to adapt to changing circumstances \citep{fabbe2008_JM}. For instance, firms that utilise advanced analytics and real-time data can make informed decisions regarding capacity adjustments, leading to significant competitive advantages \citep{aghezzaf2005_JM}.

Moreover, the integration of capacity management with supply chain strategies enhances overall performance. As noted by \cite{montgomery2020_JM}, organisations that align their capacity planning with supply chain logistics can achieve optimal performance and resource utilisation. Consequently, effective capacity management contributes to better service levels, customer satisfaction, and profitability \citep{blanchard2021_JM}.

\subsubsection*{Steps Involved in Capacity Management}
In the business context, capacity management involves several steps:

\textbf{Demand Forecasting:} The process begins with accurately predicting customer demand. \cite{babai2022_JM} stress that accurate demand forecasting is crucial in aligning supply with anticipated customer needs, thus reducing underutilisation and overutilisation of resources. This involves predicting future product or service requirements using historical data, market analysis, and trend forecasting. According to \cite{syntetos2009_JM}, incorporating advanced forecasting methods can significantly improve the accuracy of demand predictions, leading to better capacity planning outcomes.

\textbf{Capacity Evaluation:} Once demand is forecasted, businesses need to evaluate their current capacity. \cite{syntetos2009_JM} argue that continuous evaluation of capacity and capabilities is essential for improving operational performance and resilience in the face of market fluctuations. This step assesses processes, people, and technology to identify gaps between current capacity and future demand. Additionally, \cite{nia2021_JM} highlight the importance of utilizing simulation models to assess capacity in complex systems, allowing for a deeper understanding of potential bottlenecks.

\textbf{Capacity Adjustment:} \cite{wu2005_JM} highlight that timely capacity adjustments can mitigate risks and enhance responsiveness to changing market conditions. After evaluating demand and capacity, businesses may need to adjust resources by expanding, building new facilities, or introducing advanced technologies. The implementation of flexible capacity strategies, as discussed by \cite{law2007_JM}, can provide firms with the ability to scale operations up or down based on fluctuating demand.

\textbf{Performance Monitoring:} Continuous performance monitoring ensures that capacity adjustments result in desired outcomes. Regularly reviewing key performance indicators (KPIs) helps businesses determine whether their capacity management strategies are effective. According to \cite{shumsky2009_JM}, utilizing dashboard reporting tools can facilitate real-time monitoring and improve decision-making processes related to capacity management.  Moreover, leveraging advanced techniques such as Bayesian networks can enhance performance monitoring by modelling the relationships between various performance metrics, allowing organisations to identify potential issues before they escalate \citep{maleki2013_JM}. This proactive approach not only supports more informed decision-making but also fosters a culture of continuous improvement within the supply chain, leading to enhanced operational efficiency \citep{maleki2013_JM}.

\subsubsection*{Scientific Methods Used in Capacity Management}

\begin{itemize}[noitemsep,nolistsep]
\item Little's Law provides insights into the relationship between work-in-progress inventory, throughput, and lead time \citep{hopp2012_JM}.
\item Queuing theory is instrumental in analysing process flows and managing wait times in service systems \citep{gross2011_JM}.
\item Optimisation models are critical tools for efficiently allocating resources to minimise operational costs while meeting demand \citep{varian2014_JM}.
\item Simulation techniques can help in understanding complex systems and predicting the impacts of capacity adjustments \citep{fabbe2008_JM}.
\end{itemize}

\subsubsection*{Extended Strategies for Capacity Management}
\textbf{Outsourcing:} Outsourcing can improve flexibility and allow businesses to focus on core competencies, making it a cost-effective strategy for managing capacity variations \citep{quinn1999_JM}. This strategy enables organisations to adapt their capacity in line with market demands without significant capital investments.By outsourcing certain functions, companies can leverage specialised expertise and resources, which enhances their responsiveness to fluctuations in demand \citep{lee2002_JM}. This strategic decision not only minimises fixed costs but also facilitates advanced planning and scheduling, thereby improving overall supply chain performance \citep{lee2002_JM}.

\textbf{Collaboration and Alliances:} Collaboration between organisations can help capitalise on each other's strengths, achieve economies of scale, and efficiently manage complex supply chains \citep{tan2002_JM}. Joint ventures or partnerships can lead to shared resources, enhancing overall capacity. According to \cite{simatupang2008_JM}, effective collaboration design can significantly improve supply chain performance by facilitating better communication and information sharing among partners. This increased level of collaboration allows organisations to respond more swiftly to market changes and align their strategies, ultimately leading to enhanced operational efficiency and customer satisfaction.

\textbf{Subcontracting:} Subcontracting helps firms manage capacity limitations while focusing on their core strengths \citep{heikkila2002_JM}. By leveraging external expertise and resources, companies can remain agile and responsive to market changes. \cite{dainty2001_JM} highlight that subcontracting not only allows firms to access specialised skills and technology but also fosters collaborative relationships that can enhance overall supply chain performance. This approach enables firms to adapt to demand fluctuations more effectively and optimise their resource allocation, leading to improved project outcomes and customer satisfaction.

\subsubsection*{The Impact of Capacity Management}
Effective capacity management significantly reduces operational costs and enhances customer satisfaction, enabling businesses to balance costs and customer demand, thereby boosting loyalty and competitiveness \citep{Slack2010-jk_MDBJ}. Additionally, it plays a crucial role in building resilience to market disruptions \citep{kochan2018_JM}. Research by \cite{aghezzaf2005_JM} indicates that organisations proactively managing their capacity are better equipped to handle unexpected challenges. Furthermore, \cite{sazvar2021_JM} emphasise that integrating sustainability into capacity planning enhances an organisation's ability to respond to uncertainties, particularly in critical supply chains like vaccines, underscoring the strategic importance of capacity management.

\subsubsection*{Conclusion}
In conclusion, capacity management is a critical business function that involves the strategic allocation of resources to meet current and future demand. It encompasses various processes, including demand forecasting, capacity evaluation, adjustment, and performance monitoring, which collectively contribute to a business's operational efficiency and effectiveness \citep{Slack2010-jk_MDBJ}. By employing scientific principles and data-driven techniques, organisations can enhance their decision-making processes and optimise resource utilisation, ensuring long-term sustainability and growth \citep{varian2014_JM}.

The integration of capacity management with supply chain strategies amplifies its importance, as businesses that align their capacity planning with supply chain logistics can achieve superior performance and customer satisfaction \citep{blanchard2021_JM}. Additionally, the adoption of outsourcing, collaboration, and subcontracting strategies further strengthens capacity management efforts by leveraging external expertise, sharing resources, and fostering flexibility \citep{lee2002_JM, tan2002_JM, dainty2001_JM}.

In a rapidly changing business environment, effective capacity management not only reduces operational costs but also enhances resilience against market disruptions, enabling organisations to respond swiftly to fluctuations in demand and maintain competitive advantage \citep{aghezzaf2005_JM, kochan2018_JM}. As such, the significance of robust capacity management strategies cannot be overstated, as they play a vital role in achieving organisational goals and ensuring customer loyalty.

\subsection[Inventory management (Mohamed Zied Babai \& Dennis Prak)]{Inventory management\protect\footnote{This subsection was written by Mohamed Zied Babai and Dennis Prak.}}
\label{sec:Inventory_management}
Inventory management involves a set of policies and controls designed to monitor the levels of inventory -- including raw materials, components, and finished products -- across various locations within the supply chain, such as warehouses and production systems \citep{Nahmias2009-ug_ZBDP}. The goal is to maintain the right quantity of products to meet demand while minimising holding costs and preventing stockouts \citep{Cachon2023-yx_ZBDP}. From a practical standpoint, inventory management should address three main questions:
\begin{enumerate}[noitemsep,nolistsep]
\item Where to stock the product? 
\item When to place the order to replenish the stock?
\item How much to order to replenish the stock? 
\end{enumerate}

The inventory management research originated with the Economic Order Quantity (EOQ) model \citep{Harris1990-sl_ZBDP}, which optimised inventory levels for single-item, single-echelon systems under the assumption of constant demand. This model established the concept of economies of scale in inventory management. By the 1950s, advancements were made for military applications, focusing on inventory systems with stochastic demand and extending to multi-item and multi-echelon systems.

\subsubsection*{Single-echelon single-item inventory systems}
The early EOQ model, assuming constant demand, was refined by \cite{Wilson1934-sf_ZBDP}, to minimise total costs under linear inventory holding and ordering costs. Extensions included allowances for non-linear costs, imperfect items, shortages costs, etc. \citep{Rosenberg1979-zn_ZBDP,Salameh2000-yc_ZBDP}. In the 1950s, time-varying deterministic demand models emerged, considering finite time horizons and applying dynamic programming, the most well-known one being \cite{Wagner1958-qm_ZBDP}. A considerable literature was then developed to propose near-optimal solutions for this problem, often referred to as the lot sizing problem \citep[particularly]{Silver1973-uu_ZBDP}. An overview of this literature is provided by \cite{Glock2014-vk_CGTZ}.

The literature was also extended in the 1950s to consider inventory systems subject to uncertainty. Inventory control policies were then developed to deal with stochastic demand where both backordering and lost sales cases are considered. The optimality of periodic review order-up-to policies was proved in the former case \citep{Arrow1958-sp_ZBDP}. However, recognising the limitations of backordering in retail settings, later studies shifted focus to lost-sales models, which better reflect customer behaviour during stockouts \citep{Zipkin2008-yh_ZBDP,Buchanan1985-oo_ZBDP}. These models are analysed to optimise reorder points and order-up-to levels under different assumptions of lead time and demand distributions \citep{Johansen1993-xg_ZBDP} and under service-level-driven systems \citep{Federgruen1984-or_ZBDP}. 

Regarding demand modelling and forecasting in stochastic inventory systems, parametric and non-parametric approaches are used. In the former, a demand distribution is assumed, and a forecasting method is used to estimate the moments of the distribution. Distributions such as Normal, Poisson, Gamma and Negative Binomial have dominated the inventory literature for fast and slow moving items with a strong empirical evidence \citep{Syntetos2006-hz_ZBDP,Syntetos2012-qh_ZBDP,Turrini2019-xz_ZBDP}. In this case there is a plethora of forecasting methods that have been developed ranging from simple time series techniques to machine learning (ML) methods. In the non-parametric approach, no assumption is made for the demand distribution, which can be empirically determined using bootstrapping methods \citep{Hasni2019-cv_ZBDP,Hasni2019-dg_ZBDP}. More information on forecasting approaches is provided in \S \ref{sec:Forecasting}. \cite{Prak2017-vl_ZBDP} address the issue of correlated forecast errors when calculating inventory parameters. \cite{Goltsos2022-vk_ZBDP} classify the various levels of interaction and integration between the forecasting and inventory control literature.

\subsubsection*{Multi-item systems}
Most warehouses store a variety of heterogeneous SKUs. Jointly controlling these leads to new dynamics and decisions. Firstly, classification serves to manage all SKUs efficiently. Secondly, customers face the entire assortment, giving rise to new definitions of service and demand substitution between SKUs. Finally, there may be cost benefits of jointly replenishing different SKUs.  

In order to efficiently divide management effort over thousands of SKUs, these are typically divided into groups \citep{Silver2016-ju_ZBDP}. The most commonly used division is into three groups -- A, B, and C -- based on demand value (volume multiplied by price). This so-called ABC classification can be extended to an ABC-XYZ classification by adding a measure of demand regularity. XYZ classifications are often based on the coefficient of variation, whereas this does not necessarily align with forecast accuracy. Occasionally, classifications are also based on other measures, such as supply risk, lead time, or criticality \citep[e.g.,][]{Flores1987-os_ZBDP}.

Service targets are typically determined at a higher level than that of the individual SKU. Assortment-level service requirements are usually translated to the SKU level by setting the same target service level for all SKUs in a group. \cite{Teunter2010-ls_ZBDP,Teunter2017-zp_ZBDP} show that this is suboptimal and suggest a cost-based approach to determine SKU-level fill rate targets. \cite{van-Donselaar2021-kz_ZBDP} determine reorder levels for a given assortment fill rate. The so-called order fill rate \citep{Song1998-jc_ZBDP} acknowledges that customer orders may consist of multiple SKUs which must all be on stock to fulfil the order. 

Particularly in retail settings (\S\ref{sec:Retail}), a customer may buy a different product than originally intended. The product may not be in the assortment \citep[assortment-based substitution, e.g.][]{Smith2000-cg_ZBDP} or out of stock \citep[stockout-based substitution, e.g.][]{Netessine2003-fa_ZBDP}. It is typically assumed that a customer switches to their second-choice option with a certain probability and otherwise leaves, but more complex structures with multiple alternatives also exist. Joint optimisation of inventories under substitution is notoriously difficult, and for a comprehensive review we refer to \cite{Nagpal2021-zf_ZBDP}.

In practice there may be (usually fixed order) cost advantages when SKUs are replenished together, giving rise to the joint replenishment problem (JRP). Even for deterministic demand, the JRP is ${\cal NP}$-hard. \cite{Silver1976-dh_ZBDP} created an efficient heuristic that chooses the replenishment intervals of all other SKUs as multiples of some base interval. For the considerably different stochastic JRP, so-called can-order policies have been proposed \citep[e.g.,][]{Federgruen1984-he_ZBDP}. Below a certain inventory position threshold an SKU must be ordered, and below another, higher threshold, an SKU may be added to an existing order. \cite{Creemers2022-kp_ZBDP} show that can-order policies perform well. 

\subsubsection*{Multi-echelon systems }
In the 1960s, research on multi-echelon inventory systems started by dealing mainly with repairable spare parts inventory systems in military settings. A pivotal development was the Multi-Echelon Technique for Recoverable Item Control (METRIC) model by \cite{Sherbrooke1968-rp_ZBDP} that considers a single repair depot replenishing multiple stock points. The METRIC model assumes unlimited repair shop capacity, Poisson-distributed demand that can be backordered and a base stock inventory control policy. Extensions of METRIC have relaxed these assumptions to improve model realism, by considering limited repair capacity \citep{Diaz1997-ve_ZBDP}, compound Poisson demand distributions \citep{Graves1985-na_ZBDP,Costantino2017-lu_ZBDP}, lateral transshipments \citep{Lee1987-vj_ZBDP} and lost sales cases \citep{Kouki2024-bz_ZBDP}. Service restrictions and their impact on multi-echelon inventory systems under the METRIC framework have been examined by \cite{Topan2017-cg_ZBDP} and \cite{Dreyfuss2018-xm_ZBDP}. These studies investigate how limiting the availability of certain services affects overall inventory management and service levels. A discussion of maintenance aspects in OM is given in \S\ref{sec:Maintenance}.

METRIC-type models as described above deal with distribution (or divergent) systems, but multi-tier inventory systems may exhibit various structures. In assembly (or convergent) systems -- common in manufacturing settings (\S\ref{sec:Manufacturing}) -- multiple upstream stock locations converge towards fewer downstream stock locations. Raw materials or components typically have lower value than end products and may therefore be cheaper to keep in stock (upstream). After \cite{Schmidt1985-sb_ZBDP} examined the optimal policy in a finite-horizon, two-location assembly system under stochastic demand, \cite{Rosling1989-cd_ZBDP} studied a more general and infinite-horizon system, and showed equivalence with a serial system \citep{Clark1960-dc_ZBDP}, for which exact \citep{Federgruen1984-or_ZBDP} and approximate \citep{van-Houtum1991-im_ZBDP} solution procedures exist. Extensions exist, e.g., towards supplier uncertainty \citep{DeCroix2013-my_ZBDP}.

Most work discussed so far uses a stochastic service approach: each stock location keeps a safety stock so that a service level target is achieved. Stock-outs do occur and delay downstream replenishments, which complicates the analysis. \cite{Simpson1958-kt_ZBDP} described the guaranteed service approach: demand is assumed to be bounded, and every location keeps a safety stock so that a given service time to the downstream location is always guaranteed. Although bounded demand is a strong assumption, \cite{Simpson1958-kt_ZBDP} gives a possible interpretation for the bound being the maximum accepted demand. \cite{Graves2003-fq_ZBDP} contrast and illustrate the stochastic and guaranteed service approach and highlight the latter as a simple approach to tackle complex supply chains. Information asymmetries in multi-tier supply chains lead to increased upstream order variability, a phenomenon known as the bullwhip effect (see \S\ref{sec:Bullwhip_effect}). 

To the reader searching for detailed information on inventory models and their applications, we recommend \cite{Zipkin2000-vs_ZBDP}, \cite{Axsater2015-vc_ZBDP}, and \cite{Silver2016-ju_ZBDP} as three excellent examples of comprehensive inventory management textbooks.

\subsection[Warehousing (Nils Boysen \& René de Koster)]{Warehousing\protect\footnote{This subsection was written by Nils Boysen and René de Koster.}}
\label{sec:Warehousing}

The fundamental purpose of warehousing, defined as the intermediate storage of goods in a dedicated facility, is to decouple the time and place of production or supply from the time and place of consumption between successive supply chain stages \citep{bartholdi2014warehouse_NBRK}. 

By decoupling the \textit{time}, stockpiling goods can facilitate the matching of supply and demand (\S \ref{sec:Inventory_management}). Demand peaks served from stock can avoid costly excess capacities needed for just-in-time production that remain idle during the off-peak season. Warehouses can also stockpile excess supply (e.g., during harvest) to serve a stable demand. In addition to mastering predictable supply and demand peaks caused by repetitive seasonalities, inventories also protect against unpredictable events (e.g., excess wind energy and social commerce sales hype). This effect gains significance if the warehouse is also used for value-adding services (that is, late customisation of products) to reduce the impact of uncertain demand through risk pooling (\S \ref{sec:Risk_management_and_resilience}). Furthermore, decoupling the timing of production and consumption can also protect against price fluctuations of purchased goods (\S \ref{sec:Pricing}). For instance, warehouses can hold back products from the market to await price increases and store an enlarged procurement lot to realise quantity discounts. 

By decoupling the \textit{place} of production and consumption, warehousing adds flexibility to transportation processes, as a high frequency of delivery to customers (and the corresponding delivery sizes) can be accommodated more easily by a close-by warehouse, rather than by remote manufacturers (\S \ref{sec:Transportation}). On the service side, storing products in warehouses close to customers reduces delivery times and prevents split deliveries from different production sites. Instead of servicing less-than-truckload customer demands with half-empty vehicles (or at low frequency) directly from a faraway production site, a warehouse can be supplied in full truckloads, bundling the demand of multiple close-by customers. The handling of returns in a nearby warehouse promises further reductions in transportation costs. \citet{ackerman2012practical_NBRK} provides an in-depth discussion on the purposes of warehousing.

The basic process steps of warehousing are receiving, storage, order picking, shipping, and return handling \citep{gu2007research_NBRK}. Although virtually every warehouse performs these basic process steps, their specific execution varies significantly between warehousing systems. Each system consists of hardware (that is, storage devices (racks), material handling systems (\S \ref{sec:Material_handling}), and picking tools (pick-by-voice devices)), and processes defining the workflow along the hardware \citep{boysen2019warehousing_NBRK}. Warehousing systems can be categorised into picker-to-parts and parts-to-picker systems, based on how Stock Keeping Units (SKUs) and pickers come together \citep{boysen202450_NBRK}. In picker-to-parts systems, pickers move to the storage locations of the demanded SKUs. Conversely, in parts-to-picker systems, SKUs are brought to pick stations with stationary pickers using a material handling system. Picker-to-parts systems, which rely primarily on a human workforce, offer high flexibility but incur significant wage costs. On the other hand, parts-to-picker systems use fixed machinery, which, while less flexible, involves a higher investment but reduces operational expenses and can increase order throughput. In recent years, significant advancements in robotics have impacted the warehousing sector \citep{azadeh2019robotized_NBRK}. Instead of relying on fixed, inflexible hardware like conveyors, lifts, and cranes, the use of more flexible Autonomous Mobile Robots (AMRs) has introduced a third category. These robots can operate in systems that are much more flexible than traditional parts-to-picker and picker-to-parts systems. Hence, we define a third class of warehousing systems based on AMR usage. We discuss each of these three classes of warehousing systems in the following sections. 

\begin{figure}[h!]
\centering
 \includegraphics[width=0.9\textwidth]{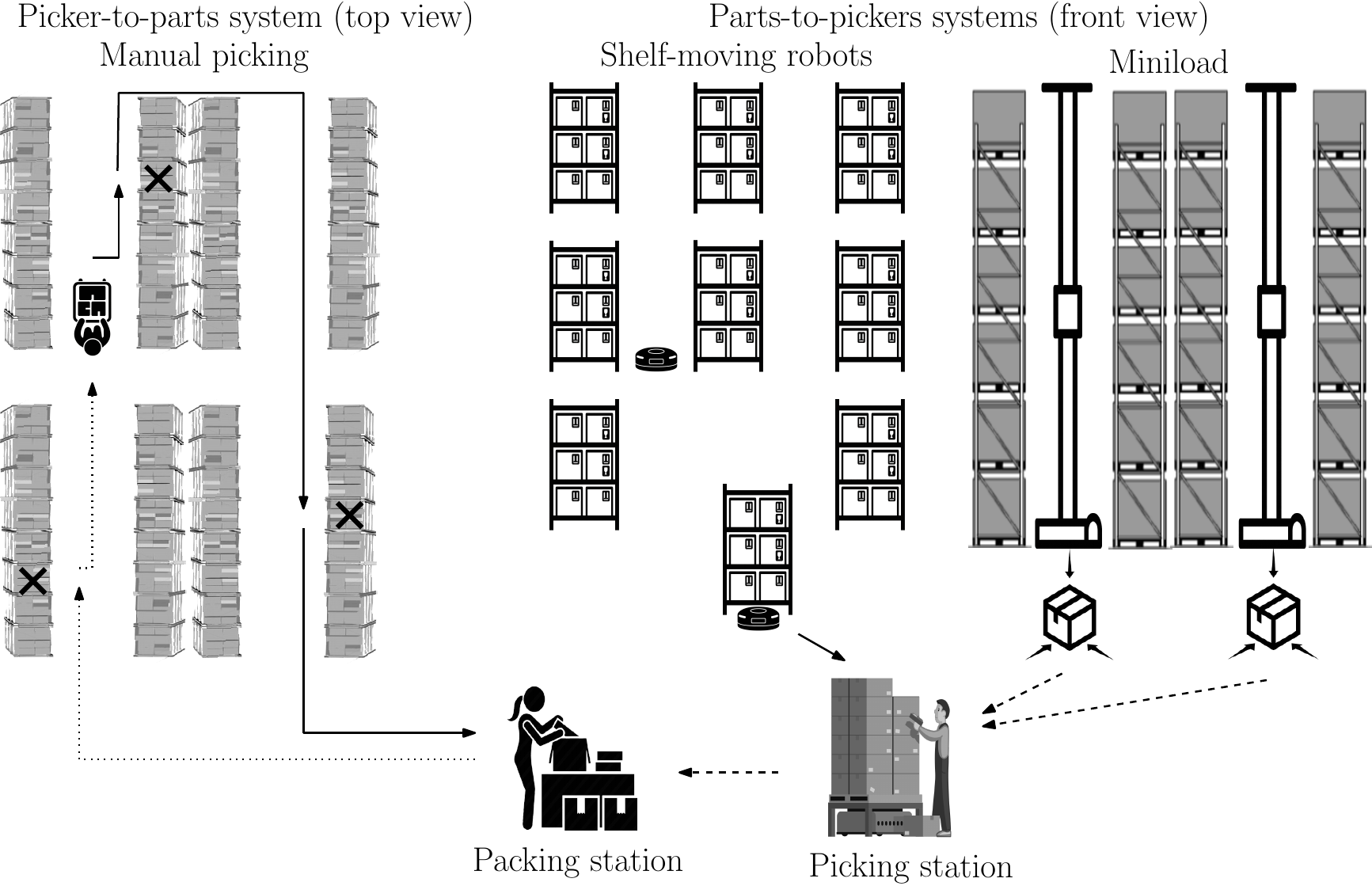}
 \caption{Warehouse with a picker-to-parts and two parts-to-picker systems.}
 \label{fig:warehouse1}
\end{figure}

Figure \ref{fig:warehouse1} gives an impression of a warehouse system consisting of a picker-to-parts and two different parts-to-picker systems: a miniload crane-based Automated Storage and Retrieval System and a system with shelf-moving robots.

\subsubsection*{Picker-to-parts systems}

\textit{System setup:} In picker-to-parts systems, pickers move toward the SKUs' storage locations to retrieve the requested number of products defined on their (nowadays electronic) pick lists. The pickers are equipped with either a manual cart or drive on a motorised cart. The SKUs are stored in racks, typically arranged to the left and right of parallel picking aisles connected by cross aisles \citep[according to][other, flying-V or fishbone-shaped rack orientations may be beneficial in unit load retrieval operations, i.e., without order picking]{GueMeller2009}. Picking tours start and end at a depot where the picked goods are collected and prepared for shipping. Often, there is just a single central depot, but some warehouses also apply decentralised systems with multiple access points to a conveyor system that connects the depot \citep{schiffer2022optimal_NBRK}. 

Especially in e-commerce, where customers order just a few items per order \citep{boysen2019warehousing_NBRK}, picking in an order-by-order manner produces excessive unproductive travel back and forth to the depot. Therefore, many warehouses apply batching and zoning to streamline their order fulfilment processes. To increase pick density per tour, batching groups multiple customer orders into extended pick lists, and zoning partitions the warehouse into smaller zones, each with a dedicated workforce. There is parallel and progressive (or sequential) zoning \citep{de2007design_NBRK}. The former promises a significant speedup because all zones start order picking on the same order set in parallel. In progressive zoning, a dedicated order bin is sent from zone to zone (typically by a conveyor), which prolongs completion times but avoids extra order consolidation. The latter is the price for batching and parallel zoning: an extra consolidation stage, often based on loop conveyors \citep{boysen2019automated_NBRK}, must sort the collected products by order.  

\textit{Decision tasks:} During layout design (\S \ref{sec:Facility_layout}), decisions on the structure of the warehouse must be made. This includes the choice of the number of picking and cross aisles and the positioning of the depot. If zoning is applied, the number of zones and their locations must be fixed. \citet{boysen202450_NBRK} provide a recent literature review of these decisions. On a subsequent level, storage assignment decides the specific storage positions where each SKU is stored. Next to the classical storage policies for unit-load storage \citep[i.e., random, closest-open-location, dedicated, full-turnover, class-based, and family grouping storage, see][]{de2007design_NBRK}, there is also scattered storage, which has established foremost in e-commerce. Scattered storage breaks up the homogeneous pallets and distributes the individual pieces per SKU over the storage racks. For hardly predictable pick lists, scattered storage increases the chance that -- whatever it is that is ordered jointly -- somewhere it is stored close together and can be picked without excessive picker travel \citep{boysen2019warehousing_NBRK}, albeit at the  expense of extra replenishment effort. On an operational level, if batching is applied, the order pool must be partitioned into pick lists, each assembled on a separate picker tour \citep[see][]{boysen202450_NBRK}. Finally, picker routing decides on the picker's tour through the warehouse for a given pick list \citep[][provide an in-depth survey on the manifold picker routing literature]{masae2020order_NBRK}.

Some papers discuss integrated modelling and solution of decision problems in picker-to-parts systems \citep[for a survey, see][]{VANGILScombinedplanning2018}. Next to integrated optimisation, the recent literature has started to emphasise the importance of human factors in system design; factors that consider human limitations and capabilities \citep{GrosseGlockNeumann2017}. These factors can be included in the planning process and focus on the well-being of human operators.  \citet{GrosseGlockJaberNeumann2015} and \citet{DeLombaert19052023} review the literature and sketch opportunities. In addition, incentive systems, goal setting, public performance feedback, gamification, and picker personality traits appear to impact performance and, in many cases, also well-being \citep[e.g.,][]{DeVries2016Aligning}.

\subsubsection*{Parts-to-picker systems}

\textit{System setup:} Parts-to-picker systems eliminate unproductive picker travel by delivering SKUs to workstations with stationary pickers: humans or automated devices (e.g., a robot arm with a vacuum gripper). The main hardware elements are an Automated Storage and Retrieval System (ASRS) where the goods are stored (often as unit loads) and retrieved, the picking workstations for product unit picking, and a conveying system connecting them. The traditional ASRS is a crane-operated (high-bay) rack filled with pallets or bins \citep{roodbergen2009survey_NBRK}, but there are also (horizontal and vertical) carousels, vertical lift modules, puzzle-based storage systems, deep-lane ASRSs, and shuttle-based storage systems \citep{azadeh2019robotized_NBRK}. Workstations range from simple racks for order bins to ergonomic stations, adaptable in height with automated bin exchange and weighing mechanism to reduce picking errors \citep{boysen2019automated_NBRK}.

\textit{Decision tasks:} During layout design, a suitable ASRS type must be identified and dimensioned, along with enough workstations for the targeted workload \citep{boysen202450_NBRK}. When deciding on the storage positions of the SKUs within the ASRS, basically the same policies are applied as already discussed for picker-to-parts systems \citep[see above and][]{roodbergen2009survey_NBRK}. Finally, order processing must coordinate the storage and retrieval of bins from the ASRS with the picking process at the picking workstations \citep{boysen2023review_NBRK}.

\subsubsection*{Robotised systems}

\begin{figure}[h!]
\centering
 \includegraphics[width=0.65\textwidth]{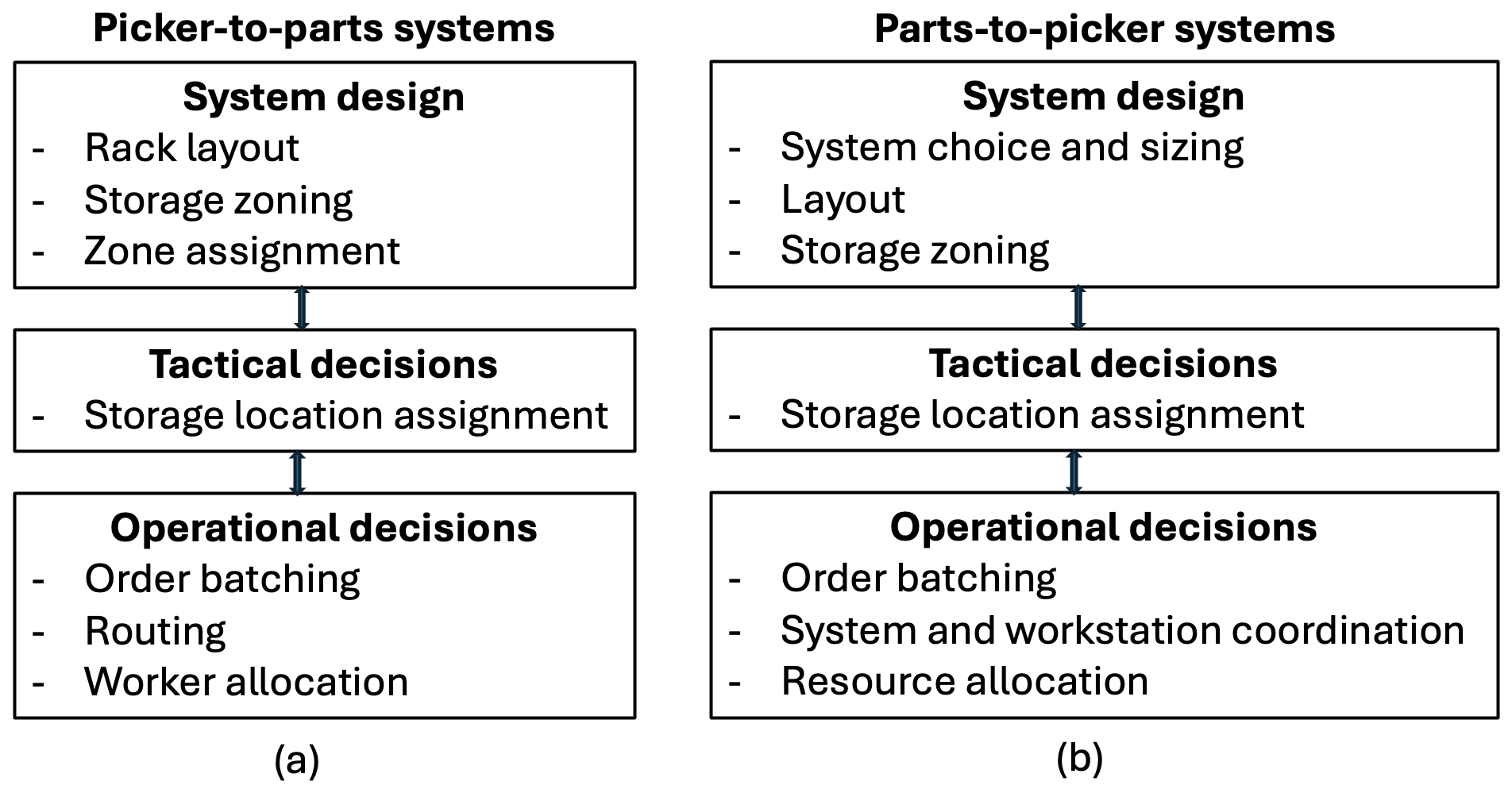}
 \caption{Overview of decisions at different hierarchical levels in (a) picker-to-parts and (b) parts-to-picker and robotised systems.}
 \label{fig:warehouse2}
\end{figure}

Driven by technological progress, like in the fields of sensors, real-time control, autonomous driving, and robotisation, AMRs have become more and more successful in recent years \citep{fragapane2021planning_NBRK}. Other than their famous predecessors, Automated Guided Vehicles (AGVs), AMRs are not bound to predefined paths; computer vision allows flexible adaptation to a changing environment. In picker-to-parts systems, AMRs that also pick, such as the Toru robot of Magazino, can substitute human pickers, at least for a restricted product assortment \citep[e.g., shoeboxes][]{boysen202450_NBRK}. For non-standard products, transport AMRs accompany human pickers through the warehouse and relieve them of the burden of transporting the picked products back to the depot \citep{Azadeh2023PSAMR2_NBRK,loffler2022picker_NBRK}. The most famous AMRs of parts-to-picker systems are (probably) shelf-moving (KIVA) robots. They drive under a rack (also denoted as inventory pod), lift it, and deliver the pod to a stationary picker \citep{weidinger2018storage_NBRK}. Other AMR representatives are storage grid-supported (Autostore) robots \citep{azadeh2019robotized_NBRK}, rack-climbing (Exotec) robots \citep{chen2022analysis_NBRK}, and multi-tote (Quicktron, Haipick) AMRs \citep{qin2024performance_NBRK} delivering bins autonomously from storage locations. Finally, there are also AMRs for order consolidation \citep{zou2021robotic_NBRK}. Equipped with a tiltable tray or small conveyor tablet, they can distribute goods between sorting destinations. The different decisions in picker-to-parts and parts-to-picker (including robotised) systems are summarised in figure \ref{fig:warehouse2}.

\subsubsection*{Conclusion}

To decouple the time and place of production from the time and place of consumption, warehouses have always been an essential part of supply chains. However, e-commerce and omnichannel retailing (\S \ref{sec:Retail}) have transformed them from a marginal element to a mission-critical supply chain stage. Hence, it seems reasonable to project that warehousing will remain a vivid field of research in the foreseeable future.

\subsection[Material handling (Veronique Limère \& Nico André Schmid)]{Material handling\protect\footnote{This subsection was written by Veronique Limère and Nico André Schmid.}}
\label{sec:Material_handling}
Material handling represents a fundamental part of operations management in any organisation or supply chain, comprising the processing of physical products in the form of parts, components, or final products within facilities such as factories, distribution centres, warehouses, cross docks, and container ports. It can be described as the activities whereby material is physically relocated from one position to another within the same facility. The objective is providing goods at the right time, in the right quantity, and to the right location where they can be retrieved for processing. This intralogistic transportation process may include additional activities such as searching, inspection, packaging, and (un)loading and can be found in several sectors such as manufacturing \citep[][also \S\ref{sec:Manufacturing}]{Kusiak.2018_VLNAC}, warehousing \citep[][also \S\ref{sec:Warehousing}]{azadeh2019robotized_NBRK}, construction \citep[][also \S\ref{sec:Construction}]{Prasad.2015_VLNAC}, container terminal operations \citep{Carlo.2014_VLNAC}, and healthcare \citep[][also \S\ref{sec:Healthcare}]{Bhosekar.2023_VLNAC}.

Manual material handling operations require significant physical and cognitive effort \citep{Soufi.2021_VLNAC} but can be supported by the implementation of appropriate material handling equipment (MHE). According to \cite{Tompkins.2010_VLNAC}, MHE can be classified into four types: (\textit{i}) containers and unitising equipment, such as pallets, totes, and cylinders; (\textit{ii}) material transport equipment, such as conveyors and industrial vehicles; (\textit{iii}) storage and retrieval equipment, such as racks or carousels; (\textit{iv}) automatic data collection and communication equipment such as tags or vision technology. MHE ranges from manual to fully automated and highly autonomous solutions such as autonomous mobile robots (AMR), also called autonomous intelligent vehicles (AIV) \citep{fragapane2021planning_NBRK,Hellmann.2019_VLNAC}, and autonomous vehicle storage and retrieval systems (AVS/RS) \citep{Marchet.2013_VLNAC}. Material handling typically incurs a high investment cost for automated solutions or a high labour cost for manual solutions \citep{Tompkins.2010_VLNAC} and effective and efficient material handling is instrumental for the operational performance of a system.

Operations managers are primarily concerned with the design and operation of material handling systems. Their design investigates both ideal facility \citep{Gue.2007_VLNAC} and process layout \citep{Schmid.2019_VLNAC} to facilitate smooth and efficient material flow while minimising handling requirements. This encompasses the selection of appropriate material handling equipment (MHE) under economic \citep{Moretti.2021_VLNAC}, human-centric (ergonomics, safety, well-being) \citep{Finnsgaard.2011_VLNAC,Hanson.2018_VLNAC}, environmental \citep{Erkayaouglu.2016_VLNAC}, or service-level \citep{Bhosekar.2021} considerations. The operation of material handling systems, on the other hand, applies principles from control theory and scheduling approaches to ensure effectiveness.

\subsubsection*{Designing Material Handling Systems: Equipment selection}
For MHE selection, frequently, the products to be handled and its packaging reduce the amount of practically feasible material handling equipment. Therefore, best practice examples are often used in the industry whereas the analytical hierarchy process is the predominant academic method to determine MHE \citep{Chan.2001_VLNAC, Chakraborty.2006_VLNAC}. Life-cycle analysis approaches can be found to assess the environmental aspects of different MHE alternatives \citep{Erkayaouglu.2016_VLNAC}. It should be noted that structured MHE selection processes are typically adjusted for the specific environment in which the equipment will be used, e.g. in warehousing \citep{azadeh2019robotized_NBRK}, hospitals \citep{Bhosekar.2023_VLNAC}, or construction \citep{Prasad.2015_VLNAC}. In most environments, one may find an array of choices spanning from labour-intensive to highly automated options as shown, e.g., by \citet{Adenipekun.2022_VLNAC} who optimise the vehicle type selection in an assembly system from an economic perspective. The final choice may be taken when considering facility layout or process design.

\subsubsection*{Designing Material Handling Systems: Facility layout and process design}
Facility design (see \S\ref{sec:Facility_layout} for more detail) concerns the placement of spaces and equipments in a facility. Therefore, it is interlinked with the selection of material handling equipment. In process-oriented facilities, where flow units follow different routes through the facility, efficient facility layout reduces the demand for material handling trips and the corresponding investments in MHE. Simplistic design improvement approaches such as CRAFT have been replaced by optimisation-based approaches that consider changes in demand over time \citep{Montreuil.1991_VLNAC} or even allow for facility redesign \cite{Mckendall.2010_VLNAC}. In contrast, flow lines and manufacturing cells require less material handling \citep{King.1980_VLNAC}. Therefore, their layout does not necessarily account for material handling considerations. Flow lines may also be designed for increased flexibility where products do not follow rigid routes but instead are transported by AGVs to resources with slack capacities using autonomous vehicles \citep{Hottenrott.2019_VLNAC}. 

The design of operational processes and alternatives impacts material handling and frequently minimises material handling efforts. One such example can be found at high variety mixed-model assembly lines (see \S\ref{sec:Assembly_line_balancing}) where part supply often requires significant material handling efforts. The assembly line feeding problem (ALFP), as initially proposed by \cite{Bozer.1992_VLNAC}, investigates increasingly more feeding process alternatives and aspects of the feeding process. Initially, these considerations focused strongly on the available storage space and fewer process alternatives \citep{Caputo.2008_VLNAC,Limere.2012_VLNAC}, more recent work considers additional aspects such as facility layout and aims to minimise material handling efforts \citep{Schmid.2022_VLNAC, Schmid.2024_VLNAC}. Another example for process alternatives can be found in order-picking operations in which zone picking may be applied in picker-to-parts systems (see \S\ref{sec:Warehousing}). If no zoning is applied, pickers may have to walk longer decreasing their picking efficiency. In contrast, zoning avoids excessive walking but often requires some material handling equipment to transport items from one zone to the next or for final consolidation \citep[i.e., progressive versus parallel zoning;][]{de2007design_NBRK}.

In summary, the combination of a facility's layout, the process choice, and MHE selection determines the efficiency of the material handling system design \citep[see also][]{Dallari.2009}.

\subsubsection*{Operation of material handling systems}
After having defined material handling systems by selecting, quantifying, and positioning the equipment those systems go into operation. We refer the reader also to \S\ref{sec:Production_and_control} and \S\ref{sec:Scheduling} where production control and scheduling are discussed more generally. Specifically for material handling, as stated above, an important objective is to provide the right good at the right time and place. To this end, one may either define certain control rules according to which systems operate or one may use real-time information and schedule activities to optimise some objective such as total or maximum tardiness. Control rules could be as simple as First-Come-First-Serve or more sophisticated. In the FCFS approach, the job that is finished first will be transported to its succeeding station, irrespective of that station's availability. More advanced control rules can be found in \cite{Klaas.2016_VLNAC}, using a simulation to derive decision rules for the effective operation of an automated storage and retrieval system. In addition, their proposed approach adapts dynamically when the system's status (utilisation or demand patterns) are evolving away from the simulation's parameters. In other environments such as assembly, manufacturing, or hospitals, material handling is frequently conducted along so-called milk runs (also called mizusumashi), which define a route that is followed periodically for replenishment. \cite{Emde.2012_VLNAC} provide an approach to define milk runs and their periodicity.

Scheduling approaches provide an alternative to control approaches. In some systems, demand may be known with certainty for multiple days in advance, allowing to schedule the part replenishment such that no material shortage occurs as described by \cite{Boysen.2011_VLNAC}. Milk runs may also be combined with a scheduling approach that determines optimised but variable deliveries frequencies, which may lead to reduced inventories at the point of consumption \citep{Emde.2017_VLNAC}. Furthermore, scheduling may allow a more flexible use of material handling equipment, possibly reducing the demand for MHE. 

The use of AMR technology impacts scheduling and control of  material handling systems significantly. These vehicles no longer need to follow predetermined paths or schedules but operate autonomously. \cite{fragapane2021planning_NBRK} give a detailed overview of how specific technological developments impact traditional planning and control decision areas.

Despite large amounts of research focusing on technical improvements and automation efforts, humans remain relevant to material handling in many enterprises \citep{Glock.2021_VLNAC}. Evidently, human-centric aspects, especially worker safety, well-being, and ergonomics, are of high relevance for operations managers in such enterprises. This relevance is even more pronounced when the workforce is ageing  \citep{Calzavara.2020_VLNAC}. According to \cite{Denis.2020_VLNAC} training for material handling workers frequently does not deliver to its goals of reduced disease and accidents. They identify a multitude of considerations for improving such training. Importantly, they state that training should address a wider range of situations and prepare operators to identify the appropriate action in their situation but also demands for additional practice during the training. Increasingly, a combination of humans and robots is operating material handling systems. Determining an effective combination, e.g., by allocating tasks to either humans or robots \citep{Boudella.2018_VLNAC}, seems relevant. 

\subsubsection*{Conclusion}
For a general overview on material handling activities and equipment, we refer to \cite{Tompkins.2010_VLNAC}, \cite{Stephens.2019_VLNAC}, and \cite{Kay.2012_VLNAC}. From a process point of view, different assembly line feeding policies and the related decision-making approaches are presented by \cite{Schmid.2019_VLNAC}. Regarding planning and control of autonomous MHE, the review article of \cite{fragapane2021planning_NBRK} gives a thorough overview and a research agenda, while for manual material handling the use of assistive devices is treated in detail in \cite{Glock.2021_VLNAC}. Finally, we believe material handling systems should be flexible and robust to variable product mixes and volumes. \cite{Klaas.2016_VLNAC} provide an excellent example for the adaptation of operating and control principles of an AS/RS when demand changes and may serve as an example for future work.

\subsection[Buffer allocation (Chuan Shi \& Stanley B. Gershwin)]{Buffer allocation\protect\footnote{This subsection was written by Chuan Shi and Stanley B. Gershwin.}}
\label{sec:Buffer_allocation}

\cite{Koen-1959_CSSBG} highlighted three challenges in the design of production lines: selecting the set of machines in the line, determining the locations of buffers, and deciding on the sizes of buffers. The third challenge is the buffer allocation problem (BAP). Efficient production line design is important in high-volume manufacturing industries.

This section explores the concepts, mathematical modelling, and recent advancements in solving the BAP. For a more comprehensive description of the topic, see the latest review articles, such as \cite{Demir-etal-2014_CSSBG}, \cite{Weiss-etal-2019_CSSBG}, and \cite{Ulas-Koyuncuoglu2024-sc_CSSBG}.

\subsubsection*{Concepts}
A production line is a set of $K$ machines connected in series and separated by $K-1$ buffers. Material flows from upstream inventory to the first machine for an operation, then to the first buffer where it waits, to the second machine, and so on, until it passes through the final machine and then goes to downstream inventory. There are two quantities associated with a buffer: its size and the time-varying inventory level (or buffer level or work-in-process or WIP).

\begin{figure}[h!]
\centering
 \includegraphics[width=0.65\textwidth]{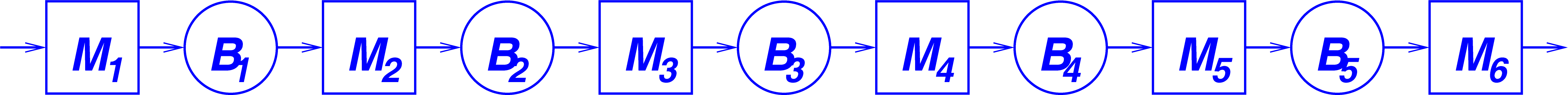}
 \caption{A Production Line Example.}
 \label{fig:line}
\end{figure}

Machines interfere with each other when material flow is disrupted by machine failures. Failures can cause neighbouring machines to be idle. If the buffer before a machine is empty, the machine will be starved and not able to work; if the buffer after it is full, the machine will be blocked and also not able to work.

The most important performance measures are the production rate or throughput, the number of items produced in a given time period divided by the length of the time period; and the average amount of inventory in each buffer. Increasing buffer sizes increases production rate. The cost of increasing buffer sizes includes the cost of the buffers and the space they occupy, the cost of holding inventory, and the increased time the parts stay in the line. The BAP seeks the buffer sizes that maximise the net benefit.

The current BAP research literature optimises analytical models of lines. It consists of two streams: one in which machines are reliable with random processing times, and one in which they are unreliable. In the first stream, each machine is modelled as an M/G/c/K queue. In the second, machines are unreliable and can fail and be repaired at random times. These models are characterised by their machines' behaviour: 
\begin{itemize}[noitemsep,nolistsep]
\item The \textit{deterministic} model: Machines' operation times are equal, deterministic, and constant; material is discrete. In most of the literature, machine failures and repairs follow geometric distributions. 
\item The \textit{exponential} model: Each machine's operation, up and down times are continuous and follow exponential distributions. Material is discrete.
\item The \textit{continuous} model: Each machine's operation, up and down times are continuous. Material is treated as a continuous fluid. 
\end{itemize}

\noindent Machines can have single or multiple failure modes \citep[e.g.,][]{Tolio-exp-decom-1999_CSSBG,TGM2002_CSSBG,Leva-Matta-Tolio-03_CSSBG}.

The machine models are simple representations of reality and the evaluation of models of lines is most often approximate. The challenge is to create models of machines and lines that are close enough to reality to be useful and to develop algorithms to predict and optimise the performance of the lines. Much of the analysis has been extended to assembly/disassembly systems, as well as constant WIP (or CONWIP) loop systems.

\subsubsection*{Mathematical modelling}
BAP is an optimisation problem where the goal is to determine the optimal buffer sizes $N_i$ for a given set of machines, subject to constraints on system parameters. Over the years, researchers have explored a variety of objective functions and constraints.

\noindent\textbf{Problem Formulation.} One classical approach minimises the total buffer size while satisfying a production rate constraint:
\begin{equation}
\label{eq:primal}
\begin{array}{rrl}
&\displaystyle\min_{\mathbf{N}} &\displaystyle\sum_{i=1}^{K-1} N_i \\
\text{Primal}\qquad&\text{subject to}& \mathbb{E}[P(\mathbf{N})] \geq P^\star\\
&&N_i\ge 0,
\end{array}
\end{equation}

\noindent where $N_i$ represents the size of the $i$th buffer, $\mathbb{E}[P(\mathbf{N})]$ represents the average production rate as a function of buffer sizes $\mathbf{N}=[N_1,\cdots,N_{K-1}]'$, and $P^\star$ is the target production rate. This is the primal problem. It is difficult to solve directly because it has a $(K-1)$-dimensional nonlinear constraint. 

To solve (\ref{eq:primal}), \cite{schor-gershwin_CSSBG} propose the following dual problem. It maximises the expected production rate for a given total buffer size $N_{\text{Total}}$:
\begin{equation}
\label{eq:dual}
\begin{array}{rrl}
&\displaystyle\max_{\mathbf{N}} &\mathbb{E}[P(\mathbf{N})]\\
\text{Dual}\qquad&\text{subject to}&\displaystyle\sum_{i=1}^{K-1} N_i = N_{\text{Total}}\\
&&N_i\ge 0.
\end{array}
\end{equation}

\noindent They solve the primal by finding the value of $N_{\text{Total}}$ such that $\mathbb{E}[P(\mathbf{N})]$ in (\ref{eq:dual}) is equal to $P^\star$ in (\ref{eq:primal}). This is easier than solving (\ref{eq:primal}) directly because (\ref{eq:dual}) has only linear constraints. The value of $N_{\text{Total}}$ is found by a one-dimensional search.

As the literature has expanded, the focus has shifted to more general forms, which encompass various other performance metrics beyond buffer size and throughput. These problems all have special cases of the following objective function \citep{Weiss-etal-2019_CSSBG}:
\begin{equation}
\displaystyle\max_{\mathbf{N}} a\mathbb{E}[P(\mathbf{N})]-\sum_{i=1}^{K-1}b_iN_i-\sum_{i=1}^{K-1}c_i\bar n_i-d\mathbb{E}[LT(\mathbf{N})]-e\mathbb{E}[WT(\mathbf{N})],
\end{equation}

\noindent where $\bar n_i$ is the average inventory level of the $i$th buffer, and $\mathbb{E}[LT(\mathbf{N})]$ and $\mathbb{E}[WT(\mathbf{N})]$ represent the average lead time and average waiting time of customer orders. For instance, \cite{Yuzukirmizi-Smith-2008_CSSBG} include the lead time in their objective function. 

\noindent\textbf{Performance Evaluation.} For these systems, the high dimensionality and the nonlinear interactions among machines and buffers make exact solutions impossible for long lines. The usual approach is to use small systems as building blocks and develop approximate numerical methods to evaluate the performance of long lines. For reliable machines, the basic models are M/M/1/K and M/G/1/K queues \citep[e.g.,][]{Smith-Cruz-2005_CSSBG}; for unreliable machines, the basic models are the two-machine one-buffer line models based on Markov chains \citep[e.g.,][]{gershwin94_CSSBG,Tan-Gershwin-2011_CSSBG,Tolio-Ratti-2018_CSSBG}. These small systems have exact or approximate closed-form solutions. Exact numerical solutions for a variety of small system structures, including production lines, can be calculated by iterated matrix multiplication using the method of \cite{BTan02_CSSBG} to generate the matrices.

For long lines, the generalised expansion method \citep{Kerbachea-Smith-1987_CSSBG,Cruz2008-wc_CSSBG} is proposed for systems with reliable machines, while decomposition and aggregation are prevailing methods to deal with unreliable machines. The decomposition method \citep{gershwin87_CSSBG,DDX_CSSBG}, and its extensions \citep[e.g.,][]{TM_CSSBG,Leva-Matta-Tolio-03_CSSBG} enabled the evaluation of different models of long lines with unreliable machines. It approximates a $K$-machine $K-1$-buffer system with $K-1$ two-machine one-buffer systems. Decomposition aims to find parameters for these two-machine systems so that, for each buffer, the material flow behaviour in the two-machine line closely matches that in the original line. The goal is for an observer of the flow of parts into and out of a buffer in a real production line can be convinced that he is observing the flow of parts in a fictitious two-machine line. The aggregation method \citep{Lim-Meerkov-Top90_CSSBG, Jacobs1995-zm_CSSBG,Li2007-cv_CSSBG} combines all the two-machine one-buffer subsystems into a single aggregated machine. It consists of a forward aggregation which merges the first subline from the start towards the end and a backward aggregation which merges the last subline into a single machine from the end towards the beginning. This is repeated until the throughput converges. Simulation is another tool for the performance evaluation of large-scale production systems. It is flexible and can be accurate but it is time-consuming \citep{Helber-etal-2011_CSSBG}.

\noindent\textbf{Optimisation techniques.} Gradient methods \citep{Helber2001-hd_CSSBG,ShiGershwin09_CSSBG} use decomposition to calculate gradients via forward difference, followed by a line search to locate the next optimal point. Although buffer sizes $N_i$s are integers, treating them as continuous variables allows gradient calculation. The optimal continuous solution is then rounded to integers \citep{Han2002-mf_CSSBG}. Meta-heuristic methods treat buffer sizes as integers. Examples include genetic algorithms \citep{Dolgui-etal-2002_CSSBG}, tabu search \citep{Demir-Tunali-Lokketangen-2011_CSSBG}, simulated annealing \citep{Spinellis-Papadopoulos-Smith-2000_CSSBG}, and degraded ceiling \citep{Nahas-AitKadi-Nourelfath-2006_CSSBG}. Hybrid algorithms that combine multiple meta-heuristics include \cite{Shi-Men-2003_CSSBG}, \cite{Dolgui2007-wk_CSSBG} and \cite{Kose-Kilincci-2015_CSSBG}. Mathematical programming techniques include the dynamic programming which breaks the problem into smaller subproblems \citep{DP04_CSSBG}. \cite{CT-2005_CSSBG} linearise decomposition equations, converting the problem into a mixed-integer linear programming problem.

\subsubsection*{Recent advancements}
Production line design involves considerations beyond buffer allocation. \cite{Traina-Gerhswin-smmso13_CSSBG} address the problem of simultaneously selecting machines and determining buffer sizes for profit maximisation. \cite{Kiesmuller-Sachs-2020_CSSBG} optimise buffer capacity and spare parts inventory. 

Many factories have very long production lines, particularly for automobile manufacturing. \cite{Shi-Gershwin-2016_CSSBG} describe a method that divides a long line into several short lines, optimises them separately, and combines their buffer distributions. \cite{Xi-etal-2019_CSSBG} decompose long series-parallel unbalanced production lines into decoupled subsystems.

Machine learning can be instrumental in developing more complex machine models. \cite{Helber-Kellenbrink-Suedbeck-2024_CSSBG} integrate an artificial neural network to replace the traditional Markov chain model in the context of two-machine lines, where the machines have significantly more than the typical two states.

\subsubsection*{Concluding remarks}
In modern manufacturing systems, the decreasing product lifetimes necessitate frequent factory reconfigurations or replacements, limiting the time available to learn and improve these systems. As a result, intuition derived from research and from careful, systematic observation of real factories becomes increasingly vital \citep{Gershwin-2017_CSSBG}. Academic research provides theoretical insights and models and offers valuable tools and methodologies that can be directly applied to improve industrial processes. Works such as \cite{Lim1990-zw_CSSBG}, \cite{Burmanetal97_CSSBG}, \cite{Liberopoulos-Tsarouhas-2002_CSSBG}, \cite{Patchong-Lemoine-Kern-2003_CSSBG}, \cite{Colledani-Scania-2010_CSSBG}, \cite{Matta-Pezzoni-Semeraro-2012_CSSBG}, and \cite{Li-2013_CSSBG} demonstrate how theoretical advancements in buffer allocation can be effectively applied to real-world production systems. For instance, these studies address practical challenges in industries such as electronics, food processing, automotive manufacturing, and continuous improvement in production lines, showcasing how operations research and engineering methods optimise throughput, reduce costs, and enhance system reliability.

\subsection[Process design (Dirk Pieter van Donk)]{Process design\protect\footnote{This subsection was written by Dirk Pieter van Donk.}}
\label{sec:Process_design}

Process design is broadly speaking developing and choosing tasks, either to be executed by equipment, machines or people (alone or combined) and arrange them to produce a product or service for the organisations customers. More specifically, according to \cite{Ellram2007-qd_DPvD} process design concerns the methods used to manufacture products, including the manufacturing strategies employed (e.g., make-to-order, assemble-to-order, make-to-stock, or just-in-time), production facilities (e.g., number, location, capacity, or specialisation), and equipment (e.g., general-purpose equipment vs. specialised equipment). As a consequence, there is a close relation with the design of a product. Hence, for some processes such as chemical ones, a specific process design will be only be able to produce one specific product. Given the above link with a product and customer, it is quite natural that process design should aim to meet the specific operations management performance criteria associated with the product and the customer being quality, speed, dependability, flexibility and cost \citep[see, for example,][]{Slack2019-hp_DPvD}. Following the developments around process design, two additions can be made. Firstly, as an additional important performance measure sustainability is considered \citep[e.g.,][]{World_Commission_on_Environment_and_Development1987-zu_DPvD, Elkington2002-kq_DPvD}. Second, while in the past product design, and process design would be considered as related but sequential processes, recent insights have argued for a concurrent approach, that argues that product, process and supply chain should be design together \citep[e.g.,][]{Fine2005-vg_DPvD,Ellram2007-qd_DPvD,Browning2023-tw_DPvD}. 

While process design might be considered as a predominantly engineering related activity, it is also related to the design of organisational and information processes. The choices made in process design being it in the physical, organisational or information design, impact operations management and supply chain management. An important contribution to the understanding of process design that links product characteristics to the process is to consider volume and variety of products. This is originally based on the work of \cite{Hayes1979-na_DPvD}. The idea behind the product-process matrix is to distinguish on one axis variety and volume and on the other process tasks and process flow. In general, the idea is that high (low) variety and low (high) volume are going together on the one axis, while diverse/complex (vs repeated/divided) process task and intermittent (vs continuous) process flow do the same on the other. Together, for manufacturing it results into five basic process types project, jobbing, batch, mass and continuous processes. We will shortly discuss these below.

\begin{itemize}[noitemsep,nolistsep]
\item Project processes can be characterised by offering large scale, unique projects, often with a duration of several years, that are highly complex, with a considerable amount of freedom for the professional workers to shape their work. Managerial challenges relate to material handling, scheduling tasks and shifting bottlenecks.
\item Jobbing processes  have also a wide variety of products with low volume, but products have more in common and usually share resources with each other. Also, products might be manufactured again, unlike project processes, although many jobs remain unique. Complexity is high, but usually the size of the product is smaller. Still, workers need considerable skills although uncertainty, as associated with project, is less. Managerial challenges are similar to the ones in projects.
\item Batch processes have some characteristics in common with jobbing, but products show less variety and are produced in batches of more than one product at a time. For very small batch sizes these processes show much similarity with jobbing, but are still different as batches of the same products will be produced regularly. Batch processes might also have just a few products that are produced in relatively large numbers. Then the use of resources might be more in a fixed order, rather than more random as in jobbing/small batches. Typical managerial challenges are the balance of different types of resources, being flexible and specifically if repetition increases maintaining workforce motivation.
\item Mass processes can be characterised by large volumes of a narrow range. Here narrow range needs to be interpreted from a production point of view as products might be different in some issues e.g., their colour, which does not affect the manufacturing process. The repetitive nature of work is high and predictable. To maintain speed managers should be sure to safeguard materials, they face huge capital investments and both mass and continuous process come with increased levels of vertical integration. 
\item Continuous processes are in fact similar to mass processes, with an even higher volume and even less variety (can be just one product), with products that are often not countable in figures but rather litres or tons. Production can be continuous 24/7, only stopping for large maintenance activities (sometimes once in several years). This process type is dominated by large investments in technology, limited manpower that mainly performs control and/or maintenance tasks rather than operational ones, and a high level of automation, including control.
\end{itemize}

The product-process matrix has been influential as it depicts clearly the specific managerial challenges for different types of operations. However, new technologies such as advanced manufacturing technology might be able to influence the range a specific product design can be used for and stretch its potential range of products \citep[e.g.,][]{Das2001-kg_DPvD}. Also, \cite{Helkio2013-vg_DPvD} have argued that the product-process matrix is somewhat simplistic and static, which makes it less useful for the dynamic changes in volume and product variety that many companies face. Similar concerns have been launched regarding contemporary developments as Industry 4.0 \citep{Tortorella2022-hp_DPvD} or additive manufacturing \citep{Eyers2022-qk_DPvD}. 

While in the above the focus is rather on the internal processes and their arrangements, the related concept of Customer Order Decoupling Point (CODP) has similarly used variety and volume of a product to decide on how products are delivered to the market in terms of specificity, delivery lead time and dependability. As such it links process characteristics (as manufacturing lead time and product characteristics) with external ones related to the customer (market demand and delivery time) while also considering information flows in a company (and associated planning processes) and between customer and manufacturer (related to the ordering process). The CODP can be defined as ``the point at which product becomes earmarked for a particular customer. Downstream of this order penetration (OP) point, customer orders drive the systems that control materials flow; upstream, forecast and plans do the driving'' \citep[p. 75]{Sharman1984-kh_DPvD}. Based on a balance between market characteristics (e.g., specificity of demand, delivery speed) and manufacturing characteristics (e.g., costs of goods, processing time) five distinct position of the CODP have been distinguished in the literature: make to stock (MTS), assemble to order (ATO), make to order (MTO), purchase and make to order (PMTO), and engineer to order (ETO). While the first one directly delivers a product form inventory upon arrival of an order, the last one starts engineering and purchasing once an order arrives. Hence there is a huge difference in delivery leadtime, but also in product specificity and the number of activities performed after ordering. \cite{Olhager2024-kr_DPvD} provide an extensive overview of the developments in the research on CODP. The positing of the CODP has consequences for a variety of operations and supply chain management issues. Some examples of subjects that have been explored include the links with various aspects of the supply chain; for example information system \citep{Giesberts1992-ng_DPvD,Olhager2010-gn_DPvD}, supply chain resilience \citep{Dittfeld2022-oo_DPvD} and supply chain integration \citep{Van_Donk2016-sx_DPvD}, internal related aspects such as capacity planning and scheduling \citep{Soman2007-rd_DPvD} and stock-keeping unit classification \citep{Van_Kampen2014-gw_DPvD} and wider aspects as operations strategy \citep{Olhager2003-jc_DPvD}. Similar to process design studies the CODP concept has been mainly studied as a static concept, while the dynamics of it are of great relevance given changes that organisations encounter in product portfolio, customer base, technology and information systems.

Process design has connections with many aspects of operations management. Some obvious cross references are operations strategy (\S \ref{sec:Operations_strategy}) and facility layout (\S \ref{sec:Facility_layout}), while the different process types each have their own specific characteristics regarding planning (\S \ref{sec:Planning}), production and control (\S \ref{sec:Production_and_control}), job design (\S \ref{sec:Job_design}) and almost all other operations management related aspects. The CODP has strong connections with buffer allocation (\S \ref{sec:Buffer_allocation}), but also with a number of supply chain management related aspects as network design (\S \ref{sec:Network_design}), outsourcing (\S \ref{sec:Outsourcing}) and forecasting (\S \ref{sec:Forecasting}).

\subsection[Work design (Tomislav Hernaus \& Matija Marić)]{Work design\protect\footnote{This subsection was written by Tomislav Hernaus and Matija Marić.}}
\label{sec:Job_design}
While managers are accountable for organisational goals and collective performance targets, employees are responsible for executing their tasks and roles. Managerial decisions about who does what in organisations are impactful and represent ongoing challenges. They should be carefully planned rather than left to chance. As a microstructural and motivational concern, work design allocates resources, assigns roles and tasks to organisational members, intrinsically (de)motivates job incumbents, and ensures coordinated workforce efforts to achieve desired individual, group, and organisational outcomes \citep{Morgeson2003-wy_THMM}. This multilayered process entails (\textit{i}) setting, dividing, and grouping tasks, (\textit{ii}) integrating tasks between and within organisational levels, and (\textit{iii}) setting authority, responsibility, roles, and interactions of employees inside and outside their organisation. Work is carried out by people as well as machines. Therefore, it is important to emphasise that work design primarily focuses on human work, while scheduling deals with task and resource allocation—often including machine operation, not just human labour (see \S\ref{sec:Scheduling}).

The theory and research on work design are extensive, especially in the field of industrial and organisational psychology. However, the study of work organisation issues extends beyond this discipline, encompassing areas such as organisational design, operations management, human resource management, sociology of work, and organisational behaviour. While disciplinary boundaries do exist, they are often blurred in practice, highlighting the interdisciplinary nature of work design. The managerial dilemmas about allocating and integrating work throughout an organisation cannot be resolved using a single lens; instead, designing and organising intertwined jobs must align with goals and strategy, structural and process design, and several HRM functions.

Within the context of operations management and supply chain management, work design is inextricably linked in a cause-and-effect relationship particularly with operations strategy (\S\ref{sec:Operations_strategy}), buffer management (\S\ref{sec:Buffer_allocation}), process design (\S\ref{sec:Process_design}), operations excellence (\S\ref{sec:Operations_excellence}), facility layout (\S\ref{sec:Facility_layout}), scheduling (\S\ref{sec:Scheduling}), digitalisation (\S\ref{sec:Digitalisation}), outsourcing (\S\ref{sec:Outsourcing}), technology management (\S\ref{sec:Technology_management}), and performance measurement (\S\ref{sec:Performance_measurement_and_benchmarking}).

It is vital to distinguish work design from job analysis to avoid confusion and misunderstanding. Job or work analysis is necessary to identify, understand, and describe work within an organisation. It serves as a baseline for the majority of HRM functions, with HR experts using job analysis when creating job descriptions for workforce planning, performance management, compensation management, and job evaluation. On the other hand, work design primarily focuses on the differentiation and integration of work within an organisation, representing a non-monetary tool for motivating employees and an optimisation tool for designing the microstructure of organisations. 

Depending on desired outcomes, we can distinguish among five main work design perspectives: motivational, mechanistic, biological, perceptual/motor, and socio-technical systems \citep[][]{Rousseau1977-ca_THMM,Campion1985-po}. The \textit{motivational perspective} aims to design meaningful jobs and promote motivation and job satisfaction \citep{Lysova2019-ij_THMM}. The \textit{mechanistic perspective} is oriented toward efficiency, while the biological perspective minimises physical stress and strain \citep{Campion1985-po}. The \textit{perceptual/motor perspective} is oriented toward human mental capabilities and limitations, primarily about the attention and concentration requirements of jobs \citep{Campion1988-oa_THMM}. Finally, the \textit{socio-technical systems perspective} considers social and technological factors, particularly emphasising the interactions between work groups and their environment \citep{Rousseau1977-ca_THMM}. The comprehensive and interdisciplinary nature of work design \citep{Campion1988-oa_THMM} highlights the importance of this workplace phenomenon. If individualised and optimised, work design may lead to positive outcomes \citep{Humphrey2007-hp_THMM}, such as increased motivation, enhanced efficiency and extra work effort, reduced stress, and a fit between job requirements and employees' capabilities (i.e., a person-job fit).

The two most common traditional approaches to work design are mechanistic and motivational. The \textit{mechanistic approach} stems from Adam Smith and proponents of Scientific Management. Organising principles such as division of labour and specialisation were (and still are) crucial for improving productivity and job performance, which consequently brought the need for changing jobs by simplifying them. It is well documented that the principles of the mechanistic approach are successfully implemented even today and have garnered much attention in operations management. On the other hand, the \textit{motivational approach} gained most of the attention in industrial and organisational psychology, organisational behaviour, and management domains. It is rooted in theories and models such as the job characteristics model \citep{Hackman1976-nf_THMM} and job demands-resources theory \citep{Bakker2023-al_THMM}. Both theories suggest that modifying work characteristics can enhance employee attitudes and motivation.

Jobs can be described and compared using work characteristics, that is, objective, stable, and universally measurable job properties \citep{Hackman1979-ds_THMM}. \cite{Demerouti2001-rz_THMM} distinguished between job demands (e.g., workload, time pressure) and job resources (e.g., autonomy, supervisor support), each type having different roles in the health impairment and motivational processes. We can also differentiate among four work design dimensions \citep{Morgeson2006-dr_THMM}: task, knowledge, social, and contextual. \textit{Task characteristics} relate to the nature and variety of tasks involved in a specific job (e.g., task variety, task significance, task identity). \textit{Knowledge characteristics} pertain to the types of knowledge, skills, and abilities required from an individual based on the tasks performed in the job (e.g., information processing, problem-solving, skill variety). \textit{Social characteristics} describe the nature and intensity of relationships with others in a specific job (e.g., task interdependence, co-worker support, interaction outside the organisation). Finally, \textit{contextual characteristics} describe physical and other aspects of the job (e.g., ergonomics, equipment use, physical demands). 

Going beyond the multiple facets of work design, an informed reader should know that organisational requirements change fast, making jobs obsolete shortly after they are designed. Thus, an essential organisational task is to monitor and redesign employees' jobs. In most cases, managers are the ones who re-think and re-assign workplace duties for subordinates. \textit{Managerial redesign} means that direct supervisors can give specific employees more autonomy over decision-making or reduce the number of tasks they have performed until now. Recently, proactive approaches to work redesign have emerged. For instance, \textit{job crafting} is the concept of employees proactively customising and altering different aspects of their work \citep{Berg2013-us_THMM}. Employees can autonomously (and under the radar) pursue new resources or demands (approach crafting) or practice avoidance crafting by reducing demands \citep{Zhang2019-xy_THMM}. \textit{Idiosyncratic deals} (i-deals) represent another proactive approach to work redesign. These personalised work arrangements negotiated between employees and supervisors \citep{Rousseau2005-bd_THMM} increase flexibility, improve morale, and boost employee loyalty. An example would be when a high performer requires shortened work days for education or personal matters, and her direct supervisor wholeheartedly grants this request for a specific time. Finally, \textit{algorithmic management} entails algorithms allocating tasks and duties to employees instead of managers. Although futuristic, this approach \citep[i.e., ``using a computational formula that autonomously makes decisions based on statistical models or decision rules without explicit human intervention'';][]{Parent-Rocheleau2022-io_THMM} has already been applied (e.g., ridesharing, delivery services) with negative consequences for work design (decreased job resources and increased job demands).

Content-wise, each job can be redesigned using the four most common work redesign techniques: job enrichment, job enlargement, job rotation, and job simplification. \textit{Job enrichment} is an attempt to make the job more enjoyable for job incumbents by introducing new tasks and responsibilities \citep{Cimini2023-if_THMM}. The idea behind such a technique is to offer learning and growth opportunities and training for employees with potential KSAs so that they can perform in more ambitious jobs. Job enrichment is particularly effective for employees with passive or low-stress jobs. \textit{Job enlargement} involves adding new tasks that are similar to the existing ones. This strategy does not imply adding new responsibilities. It is more appropriate if organisations need to allocate specific tasks to employees who are already competent in that area and lack the ambition or KSAs to occupy more responsible job positions. \textit{Job simplification} implies reducing tasks and duties an employee is expected to perform. While it boosts efficiency, it is also an effective method for alleviating the strain on employees in highly stressful roles, helping to prevent burnout and maintain work engagement. Finally, \textit{job rotation} entails switching people between different tasks or jobs. It is primarily used to prevent boredom and reduce injuries and exhaustion, but it also helps build KSAs and an understanding of organisational processes.

Since work design is especially prominent in the digital era, it remains to be seen how the nature of work (i.e., employee tasks and roles) will change. Technological advancement, the introduction of collaborative robots, and the use of AI and algorithmic management create uncertainty and new avenues in work design \citep{Huang2019-rm_THMM,Otting2022-ro_THMM,Parker2022-mb_THMM,Rogiers2024-dv_THMM}. The existing knowledge of work design \citep[e.g.,][]{Parker1998-gd_THMM,Parker2017-cy_THMM} will not become obsolete yet needs to be upgraded. We witness two very different streams in the literature. The first one predicts that automation would be used to complement human efforts by focusing on manual routinised tasks, which would lead to fewer low-skilled manual jobs but also increase the number of high-skilled jobs for operators with more decision-making autonomy, coordination, planning, and job complexity \citep{Cagliano2019-zg_THMM,Vereycken2021-ln_THMM}. The second scenario predicts that technologies such as automation will lead to manufacturing controlled by cyber-physical systems, which will consequently change the nature of jobs for the worse by giving operators less autonomy, simplified tasks, and more organisational procedures \citep{Cagliano2019-zg_THMM,Vereycken2021-ln_THMM,Waschull2020-vp_THMM}. Instead of rolling a die, managers should continue to upskill or cross-skill themselves and their workforce and apply work redesign strategies to meet the changing business and work requirements.

\subsection[Operations excellence (Maneesh Kumar \& Andrea Chiarini)]{Operations excellence\protect\footnote{This subsection was written by Maneesh Kumar and Andrea Chiarini.}}
\label{sec:Operations_excellence}
Organisations increasingly pursue Operational Excellence (OPEX) to gain a competitive edge by implementing business improvement initiatives like Total Quality Management (TQM), Lean, Six Sigma, and Lean Six Sigma (LSS). These approaches focus on improving processes, reducing inefficiencies, and enhancing customer satisfaction across industries \citep{Womack2003-zk_MKAC,Shah2007-lu_MKAC,Antony2023-zg_MKAC,Sunder-M2024-to_MKAC}. While Lean and Six Sigma methodologies have been applied beyond manufacturing to sectors like healthcare and finance \citep[see, for example,]{Browning2021-lg_MKAC}, a narrow focus on efficiency limits long-term growth. A more holistic, systems-based view that includes social dimensions is essential for sustainable improvement \citep{Browning2021-lg_MKAC}. While some researchers view LSS as a strategic approach \citep{Arnheiter2005-dx_MKAC,George2002-ir_MKAC}, many regard it as one of the most effective frameworks within operational excellence \citep{Chiarini2011-mj_MKAC,Jaeger2014-xk_MKAC,Antony2023-zg_MKAC}. 

While improvement methodologies like Lean and Six Sigma are essential, they alone are insufficient to sustain a competitive advantage \citep{Sunder-M2024-to_MKAC}. A narrow focus on efficiency limits a firm's ability to explore broader opportunities \citep{Luz-Tortorella2022-wp_MKAC,Chiarini2021-ht_MKAC}. Researchers emphasise the need for a more holistic, systems-based view of OPEX that incorporates social factors for sustainable improvement \citep{Browning2021-lg_MKAC}. Just as Lean and Six Sigma have evolved in definition \citep{Hopp2021-xi_MKAC,Cusumano2021-ba_MKAC,Browning2021-lg_MKAC,Kumar2008-we_MKAC,Kumar2011-es_MKAC}, OPEX also requires a broader, more integrated perspective, as outlined in table \ref{tab:opsexc}.

\begin{table}
\small
\centering
\caption{Defining Operational Excellence.}\vspace{0.25cm}
\begin{tabular}{p{0.2\linewidth}  p{0.7\linewidth}}
\hline
Authors & Definition of Operational Excellence \\
\hline
\cite{Treacy1995-kh_MKAC}	& Operational Excellence is the \textbf{strategy} for organisations striving to \textbf{deliver a combination of quality, price, ease of purchase, and service} that surpasses that of any other organisation in their market or industry. \\
\cite{Dahlgaard1999-vu_MKAC}	 & Operational Excellence is defined in terms of 4Ps: (\textit{i}) excellent \textbf{people}, who establish, (\textit{ii}) excellent \textbf{partnerships} (with suppliers, customers and society) in order to achieve, (\textit{iii}) excellent \textbf{processes} (key business processes and management processes) to produce, and (\textit{iv}) excellent \textbf{products}, which are able to delight the customers. \\
\cite{Sousa2001-ph_MKAC} & Describe OPEX as a set of \textbf{mutually reinforcing principles} (e.g., continuous improvement and process management) alongside tools and techniques (e.g., control charts and Pareto diagrams).  \\
\cite{Bigelow2002-ed_MKAC} &	OPEX represents a \textbf{structured and systematic approach to enhancing business and operational processes} to drive successful strategy implementation. \\
\citeauthor{Hammer2004-rl_MKAC} (\citeyear{Hammer2004-rl_MKAC}, p. 85) &	The term operational excellence, or operational improvement,
``refers to \textbf{achieving high performance via existing modes of operation}: ensuring that work is done as it ought to be to reduce errors, costs and delays but without fundamentally changing how that work gets accomplished''. \\
\cite{Spear2010-qy_MKAC} &	OPEX is all about using a \textbf{systematic problem-solving approach to keep learning and improving} within the company's internal and external environment. \\
\cite{Carvalho2019-yu_MKAC}	& Operational Excellence should not be considered \textbf{an approach to driving change}, but rather as \textbf{a means of providing tools and a framework} for individuals within the organisation to implement it. \\
\cite{Found2017-my_MKAC} &	The Boston Scientific Strategic Operational Excellence Model aligns vision with results through products, technologies, and partnerships, emphasising skills development and a culture of improvement. It employs interconnected tools, systems, and metrics, driven by strategy, values, leadership, and change management. The model is dynamic, incorporating continuous improvement and a systems approach to Operational Excellence. \\
\cite{Edgeman2019-ae_MKAC} &	The Shingo Institute defines Operational Excellence through five key concepts for the Shingo Prize: (\textit{i}) OPEX focuses on results and behaviours, (\textit{ii}) Ideal behaviours stem from guiding principles, (\textit{iii}) Principles are the foundation for a sustainable culture, (\textit{iv}) Aligning management systems fosters ideal behaviours, and (\textit{v}) Tools like Lean and Six Sigma are enablers that should be strategically integrated to drive behaviours and achieve results. \\
\cite{Browning2021-lg_MKAC}	& When aiming for operational excellence, it is super important to see operations as processes and systems, especially at higher levels in organisations. It is also crucial to consider this same mindset when defining value. This means looking beyond just value-added actions and considering their network of interactions as well. \\

\hline
\end{tabular}
\label{tab:opsexc}
\end{table}

The key themes of OPEX emphasise process improvement, systematic approaches, continuous learning, and the crucial role of culture and people. Successful organisations constantly refine their processes to reduce inefficiencies and sustain improvement through structured methodologies like Lean and Six Sigma \citep{Treacy1995-kh_MKAC,Sousa2001-ph_MKAC}. A strong culture of continuous learning and people development is critical for long-term success \citep{Spear2010-qy_MKAC,Found2017-my_MKAC}. OPEX initiatives must align with organisational goals and drive behaviours that support strategic objectives, ensuring processes, tools, and culture are integrated effectively \citep{Edgeman2019-ae_MKAC}.

Quality awards like the Malcolm Baldrige National Quality Award (MBNQA) and the European Foundation for Quality Management (EFQM) recognise operational excellence using defined scoring systems \citep{Jaeger2014-xk_MKAC}. These frameworks guide organisations toward business excellence through self-assessment. However, their applicability to small and medium-sized enterprises (SMEs) is debated, as the significant preparation and execution efforts required are challenging for SMEs with limited time and resources \citep{Rusjan2005-wv_MKAC}.

Agility is also considered an indicator of OPEX in a dynamic and unstable market environment \citep{Vinodh2010-wj_MKAC,Carvalho2019-yu_MKAC,Carvalho2021-ku_MKAC}. The semblance between high performance and maintaining agility when operating in a changing external environment is also defined as operational excellence, which can only be realised through the effective integration of social and technical aspects with a focus on cultivating organisational culture and agility \citep{Carvalho2019-yu_MKAC,Carvalho2021-ku_MKAC}. This approach to operational excellence not only enables organisations to stay resilient and responsive to evolving market demands \citep{Saleh2017-ly_MKAC} but also fosters the development of technical capabilities and a supportive cultural orientation that sustains operational excellence over time \citep{Carvalho2019-yu_MKAC,Carvalho2021-ku_MKAC}.

The integration of operational excellence practices, such as Lean, with Industry 4.0 base technologies (e.g., cloud computing, IoT, big data, and machine learning), enables end-to-end supply chain collaboration and improved performance \citep{Frank2019-ru_MKAC,Chiarini2021-ht_MKAC,Luz-Tortorella2022-wp_MKAC}. Early research highlighted the synergy between Lean and I4.0, giving rise to concepts like Lean 4.0 and Digital Lean \citep{Kolberg2017-ee_MKAC,Buer2018-yr_MKAC}. Combining these frameworks is expected to deliver superior results compared to using them in isolation, offering companies a powerful tool for boosting efficiency and effectiveness \citep{Chiarini2021-ht_MKAC,Januszek2022-ig_MKAC}. However, companies must approach digital transformation with caution, focusing on developing capabilities for real-time data processing and informed decision-making, rather than overinvesting in costly technologies \citep{Koenig2019-hh_MKAC,Koenig2020-zz_MKAC,Gupta2023-fm_MKAC,Gremyr2022-ru_ABJ}. \cite{Koenig2019-hh_MKAC,Koenig2020-zz_MKAC}  and \cite{Gupta2023-fm_MKAC} demonstrated the value of frugal innovation in airport operations, showing that targeted, real-time monitoring of critical quality factors can be more effective than a complete technological overhaul. The majority of literature fails to emphasise the role of softer aspects such as the role of the leadership team and the impact on employees to embrace the technology, which are critical for managing digitalisation induced changes in the workplace \citep{Tortorella2020-gq_MKAC,Tortorella2021-vk_MKAC,van-Dun2023-av_VGABKA,Januszek2022-ig_MKAC}. 

The first step in operational excellence (OPEX) is understanding value from the customer’s perspective, balancing both efficiency and effectiveness \citep{Browning2021-lg_MKAC}. A reductionist approach that separates value-added from non-value-added activities limits improvement potential. Value depends on the quality of inputs -- products, processes, and information -- used to create it \citep{Browning2003-uk_MKAC}. Focusing solely on cost reduction risks neglecting quality and customer satisfaction, leading to suboptimal outcomes \citep{Hines2008-tn_MKAC}. Sustainable improvement requires balancing efficiency and effectiveness; overemphasising efficiency promotes short-term thinking and negative behaviours \citep{Chiarini2021-ht_MKAC,Chiarini2022-si_ABJ}.

Researchers consistently highlight the importance of softer, socio-cultural factors as critical to the success of OPEX initiatives, particularly in embedding a culture of continuous improvement within organisations and their supply chains \citep{Hines2008-tn_MKAC,Found2018-wm_MKAC,Cusumano2021-ba_MKAC,Tortorella2021-vk_MKAC,Antony2023-zg_MKAC,Arumugam2024-mz_MKAC}. For example, \cite{Hines2008-tn_MKAC} used the iceberg model to emphasise that beneath-the-surface factors such as strategy alignment, leadership, behaviour, and employee engagement are essential for sustaining continuous improvement. However, these often get overlooked in favour of tools, techniques, and technology, despite their fundamental role in driving long-term success. \cite{Antony2023-zg_MKAC} further highlighted key soft factors such as employee involvement, training, leadership, and communication as essential for successful OPEX implementation. Additionally, \cite{van-Dun2023-av_VGABKA} showed that transformational leadership and emotional intelligence play a critical role in navigating digitalisation changes.

Based on table \ref{tab:opsexc} and the scope of OPEX, we define Operational Excellence as the strategic pursuit of improving end-to-end processes (from customer order to delivery) through a combination of people development, systematic problem-solving, technology and continuous learning. It focuses on aligning organisational culture, behaviours, and tools with strategic goals to create sustainable improvement, efficiency, and value creation. Operationally excellent organisations are built on a foundation of well-defined principles, empowered people, and a culture of continuous improvement. For deeper insights on OPEX, see works by \cite{Found2018-wm_MKAC}, \cite{Chiarini2021-ht_MKAC,Chiarini2022-si_ABJ}, \cite{Tortorella2020-gq_MKAC,Tortorella2021-vk_MKAC,Luz-Tortorella2022-wp_MKAC}, and \cite{Sunder-M2024-to_MKAC}. For critiques and future research on Lean and OPEX, refer to \cite{Hopp2021-xi_MKAC}, \cite{Cusumano2021-ba_MKAC}, and \cite{Browning2021-lg_MKAC}.

\subsection[Quality management (Alistair Brandon-Jones)]{Quality management\protect\footnote{This subsection was written by Alistair Brandon-Jones.}}
\label{sec:Quality_management}
Quality Management plays a central role in operation management. It has evolved from a relatively routine activity focused largely on error prevention to an integrative organisational philosophy that seeks to deliver operations improvement and enhanced customer satisfaction. 

There are two key perspectives on the term `quality'. Dominant in early quality management discourse is the \textit{product-based approach}, where quality reflects differences in measurable attributes of the product \citep{Garvin1984-mq_ABJ}. Taking this approach, quality is achieved by meeting customer expectations through the consistent conformance to a predetermined set of specifications. The idea of `consistent' implies quality management is used to design and run processes in such a way that ensure product or service conformance. In contrast to this relatively objective and mechanistic view of quality, the \textit{customer's view of quality} explicitly acknowledges that expectations are shaped by individual needs and experiences \citep{Slack2024-nn_ABJ}. As such, customers will inevitably perceive the same product or service in different ways. Furthermore, when they are not able to assess the `technical' specification of a product or service, customers adopt surrogate measures to assess quality. 

In seeking to combine these alternative views of quality, the disconfirmation perspective (or gap perspective) has emerged as the dominant orthodoxy \citep{Berry2004-fn_ABJ,Cadotte1987-fu_ABJ,Oliver1980-pr_ABJ}. This treats quality as a function of the gap between the customer's expectations of the product or service and their perception of its performance. Disconfirmation is positive when performance exceeds expectations and negative when it falls short of expectations \citep{Ganesh2000-bl_ABJ}. Within the disconfirmation process, both \textit{assimilation effects} and \textit{contrast effects} \citep{Oliver1999-zp_ABJ} play a significant role.  Assimilation effects speak to the way that previously held expectations anchor quality assessments and are particularly important when there is performance ambiguity \citep{Ganesh2000-bl_ABJ}. Contrast effects concern the magnification of quality assessment whereby customers have a propensity to over-emphasise the scale of their positive or negative performance assessments \citep{Brandon-Jones2006-pk_ABJ}. 

In operationalising quality, the `Nordic' perspective \citep{Gronroos1984-lg_ABJ} defines quality dimensions in global terms. \textit{Technical quality} refers to what a customer receives, whilst \textit{functional quality} is concerned with how the product or service is delivered. \cite{Rust1994-ta_ABJ} build on this perspective with their three-component model including \textit{service product}, \textit{service delivery}, and \textit{service environment}. In contrast, the `American' perspective applies terms to describe the various aspects of the service encounter (e.g., reliability, responsiveness, assurance, empathy, and tangibles) and has been applied extensively through applications of the SERVQUAL scale \citep{Parasuraman1988-ul_ABJ}. 

The `sand cone' model or cumulative capability perspective \citep{Ferdows1990-ti_ABJ} provides a useful view on the role of quality in driving operations improvement. The model states that quality should be the priority for any organisation as it forms the foundation upon which other aspects of performance – dependability, speed, flexibility and cost – can be layered cumulatively (rather than sequentially). Only when a minimum quality threshold has been reached can the operation move onto improving dependability. To achieve improvements to dependability, requires further improvements to quality. In turn, once a sufficient level of dependability has been achieved, the focus can shift towards improving speed, then flexibility, and finally cost of the operation. All the time, attention on the `lower layers' of the Sand cone must be maintained.

Quality management has evolved significantly over the last hundred years. In the late 1800's, the focus of quality management (though the term was not really established) was essentially on enabling the transition towards mechanised industrial systems and production volume. As such, when present, quality management predominantly involved inspection to monitor product quality \citep{Watson2019-gi_ABJ} and to identify defects. The shift in focus towards productivity at the turn of the 20th century resulted in a subtle shift of emphasis away from simply identifying defects and towards collecting data to identify sources of variation in products and processes, often using statistical process control \citep{Chiarini2022-si_ABJ}. Furthermore, there was an increase in the number of dedicated individuals tasked with assuring product quality and the establishment of formal quality functions within organisations \citep{Sousa2002-hx_ABJ}. 

During the third industrial revolution, customers began demanding higher product and service quality in response to greater per capita income \citep{Chiarini2022-si_ABJ}. This necessitated a move beyond defect detection and rejection to a focus on defect prevention. This period witnessed growing attention on quality management and the flourishing of Total Quality Management (TQM) as well as the related improvement philosophy of Lean (see \S\ref{sec:Operations_excellence} and \S\ref{sec:Lean_and_agile}). Based on the core writings of Philip Crosby, W. Edwards Deming, Joseph Juran and Kaoru Ishikawa, TQM gained huge traction across many business sectors during the 1980s and 1990s \citep{Hackman1995-dd_ABJ,Sousa2002-hx_ABJ}. Whilst the popularity of TQM has declined in the last twenty years, its general precepts and principles remain the dominant mode of quality management in contemporary organisations \citep{Slack2024-nn_ABJ}. In some ways, TQM merely codified the way in which quality-related practices had progressed up to the 1980s. Yet it also extended the established ideas of inspection, quality control and quality assurance with a more strategic orientation to become ``an integrative management philosophy aimed at the continuous improvement of performance'' \citep{Ebrahimi2013-hn_ABJ}. Here, the \textit{voice of the customer} (also referred to a \textit{customer centricity}) and \textit{supplier involvement} were increasingly recognised as key elements of quality management \citep{Flynn1994-cd_ABJ}. The recognition that everyone within the organisation is an internal customer receiving products and services largely from internal suppliers reinforced the idea that any errors within the organisation ultimately impact on the end customer \citep{Slack2024-nn_ABJ,Sousa2002-hx_ABJ}. 

Quality management considers key \textit{costs of quality} – prevention costs, appraisal costs, internal failure costs and external failure costs. Prevention costs are those incurred trying to stop errors occurring. Appraisal costs are those associated with quality control in order to detect errors (for example, setting up SPC systems, input and output inspection, quality audits and customer surveys). Internal failure costs are those associated with addressing failures such as reworking, material waste, lost capacity, and work stress when dealing with errors. Finally, external failure costs are incurred when a defective product or service is experienced by the customer. Such costs may involve replacing products or re-delivering services, damage to organisational reputation, compensation and litigation. In traditional quality management, the assumption is that there exists an optimum level of quality effort that will minimise the total cost of quality. Beyond this optimum level, prevention and appraisal costs combined are greater than internal and external failure costs combined. In contrast, TQM proponents argue that prevention and appraisal costs are significantly lower, and the costs of internal and external failure are significantly higher, than implied by traditional quality management. Therefore, TQM refutes the concept of an optimum quality threshold and instead places significant emphasis on managing prevention and appraisal to ensure product and service failure are kept to an absolute minimum.

We are now witnessing the fourth industrial revolution with its significant emphasis on servitisation, mass customisation and digital offerings \citep{Gremyr2022-ru_ABJ,Kohtamaki2021-gb_ABJ}. Furthermore, the continued rise of sustainability agendas across a wide range of sectors has increased attention on the potential ways in which quality management can support environmental and social dimensions of performance improvement \citep{Siva2016-uq_ABJ}. Arguably, this has necessitated the shift of quality functions away from a policing role to a more proactive role that involves greater levels of collaboration with both internal functions and external stakeholders, such as customers, suppliers, competitors, and regulators. Furthermore, it requires an even greater understanding of the subjective views of customers to inform the actions taken within quality management, for example in evaluating personalised customer feedback and in collecting customer data on products and services in use \citep{Fundin2006-qv_ABJ}. 

Several bodies have been established to encourage the adoption of quality management practises. Arguably, the most well-known of these is The Deming Prize, established in 1951 in honour of W Edwards Deming for his key role in the regeneration of Japan after the Second World War and his broader contributions to the quality improvement movement \citep{Anderson1994-pi_ABJ}. This prize is awarded to organisations who have demonstrated leadership in `company-wide quality control'. In 1987, The Malcolm Baldrige National Quality Award was established to encourage the adoption of quality improvement in American organisations. The award had the effect of raising awareness amongst senior executives concerning the value of quality management and of the key role that employees play in driving operations improvement \citep{Stanley2001-rg_ABJ}. The European Foundation for Quality Management (EFQM) Excellence Award (now known as the EFQM Global Award), was first awarded in 1992. The award was established to recognise European businesses who demonstrated excellence in operations improvement. 

\subsection[Production control (Mélanie Despeisse \& Björn Johansson)]{Production control\protect\footnote{This subsection was written by Mélanie Despeisse and Björn Johansson.}}
\label{sec:Production_and_control}
Production control is a crucial subset of activities within the broader topic of operations management. It focuses on execution, monitoring and optimisation of resources to ensure process efficiency, product quality and timely delivery to fulfil customers' demands \citep{Slack2010-jk_MDBJ}. Production control requires inputs from many other functions such as scheduling (\S\ref{sec:Scheduling}) and planning (\S\ref{sec:Planning}), but also plays a role in supporting inventory management (\S\ref{sec:Inventory_management}), capacity management (\S\ref{sec:Capacity_management}), material handling (\S\ref{sec:Material_handling}), quality management (\S\ref{sec:Quality_management}), operations excellence, and assembly line balancing (\S\ref{sec:Assembly_line_balancing}), amongst others. Production planning and control are often considered together, but they differ in that planning is about allocating resource according to what is expected to happen, while control is about executing the plan, monitoring the operations, and adapting to the actual conditions and unexpected changes. However, production planning and control are intricately connected as feedback loops between the two are often necessary to ensure successful product delivery. The line between the two are increasingly blurred with the rise of automation and more integrated systems with the increased digital maturity of manufacturing companies (digitisation \S\ref{sec:Digitalisation}), often combining planning and control functionalities as a single activity. 

Traditionally, production control focuses on internal information flows and can generally be categorised as push (e.g., make-to-stock, driven by demand forecast) or pull \citep[e.g., make-to-order, triggered by customer demand;][]{Lodding2011-nf_MDBJ,Thurer2024-wy_MDBJ}. With push control, the flow of materials is pushed through successive processes from suppliers to inventory and then queues, workstations, buffers, etc. until the product is complete and ready to be delivered to the customer. In pull systems, production is triggered by an order (customer demand) which pulls the materials through the processes with the aim to minimise inventory and overproduction. In modern production control systems, hybrid systems are often used to combine push and pull logic. Common hybrid approaches include drum-buffer-rope and constant work in progress \citep[CONWIP;][]{Spearman1990-pc_MDBJ}. Alternative Material Flow Control (MFC) methods have been proposed, such as order generation, order release and production authorisation \citep{Thurer2024-wy_MDBJ}. The dominant trend remains with pull approaches as Lean, Kanban and Just-In-Time (JIT) systems have become established best practices for cost and waste reduction. This is further reinforced with the increased integration of external information flows with the rise of customisation, penalisation and engineer-to-order models. 

To facilitate production control, a structured data exchange network is needed where resources and materials are documented and linked to what will happen when and where in the production chain \citep{Givehchi2017-co_MDBJ}. This interoperable network can be structured in several ways often categorised as centralised, distributed or hierarchical \citep{Kasper2024-pd_MDBJ}. Traditionally, a hierarchical network with a tree-like structure has been used and is still in use for most production facilities. The distributed network -- also known as decentralised, holonic or heterarchical control systems which resemble a spider web with resources connected to each other directly without hierarchy -- has been under development for a long time \citep{Bongaerts2000-lc_MDBJ,Cupek2016-tq_MDBJ,Mcfarlane2000-hu_MDBJ}. Industry 4.0 developments, such as RAMI 4.0 \citep{Hankel2015-wk_MDBJ} and GAIA-X \citep{Eggers2020-pl_MDBJ}, are examples of standardisation developments and facilitators for a distributed control system.

The ISA-95 standard defines different levels for enterprise-control system integration \citep{International-Society-of-Automation2010-ev_MDBJ}. Production control requires management tools ranging from IT systems at level 3-4 with Product Lifecycle Management (PLM), Enterprise Resource Planning (ERP), Manufacturing execution systems (MES) and Material Requirements Planning (MRP), all the way to actual production control with operational technology (OT) systems at level 0-2 with supervisory control and data acquisition (SCADA) systems, distributed control system (DCS), programmable logic controllers (PLC), and physical systems with sensors and actuators. While the top-level structure is well-defined \citep{Cupek2016-tq_MDBJ}, the lower levels of detailed control on PLC level are not standardised to give manufacturers the freedom to choose appropriate approaches meeting their specific needs and conditions.

With Industry 4.0 developments and implementations, increased digitisation and Industrial Internet of Things (IIoT) are becoming more prevalent \citep{Kolberg2015-py_MDBJ}. However, the amount and richness of data can be overwhelming, especially for SMEs \citep{Moeuf2018-gq_MDBJ}. Machine learning and predictive methods used in simulation and optimisation models are promising ways to use the data for decision making and to increase production efficiency, flexibility and resilience \citep{Martins2020-ak_MDBJ}. 
For example, using a predictive control strategy to adapt to unpredictable changes in material inputs, such as in pharmaceutical manufacturing, can help maintain consistent product quality and yields \citep{Eslami2024-zw}.
Digital twins for real-time monitoring and closed-loop feedback systems can address the challenges of adapting to rapidly changing manufacturing environments. 
For example at a process level, machine learning has been used for in-situ monitoring of metal additive manufacturing to take evasive or corrective actions when detecting avoidable defects and anomalies \citep{Gunasegaram2024-se}. At the manufacturing systems level, the use of real-time control methods to adapt operations to variable renewable energy supply and to support more self-sufficient and resilient manufacturing systems \citep{Beier2017-se}.
The ISO 23247 standardisation effort on ‘Automation systems and integration -- Digital twin framework for manufacturing’ are beneficial for structuring production planning and control systems \citep{International-Organization-for-Standardization2021-wg_MDBJ,Shao2023-kw_MDBJ}, including, for example, Part 1: Overview and general principles, Part 2: Reference architecture, Part 3: Digital representation of manufacturing elements, and Part 4: Information exchange. Additional validation and more detailed specification work including validation with demonstrators are ongoing as of now. Blockchain-like technologies are also on the rise to facilitate multi-stakeholder interoperability across value chains to enable traceability and transparency in production processes and along these value chains. 

Numerous metrics and key performance indicators (KPIs) can be used to evaluate production control performance; e.g., cost, reliability, flexibility, productivity, speed, lead time, throughput, quality, overall equipment effectiveness (OEE), etc. \citep{Bueno2020-bk_MDBJ,Wheelwright1978-je_MDBJ}. Improvement approaches, such as Lean Six Sigma and Total Quality Management (TQM) are also effective to monitor and improve processes and production controls \citep{Bendell2006-jn_MDBJ,Moeuf2018-gq_MDBJ}. In this respect, it is important to measure the right aspects to achieve more holistic improvements accounting for increased product variety, demand variability, product and process complexity, and others \citep{Rahmani2022-la_MDBJ}. Therefore, additional performance dimensions need to be further explored and developed as they are often overlooked under the Industry 4.0 paradigm, including environmental impacts \citep{Despeisse2012-ki_MDBJ}, knowledge discovery and generation, and human factors \citep{Usuga-Cadavid2020-aq_MDBJ}. 

Focusing on sustainability dimensions, the environmental and social dimensions are less well-addressed than the economic one, largely related to productivity, cost and quality. Various methods have been developed to include aspect of environmental management in production control, for example energy efficiency and waste minimisation \citep{Dai2013-dd_MDBJ,de-Ron1998-sg_MDBJ,Matsunaga2022-ln_MDBJ}. Looking at strategies for production controls accounting for external information flows, new methods have been developed to support circular economy through production planning and control of remanufacturing operations \citep{Dev2020-zc_MDBJ,Guide2000-wv_MDBJ}. While automation is increasing, it remains essential to keep the human in the loop \citep{Cimini2020-zc_MDBJ,Pinzone2020-sz_MDBJ}. Production managers and operators have key roles to play in decision making and problem solving related to production control. Challenges remain with skills and education, as well as resistance to change, teamwork and other non-technical issues. This is in line with the upcoming trends guiding future developments in production control models and technologies, promising to be more digitalised, resilient, sustainable and human-centric under the Industry 5.0 paradigm. 

\subsection[Facility layout (Andrew Kusiak \& Yong-Hong Kuo)]{Facility layout\protect\footnote{This subsection was written by Andrew Kusiak and Yong-Hong Kuo.}}
\label{sec:Facility_layout}
Facility layout applies to spatial arrangement of spaces (e.g., office spaces) and objects (e.g., machine tools). It has seen a tremendous coverage in the literature on manufacturing and other applications \citep{Reed1961-eo_AKYHK,Snow1961-aw_AKYHK}. Any manufacturing and service organisation includes facilities, workspaces, equipment, and furniture which need to be organised to meet the business objectives. The high computational complexity of the facility layout problem might have contributed to the diversity of models and algorithms that date back to 1950s \citep{Koopmans1957-zi_AKYHK}.
The facility layout problem (FLP) has seen many applications and models, some having distinct properties reflected in the objective functions and constraints and names such as: 
\begin{itemize}[noitemsep,nolistsep]
\item Facility layout problem: Facilities may change shapes are located at different sites usually in predefined area in two-dimensional \citep{Irani2000-ym_AKYHK} or three-dimensional space \citep{Kalita2024-uv_AKYHK}. 
\item Machine layout problem \citep{Zuo2019-fs_AKYHK}: Machine tools have a fixed footprint while facilities may change their shape.
\item Cutting stock problem \citep{Hadj-Salem2023-gp_AKYHK}: Arranging patterns to be extracted usually from a sheet of material (e.g., metal, wood, plastic). The literature reports models in 1, 1.5, and 2 dimensions.
\item Wind turbine siting problem \citep{Kusiak2010-fa_AKYHK}: Locating wind turbines at different sites, where the turbines are optimally located subject to constraints resembling those encountered in manufacturing applications. 
\item Facility location problem \citep{Zhang2023-sc_AKYHK}: Locating facilities (e.g., manufacturing plants, warehouses, stores, restaurants) at different sites.  Though most often it is solved in two dimensions, other cases are possible.
\end{itemize}

\subsubsection*{Modelling the facility layout problems}
FLPs are frequently applied to positioning departments in a facility.  A common objective is to minimise the overall cost associated with inter-departmental logistics within the facility. The cost is often related to the transportation of materials, parts, or products, and is directly proportional the distances travelled between departments \citep{Anjos2017-hk_AKYHK}. In many settings, the facility and its departments are rectangular in shape. The dimension of the facility is given and fixed, while those of the departments may vary. Typical constraints in FLPs can be categorised as: (\textit{i}) requirements related to department shapes (e.g., specified areas, dimensions, and aspect ratios) and (\textit{ii}) requirements concerning the placement of departments (e.g., non-overlapping departments and pre-specified locations); see also \cite{Anjos2017-hk_AKYHK}.

A widely studied class of FLP is the single and double row layout problem \citep{Heragu1991-lb_AKYHK}. The single row FLP involves departments with the known length. The objective is to optimally position the departments along this row to minimise the weighted sum of flow distances. This single-row FLP was discussed in, e.g., \cite{Ravi-Kumar1995-za_AKYHK} and \cite{Maier2023-ue_AKYHK}. The row arrangement concept can be expanded to include the double-row FLP \citep{Heragu1988-gw_AKYHK,Chung2010-ws_AKYHK}. In the double-row FLP, two rows of facilities are placed on the opposite sides of an aisle, and the the departments are to be arrange using the same objective as in the single-row FLP. The problem can be further generalised to the multi-row FLP \citep{Wan2022-eq_AKYHK}, which deals with layouts across multiple rows. Additionally, loop layouts \citep{Cheng1998-wi_AKYHK} and open-field layouts \citep{Yang2005-af_AKYHK} are specific facility configurations that have been explored in the literature.

Another popular class of FLPs is the unequal-area FLPs \citep[UAFLPs;][]{Armour1963-wm_AKYHK,Montreuil1991-hi_AKYHK,Tate1995-ld_AKYHK}. Similar to the row FLP, the height and the width of the facility are given and fixed. The UAFLP is to determine the positions and dimensions of departments within a facility while ensuring adherence to the constraints on the ranges of department widths and heights (and/or aspect ratios). Additionally, the non-overlapping condition needs to be met.

Row FLPs and UAFLPs focus on the layout within a single floor. However, when FLPs require the arrangement of departments across multiple floors, a new class of problems, multi-floor FLPs \citep[MFFLPs;][]{Johnson1982-na_AKYHK,Meller1997-dh_AKYHK}, is introduced. In MFFLPs, the dimensions of the facility’s floors and the areas of the departments are pre-defined. MFFLPs require additional factors to be considered, which includes the consideration of floor heights and/or the utilisation of an elevator, capturing vertical travel costs in the planning process.

\subsubsection*{Solving the facility layout problems}
FLPs are known to be ${\cal NP}$-hard \citep{Anjos2017-hk_AKYHK}. There have been significant research efforts to devise computationally efficient algorithms for solving these complex problems \citep{Kusiak1987-de_AKYHK}. In the literature and industrial applications, two primary approaches have been adopted to tackle FLPs, mathematical programming and heuristics. 

\subsubsection*{Mathematical programming}
Most FLPs are formulated as mixed-integer linear programs (MILPs) or semidefinite programs (SDPs) involving rigorous mathematical representations of the problem. A crucial aspect that FLPs account for is the cost associated with the flows within the facility. To capture the distances and flows across the departments, one needs to figure out the relative position of each department in relation to the other units. For instance, for a single-row FLP, one may need to determine the placement of department \textit{i} with respect to department \textit{j}. Concepts such as permutation and betweenness need to be incorporated in the formulation \citep{Anjos2017-hk_AKYHK}. In a MILP formulation, binary variables taking values of 0 and 1 can be utilised to indicate whether one department is positioned between two others, with the use of a set of linear constraints \citep{Amaral2009-kn_AKYHK}. One the other hand, an SDP may require binary variables with values of -1 and 1 to indicate if one department is positioned right or left to another department, complemented by a set of quadratic constraints \citep{Anjos2012-us_AKYHK}. For more complex cases of FLPs such as the multi-row FLP and MFFLP, additional binary variables and constraints are introduced to indicate the placement of a department in a specific row or floor, while ensuring that the constraints related to the shape and position of each department are respected. The reader is referred to \cite{Kusiak1987-de_AKYHK} and \cite{Anjos2017-hk_AKYHK} for detailed discussions on the MILP and SDP formulations of FLPs.

When using exact solution techniques such as the branch-and-bound algorithm to solve FLPs, using valid inequality constraints can significantly reduce the computational time. For instance, in the context of the row FLP, researchers have derived valid inequalities to strengthen formulations or offer polyhedral results \citep{Amaral2008-ov_AKYHK,Amaral2009-kn_AKYHK,Amaral2013-of_AKYHK}. Similarly, for UAFLPs, studies have investigated the use of valid inequalities to enhance the computational performance \citep{Meller1998-jo_AKYHK,Sherali2003-jn_AKYHK,Konak2006-ga_AKYHK}.

\subsubsection*{Heuristic algorithms}
As FLPs are ${\cal NP}$-hard, optimal algorithms face computational challenges when tackling large-scale instances. In such situations, heuristics are usually adopted to obtain near-optimal solutions. A variety of heuristics, such as ant colony optimisation \citep{Kulturel-Konak2011-sw_AKYHK,Zouein2022-vv_AKYHK}, genetic algorithms \citep{Kulturel-Konak2013-qs_AKYHK}, GRASP \citep{Wan2022-eq_AKYHK}, simulated annealing \citep{Meller1996-is_AKYHK}, Tabu search \citep{Kulturel-Konak2012-xd_AKYHK}, and variable neighbourhood search \citep{Herran2021-if_AKYHK}, have been utilised to derive high-quality solutions for FLPs within reasonable computational times. In recent years, due to the advances in computing and data science, applications of reinforcement learning have emerged as new approaches for solving FLPs \citep{Yan2022-cm_AKYHK,Heinbach2023-cr_AKYHK,Zhang2023-sc_AKYHK,Kaven2024-ra_AKYHK}.

\subsection[Machining, assembly and disassembly line balancing (Olga Battaïa \& Alexandre Dolgui)]{Machining, assembly and disassembly line balancing\protect\footnote{This subsection was written by Olga Battaïa and Alexandre Dolgui.}}
\label{sec:Assembly_line_balancing}
Line balancing is a well-known combinatorial optimisation problem, proved to be ${\cal NP}$-hard in general. Initially developed for manual assembly environments, the core problem has been adapted to accommodate robotic, machining, and disassembly settings 
\citep{Battaia2013-tr_OBAD,Battaia2022-qd_OBAD}. Despite the diversity of industrial environments and line configurations, the mathematical models employed usually share similar types of decisions, constraints and objective functions.

Generally, the goal of line balancing is to determine how to manufacture a product within the production line, which consists of a set of workstations \citep{Scholl1999-hd_OBAD}. This involves assigning tasks to these workstations to meet specific requirements and optimise one or more objectives. By the time the line balancing step is reached, the layout of the production line is typically already established from a previous decision. Each layout imposes specific constraints that must be adhered to when assigning tasks to the workstations.

One of the earliest formulations of this problem was proposed by \cite{Salveson1955-ta_OBAD} for manual assembly lines. It involves assigning a set of tasks $I$ to a sequence of linearly ordered workstations $M$ in such a way that the sum of durations of the tasks assigned to the same workstation does not exceed a given takt time and the precedence constraints between the tasks are respected. The objective function for this first formulation was to minimise the number of required workstations for the assignment of all given tasks. 

This problem, known as the Simple Assembly Line Balancing Problem (SALBP) following the definition by \cite{Baybars1986-kw_OBAD}, was shown to be ${\cal NP}$-hard in general \citep{Wee1982-dl_OBAD}. Early studies were dedicated to the development of efficient solution techniques for this simplified problem \citep{Gutjahr1964-jj_OBAD,Arcus*1965-tu_OBAD,Baybars1986-kw_OBAD,Hoffmann1990-rm_OBAD}. 

SALBP represents a basic version of industrial line balancing problems existing in various manufacturing environments, such as machining, assembly and disassembly. Over the decades, more precise problem formulations have been developed to address the specific needs of various industrial contexts:
\begin{itemize}[noitemsep,nolistsep]
\item \textit{Machining}: the first formulation for machining environment was developed for Transfer lines and named Transfer Line Balancing Problem \citep{Dolgui1999-kd_OBAD}. Typically, in this context, there are fewer precedence relationships between tasks compared to assembly or disassembly processes. However, numerous compatibility constraints may determine which tasks must be performed together at the same workstation due to tolerance requirements, or conversely, which operations cannot be performed together due to fixturing limitations. These lines are generally automated. For further details, see \citep{Guschinskaya2009-rd_OBAD,Battaia2024-xv_OBAD}.
\item \textit{Assembly}: the various configurations of assembly lines were addressed ranging from fully manual, with one or multiple workers assigned to each workstation, eventually collaborating with cobots, to fully automated systems \citep{Scholl2006-on_OBAD,Boysen2007-dc_OBAD,Boysen2022-aj_OBAD}.
\item \textit{Disassembly}: The development of the circular economy has intensified the need for dis-assembly. The first formulation of Disassembly Line balancing Problem is due to (McGovern and Gupta, 2007)\citep{Mcgovern2007-dc_OBAD}. Similar to Salveson's formulation, this model has since been extended to address a range of industrial configurations, from fully manual to fully automated systems \citep{Mcgovern2011-dq_OBAD,Ozceylan2019-kq_OBAD,He2024-ax_OBAD}.
\end{itemize}

All existing formulations for diverse line balancing problems can be classified according to common features such as variety of products treated in the line, layout of the line to be balanced, number of lines to be balanced, task characteristics (attributes), workstation characteristics (attributes), constraints to be observed, and objective function to be optimised \citep{Battaia2013-tr_OBAD}. 

\textbf{Variety of products treated in the line.} Based on this criterion, the following types of production lines are commonly identified in the literature \citep{Scholl1999-hd_OBAD}:
\begin{itemize}[noitemsep,nolistsep]
\item \textit{Single-model lines}: A single homogeneous product is manufactured on the line \citep{Otto2013-sj_OBAD}. During each cycle time, every workstation performs the same set of tasks assigned to it. Consequently, the line-balancing problem deals with the decision how to distribute the whole set of tasks required for the product completion among the workstations.
\item \textit{Mixed-model lines}: Multiple models of products from a basic product family are produced simultaneously \citep{Zeltzer2017-ee_OBAD}. These models share similar core processes, differing only in certain attributes or optional features. As a result, the full set of tasks must be divided among the workstations, with each workstation assigned a subset of tasks for each model. The number of subsets associated with each workstation corresponds to the number of models being produced.
\item \textit{Multi-model lines}: Several distinct products are produced in separate batches \citep{Pereira2018-tm_OBAD}. In this scenario, the line can be rebalanced for each batch, allowing for setup times between them and an additional problem of sequencing of lots should be also solved if there are setup times.
\end{itemize}

\textbf{Line layout.} According to this criterion, the following types of production lines are commonly distinguished in the literature:

\begin{itemize}[noitemsep,nolistsep]
\item \textit{Basic straight lines}: In this layout, each workpiece moves through a series of workstations in the order they are set up \citep{Scholl1999-hd_OBAD}. A specific set of tasks is assigned to each workstation, and these tasks are performed sequentially within each station.
\item \textit{Straight lines with multiple workplaces}: Workstations are arranged in a straight line but each workstation includes several parallel \citep{Lovato2023-qr_OBAD}, sequential \citep{Battaia2012-jy_OBAD}, or mixed-activation workplaces \citep{Dolgui2009-kk_OBAD}. These workplaces enable workers or equipment to operate on each workpiece simultaneously, sequentially, or in a series-parallel manner.
\item \textit{U-shaped lines}: These lines feature both the entry and exit points at the same location, and they are usually manual. Workers positioned between the two legs of the line can move between them allowing them to work on two or more workpieces during the same takt time \citep{Fattahi2015-gi_OBAD}. 
\item \textit{Lines with circular transfer}: Workstations are arranged around a rotating table which facilitates the loading, unloading, and transfer of parts between stations \citep{Battaia2023-az_OBAD}. The process can be multi-turn \citep{Battini2007-fm_OBAD}.
\item \textit{Asymmetric lines} are designed to delay product differentiation, maintaining a common line configuration for as long as possible \citep{Ko2008-ud_OBAD,AlGeddawy2010-ep_OBAD}. 
\end{itemize}

\textbf{Number of lines to be balanced.} Based on this criterion, the following types of production situations are commonly identified in the literature:

\begin{itemize}[noitemsep,nolistsep]
\item \textit{Independent lines}: These lines can be identical or different and may produce the same product or different ones \citep{Aguilar2020-th_OBAD}.
\item \textit{Lines with multiline workstations}: In this configuration, workstations or workers per-form tasks across more than one line \citep{Ozcan2010-dl_OBAD}. While this setup can reduce the total number of required workstations, a breakdown at a multiline workstation can impact production on multiple lines.
\item \textit{Parallel lines with crossover workstations}: These lines have crossovers that allow workpieces to be transferred from one line to another in case of a failure \citep{Spicer2002-wo_OBAD}.
\end{itemize}

\textbf{Tasks attributes.} Task attributes such task time, cost, may have constant/uncertain/dynamic values or be functions of partial assignment solutions. 

\textbf{Workstations and their attributes.} Workstation can have scalar (e.g. the number of identical parallel workstations/machines/workers associated with a workstation) or vectorial attributes which indicate the repartition of the tasks assigned to the same workstation to its resources (equipment, operators, etc).

\textbf{Problem constraints.} Constraints are used to distinguish feasible assignments of tasks to workstations. They can have technological, economical or logical origins. The most frequently modelled constraints include: assignment constraints, precedence constraints, inclusion/conjunctive/positive zoning constraints, exclusion/disjunctive/negative zoning constraints, distance and positional constraints, correspondence constraints (matching of task and workstation attributes) and constraints on performance measures calculated for workstations (takt time constraint, workstation space utilisation,  ergonomic constraints, energy consumption constraints, workstation reliability, etc).

\textbf{Objective functions.} An objective function is used to evaluate the quality of feasible solutions and choose the best one, the most common objective functions are:  the number of workstations, line takt time, line efficiency, system utilisation, line smoothness, line cost or profit, line reliability, energy consumption. Frequently, multiple objective functions are used \citep{Laili2020-gu_OBAD}.

\textbf{Hybridisation with other optimisation problems.} Making several decisions simultaneously along with line balancing leads to better final line performance and effectiveness, the most popular hybridisations in the literature are with task sequencing at each workstation \citep{Ozcan2019-sc_OBAD}, with process planning \citep{Battaia2017-cb_OBAD,Bentaha2023-rr_OBAD}, and with line design \citep{Battaia2024-xv_OBAD}.

\textbf{Solution methods.} From the perspective of the quality of the results, solution methods are of-ten divided into exact and approximate. More details on existing solution methods can be found in \cite{Sewell2012-il_OBAD}, \cite{Morrison2014-aa_OBAD}, \cite{Pape2015-os_OBAD}, \cite{Bukchin2018-mo_OBAD}, and \cite{Boysen2022-aj_OBAD}. To assess more precisely the performances of the line, simulation can be used to evaluate its dynamic behaviour \citep{Cortes2010-wd_OBAD}.

\subsection[Planning (Christoph H. Glock \& Ting Zheng)]{Planning\protect\footnote{This subsection was written by Christoph H. Glock and Ting Zheng.}}
\label{sec:Planning}
Planning is the process of developing and implementing strategic and tactical measures to reach specific goals. It entails defining objectives, constraints, and using algorithms to optimise the coordination of material requirements, demand, production, and distribution. Effective planning aligns resources with strategic objectives, thereby optimising operational efficiency and enhancing customer satisfaction while minimising associated risks. Comprehensive planning requires the integration of suppliers, the enterprise, and customers. However, achieving true optimisation and integrated planning is often impractical due to the complexities introduced by numerous concurrent planning tasks \citep{Fleischmann2003-fk_CGTZ}. This section provides an overview of planning problems that occur in operations management, exploring hierarchical planning and advanced planning systems (APSs) together with their corresponding planning modules. We also address collaboration and information sharing as well as planning enabled by digital technologies. 

\begin{figure}[h!]
\centering
 \includegraphics[width=0.65\textwidth]{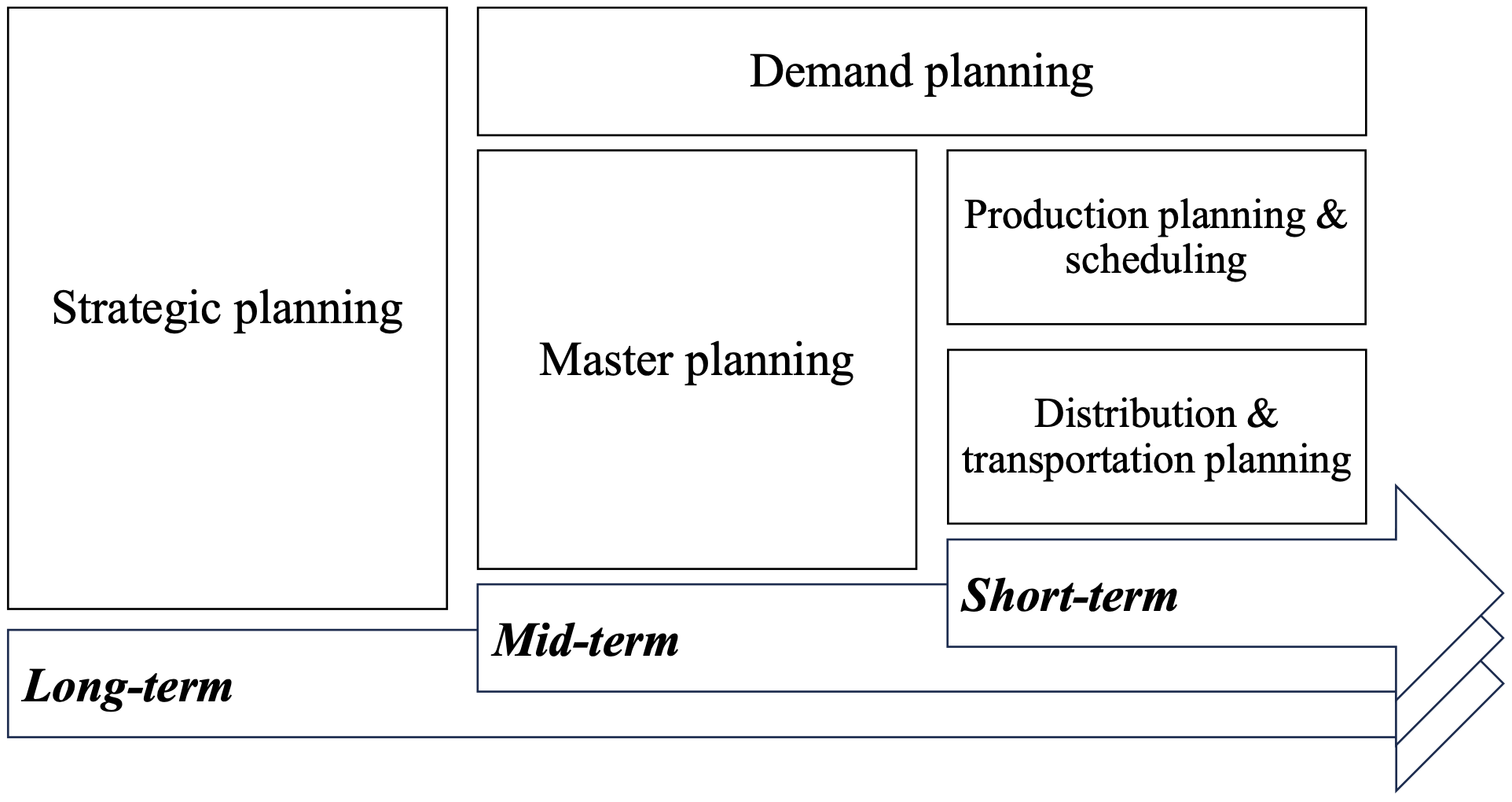}
 \caption{Hierarchical planning concept.}
 \label{fig:planning}
\end{figure}

Hierarchical planning offers a practical solution by segmenting the overall planning task into manageable subtasks \citep{Hax1973-uv_CGTZ}, organising planning tasks into three tiers: long-term, mid-term, and short-term planning \citep{Anthony1965-jp_CGTZ}. Long-term planning, also known as strategic planning or business planning, focuses on a multi-year horizon and shapes the company's strategic direction, influencing decisions related to supplier selection, production capacity, and distribution structure \citep{Stadtler2005-ax_CGTZ}. Mid-term planning comprises demand planning and master planning. It spans several months to a year and addresses aggregate-level issues such as production quantities, inventory levels, workforce requirements, and distribution channels \citep{Pereira2020-on_CGTZ}. Short-term planning, which includes scheduling and distribution planning, is conducted on a weekly or daily basis. It involves decisions regarding purchase quantities, production lot sizes, transportation modes, and vehicle routes \citep[see also \S\ref{sec:Scheduling}]{Bitran1993-nd_CGTZ}. These planning tasks are intricately linked, with a continuous flow of information across different levels, necessitating coordination to ensure alignment and integration of planning efforts across the hierarchical spectrum \citep{Fleischmann2003-fk_CGTZ,Ozdamar1999-yr_CGTZ}. The hierarchical planning concept is illustrated in figure \ref{fig:planning} based on \cite{Meyr2002-tj_CGTZ} and \cite{Rohde2005-dm_CGTZ}.

While various algorithms exist to manage planning tasks, addressing the entire task at once is often not possible. APSs provide a robust framework that integrates different modules and algorithms to tackle real-world planning challenges \citep{Meyr2002-tj_CGTZ}. As decision support systems, APSs extract data from enterprise resource planning (ERP) systems, generate high-quality solutions, and relay decisions back to the ERP for execution \citep{Zoryk-Schalla2004-ou_CGTZ}. Numerous vendors offer APS solutions, including Oracle, SAP, i2, and AspenTech \citep{Jonsson2007-xn_CGTZ}. An APS typically comprises several of the modules illustrated in figure \ref{fig:planning} \citep{Meyr2002-tj_CGTZ}. These modules operate based on both constructional and organisational hierarchies, effectively addressing complex planning problems by decomposing them into simpler sub-problems and coordinating across distinct functional areas \citep{Fleischmann2003-fk_CGTZ}.

Strategic planning sets the foundation for tactical and operational planning, occurring over a long-term horizon. Once established, strategies are generally costly to modify \citep{Colotla2003-vn_CGTZ}. The strategic planning module within APSs utilises problem-oriented heuristics for decision-making, providing rapid evaluations based on qualitative and quantitative criteria \citep{Jonsson2007-xn_CGTZ}.

Master (or tactical) planning addresses mid-term issues covering a planning horizon from a season to a year \citep{Rohde2005-dm_CGTZ}. It serves as a conduit between strategic and short-term operational planning, utilising the infrastructure defined by strategic planning to optimise revenues and minimise costs while setting goals for operational activities. Typical decision variables in master planning include production, inventory, transport, sales, supply and backlog quantities, estimated hirings and lay-offs, as well as subcontracting and overtime, constrained by capacity limitations \citep{Jonsson2007-xn_CGTZ}. APSs employ linear programming or mixed-integer programming optimisers for solving master planning problems, although increasing model complexity may necessitate the use of heuristics or metaheuristics \citep{Silver1998-go_CGTZ,Venkataraman1996-dj_CGTZ}.

Demand planning considers both mid-term and short-term horizons. It comprises demand forecasting, what-if-analyses/simulation, and safety stock determination \citep{Wagner2002-ju_CGTZ}. Forecasting may employ top-down, bottom-up, or middle-out approaches to ensure consistency across various dimensions, such as product and time (year, month, week). Single-level forecasting can use time series models, explanatory models, Bayesian models or the Delphi-method \citep[][see also \S \ref{sec:Forecasting}]{Box1976_SMD,Fleischmann2003-fk_CGTZ,Silver1998-go_CGTZ}. What-if-analyses/simulation help strategise the introduction of new products or promotions. Safety stocks are finally calculated to buffer against forecasting errors, lead time variability, or other uncertainties \citep{Graves2000-hh_CGTZ}.

Turning to short-term planning, production planning and scheduling aims to optimally allocate resources within a single plant, featuring detailed daily minute-by-minute plans. It encompasses two levels: aggregate production planning (APP) and short-term scheduling. APP, similar to master planning, determines production quantities, overtime work, subcontract quantities, inventory levels, backlog quantities, external purchases, and aggregated lot-sizes over a finite time horizon \citep{Nam1992-pe_CGTZ}. The result is then disaggregated into detailed lot-sizing \citep{Glock2014-vk_CGTZ} and machine schedules \citep{Nelson1967-zf_CGTZ}, aiming to fully utilise available resources. Material requirements planning (MRP-I) and its extension, manufacturing resource planning (MRP-II), are widely used for APP. While MRP-I assumes infinite capacities, MRP-II incorporates capacity constraints and integrates functions such as aggregate planning, capacity checks, and short-term scheduling \citep{Harhen1988-py_CGTZ}. MRP-II requires a step-by-step process to plan materials and capacities, and it is sometimes difficult to generate plans that consider volatility. APSs enhance these capabilities by providing adaptive plans that apply linear programming and mixed-integer programming and allow for the visualisation of planning results. This integration of adaptive planning through APSs bridges the gap between material and capacity planning and the operational execution of production, ensuring that production control processes are more responsive and aligned with real-time changes in demand and available resources (\S\ref{sec:Production_and_control}).

It is crucial to recognise that planning extends beyond a company-centric perspective to include an inter-organisational dimension. In this regard, distribution and transportation planning determine optimal product flows \citep{Fahimnia2018-qe_CGTZ}. In a typical distribution network, products from different factories are transported to distribution centres, where they are bundled for long-distance transportation to regional warehouses or transshipment points. In procurement networks, materials from suppliers are collected at regional warehouses before being shipped to factories \citep{Amiri2006-sh_CGTZ}. APSs utilise information from demand planning and master planning to address the strategic distribution network configuration, mid-term plans (e.g., shipment frequencies), and short-term transportation plans (e.g., routes, shipment quantities, vehicle loading). The distribution planning module generally focuses on longer-term planning issues, while the transportation module handles daily transport activities \citep{Fleischmann2003-fk_CGTZ}.

Using advanced information and communication technology enhances information sharing and processing, enabling collaborative planning, forecasting, and replenishment (CPFR). In the mid-1990s, practitioners introduced the CPFR concept to facilitate the exchange of strategic decisions, demand forecasts, and production plans between retailers and manufacturers \citep{HollmannUnknown-bm_CGTZ}. CPFR has since found widespread application in industries such as food, apparel, and manufacturing \citep{Fliedner2003-ha_CGTZ}. Through collaborative business plan development, real-time data sharing, and future demand forecasting, stakeholders along the supply chain can synchronise inventory replenishment processes, thereby reducing inventory costs while enhancing product availability \citep{Barratt2001-ly_CGTZ}. It is important to note that the use of advanced information and communication technology is essential for facilitating CPFR.

The increasing computing power and advancement in digital technologies such as the Internet of Things (IoT), big data, cloud computing, and artificial intelligence (AI) are further revolutionising planning capabilities. For example, big data and machine learning can uncover complex patterns from vast datasets, enhancing demand forecasting accuracy \citep{Carbonneau2008-wn_CGTZ,Seyedan2020-wa_CGTZ}. Incorporating IoT and machine learning into planning processes allows for real-time data collection and facilitates better informed decisions in production, inventory control, and vehicle routing \citep{Bai2023-xw_CGTZ,Rolf2023-ie_CGTZ}. The application of machine learning is particularly promising for addressing problems with uncertainty or nonlinear properties \citep{Esteso2023-cr_CGTZ}. As AI technology continues to evolve, it is expected to provide more visible, transparent, and intelligent planning solutions \citep{Jackson2024-hq_CGTZ,Rolf2023-ie_CGTZ}.

Planning problems are often complex due to the interplay of multiple objectives, constraints, time horizons, and decision layers. To address these challenges, numerous mathematical modelling approaches have been developed, ranging from mixed-integer (linear) programming \citep[e.g.,][]{Bilgen2007-ve_CHGTZ,Dogan1999-fu_CHGTZ} to fuzzy programming \citep[e.g.,][]{Peidro2010-ki_CHGTZ}, or stochastic programming \citep[e.g.,][]{Sabri2000-la_CHGTZ}, aiming to obtain optimal or near-optimal solutions. The richness of this modelling landscape reflects the diversity of planning environments, each requiring tailored formulations to capture domain-specific characteristics. Reviews by \cite{Fleckenstein2023-zs_CHGTZ}, \cite{Gelders1981-mv_CHGTZ}, and \cite{Mula2010-ji_CHGTZ} provide valuable overviews of optimisation-based models that support planning.

\subsection[Scheduling (Rainer Kolisch \& Dirk Briskorn)]{Scheduling\protect\footnote{This subsection was written by Rainer Kolisch and Dirk Briskorn.}}
\label{sec:Scheduling}
Scheduling is the allocation of \textit{resources} to \textit{tasks} over time. The result is a \textit{schedule}, which gives information on the \textit{start time} and the \textit{completion time} of each task as well as the times when resources are processing (which) tasks throughout the planning horizon. Synonyms for tasks in the literature are jobs or activities. Scheduling is one of the major and classical planning problems in \textit{operations management}, see, e.g., \S\ref{sec:Warehousing}, \S\ref{sec:Production_and_control}, \S\ref{sec:Project_management}, \S\ref{sec:New_productservice_development}. Recently, scheduling is also addressed in  \textit{supply chain management}, see \citet{Amaro/P:08_RKDB}, \citet{Chen/H:22_RKDB}, \citet{Kreipl/P:04_RKDB}, and \citet{Sawik:11_RKDB}. In most cases, scheduling is a \textit{short- to mid-term planning problem}, where the planning horizon stretches from a few hours to several months. Scheduling is done in manufacturing and services, see, e.g., \citet{Pinedo2009_RKDB}. In \textit{manufacturing}, tasks relate to physical products, which have to be processed in production facilities. In \textit{services}, the main fields for scheduling problems are in transportation (e.g., trucks, which have to be (un)loaded at terminal gates, see \citeauthor{Boysen2010_RKDB}, \citeyear{Boysen2010_RKDB}, ships which have to be (un)loaded at berths, see \citeauthor{BIERWIRTH2010615_RKDB}, \citeyear{BIERWIRTH2010615_RKDB}, and \citeauthor{BIERWIRTH2015675_RKDB}, \citeyear{BIERWIRTH2015675_RKDB}, containers, which have to be moved by cranes, see \citeauthor{Boysen2017_RKDB}, \citeyear{Boysen2017_RKDB}, aircrafts, which have to use runways for starting and landing, see \citeauthor{IKLI2021105336_RKDB}, \citeyear{IKLI2021105336_RKDB}), healthcare (e.g, patients, which have to be treated by doctors, see \citeauthor{Hall:12_RKDB}, \citeyear{Hall:12_RKDB}, and \citeauthor{Youn/G/P:22_RKDB}, \citeyear{Youn/G/P:22_RKDB}), sports \citep[see][]{Kendall/K/R/U:10_RKDB,Ribeiro:13_RKDB}), and workforce scheduling \citep[see][]{VanddenBergh/B/B/D/B:13_RKDB}. See also \citet{Pinedo2009_RKDB} and \citet{Pinedo/Z/Z:15_RKDB} for a survey of scheduling in services. 

For a general overview over scheduling as a field we refer to \citet{BLAZEWICZ2019_RKDB}, \citet{Brucker2006_RKDB}, \citet{Brucker2012_RKDB}, \citet{LEUNG2004_RKDB}, \citet{Pinedo2009_RKDB,Pinedo2022_RKDB}, and \citet{Schwindt2015a_RKDB,Schwindt2015b_RKDB}.

\subsubsection*{Tasks}
While being processed a task occupies resources. In the most simple case the resource is occupied completely and cannot handle any other tasks in parallel. However, tasks also might seize a fraction of a resource's capacity and, if it is sufficiently small, the resource might handle other tasks in parallel. A task's \textit{processing time} specifies how long it occupies the resource. In most cases, the processing time is a given parameter. In more involved settings the decision maker might have the opportunity to decide on the processing time \citep{WeglarzJMW:2011_RKDB,SHABTAY20071643_RKDB,Gawiejnowicz2020_RKDB}. Once the processing of a task has started it may or may not be \textit{preempted}. However, processing a task is only allowed within a time interval between its \textit{release date} and its \textit{deadline}. In (quite common) special cases, release dates and deadlines may not impose any relevant restrictions and we can think of them as non-existent.

\subsubsection*{Precedence constraints}
The times when tasks are processed are interdependent through the joint need for scarce resources. Hence, processing one task might restrict the freedom when to process another tasks. Another form of interdependency is precedence constraint between task pairs. The most common precedence constraint is a minimal time lag of 0 between the completion time of the {\it predecessor task} and the start time of the {\it successor task}. In its most general from, precedence constraints impose minimal or maximal time lags on the start and/or completion times of two tasks \citep{Neumann/S/Z:03_RKDB}. General precedence constraints are powerful modelling tools, allowing to depict many settings such as two tasks, which must not be processed in parallel or are partially processed in parallel, or completion of two tasks within a given time frame, just to name a couple \citep{HARTMANN20221_RKDB}.

\subsubsection*{Resources}
There are two predominant generic concepts in the literature: machine scheduling and project scheduling. In \textit{machine scheduling} resources are typically fully occupied while processing one task. Since in this case one machine does process no more than one task at a time, we often simply seek a sequence of tasks (potentially with idle time in between consecutive tasks) on each machine. Different settings with multiple machines differ in the relation of machines. In settings with \textit{parallel machines} we consider alternative resources where each task requires to be processed by one machine only. These alternative resources might differ in their processing capabilities, e.\,g., in their speeds, or not  \citep{Pinedo2022_RKDB}. In \textit{shop settings}, each task needs to be processed by each machine. The order in which machines are visited is given and identical for all tasks in a \textit{flow shop} setting  \citep{ROSSIT2018143_RKDB} given and individual in a \textit{job shop} and \textit{flexible job shop} setting \citep{XIONG2022105731_RKDB,Dauzeres-Peres/D/S/T:24_RKDB}, and to be determined for each task as part of the decision process in an \textit{open shop} setting \citep{AHMADIAN2021399_RKDB}. In \textit{project scheduling} resources are capable of processing multiple tasks at a time \citep{HARTMANN20101_RKDB,HARTMANN20221_RKDB,Kolisch/P:01_RKDB}. Processing a task induces a certain load of capacity for each resource. As far as capacities are concerned, a set of tasks can be handled in parallel if the total load does not exceed the capacity of any resource. With respect to the capacity of resources there are different concepts, which can be combined. The most frequent one are \textit{renewable resources}, where the capacity of a resource is available in each period anew. The capacities of non-renewable resources are  available only once for all periods of the planning horizon \citep{WeglarzJMW:2011_RKDB} while the capacities of partially renewable resources are only available for subsets of the periods of the planning horizon \citep{Alvarz-Valdes/T/V:15_RKDB}. In most settings capacities of resources are given and might by dynamic over time or not \citep{HARTMANN2015_RKDB}. However, there are problems where the capacity of resources has to be determined as part of the decision problem, e.g. the \textit{resource investment problem} and the \textit{resource levelling problems}, see \citet{Rieck/Z/G:12_RKDB} and Part V of \citet{Schwindt2015a_RKDB}. 

\subsubsection*{Objectives}
Naturally, finding an arbitrary feasible schedule is often not sufficient (although this might prove challenging enough) but a good or even optimal one with respect to a given objective is to be determined. While we can find a huge variety of objectives in the literature, we restrict ourselves to classic objectives in scheduling. We, first, address \textit{time objectives} and consider \textit{cost objectives} afterwards. The classical and most prominent scheduling objective is to minimise the so-called \textit{makespan}, which is the last completion time among all tasks. Instead of focusing on the bottleneck task which completes last, we also might be interested in minimising the \textit{sum of completion times} of all tasks which is equivalent to minimising the average completion time among tasks. Several objectives involve a due date associated with each task. As opposed to deadlines implied by time windows, due dates may be violated which is, however, penalised in terms of the objective. The \textit{lateness} of a task is defined as the difference between the task's completion time and its due date. A positive lateness is termed as \textit{tardiness} and the absolute value of negative lateness as \textit{earliness}. The most common objective concerning lateness is to minimise the \textit{maximum lateness} among tasks. Further objectives related to lateness are to minimise the \textit{sum of tardiness} of tasks or the \textit{number of tardy tasks}. For sum of completion times, sum of tardiness, and number of tardy jobs, also weighted versions are prominent where each task has a \textit{weight}, reflecting its impact on the objective value. Minimising the weighted sum of earliness and tardiness of tasks is an important objective for just-in-time manufacturing \citep[e.g.,][]{Jozefowska:07_RKDB}. In project scheduling, cost objectives are addressed in particular when the availability of capacities is to be decided. In the resource investment problem, the total \textit{cost for buying capacity} is to be minimised while in the resource renting problem, the \textit{renting cost}, potentially reflecting setup costs and linear costs, are to be minimised. Furthermore, the \textit{net present value} objective seeks to maximise the the sum of discounted task costs, where tasks costs are incurred at the start or completion time of tasks \citep{Gu/S/S/W/C:15_RKDB}. Of course it might be interesting to consider not only a single objective but to account for multiple objectives \citep{Kindt/B:06_RKDB}.

\subsubsection*{Solution methods}
Scheduling problems are combinatorial in nature and the vast majority is ${\cal NP}$-hard. Hence, classical methods for solving scheduling problems exactly are mixed-integer linear programs (MIPs), branch-and-bound, and dynamic programming \citep{Agnetis2025-od}. Constraint programming \citep[CP;][]{Baptiste2012-bg} is a solution approach, which is particularly suited to scheduling problems. Recent results show that for certain scheduling problems, CP is superior to MIPs \citep{Naderi2023-vy}. Additionally, combining MIP and CP is a promising solution strategy \citep{Pohl2022-lx}. Heuristics such as priority rules and metaheuristics are used to solve large problem instances \citep{Agnetis2025-od,Ruiz2018-uz}. Furthermore, matheuristics and CP heuristics successfully combine heuristic techniques with MIP- and CP-approaches \citep{Agnetis2025-od}. Recent research investigates machine learning methods, most prominently reinforcement learning and deep reinforcement learning \citep{Kayhan2021-fa,Khadivi2025-uz}. For the latter, graph neural networks are of particular interest, since the precedence constraints between tasks are naturally represented by graphs \citep{Smit2025-dm}.

\subsubsection*{Beyond deterministic offline scheduling}
While almost all scheduling problems revolve around the components introduced above, we can distinguish different problem classes according to the nature of the information available for decision making. While in \textit{deterministic scheduling problems} the parameters involved are not subject to any kind of stochasticity but specific numerical values are known, in \textit{stochastic scheduling problems} one or more parameters are stochastic \citep{CaiWuZhou2014_RKDB}. When we encounter stochastic parameters, \textit{robustness} of schedules becomes an interesting feature \citep{KOUVELIS1997_RKDB}. Robustness can be addressed with respect to feasibility of the schedule and/or the achieved objective value. In \textit{offline problems} all parameters (stochastic or not) are revealed beforehand while in \textit{online problems} some or maybe even all tasks or their parameters are revealed over time \citep{RobertVivien2009_RKDB,Albers2009_RKDB}. In stochastic settings or in online settings the goal often is not to determine a full schedule immediately. Instead, a policy is determined which constructs a schedule over time when tasks or their parameters are revealed.

\subsection[Project management (Jens K. Roehrich \& Tyson R. Browning)]{Project management\protect\footnote{This subsection was written by Jens K. Roehrich and Tyson R. Browning.}}
\label{sec:Project_management}

The work done by individuals, teams, and organisations falls along a spectrum, from repetition of identical tasks to the invention of new ones \citep{Hayes1984-ts_JRTB,Browning2017-sw_JRTB}. While operations management (OM; \S \ref{sec:Operations_Management}) pertains to the management of all such work, project management (PM) is a subset dealing with the latter end of the spectrum -- where a ``temporary'' set of tasks must be ``undertaken to create a unique product, service, or result'' \citep[][p. 245]{Pmi2021-ki_JRTB}. That is, PM is a discipline that involves planning, organising, securing, managing, leading, and controlling resources to achieve specific goals within a defined timeline \citep{Davies2005-kc_JRTB,Browning2017-sw_JRTB}. It is a structured approach to guide a project across its lifecycle from its inception through to its completion \citep{Meredith2021-xg_JRTB}. Because PM involves so many aspects of management, PM techniques naturally draw upon numerous areas of general management theory and other management areas \citep[e.g., innovation;][]{Davies2018-os_JRTB}, yet key aspects of PM theory tend to emerge from the temporary and unique characteristics of project work.

The importance of PM to public, private, and not-for-profit organisations as well as whole industries and governments cannot be overstated, given its significant economic and social value creation \citep{Roehrich2024-sn_JRTB}. Economically, effective PM -- characterised by ensuring that resources are utilised efficiently, objectives are met within budget and on schedule, and the quality of the deliverables is maintained -- strongly contributes to higher productivity, reduced costs, and enhanced profitability. From a societal and social value perspective, PM may lead to improved infrastructure, better services (\S \ref{sec:Servitisation}), and overall societal advancement by, for example, developing skillsets through apprenticeships or upskilling small- and medium-sized enterprises (SMEs; \S \ref{sec:Small_and_medium-sized_enterprises}) in deprived regions \citep{Kalra2019-bs_JRTB}.

Conversely, poor PM may have severe consequences, including cost overruns, delayed timelines, and subpar quality, resulting in an organisation’s or government’s financial losses, wasted resources, and reputational damages \citep{Bendoly2007-zl_JRTB,Maylor2018-jw_JRTB}. High-profile project failures such as the Denver International Airport’s baggage handling system serve as reminders of the implications of poor PM. These project failures are often caused by poor planning, miscommunication or lack of collaboration between (key) stakeholders, and/or scope changes, highlighting the vital need for robust and continuous (across the project lifecycle) PM practices by all project stakeholders -- individuals, groups, and/or organisations that have an interest in or are affected by the outcome of a project \citep{Donaldson1995-sw_JRTB}.

The history of traditional PM tools and frameworks such as the Gantt chart and Work Breakdown Structure \citep{Gantt1919-pl_JRTB,Miller2008-rv_JRTB} traces back to the early 20th century, with the advent of complex industrial and engineering projects necessitating more systematic approaches. However, it could be argued that early roots can be seen in the construction of the Egyptian pyramids and the Great Wall of China, where rudimentary PM principles (rather than tools) were employed \citep{Cleland2004-vf_JRTB}. Modern PM began to take shape during the 1950s with the development of methodologies like the Critical Path Method \citep[CPM;][]{Meredith2021-xg_JRTB} and Program Evaluation and Review Technique \citep[PERT;][]{Malcolm1959-yn_JRTB}. These methodologies formalised the way the `triple constraint' of projects -- time, cost, and quality -- was planned and managed. 

Over the last decades, PM principles have expanded beyond engineering and construction to other sectors such as healthcare and information technology, often spanning public and private organisations \citep{Chakkol2018-ar_JRTB,Roehrich2022-cn_JRTB}. Prior PM scholars have drawn attention to the fact that we are experiencing an ongoing `projectification' of work \citep{Midler1995-ae_JRTB}, where economic activities are increasingly delivered via projects. The establishment of professional bodies, such as the Project Management Institute (PMI) and the Association for Project Management (APM), and the emergence of PM degrees, modules, courses, and programmes at (higher) educational institutions further normalised the discipline, offering standardised guidelines, certifications, and a shared body of knowledge. PMI’s Project Management Body of Knowledge (PMBOK) Guide has provided a framework that has since been continuously updated since the 1980s \citep{Pmi2021-ki_JRTB}. This development was also further supported by an increasing number of academic scholars researching various aspects of PM \citep[e.g.,][]{Morris1994-qw_JRTB,Davies2005-kc_JRTB,Soderlund2011-xg_JRTB}. 

Technically, PM encompasses various models, frameworks, and tools designed to optimise project execution. Most such tools assume that project activities and their relationships can be specified in advance and thus modelled as an activity network, from which schedules, resource-loading profiles, and other decision-support views may be derived and optimised \citep{Browning2010-iq_JRTB}. However, scholars and practitioners have recently questioned the underlying assumptions of project predictability that enables detailed planning and optimisation. Some aspects of unpredictability can be anticipated, such as the typical failure modes causing rework \citep[e.g.,][]{Eppinger2012-bc_JRTB} and other project risks \citep[e.g.,][]{Hillson2020-ay_JRTB}, whereas others, called ``unknown unknowns'', can render some detailed planning moot \citep[e.g.,][]{Ramasesh2014-kc_JRTB}. On the other hand, Agile PM methodology, which emerged in the software industry, promotes iterative development, flexibility, and customer collaboration \citep[e.g.,][]{Highsmith2009-ih_JRTB}. Agile methods, such as Scrum, focus on delivering incremental value and adapting to changing requirements. Agile’s iterative cycles, known as sprints, allow teams to review progress and replan between sprints in response to evolving customer needs. 

Apart from the more technical considerations of PM, behavioural, sociocultural, and sociopolitical aspects of PM are equally crucial. Here, effective interpersonal and interorganisational relationships are the bedrock of successful project execution, including integrated project teams \citep{Roehrich2019-nw_JRTB} and relational coordination and cooperation \citep{Roehrich2023-nt_JRTB,Taubeneder2024-am_JRTB}. Stakeholder management (\S\ref{sec:Stakeholder_management}) is vital to ensure that all parties involved are aligned with the project's goals, and their expectations are managed appropriately from project front-end \citep{A-Lewis2023-dx_JRTB} to transition to the operations phase \citep{Zerjav2018-jt_JRTB}. Dynamics between members of the project team also play a critical role in project success. Project managers and leaders must possess strong interpersonal skills, including communication, negotiation, and empathy, to effectively lead diverse teams to foster a collaborative environment, manage conflicts constructively, use nudges to drive performance, and ensure learning across projects \citep[e.g.,][]{Brady2004-iz_JRTB,Mishra2015-ih_JRTB,Bukoye2022-iv_JRTB}. Power dynamics (between team members, and/or organisations), as well as (organisational) politics and culture can influence project outcomes significantly \citep[e.g.,][]{Clegg2004-lh_JRTB}. Project managers and teams must navigate these complexities, balancing the interests of different stakeholders while maintaining focus on the project goals and objectives. 

Emerging themes in current PM research highlight the growing complexity and scale of projects \citep{Harrison2024-xm_JRTB}. Large interorganisational (or mega-) projects \citep[LIPs;][]{Flyvbjerg2014-ut_JRTB,Denicol2020-bp_JRTB,Roehrich2024-sn_JRTB} are deployed for infrastructure developments and major technological implementations, and bring significant economic and social value, but also pose substantial challenges such as managing cooperation between large, often competing, firms \citep{Taubeneder2024-am_JRTB} or public and private organisations with different objectives \citep[e.g.,][]{Roehrich2022-cn_JRTB}. These projects require sophisticated management techniques to address their scale, complexity, uncertainty, resource constraints, risks, and impact \citep{Pich2002-tf_JRTB,Geraldi2011-ht_JRTB,Ramasesh2014-kc_JRTB,Browning2019-wb_JRTB}. 

The role of emerging technologies (\S\ref{sec:Emerging_technologies}), digitisation (\S\ref{sec:Digitalisation}), and data in PM are other critical areas of research and practice. Advances in data analytics, artificial intelligence (AI), and digital tools are transforming PM practices \citep{Whyte2019-xz_JRTB,Muller2024-wy_JRTB}. Data analytics enables project managers to derive insights from project data, facilitating better decision-making and risk management. AI-powered tools can automate routine tasks, such as scheduling and progress tracking, allowing project managers to focus on strategic activities. Similarly, Building Information Modelling (BIM) is revolutionising PM in the construction industry by providing detailed 3D models that integrate design, construction, and operational information \citep{Volk2014-lr_JRTB}.

In conclusion, PM is a vital discipline that encompasses technical, sociocultural, and emerging aspects to ensure the successful completion of projects. From its historical roots to its modern methodologies and tools, PM continues to evolve, addressing the growing complexities of contemporary projects. Effective PM drives economic and social value, mitigates risks, and ensures that resources are utilised optimally to achieve the desired outcomes. Through continuous innovation and adaptation, PM will remain a critical enabler of progress and development for individuals, organisations, industries, and governments. However, PM is fraught with many interesting challenges and open questions, leaving plenty of opportunities for exciting and impactful research at the intersection of academia, practice, and policy.

\subsection[Programme management (Andrew Davies \& Juliano Denicol)]{Programme management\protect\footnote{This subsection was written by Andrew Davies and Juliano Denicol.}}
\label{sec:Programme_management}
Programme management is often positioned in the project management literature, highlighting the management of projects, programmes and portfolios \citep{Morris2011-yf_ADJD,Morris2013-dj_ADJD}. Project management is often concerned with the achievement of traditional performance metrics (time, cost, quality), whilst programme management incorporates connections with a more strategic level of organisations, external stakeholders, and wider society \citep{Artto2009-rz_ADJD,Rijke2014-bx_ADJD}. The programme management domain is the integration of multiple temporary organisations (projects), each with specific temporalities within the programme life cycle, coordinated as an integral part for the successful delivery of the entire programme \citep{Lycett2004-qe_ADJD,Pellegrinelli2011-su_ADJD,Pollack2022-qe_ADJD}. In summary, a programme cannot be delivered without the completion of all its interdependent projects, and programme management unlocks outcomes that would be impossible if projects are managed in isolation. The coordination of resources at programme level includes strategic topics that often go beyond the technical dimension, moving into the interfaces with organisational, institutional, and political spheres \citep{Maylor2006-nh_ADJD,Rijke2014-bx_ADJD}. Therefore, programme management should not be conceptualised simply as an expansion and amplification of project management. Successful project managers are often challenged when promoted to lead programmes, where they face strategic topics, high ambiguity, amplification of political dynamics, personal interfaces with internal and external stakeholders, and decisions with less information. Programme managers need to have constant external awareness and have the skills to navigate topics in close negotiations with the strategic and political levels. 

\subsubsection*{Dynamics of programme management}
The decomposition of a programme in multiple projects, each project in sub-projects and systems, provides the technical and hierarchical structure of reporting and integration \citep{Davies2009-cp_ADJD,Whyte2021-hx_ADJD}. Although essential to the programme, the technical integration (inward-looking) is only one part of programme management and needs to be complemented by the organisational and institutional engagement and integration (outward-looking). Programme management needs to account for multiple relationships within and outside the focal/client organisation (contracting authority), creating alignments that will enable and shield the technical delivery of projects from external disturbances \citep{Denicol2020-bp_JRTB}. The establishment and development of key relationships with internal and external stakeholders will bring a degree of stability to an environment that is intrinsically changing at fast pace \citep{Muruganandan2022-yx_ADJD}. The clarity of purpose and objectives tend to be an enabler if the relationships and alignments are in place, contributing to collectively achieve the outcomes of the programme.

The coordination at programme level is of utmost importance, as only the focal entity (e.g., client) or appointed delivery partner (e.g., programme management organisation) would have the overall visibility of multiple projects \citep{Castaner2020-bb_ADJD,Denicol2021-ds_ADJD,Oliveira2017-lz_ADJD,Zani2024-ol_ADJD}. The teams and organisations (often Joint Ventures) delivering the individual projects (e.g., part of a new product development, one section of a railway line, a stadium within the Olympic Park) will be entirely focused and driven to achieve their mission, yet with little or no detailed visibility of other projects happening in parallel or sequentially as part of the programme. Considering the fragmentation of the supply network and potential low profit margins, organisations tend to allocate resources to maximise the delivery of their own specific parts of the contract \citep{Denicol2020-jv_ADJD}. This follows a commercial logic often connected with the cost of managing transactions rather than overall value, which brings a clear tension between best for individual projects (and organisations delivering them), and best for the overall programme. 

Programme management will have an interface with the strategy of the focal organisation \citep{Denicol2021-ds_ADJD,Gulati2012-js_ADJD}, therefore it is essential to analyse the strategy of organisations (and their maturity levels) to assess their suitability and capability gap to embark in the journey of delivering a major programme. The organisational strategy is often implemented through portfolios, programmes, and projects – building upon teams, and individuals. Programmes might be organised as a one-off \textit{major project} or ongoing \textit{programmes of work}. Programmes of work are often smaller and by inherent characteristics (e.g., maintenance) might offer more opportunities for repetition and standardisation. Major projects are on the other end of the spectrum, often bringing dimensions of complexity, novelty and uncertainty, yet organised and managed as a programme, with clients decomposing the scope in sub-projects and engaging several tier one contractors \citep{Denicol2020-bp_JRTB}. Given the evolution of scope to build the asset, tier one contractors will be appointed according to their expertise throughout the project’s lifecycle (e.g., from professional service firms in planning, to integrators of systems and sub-systems in execution). Structurally, they might be embedded in a permanent organisation as part of their capital projects directorate or delivered outside through a new entity created with the purpose of being responsible for the programme and act independent from the parent/permanent organisation \citep{Stefano2023-bw_ADJD}. The dynamics of multiple projects within a programme deserves further attention and scientific exploration, as they might happen concurrently or sequentially embedded in multiple networks.

Programme structures have more permanency than project-related boundaries, which are organised within the programme. Different projects will have multiple temporalities (i.e., projects within phase 1 of a railway line), moving to sequential phases (Phase 1 and Phase 2), then a meta level of corporate entity coordinating the entire programme \citep[i.e., the client organisation;][]{Davies2014-cn_ADJD}. This entity can be a temporary \citep{Bakker2010-mi_ADJD,Burke2016-mu_ADJD}, semi-permanent, quasi firm, or virtual firm, which is responsible for setting the scene for multiple levels of programme management. There is significant interest in how different organisations (e.g., clients, delivery partners, tier 1s, tier 2s) are creating programme management strategies \citep{Denicol2022-jw_ADJD} and the connections and implications to the organisational system across the multi-level structure.

Programmes are delivered by a nested network of suppliers, ultimately procured by a client and resource orchestrator \citep{Roehrich2023-nt_JRTB,Stefano2023-bw_ADJD}. The focal organisation acts as the coordinator and integrator of a large network of organisations involved in multiple projects that form the programme \citep{Denicol2021-ds_ADJD}. The organisational/enterprise level is essential to unlock the potential of programme management. Choices related to organisational design is one of the most powerful strategic levers available to the leadership of contemporary organisations \citep{Browning2009-iz,George2023-kt_ADJD,Gulati2012-js_ADJD,Zani2024-mf_ADJD}. The capabilities of the enterprise in charge of programme management are underexplored and present an opportunity for further scientific investigation. 

A well designed and resourced focal entity would provide the conditions to unlock the potential of programme management, maximising the performance within and across different projects of the programme \citep{Denicol2020-zl_ADJD}. A capable focal entity would be able to identify pockets of performance across the supply chain, find synergies between project boundaries, and add value from interactions with the institutional level. Such dynamics create a tension of intra- and inter-organisational levels. First, there is an emphasis on the capabilities of the client/contracting authority (intra-organisational), which creates the conditions for the focal entity to establish the rules and design the structure and relationships with the programme network (inter-organisational); see \cite{Provan2007-zb_ADJD}, \cite{Roehrich2024-sn_JRTB}, \cite{Sydow2018-ks_ADJD}, and \cite{Zhang2024-sb_ADJD}. Second, different organisations of the programme supply network will have to implement their own programme management strategies, as tier one contractors will subcontract parts of their scope (products and services) to specialist suppliers in lower tiers \citep{Denicol2021-ds_ADJD,Stefano2023-bw_ADJD}.

\subsubsection*{Future research avenues}
Future research is needed to investigate how the different programme management at multiple levels might be coordinated and integrated, vertically as part of individual projects and horizontally as part of the programme \citep{Denicol2022-jw_ADJD}. The capabilities of the focal entity to effectively coordinate and integrate a large network of suppliers, the information and knowledge flows across multiple actors, are the foundation to enable systemic and programmatic actions.  Avenues that can support future investigation relate to the design of the focal organisation, the necessary capabilities, and models of engagement with delivery partners that might provide complementary capacity and capability at client side \citep{Zani2024-mf_ADJD}. Researchers might be interested in exploring the role of integrated teams within the programme organisation, trajectories that may lead to high-performing teams, and the organisational impact in designing and managing the programme supply network \citep{Roehrich2024-sn_JRTB,Zhang2024-sb_ADJD}.

Researchers could expand the boundaries of public and private enterprises, by exploring the role of public procurement in the creation of the inter-organisational network \citep{Stefano2023-bw_ADJD}. Procurement models and strategies could be examined to push the boundaries of our understanding regarding standardisation and novelty across different parts of the programme. This line of investigation would unlock synergies and potential strategies to integrate suppliers involved in delivering the programme, with the network of partners hired by the focal entity in their ongoing operational activities. 

\subsection[New product/service development (Anders Haug \& Ewout Reitsma)]{New product/service development\protect\footnote{This subsection was written by Anders Haug and Ewout Reitsma.}}
\label{sec:New_productservice_development}
The Product Development and Management Association defines a new product as ``a product (either a good or service) new to the firm marketing it'' and new product development (NPD) as ``the overall process of strategy, organisation, concept generation, product and marketing plan creation and evaluation, and commercialisation of a new product'' \citep[][p. 458]{Kahn2012-oo_AHER}. While it should be noted that there are parts of the NPD literature that focus on physical products or services only, this section applies a more general perspective. NPD research has resulted in many papers and books across various fields, such as marketing, innovation, operations management, and supply chain management. This section discusses some of the major topics across NPD research: (1) NPD and firm strategy, (2) portfolio management for NPD, (3) the NPD process, (4) product architecture, (5) customer involvement, (6) supplier involvement, (7) cross-functional collaboration, (8) the use of technology, and (9) servitisation.

Research on NPD and firm strategy has explored how firm characteristics and strategies affect NPD. A prominent example is the use of the concept of `dynamic capabilities', which describe a ``firm's ability to integrate, build, and reconfigure internal and external competencies to address rapidly changing environments'' \citep[][p. 516]{Teece1997-od_AHER}. Specifically, by reconfiguring NPD capabilities according to market and technology needs, firms may increase their ability to deal with turbulent environments and rapidly changing technologies \citep{Pavlou2011-kv_AHER}, as well as improve NPD performance \citep{Bruni2009-me_AHER}. NPD performance has also been linked to different strategic orientations. For instance, \cite{Narver2004-el_AHER} found that firms need both responsive and proactive market orientation to create and sustain new product success. Similarly, \cite{Calantone2003-lu_AHER} found that firm innovativeness, market orientation, and top management risk-taking relate positively to NPD speed, which is particularly important in highly turbulent environments. In recent decades, research has increasingly focused on green (or sustainable) NPD. Such research focuses on topics such as the conditions for green NPD \citep{Yalabik2011-zj_AHER}, the role of (green) dynamic capabilities \citep{Chen2013-tg_AHER}, and the link to green process innovation \citep{Xie2019-uz_AHER}.

Portfolio management for NPD is the manifestation of the firm’s strategy and dictates NPD investments \citep{Cooper2023-lp_AHER}. It involves the evaluation, selection, and prioritisation of new projects, as well as the acceleration, termination, or downgrading of existing ones. Research on this topic has explored how to increase the NPD project portfolio value, identify a profitable mix of projects, and strategically allocate NPD resources \citep{Cooper2001-lv_AHER}. In this context, several tools have been introduced for portfolio reviews, including the value-based scorecard, the expected commercial value method, the productivity index, and bubble diagrams \citep{Cooper1997-sb_AHER,Cooper2023-lp_AHER,Si2022-pb_AHER}. For a review of research on portfolio management for NPD, see \cite{Si2022-pb_AHER}.

Research on NPD processes has investigated the content and sequence of NPD activities, such as planning, product design, testing and validation, and commercialisation. This has led to the introduction of the stage-gate model, which is a popular approach for managing NPD processes from idea to launch \citep{Cooper2019-fr_AHER}. It provides an overview of NPD activities, where ‘stages’ reflect the sequence of these activities, and ‘gates’ act as quality control checkpoints. In this context, NPD lead time can be reduced by making NPD activities overlap \citep{Cooper2014-aa_AHER}. This, however, requires a detailed representation of the information exchanges necessary between individual activities, which can be created with a design structure matrix \citep{Browning2001-nn_AHER,Eppinger1994-kh_AHER}. It should also be noted that the stage-gate model can be combined with strategies such as agile, lean, design thinking, and open innovation \citep{Cocchi2024-yo_AHER,Cooper2019-fr_AHER}. For a comprehensive description of NPD processes, see the textbook by \cite{Ulrich2019-mb_AHER}.

Another area of NPD research focuses on the choices pertaining to product architecture. A central conceptualisation in this research is the five categories of architectural modularity by \cite{Ulrich1991-dg_AHER}: (1) component swapping, (2) component sharing, (3) fabricate-to-fit, (4) bus, and (5) sectional modularity, which \cite{Pine1993-jm_AHER} extended with a sixth type named ‘mix modularity’. Later, \cite{Ulrich1995-yb_AHER} provided another oft-cited categorisation of product architectures by conceptualising three distinct types of product modularity – slot, bus, and sectional – which are contrasted to a non-modular (integral) architecture. The use of modular product architectures has often been associated with increased operational performance \citep{Salvador2002-df_AHER}. Furthermore, modularisation allows firms to pursue a ‘mass customisation’ strategy, which aims to target individual customer requirements while having near-mass production efficiency \cite{Pine1993-jm_AHER}. For a review of the literature on modular product design, see \cite{Bonvoisin2016-zy_AHER}.

Since customers may contribute to the generation of new ideas and knowledge \citep{Chang2016-pt_AHER}, research has also focused on customer involvement (CI) in NPD. CI is often associated with the concept of ‘open innovation’, which describes a strategy of integrating ideas and knowledge from external sources to improve innovation performance \citep{Chesbrough2003-gm_AHER}. Research on CI has shown that customers can take different roles in NPD, including that of information providers and codevelopers \citep{Wang2020-gk_AHER}. In this context, it should be noted that excessive or poorly managed CI can negatively impact NPD performance. For instance, customers may be unable to articulate needs for advanced, technology-based products or be unable to conceptualise ideas beyond their own experience \citep{Knudsen2007-iq_AHER}. For a meta-analysis of the contextual factors moderating the relationship between CI and NPD performance, see \cite{Chang2016-pt_AHER}.

Research on supplier involvement (SI) in NPD gained traction with the observation that Japanese automakers outperformed their Western counterparts in time-to-market and NPD cost through SI \citep{Clark1989-mn_AHER}. Research has shown that SI can take different forms, depending on the level of supplier design responsibility and the moment of SI. By accessing – rather than acquiring – supplier knowledge, firms can expect higher NPD efficiency and effectiveness \citep{Suurmond2020-bo_AHER}. For instance, SI can have a positive effect on product design quality and time-to-customer targets \citep{Ragatz2002-ag_AHER,Takeishi2001-qr_AHER}. However, research has also shown that SI may involve risks such as the loss of proprietary information \citep{Wagner2006-ui_AHER,Yan2020-yh_AHER} and suppliers being uncooperative or unmotivated \citep{Primo2002-de_AHER}. For a meta-analysis of the relationship between SI and performance outcomes, see \cite{Suurmond2020-bo_AHER}.

NPD research has also explored the role of cross-functional collaboration. Such studies have, for instance, revealed the potential benefits of increased collaboration between the research and development function and other internal functions such as marketing, purchasing, logistics and manufacturing. While marketing and manufacturing involvement tends to slow down NPD, it increases a product’s competitive advantage and a project’s return on investment \citep{Swink2007-fc_AHER}. Furthermore, since the purchasing function has become the common interface with suppliers, it can contribute to NPD by identifying innovation potentials from the supply market and facilitating the absorption of supplier knowledge \citep{Picaud-Bello2022-no_AHER}. In this context, \cite{Reitsma2023-td_AHER} provided examples of cross-functional NPD activities, while \cite{Yao2019-gr_AHER} discussed decision-support tools that can support such activities. It should be noted that functional diversity in NPD may decrease NPD speed; therefore, it may be desirable to keep the core NPD team small and integrated \citep{Chen2010-qd_AHER}. Furthermore, it is highly important that project managers can detect and resolve knowledge-sharing-related problems in NPD teams \citep{Haug2023-zv_AHER}. For a meta-analysis of how cross-functional collaboration may have an impact on NPD success, see \cite{Troy2008-ps_AHER}.

Another area of NPD research focuses on the role of technology. This research has shown that information technology (IT) can have a positive effect on NPD performance \citep{Mauerhoefer2017-go_AHER}, for instance, through facilitating collaboration with customers and suppliers in NPD \citep{Zheng2019-ts_AHER}. In this context, the use of social media (particularly blogs and forums) can facilitate knowledge sharing with customers and suppliers and enhance NPD performance \citep{Cheng2018-wt_AHER}. More recently, the concept of `Industry 4.0' (I4.0) \citep{Lasi2014-tm_AHER} has been studied extensively in relation to NPD. One of the most discussed I4.0 technologies is `additive manufacturing' (or 3D printing), which allows fast production of prototypes and manufacturing of highly complex components \citep{Huang2013-sc_AHER}. Another often discussed I4.0 technology is `digital twins', which allow the creation of a digital model of possible or actual real-world objects to be used for simulation, testing, monitoring, and maintenance \citep{Lo2021-gd_AHER}. For an overview of the potential applications of artificial intelligence across NPD activities, see \cite{Cooper2023-lp_AHER}.

Finally, in recent decades, the concept of ‘servitisation’ has drawn increasing interest, which has been defined as ``a process of building revenue streams for manufacturers from services'' \citep[][p. 257]{Baines2017-el_AHER}. In this context, the review by \cite{Raddats2019-mb_AHER} identified five main themes in the servitisation literature: (1) service offerings, (2) strategy and structure, (3) motivations and performance, (4) resources and capabilities, and (5) service development, sales, and delivery. The literature has also identified several aims of pursuing a servitisation strategy, including increased revenue or profit, improved response to customer needs, improved product innovation, new revenue streams, higher customer loyalty, and increased barriers to competition \citep[see the review by][]{Baines2017-el_AHER}.

As demonstrated by this section, NPD research covers many topics across different research fields. Thus, when searching for NPD literature, we advise researchers to craft their search strategy carefully to avoid excluding potentially relevant research.

\subsection[Pricing (Hubert Pun \& Salar Ghamat)]{Pricing\protect\footnote{This subsection was written by Hubert Pun and Salar Ghamat.}}
\label{sec:Pricing}
Pricing is a crucial element in operations and supply chain management, influencing firms' strategic and operational decisions. It extends beyond the act of setting prices, requiring firms to navigate a complex landscape of considerations that affect overall supply chain efficiency and market positioning. This section explores various pricing issues relevant to supply chain management, reflecting critical considerations in the literature. By examining these topics, we provide insights into how firms can address complex pricing challenges to optimise performance, enhance market positioning, and improve coordination across supply chains, offering a comprehensive understanding of the factors that shape pricing strategies.

Price competition (Bertrand competition) and quantity competition (Cournot competition) are two models for competitive market. In price competition, firms set prices based on competitors, often leading to price wars that drive prices to marginal costs. Quantity competition involves setting production levels based on competitors' output, generally resulting in higher prices and profits. However, price competition can also lead to higher prices and lower consumer welfare when essential inputs are controlled by a vertically integrated competitor with upstream market power \citep{Arya2008-vm_HPSG}. \cite{Hirose2019-ay_HPSG} find that price competition enhances profits and welfare under public leadership, while private leadership sees higher profits with price competition but greater welfare with quantity competition. \cite{Singh1984-su_HPSG} suggest firms choose price contracts when products are complements and quantity contracts when they are substitutes. When the firms choose either a price to charge or a quantity to produce, \cite{Klemperer1986-uu_HPSG} show that both firms would make the same choice in the presence of demand uncertainty. \cite{Miller2001-sb_HPSG} argue that price and quantity competition are equivalent when managers, incentivised by owners, decide on the competition type. 

Beyond the impact of competition on pricing, double marginalisation occurs when both the manufacturer and retailer add markups to the product price, leading to a higher final consumer price. This happens because each firm maximises its own profit without considering the overall efficiency of the supply chain. As a result, the combined markup reduces total demand and overall supply chain profitability. \cite{Heese2007-kw_HPSG} explores double marginalisation in the context of inventory record inaccuracy, showing that such inaccuracies worsen inefficiencies in decentralised supply chains. \cite{Dellarocas2012-pw_HPSG} highlights that pay-per-action advertising, as used by companies like Google and Microsoft, can lead to double marginalisation, distorting product prices and reducing consumer welfare. \cite{Li2013-wx_HPSG} examine supply chains with uncertainty in input levels, finding that double marginalisation occurs under wholesale price contracts, and propose a coordination mechanism for random demand and yield. \cite{Chen2016-nx_HPSG} analyse supply chains with long lead times and short selling seasons, where poor inventory cost distribution between supplier and retailer leads to incentive misalignment, suggesting a coordination contract with inventory subsidising to address these issues. 

Pricing strategies are heavily influenced by power dynamics among stakeholders significantly impacting final pricing and supply chain profitability, often favouring stronger firms while disadvantaging weaker ones \citep{Luo2017-px_HPSG,Davis2018-pt_HPSG}. Models like Nash bargaining and Stackelberg illustrate how power dynamics shape pricing decisions. In Nash bargaining, both parties negotiate simultaneously, with outcomes reflecting their bargaining strengths. In Stackelberg model, featuring a leader-follower dynamic, the leader sets the price first, followed by the followers pricing decision. This sequential decision-making creates different equilibrium prices compared to simultaneous negotiations, as the leader's actions influence the follower's pricing decisions \citep{Wang2013-ye_HPSG}.

\cite{Leider2016-fa_HPSG} show that when multiple retailers compete for a single supplier, their bargaining power shapes pricing dynamics. Retailers with greater power can secure better wholesale prices, influencing their pricing strategies and competitiveness. Conversely, suppliers may use strategies to mitigate pricing risks, such as contracts that share inventory risks or incentivise retailers to maintain certain pricing levels \citep{Cachon2005-yw_HPSG}. 

Price discrimination is a multifaceted strategy that encompasses various forms—first, second, and third degree. Effective implementation of these strategies helps firms improve their competitive position by navigating consumer behaviour and market dynamics. First-degree price discrimination, or perfect price discrimination, charges each consumer the maximum they are willing to pay, often enabled by detailed consumer data that allows personalised pricing. \cite{Choudhary2005-fm_HPSG} demonstrate how firms can use customer insights to enhance competitive strategies and maximise profits through personalised pricing.

Second-degree price discrimination involves varying prices based on quantity consumed or product version, allowing firms to capture consumer surplus through different pricing tiers. This approach is effective in markets with diverse consumer preferences, enabling segmentation and pricing optimisation based on demand elasticity. \cite{Besanko2003-wf_HPSG} highlight how competitive price discrimination influences pricing strategies, market dynamics, and profitability in oligopolistic settings.

Third-degree price discrimination, which charges different prices to distinct consumer groups based on observable characteristics, is common in supply chains. This strategy maximises revenue across various market segments. \cite{Luo2017-px_HPSG} demonstrate how optimal pricing policies can be applied under different supply chain power structures, emphasising the importance of understanding consumer demand intensity for effective implementation.

Price discrimination impacts not only revenue but also market competition. \cite{Candogan2012-dg_HPSG} show that optimal pricing in networks can affect consumer surplus and market equilibrium, demonstrating the link between pricing strategies and network externalities. \cite{Montes2019-hq_HPSG} emphasise the role of consumer privacy and information asymmetry in competitive pricing, highlighting the need for firms to balance the use of consumer data with privacy considerations.

Firms often adjust prices to reflect changing market dynamics, making a multi-period model effective for optimising long-term outcomes. In a multi-period model, firms can choose committed pricing, setting a fixed price throughout all periods to signal a stable market strategy, or dynamic pricing, where prices are adjusted based on changes in demand, competition, or costs. The choice between these strategies depends on the market environment, competitive dynamics, and the firm’s long-term goals. For example, \cite{Monahan2004-nc_HPSG} explore dynamic pricing for products with random demand and fixed supply, framing it as a price-setting newsvendor problem and offering insights into optimal pricing, procurement, and profits. \cite{Levin2009-br_HPSG} develop a model for oligopolistic firms selling differentiated perishable goods, showing that consumer strategic behaviour can impact revenues and that firms might benefit more from limiting consumer information than fully responding to it. \cite{Wang2014-qj_HPSG} study a duopoly with demand uncertainty, finding that firms with greater capacity tend to commit to fixed pricing, while those with less capacity prefer dynamic pricing. \cite{Cohen2020-wp_HPSG} examine pricing for differentiated products, where firms adjust prices dynamically based on past sales data. 

Pricing strategies are influenced by information asymmetry between firms, leading to inefficiencies and suboptimal pricing. \cite{Corbett2004-mh_HPSG} show that specific contract types, like two-part tariff, can mitigate inefficiencies by acting as signalling mechanisms, allowing suppliers to convey their private information to buyers and thus influence pricing decisions. \cite{Mukhopadhyay2008-bw_HPSG} emphasise the importance of optimal contract design in the presence of information asymmetry, proposing models that account for the incomplete information. \cite{Li2015-jt_HPSG} extend this discussion by investigating supplier encroachment and its implications for nonlinear pricing strategies under information asymmetry. \cite{Zhang2015-xq_HPSG} introduce a product life cycle perspective on pricing strategies under information asymmetry, suggesting that manufacturers must adapt their pricing approaches in response to consumer expectations and purchase timing. Furthermore, \cite{Narayanan2004-fn_HPSG} study the impact of strategic information sharing on pricing, highlighting how deliberately shared information can influence competitive dynamics and lead to more efficient pricing.

Integration of technology into pricing has garnered significant scholarly attention, particularly with the advent of data-driven pricing models and artificial intelligence (AI). The emergence of e-business technologies has transformed traditional pricing mechanisms, enabling firms to adopt dynamic pricing strategies that respond in real-time to market conditions and consumer behaviour. However, while e-business technologies have been widely adopted for coordination, the adoption of price determination technologies remains relatively low \citep{Johnson2007-rd_HPSG}. \citeauthor{brynjolfsson2000frictionless_SFRTLBNO}'s (\citeyear{brynjolfsson2000frictionless_SFRTLBNO}) seminal work illustrates how the Internet has facilitated a competitive pricing environment, leading to lower prices for consumers compared to conventional retail settings. Firms use data analytics to analyse large datasets, predict consumer behaviour, and dynamically optimise pricing. This approach enables firms to implement price discrimination strategies effectively. \cite{Li2017-yv_HPSG} highlight AI’s role in pricing, showing how sensor data in perishable food supply chains can enhance pricing accuracy.

When setting prices, businesses must also consider psychological factors such as fairness concerns, which influence consumer perceptions of price justification and transparency, affecting satisfaction and loyalty. \cite{Ho2014-ru_HPSG} demonstrate that in a one-supplier, two-retailer supply chain, fairness concerns can lower wholesale prices but also lead to complex outcomes, such as higher prices set by retailers. \cite{Nie2017-vp_HPSG} show that such supply chain requires fixed fees rather than quantity discount contracts for coordination. Risk and loss aversion are also crucial psychological factors, with players favouring predictable pricing. Firms must carefully manage and communicate price changes, particularly when consumers are sensitive to increases. \cite{Pun2015-pg_HPSG} show that economic incentives or enforcement can motivate suppliers' subcontracting decisions under risk aversion. \cite{Choi2018-zu_HPSG} examines high-risk fashion supply chains, finding that make-to-order quick response systems can achieve Pareto improvements for risk-averse retailers. \cite{Zhang2022-qn_HPSG} further explore the coordination of Pareto optimality for risk-averse supply chain members using mean-variance and mean-downside-risk objectives.

For a comprehensive understanding of pricing strategies in supply chain management, readers are directed to foundational texts like ``Pricing and Revenue Optimisation'' by \cite{Phillips2021-ub_HPSG} and `The Theory of Industrial Organisation'' by \cite{Tirole1988-fq_HPSG} that offer valuable insights into this complex field.

\subsection[Maintenance (Rommert Dekker \& Geert-Jan van Houtum)]{Maintenance\protect\footnote{This subsection was written by Rommert Dekker and Geert-Jan van Houtum.}}
\label{sec:Maintenance}
In this section we discuss maintenance and its role in operations management. Maintenance can be defined as all actions taken to keep a system into a state in which it can function as prescribed. We identify different types and indicate when these are appropriate. Next we present theory from different angles, both from a management and from an optimisation point of view. In the sequel we use the following terminology: \textit{system}, \textit{unit} and \textit{component}. A system consists of units and units consist of components. Maintenance actions can be executed at each of these levels. A \textit{failure} indicates the event by which a system, unit, or component is no longer able to function.

\subsubsection*{Typology of Maintenance}
Several types of maintenance can be distinguished. Corrective maintenance (CM) is initiated by failures and intends to bring a system back into a working condition. Preventive Maintenance (PM) refers to actions which improves a system's state and is carried out to prevent failures, reduce energy use or improve the quality of the output. PM is generally done based on calendar time or usage of a system, unit or component (e.g., kilometres in the case of a car). Inspection-based maintenance (IBM) refers to inspecting a system, unit or component and replacing/repairing it in case it is in a bad state. Condition-based maintenance (CBM) can be applied when a system, unit or component has a clear \textit{degradation process} that describes the health status. The state of this degradation process is monitored continuously or at least with a high frequency (e.g., every day or week). When a certain state has been reached or passed, then a repair or replacement is executed. It is also possible that there is no clear degradation process, but that all kinds of data is collected and that failures/faults can be predicted via a certain data analysis/AI method. If maintenance is based on such predictions, we refer to it as predictive maintenance (PDM). 

Maintenance varies depending on the environment it is applied to. In case of manufacturing, systems have a clear output which facilitates assessing the value of maintenance. In transport systems, companies operate fleets (e.g., trucks and planes) which allows to standardise maintenance. In case of civil infrastructure, lifetimes are long and deterioration rates are low. These systems often have indirect use, which complicates the justification of maintenance. On one hand there are many identical systems like roads and railway tracks, but also many unique systems like bridges and buildings. For consumer products maintenance has become less important as replacing by a new product is often cheaper. From a sustainability viewpoint there is a renewed emphasis on repairing products, yet labour costs are quite high.  

\textbf{Role of Maintenance in Operations Management.} Maintenance can be seen as a necessary evil with yearly costs of 2-10\% of the total capital investment. It requires specific manpower and may tie to specific companies. Maintenance is difficult to manage as the effect is only seen on a longer term, but then deterioration may be irreversible. Companies may have an own maintenance department or may (partially) outsource maintenance. Original Equipment Manufacturers (OEMs) typically provide services, like advice and spare parts for maintenance. In several industries, it is common that OEMs offer full-service maintenance contracts to customers. Under these contracts, the OEM is responsible for all maintenance actions and for realising the required system uptime percentages. Customers may also lease systems and pay for the use of the system and the uptime percentage (see also \S\ref{sec:Servitisation}). For many OEMs, the revenue generated by after sales service is more than 25\% of their total revenues and often the profit percentage on after sales activities is higher than on the manufacturing activities \citep{2006Deloitte_RDGJH}. For very advanced systems, advanced knowledge and specialised service engineers are needed, and then it is attractive for OEMs to offer full-service maintenance contracts. There is limited competition in that case and it leads to close relationships with customers. 
  
\subsubsection*{Optimisation models}
Optimisation models balance maintenance work in terms of manhours and materials against reduced downtime and increased efficiency. In particular the aim is to avoid failures by doing maintenance preventively. The ideal is to do the maintenance just before the failure, but it is a challenge to predict failures correctly.

\textbf{Single component.} Originally one let systems (and underlying units and components) run until they broke down and CM has to be done. This is called run-to-failure. About half a century ago one introduced PM. Extensive maintenance schemes were developed to make sure that no critical failures could happen. This is still the case in airplane maintenance. At the same time the first optimisation models were developed by \cite{doi:10.1137/1.9781611971194_RDGJH} to determine the best time for PM. The models use statistical distributions of the time to failure. Especially the Weibull distribution was and still is popular. In the simplest case one does PM at fixed intervals, regardless of failures in between. This is a so-called block replacement policy (BRP) and it requires the evaluation of the so-called renewal function. In age replacement policy (ARP) one does maintenance when the component reaches a critical age, while the age is also reset in case of failure replacement. Evaluation of an ARP policy is much easier and it can be shown to outperform BRP.

Many variants of these models were developed and analysed. A recent review is given by \cite{de2020review_RDGJH}.  We like to mention the Modified Block Replacement model, where one does a preventive replacement at fixed time-intervals, but skips one if CM has recently been done \citep{https://doi.org/10.1002/nav.3800230103_RDGJH}. While in the basic ARP and BRP models maintenance costs are the same at all times, a stream of models appeared which considers  occasions at which PM can be done at much lower costs (so-called opportunity maintenance, see e.g. \cite{ab2014opportunistic_RDGJH} for a review). 

A major bottleneck in using these models is the difficulty to obtain appropriate lifetime distributions \citep{DEKKER1996229_RDGJH}. On one hand because there are not always indications of wear out, justifying PM, and on the other hand not many failures occur. Moreover, these models do not use indications that failures are upcoming. 

In the early nineties, ideas on CBM were coming up. In his last lecture, \cite{1991Geraerds_RDGJH} gave a list of trends for the field of maintenance. He stated that: ``Periodic, usage based maintenance will considerably decrease, while condition-based maintenance will start to dominate''. And: ``Together with the automation of condition-based maintenance, one will see a centralisation of inspections and analysis of the measurements (remote monitoring and remote analysis)''.
He was right with seeing the trend that IBM, CBM, and PDM would become more important. IBM came up first \citep[see][]{2005Jardine_RDGJH}. For IBM, the basis is already to know the degradation modes, which may be identified by executing a Failure Mode and Effect Analysis (FMEA). 
Next, you try to describe the degradation processes, e.g. by stochastic processes. Many different processes are being applied \citep{2011Si_RDGJH}. A model/process that requires limited data and has been applied a lot is the Delay Time concept introduced by \cite{1982Christer_RDGJH}. For CBM and PDM, it is required to collect many data, and they came up together with the increased use of sensor technologies and the Internet of Things (IoT); see \cite{2021Tortorella_RDGJH}. These forms of maintenance are studied extensively in the current literature; see \cite{de2020review_RDGJH}.

\textbf{Multiple components.} Systems often consist of many identical or nonidentical components. Several types of relations may exist between components effecting maintenance (see \citeauthor{dekker1997review_RDGJH}, \citeyear{dekker1997review_RDGJH}, and \citeauthor{Nicolai2008_RDGJH}, \citeyear{Nicolai2008_RDGJH}, for reviews). In \textit{structural dependence} maintenance of one component has to be accompanied by maintenance of other components. In \textit{stochastic dependence} the condition of a component influences the deterioration of other components. Next in \textit{economic dependence} maintaining multiple components may be cheaper or more expensive than individual ones. The first is the case when there are large logistical costs, e.g. road maintenance \citep{golabi1982statewide_RDGJH}. The latter may be the case for redundant systems, like a 1-out-of-2 system, where one component can be maintained without shutting down the system, but two would require system shutdown.

In economic dependence one may do a direct grouping of components into a fixed maintenance package, or an indirect grouping. The block replacement model is very suited for the first as it allows easy coordination. In case of age replacement one can only anticipate on opportunities for combination and that is indirect grouping.

In case CBM is applied to multiple components of a system, while a fixed maintenance schedule is followed for other components, coordinating maintenance is much more difficult as maintenance can be initiated at many different moments. Such systems are studied as well and in these studies also the trade-off between total maintenance costs and system availability is studied; see, e.g., \cite{2024Kivanc_RDGJH}.

\textbf{Maintenance and production.} There are several relations between maintenance and production \citep[see][for an overview]{Budai2008_RDGJH}. If maintenance can be done during production or if production is regularly stopped for other reasons, then maintenance work can be fit within the production schedule. A nice example is the night maintenance which is done on the Dutch railway track, when no trains are running. Yet often maintenance needs a shutdown of production to carry out the work in a safe environment. Especially if production increases and PM is delayed one sees a growing conflict. This is for example visible in the maintenance of civil infrastructure where bridges need to be replaced, but closing roads gives very much inconvenience to road users.

A quite popular recent topic in maintenance optimisation is the maintenance of offshore wind turbines. On one hand logistic costs are very high as maintenance personnel has to be shipped to offshore locations, while shutting down a wind turbines for PM or CM induces high costs. A recent review is given by \cite{REN2021110886_RDGJH}.

\subsection[Servitisation (Martin Spring \& Andreas Schroeder)]{Servitisation\protect\footnote{This subsection was written by Martin Spring and Andreas Schroeder.}}
\label{sec:Servitisation}
Servitisation describes a strategic transformation where manufacturers expand from a product-focused business to a service-focused business. It is a long-term transformation process that involves significant changes to the manufacturer but also its customers and the wider delivery network. The majority of servitisation research is focused on the transformation of manufacturers in a business-to-business (B2B) industrial context. Prominent examples include manufacturers of high capital value assets such as aerospace engines, trains, and trucks \citep{Baines2013-ot_MSAS}, but the approach is increasingly also applied to lower unit-cost items such as commercial vehicle tyres, and to some consumer offerings. 

The approach draws in part on the tradition of industrial ecology, where the motivation for moving to a service-based model - whereby users pay for access to assets rather than buying them outright – was to reduce material use. This environmentally motivated form of servitisation became known as Product-Service Systems (Mont, 2002). A second motivation \citep{Neely2008-ty_MSAS} came from a strategic agenda for manufacturers, especially of capital equipment, to ``move downstream'' \citep{Wise1999-kd_MSAS} in order to capture profits from the assets in use, as well as or instead of the sale of the asset, and to lock in customers with an offering differentiated by the service element. This shift along the supply chain helps customers to concentrate on serving their own customers, rather than expending effort on managing maintenance and management of their process equipment or other capital assets. The manufacturer’s technological knowledge of their product allows them effectively to provide services closely related to the product, and their intimate knowledge of contexts of use can allow them to improve product designs \citep{Davies2004-cn_MSAS}.

\subsubsection*{Servitisation as process and outcome}
The term servitisation refers to the process of strategic transformation away from a purely product-based offering. A related question is the extent of the process, more specifically how and to what degree services form part of the overall offering. A popular framework \citep{Baines2013-ot_MSAS} for describing this differentiates between base, intermediate, and advanced services:
\begin{itemize}[noitemsep,nolistsep]
    \item Base services focus on spare part provision and repair.
    \item Intermediate services focus on condition maintenance, such as scheduled overhauls and upgrades, condition monitoring and operator training.
    \item Advanced services see manufacturers committing to providing the outcome of the use of the product, for example when a production machine manufacturer enters into an outcome-based contract with its customer and is paid per pack produced. 
\end{itemize}

Advanced services create critical operational implications, as asset ownership often remains with the manufacturer, and the manufacturer takes on more risk. For the manufacturer, the advantages are that they can overcome commoditisation of the product, create customer stickiness, capitalise on product quality, and smooth the revenue streams by shifting from occasional capital sales to continuous sales for services provided. The advantages and disadvantages for the customer mirror those of the manufacturer: they can focus on their own activities, reduce capital employed and reduce risk, but can become locked in to one supplier and lose important technical capabilities.

For both sides, manufacturer and customer, servitisation and particularly advanced services offerings forge a higher mutual dependence than would emerge in conventional product-based commercial relationships. For the customer to become so dependent on the manufacturer’s capabilities and the manufacturer taking on so much of its customer's risk requires high levels of trust and transparency, but also extensive relationship management. Yet, although much existing research examines these phenomena in particular cases, we still lack econometric large-scale studies that show a complete picture of the amount and circumstances of the benefits. 

\subsubsection*{Digital servitisation}
Research in servitisation began to gain momentum in the early 2000s: for the first 10-15 years, the focus was mostly on defining servitisation strategies and understanding the transition challenges for manufacturers. More recently, servitisation research has been dominated by work on the intersection of digital technologies and servitisation \citep{Gebauer2020-lp_MSAS}. One perspective is that digital connectivity in products enables manufacturers to deliver outcome-based advanced services \citep{Schroeder2020-jp_MSAS}. Alternatively, it can be argued that servitisation provides a pathway by which to profit from digitisation [ref]. This intersection has seen servitisation research combine streams from operations management and information systems research, with prominent areas including the alignment between these domains \citep{Kohtamaki2022-sp_MSAS}, the affordances that are required to leverage the opportunities digitisation offers \citep{Naik2020-sm_MSAS}, and the strategic implications the digital twin and the industrial metaverse will create for the future of servitisation \citep{Bertoni2022-cn_MSAS}. 

\subsubsection*{Critical servitisation research areas}
Because servitisation is an empirical phenomenon rather than a sub-discipline of operations management, it has practical and research implications that cut across various areas of operations, as well as across industrial marketing, information systems and accounting \& finance. We identify some existing and promising research areas, which connect to some of the other operations themes in the present paper. 

Servitisation is essentially an innovation of the manufacturer’s value proposition and business model. This presents challenges for manufacturers who are accustomed to product innovation, but not to innovation in service offerings and processes, and the organisational changes that accompany the shift to service. Early research asked whether manufacturing and service provision should be separated \citep{Oliva2003-bj_MSAS}; more recently, researchers have attempted to understand whether manufacturers should simultaneously innovate product and business model, or concentrate on one \citep{Visnjic2016-xw_MSAS}. Innovation in the product itself is transformed by servitisation, as the product must be designed for service, and increasingly must incorporate digital functionality; this radically transforms the capabilities required and the financial basis on which product innovations are evaluated \citep{Solem2022-yp_MSAS} and, because the manufacturer often retains ownership, can create opportunities for more daring product innovations. Designing and resourcing service delivery is often a major challenge for manufacturers, especially when it comes to delivering service at scale \citep{Sklyar2019-hy_MSAS} Most research on the innovation associated with servitisation has been based on larger firms, but recent research \citep{Kolagar2022-kf_MSAS} demonstrates that SMEs can be especially adept at shifting to service, due to their flexibility and close customer relationships. 

Supply chains and networks are becoming a more important area of concern in servitisation research and practice. Much early research concentrated on the changes required within the manufacturing firm and, to a lesser extent, the customer \citep{Shen2023-il_MSAS}. Some earlier studies have addressed the role of third parties such as dealers \citep{Karatzas2017-vg_MSAS} and suppliers \citep{Johnson2008-sz_MSAS} but the wider network is now becoming increasingly important as more divergent technologies, especially digital technologies, play a greater part \citep{Marcon2022-ee_MSAS}, and the importance of actors such as asset finance providers becomes clearer. Important current research issues in this area include the alignment in the network and the balancing of value capture.

Sustainability was one of the early motivations for a shift to service \citep{Mont2002-wf_MSAS,Stahel1981-mr_MSAS} but, as industrial adoption took hold, sustainability concerns became less prominent. Recently, however, as the climate emergency makes sustainability an almost universal mainstream business objective, sustainability has become a central rationale for servitisation \citep{Rabetino2024-ll_MSAS}. Manufacturers who retain ownership of products under outcome-based contracts can be incentivised, and have the capabilities, to monitor the product in use and provide advisory services for the efficient and sustainable consumption of the service. Research is concerned with evidencing the environmental benefits and exploring how they can be further enhanced. The notion of `Scope 3' of the Greenhouse Gas Protocol provides additional impetus and structure to such initiatives. Shifts in ownership and responsibility within servitisation can also promote the adoption of circular economy approaches \citep{Spring2017-nm_MSAS}.

\subsubsection*{Conclusion}
Although the areas discussed have been the most active in servitisation research, as a novel empirical setting, servitisation builds on many core OM topic areas, such as: managing capacity in supporting a distributed fleet of capital assets; quality management and assurance; job design for service-delivery staff. Yet, in a lot of servitisation research these core OM topics are not yet sufficiently integrated despite their practical importance and theoretical relevance for understanding success and failure of servitisation. Thus, as well as developing further the distinctive themes examined earlier, researchers can use servitisation as an empirical context for research into classic operations themes using a variety of methods. \cite{Baines2024-qg_MSAS} provides an accessible overview of the field, and recent review papers present the state-of-the art in key areas of servitisation research related to sustainability \citep{Rabetino2024-ll_MSAS} and digitalisation \citep{Shen2023-il_MSAS}. 

\clearpage

\section{Supply Chain Management}
\label{sec:Supply_Chain_Management}

\subsection[Supply chain strategy (Steven J. Day \& Janet Godsell)]{Supply chain strategy\protect\footnote{This subsection was written by Steven J. Day and Janet Godsell. The author names are listed alphabetically as both contributed equally in the writing of this subsection.}}
\label{sec:Supply_chain_strategy}
The origins and purpose of supply chain strategy are best understood through early research on strategy formulation and implementation, as detailed in the dominant process model linking corporate and functional strategies. \cite{Hill1985-zy_JGSD} and \cite{Leong1990-db_JGSD} observed that a firm starts with an overarching corporate strategy and objectives. Strategic business units then develop marketing strategies to meet these objectives. Functions like new product development assess product competitiveness, while manufacturing sets up processes for production. During implementation, manufacturing, logistics, procurement, and other functions build the necessary physical and digital infrastructure. Information continuously flows between corporate, business unit, and functional leadership, making strategy formulation and implementation a dynamic, ongoing process aimed at aligning objectives and actions across different levels.

Supply chain strategy emerged from the challenge of aligning the diverse strategies of functions, particularly as manufacturing, logistics, and procurement. Adopting an end-to-end perspective that encompasses the firm's entire network -- including suppliers and customers -- enhances alignment between corporate strategy and integrated functions and prevents functions working at cross-purposes. This approach enables comprehensive strategy formulation and implementation through the diverse capabilities and connections of the different functions closely involved with demand satisfaction. Although academic definitions of supply chains and supply chain management are not universally agreed upon \citep{Burgess2006-ku_JGSD}, the Supply Chain Operations Reference (SCOR)  model effectively illustrates the scope of supply chain strategy. The SCOR model outlines key activities -- plan, source, make, deliver, return, and enable -- that address customer demand. Consequently, supply chain strategy focuses on satisfying this demand by coordinating functions such as manufacturing, logistics, and procurement to execute SCOR model activities effectively.

\begin{figure}
\centering
{\includegraphics[width=0.95\linewidth]{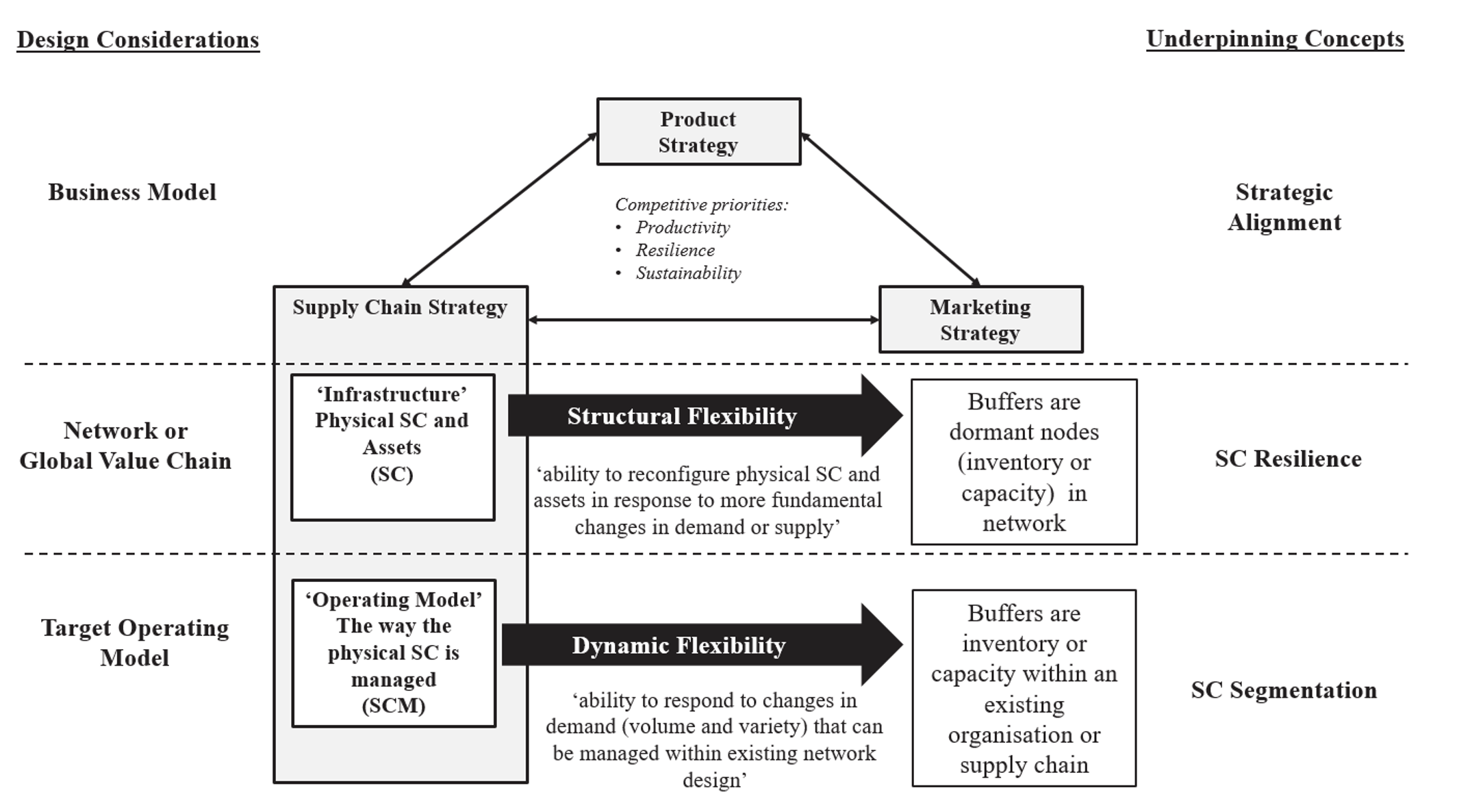}}
\caption{Supply chain strategy diagram.}
\label{fig:scstrategy}
\end{figure}

Taking a holistic view, supply chain management is one of the three core business processes identified by \cite{Srivastava1999-ow_JGSD}, alongside new product development and customer relationship management. For effective supply chain strategy, alignment with product and marketing strategies is essential, ensuring coherence among all three \citep{Godsell2010-hz_JGSD}, as shown in the supply chain strategy diagram (see figure \ref{fig:scstrategy}). Traditionally, supply chains have been tasked with supporting corporate growth objectives by cutting costs through productivity improvements. However, this focus can undermine other goals; overly lean systems narrow flows of products and reduced inventory and capacity buffers may struggle with resilience and absorbing variability in supply and demand. Additionally, cost-cutting measures can lead to short-term decision-making, such as reliance on carbon-intensive energy whose externality costs are increasingly redirected at supply chains by regulation. An example is the European construction industries' reliance on imports from countries reliant on coal for energy, the externality costs of which are redirected to importers through the European Union's carbon border adjustment mechanism (CBAM). Supply chain strategy thus must go beyond productivity to also balance resilience and sustainability while aligning with overall corporate, product, and marketing objectives to satisfy diverse customer expectations \citep{Ketchen2007-lv_JGSD}. One of the best examples of a firm consistenly aligning product, marketing, and supply chain strategy with this goal would be Apple and particularly its iPhone product line.

Once such alignment has been achieved, the supply chain strategy is configured through two primary components: a) the infrastructure and b) the operating model. While the infrastructure is developed through network design (see \S\ref{sec:Network_design} and \S\ref{sec:Outsourcing}) and reflects the physical supply chain and its assets, the operating model is the outcome of the sum of various supply chain management design decisions. When correctly configured and aligned with the overall supply chain strategy, supply chains can achieve more balanced outcomes and avoid productivity initiatives coming at the expense of resilience and sustainability.

The concepts of dynamic and structural flexibility \citep[][see also figure \ref{fig:scstrategy}]{Christopher2011-wv_MMMK} are key to understanding how such a balance can be struck. Firms are able to build a certain degree of flexibility into their supply chains via specifically configured operating models that, through segmentation, account for the particular features of customers, products, and the underlying supply chains. Again the origins of supply chain strategy in manufacturing strategy shine through here as such operating models and their underlying segmentation concepts are informed by the product-process matrix \citep{Hayes1979-na_DPvD} and the idea that a supply chain can be optimised around a product \citep{Fisher1997-lt_UR}. Coupled with a nuanced understanding of customer demand acquired through demand profiling \citep{Godsell2011-ay_JGSD}, firms can then consider what proportion of demand can be met through a lean, cost-efficient solution and how much short-term agility is necessary to cope with ongoing demand variations \citep[][see also \S\ref{sec:Lean_and_agile}]{Naylor1999-cn_JGSD}. Zara serves as an illustrative example here; while basic products with a predictable demand profile (e.g., white t-shirts) are provided through a lean supply chain, fashion products with significant demand uncertainty are supplied through an agile solution closer to the end-consumer. Firms like Zara thus continuously run parallel supply chain configurations depending on products' demand profiles. Other firms have also found ways to frequently assemble and disassemble supply chains for planned but non-continuous surge demand periods. For example, grocery retailers such as Aldi and Lidl temporarily offer products and create large supply chains for the occasion that are scheduled to be scaled down or wound down entirely when new products are cycled in. In practice, lean is the preferred SC strategy, when demand is stable and predictable. However, when demand is more unpredictable, then a more agile strategy is preferred. Both strategies require buffers (inventory, manufacturing capacity or time) to be held to buffer against the uncertainty, with larger buffers required as unpredictability increases. Dynamic flexibility is a reflection of the agility of the supply chain and particularly its ability to respond rapidly to variations in volume and mix is thus an important but often neglected strategic objective of the operating model.

There are also limits to what can done by tweaking the operating model; as \citeauthor{Christopher2011-wv_MMMK} (\citeyear{Christopher2011-wv_MMMK}, p. 64) state, dynamic flexibility achieved through clever management only allows firms to respond and ``cope with certain shifts in demand and technology, but only within the set structure of their existing supply chain design [\dots] to meet the challenges of a turbulent business environment, we need structural flexibility that builds flexible options into the design of supply chains''. Such structural flexibility, achieved through network design, can be created through the strategic configuration of infrastructure. Strategic inventory and capacity buffers, dual sourcing, asset sharing, postponement, rapid manufacture, outsourcing, and other approaches can build a supply chain’s ability to adapt to fundamental change. Such fundamental change may be caused by various human-made and natural factors; its primary effect is the shifting of the `centre of gravity' between supply and demand locations. Greater consolidation in the infrastructure with fewer, more centralised assets likely far removed from either suppliers and/or customers may improve efficiency but also lowers structural flexibility, which would benefit from a more distributed infrastructure that can adjust to a changing ‘centre of gravity’. Without structural flexibility built into the supply chain through purposeful network design, change cannot be adjusted to as the existing infrastructure lacks the levers that would allow for a reconfiguration of flows.

While supply chain strategy aims for an end-to-end approach, practical execution is challenging due to limited visibility and growing complexity in coordinating activities across numerous firms. Consequently, many firms adopt a more focused approach. \citeauthor{Frohlich2001-jr_JGSD}'s (\citeyear{Frohlich2001-jr_JGSD}) `arcs of integration' concept is valuable here; it reflects how firms integrate activities with suppliers and customers. Firms implementing a new supply chain strategy might begin with a narrow, inward-facing arc with minimal integration or a peripheral arc that integrates selectively. Over time, they may expand integration by enhancing access to planning systems, sharing inventory information, and using common logistics resources. \cite{Frohlich2001-jr_JGSD} and subsequent research indicate that while many firms focus primarily on either supplier or customer integration, a more comprehensive outward-facing arc, i.e., integrating with both, tends to yield the highest performance improvements. However, as integration extends to more tiers of suppliers and customers, firms must evaluate when the costs of further integration exceed the anticipated benefits.

Addressing evolving challenges, there is an urgent need to balance social, environmental, and economic outcomes (\S\ref{sec:Environmental_sustainability} and \S\ref{sec:Ethical_sustainability}). Long-term, we must establish a solid social foundation that meets basic human needs while maintaining an ecological ceiling to mitigate the adverse effects of economic activity, as outlined in \citeauthor{Raworth2017-cy_JGSD}'s (\citeyear{Raworth2017-cy_JGSD}) doughnut economics. To achieve this, promising macro-level ideas are being translated into actionable strategies at the meso-level of supply chains. Concepts such as the circular economy (\S\ref{sec:Circular_economy}) and its meso-level components of closed-loop supply chain management and cradle-to-cradle design offer valuable approaches. Additionally, many supply chains still have untapped opportunities for improvement through simpler, underutilised practices. Crucially, firms should better integrate consumers into the supply chain, as they influence demand patterns and contribute to preventable social, environmental, and economic costs. Reducing inventory and capacity buffers through a closer connection with consumers can thus help decrease the overall supply chain footprint. Digitalisation and emerging technologies (see\S\ref{sec:Digitalisation} and \S\ref{sec:Emerging_technologies}) will be key in facilitating this transition toward more balanced supply chains and will play a greater role at the formulation stage of supply chain strategy.

Beyond the literature referred to above, we direct the reader to several reviews that offer more insight into particular topics within the supply chain strategy literature, such as flexibility and resilience \citep{Stevenson2007-yf_JGSD,Tukamuhabwa2015-hj_WPLD} and integration \citep{Ataseven2017-ib_JGSD}. Formal theories relevant to supply chain strategy are scarce, but \cite{Schmenner2001_SMD} develops a seminal perspective with the Theory of Swift, Even Flow. Given the cross-functional scope of supply chain strategy, contributions from other literatures, particularly strategic management \citep{Hitt2011-bx_JGSD,Ketchen2020-tn_JGSD} and marketing \citep[e.g.,][]{Juttner2007-wo_JGSD,Kozlenkova2015-px_JGSD}, are also relevant. Textbooks by \cite{Chopra2021-lr_JGSD} and \cite{Cohen2013-ni_JGSD} provide structured, accessible, and illustrated insight. Practitioner books, such as \cite{Sarkar2017-sd_JGSD} and \cite{Handfield2022-gb_JGSD}, offer perspectives on emerging developments. 

\subsection[Digitalisation (Christoph G. Schmidt)]{Digitalisation\protect\footnote{This subsection was written by Christoph G. Schmidt.}}
\label{sec:Digitalisation}
Digitalisation has fundamentally reshaped how individuals interact, how societies function, and how organisations operate. In a business context, digitalisation is understood as using digital technologies to enhance and enable value creation.  Digital technologies are electronic tools and systems that generate, store, process, or exchange data. Traditional examples of digital technologies include computers, smartphones, and the Internet. Emerging digital technologies (\S\ref{sec:Emerging_technologies}) refer to, for instance, the Internet of Things (IoT), Artificial Intelligence (AI), blockchain, drones, robotics, and cloud computing \citep[e.g.,][]{Choi2022-qh_MALMA, Maghazei2022_CGS, Kloeckner2022_CGS}. 

Firms' digitalisation efforts increasingly include their supply chain partners. Modern supply networks (\S\ref{sec:Network_design}) are characterised by complex global structures involving diverse actors worldwide. As emerging digital technologies (\S\ref{sec:Emerging_technologies}) enable seamless coordination and real-time information sharing, they exhibit a ``natural fit'' with the challenges associated with complex supply networks. The firm's digitalisation efforts can enhance communication, collaboration, and visibility across these interconnected relationships \citep[e.g.,][]{Choi2022-qh_MALMA, OlsenTomlin2020_CGS, hastig2020blockchain_CGS}. 

Digitalisation impacts firms and supply chains on three levels \citep[see][]{baiyere2023digital_CGS}. The data level (i.e., digitisation) involves converting analog data into digital formats for easier management, access, and use. At the process level, digitalisation optimises processes within and across firm boundaries, enhancing efficiency and effectiveness. At the business level, new technologies can also potentially shift business models and practices at the business level (i.e., digital transformation). 
Overall, digitalisation transforms how firms and their supply chains operate, collaborate, and compete in a global digital economy. 

\subsubsection*{The role of data}
The role of data in supply chain management has radically changed over time. Digitisation vastly increases the amount of data generated by the firm and its supply chain partners. This data includes a wide variety of types, such as real-time sensor data, transactional data from supply chain exchanges, customer-related data, as well as information on production (\S\ref{sec:Production_and_control}), inventory levels (\S\ref{sec:Inventory_management}), and maintenance requirements (\S\ref{sec:Maintenance}). Additionally, data from external sources like market trends, weather conditions, and geopolitical events can also be integrated to enhance decision-making across the supply chain \citep[e.g.,][]{colombari2023interplay_CGS, xu2023data_CGS}. 

Research explores how firms can use the new data and the associated financial and operational performance implications. Advances in (big) data analytics, utilising machine learning algorithms and artificial intelligence approaches, allow firms to leverage these data \citep[e.g.,][]{Sanders2018_CGS, TsanMingChoi2018_CGS}. 
Data-driven operations and supply chain management offers numerous benefits, such as enhanced visibility, real-time decision-making, predictive analytics, and improved customer satisfaction \citep[e.g.,][]{LiLi2023_CGS, OlsenTomlin2020_CGS}. 
Current findings generally suggest that utilising big data analytics can translate into increased firm performance \citep[e.g.,][]{LiLi2023_CGS, Cohen2018_CGS}.

Various research streams examine how the availability of new data types and higher volumes influences decision-making in operations and supply chain management. Specifically, related studies cover pricing, inventory, scheduling, or maintenance decisions \citep[e.g.,][]{LiLi2023_CGS, xu2023data_CGS, van2022data_CGS}. In addition, research also focuses specifically on the opportunities and challenges of using specific types and data sources. For example, weather forecasts are used to optimise logistics and inventory, and social media data provides valuable insights into consumer sentiment and trends \citep[e.g.,][]{Schmidt2023ant_CGS, roth2023sellin_CGS, agnihotri2022utilizing_CGS}. \citet{huang2020social_CGS} provides an overview of the social media application areas. Overall, data has become a strategic asset in supply chain management, essential for decision-making and optimisation.

One of the major insights in research and practice is that the increasing reliance on data in operations and supply chain management presents clear downsides. The threats to a firm's digital infrastructure and data integrity are operational risks, frequently referred to as \textit{cyber risk} \citep{KumarMookerjee2018_CGS}. On the one hand, globally integrated IT infrastructures exhibit new vulnerabilities, and cybersecurity and information systems security threats pose significant risks. Materialised risks can result in data breaches or supply chain disruptions with substantial negative performance implications for firms and their supply chain partners \citep[e.g.,][]{Foerderer2022ms_CGS, durowoju2021supply_CGS}. Investing in operational and supply chain resilience related to information security, for example, through secure data sharing or advanced encryption, can partially protect against these threats \citep[e.g.,][]{Ampel2024_CGS}. \citet{Kumar2022-ib_WPLD} elaborate on the growing importance of cybersecurity in operations and supply chain management. 

On the other hand, there are increasing concerns regarding the handling and exposure of data, highlighting the importance of data privacy and confidentiality \citep[e.g.,][]{Fainmesser2022ms_CGS, Massimo2018_CGS}. Customers are more sensitive to data breaches, understood as the unintended exposure of confidential information, and these incidents have received increasing attention from customers and policymakers. \cite{Massimo2018_CGS} outline data privacy and confidentiality issues and discuss their importance for operations and supply chain management research and practice. 

Researchers continue to use established methods like case studies, surveys, regression analyses, event studies, and various analytical approaches to explore how firms utilise new digital technologies. However, new data sources also allow for innovative research approaches. For instance, textual data is now analysed using machine learning techniques to extract topics, identify similarities, or measure sentiment, providing fresh insights into supply chain dynamics and decision-making processes \citep[e.g.,][]{Frankel2022ms_CGS, BaoDatta2014_CGS}.

\subsubsection*{Digital technologies in supply chain management}
Supply chain management is a promising and prominent area for digitalisation due to its complex and interconnected nature. As global supply chains face an increasing need for efficiency, transparency, sustainability, and resilience, digitisation becomes a key driver for innovation and competitive advantage \citep[][]{Kumar2022-ib_WPLD, Choi2022-qh_MALMA}.

Digitalisation efforts increase supply chain \textit{efficiency} by automating or supporting production, logistics, or retailing processes across the supply chain \citep[e.g.,][]{pournader2020blockchain_CGS}. Enhanced data sharing, for example, fosters better coordination among partners, optimises inventory management, and reduces lead times, resulting in more streamlined supply chains \citep[e.g.,][]{Choi2022-qh_MALMA}. Digital technologies also increase supply chain transparency, which is understood as the ability to access and share accurate information across the supply chain, improving collaboration and coordination between supply chain partners \citep[e.g.,][]{kalaiarasan2022abcde_CGS, dolgui20225g_CGS}. A major development is the global importance of transparent supply chains, as they are a prerequisite to clear customer and investor communication, for example, regarding product sourcing and quality and increasing regulatory requirements \citep[e.g.,][]{hastig2020blockchain_CGS}. Digital technologies, such as IoT, RFID, and blockchain, have been shown to improve visibility and transparency across supply chains \citep[e.g.,][]{kalaiarasan2022abcde_CGS, hastig2020blockchain_CGS}.

Digital technologies, such as AI and predictive analytics, can strengthen \textit{risk management} and enhance \textit{resilience} in supply chains \citep[e.g.,][]{huang2023impact_CGS, pettit2019evolution_CGS}. New tools allow firms to identify potential risks early, assess their impact, and implement proactive mitigation strategies while also increasing firm operational and financial performance \citep[e.g.,][]{zhao2023impact_CGS, BaoDatta2014_CGS}. By leveraging new technologies, firms can also better withstand and adapt to both global and local disruptions, ensuring stability and continuity \citep[e.g.,][]{Ivanov2024-my_DIAD}. Digitalisation can also significantly enhance \textit{sustainability} efforts in supply chains by, for example, optimising transportation routes, reducing energy consumption, and minimising emissions \citep[e.g.,][]{tsolakis2022towards_CGS, pournader2020blockchain_CGS}. It also supports better resource use and waste management through continuous production monitoring and improved supply tracking \citep[e.g.,][]{Schilling2024-xs_SSSAQ, OlsenTomlin2020_CGS}.  \cite{Schilling2024-xs_SSSAQ} provide an overview of the challenges and opportunities of digitalisation efforts for sustainable supply chain management. 

Firms are increasingly pushing \textit{digital transformation} initiatives involving their supply chain partners. Supply chains move beyond traditional systems, and firms drive the development of new business models, enabled by integrating advanced digital technologies such as AI, IoT, blockchain, and big data analytics \citep[e.g.,][]{gokalp2022digital_CGS, Choi2022-qh_MALMA}. New business models relate to platform-based ecosystems, and on-demand manufacturing \citep[e.g.,][]{lerch2024manufacturers_CGS, Stark2023_CGS}, all of which offer greater flexibility and scalability. Digital transformation creates significant changes in organisational structures, roles, and capabilities and requires that firms build digital competencies across their workforce to stay competitive in a global market \citep[][]{Schilling2024-xs_SSSAQ, gokalp2022digital_CGS}. In conclusion, digitalisation is transforming operations and supply chain management in various ways, bringing both opportunities and challenges. 

\subsection[Network design (Piera Centobelli)]{Network design\protect\footnote{This subsection was written by Piera Centobelli.}}
\label{sec:Network_design}
A supply chain network is a complex system interconnecting multiple individual supply chains. This includes both supplier and consumer supply chains, as well as a company's internal supply chain. The design of this network determines how resources, information, and goods flow from raw material suppliers to end consumers. In today's globalised economy, supply chain networks are essential for the efficient production and distribution of goods and services. The design of a supply chain network involves strategic planning to ensure that the entire system operates efficiently, minimises costs, and mitigates risks.

In the current landscape, several pressing issues impact supply chain network design. Some challenges are supply chain disruptions, which have been exacerbated by events such as the COVID-19 pandemic and ongoing geopolitical tensions. These disruptions have underscored the vulnerability of global supply chains, leading to delays, shortages, and increased costs. As a result, many companies are reassessing their supply chain networks to enhance resilience by diversifying their supplier base, increasing local sourcing, and establishing strategic reserves. The design of resilient supply chains should support disruptions, whether from natural disasters, economic shifts, or pandemics. For example, one study proposes a bio-inspired framework for designing resilient and sustainable resource and infrastructure networks, drawing on ecological principles to balance redundancy and efficiency \citep{Chatterjee2020-xl_PC}. Another research explores the use of probabilistic-stochastic programming to create a robust humanitarian logistics network, emphasising the importance of scenario-based planning in the face of unpredictable disruptions like earthquakes or pandemics \citep{Nezhadroshan2021-lm_PC}. Similarly, a study on earthquake relief logistics during COVID-19 highlights the challenges of simultaneous disasters, proposing a mathematical model that considers facility failures and multi-trip vehicle routing to enhance reliability and timely delivery \citep{Abbasi2023-sx_PC}.

The stream of research focused on risk management highlights that strategies such as diversifying supply sources, adopting multi-sourcing approaches, and maintaining safety stocks are effective strategies in mitigating risks \citep{Esmizadeh2021-xh_PC,Schatter2024-gr_PC,Vieira2023-fs_PC}. However, these approaches often lead to increased operational costs, necessitating a balance between risk mitigation and cost efficiency based on a company's specific risk acceptance and business objectives.

Another significant issue is the growing emphasis on sustainability and environmental concerns. There is an increasing pressure on companies to minimise their carbon footprints, which has driven a shift towards greener logistics, sustainable sourcing practices, and the adoption of circular supply chains. While integrating sustainability into supply chain design can introduce additional complexity and costs, it is crucial for long-term viability and compliance with evolving environmental regulations. Some studies have investigated the multi-period, multi-echelon, multi-product, and multi-modal sustainable supply chain network design problem developing a Mixed-Integer Linear Programming model that explicitly considers the environmental footprint and social responsibilities \citep{Guo2021-aw_PC}. Several studies address the integration of environmental, economic, and social objectives into supply chain design. \cite{Jiang2018-ov_PC} introduced a multi-objective mixed-integer linear programming model to design a sustainable supply chain network that minimises costs while reducing carbon emissions and improving social outcomes, such as local employment. Another research discusses the use of graph theory in social network analysis to explore the relationships within local food systems, revealing how diversified marketing practices contribute to system resilience and sustainability \citep{Brinkley2018-oz_PC}. Additionally, a study of the bioenergy supply chain emphasises the importance of designing networks that balance economic viability with environmental impacts, particularly in the context of transitioning to renewable energy sources \citep{Rafique2021-ce_PC}.

The literature underscores the transformative impact of digital technologies on supply chains. The digital transformation and integration of advanced technologies such as artificial intelligence (AI), the Internet of Things (IoT), and blockchain are reshaping supply chain management \citep{Ghomi-Avili2023-hj_PC}. \cite{Modares2023-aw_PC} assessed the use of IoT and blockchain to optimise supply chain management by improving information exchange between suppliers and manufacturers, reducing waste and returns. A multi-criteria decision-making model, coupled with genetic algorithms and particle swarm optimisation, is used to select optimal suppliers, with the approach validated through a food supply chain case study. These technologies enhance visibility, improve decision-making processes, and enable real-time tracking of goods and information. However, incorporating these technologies into existing supply chain networks presents challenges, including a significant investment and the complexity associated with the system integration. Also, other challenges emerge, such as data security concerns, the skills gap in the workforce that must be addressed to fully realise the benefits of digitalised supply chains. \cite{Schulz2019-mp_PC} investigated the role of blockchain technology in enhancing data security within Industry 4.0's interconnected supply chains. By integrating blockchain with the industrial Internet of Things, the study addresses critical concerns such as data integrity, access control, and transaction security, proposing a decentralised information processing system to support the secure and efficient management of self-organised production networks. On the other hand, \cite{Ronchini2023-vk_PC} examined the impact of additive manufacturing adoption on upstream supply chain design, revealing that decisions to make, buy, or vertically integrate depend on factors like additive manufacturing (AM) experience, application type, and the need for production control. High initial investments and a lack of skills emerged as significant barriers, influencing supply chain restructuring and the make-or-buy decision.

Additionally, the debate between globalisation and rationalisation is influencing supply chain network design. While globalisation has been a dominant trend, recent shifts towards rationalisation are driven by factors such as trade wars, rising tariffs, and the need for greater supply chain resilience. Companies are increasingly focusing on regional hubs rather than extensive global networks, which can reduce risks associated with long-distance logistics but may also limit economies of scale. One paper proposed and examined the impact of the Regional Comprehensive Economic Partnership agreement on supply chain resilience, using a fuzzy optimisation model to address demand uncertainty and evaluate the effectiveness of resilience strategies like multiple sourcing and capacity redundancy \citep{Cheng2024-dn_PC}. Another study analyses the challenges faced by multinational companies in managing complex production networks, proposing a stochastic programming model to optimise the allocation of production stages to suppliers under uncertain demand conditions \citep{Zhen2016-ks_PC}.

Although the literature identifies certain common themes and shared issues, there are undoubtedly sector-specific traits that necessitate special attention. For instance, the challenge of managing perishable goods presents unique logistical and supply chain difficulties that are not as prevalent in other industries \citep{Biuki2020-kd_PC,Gong2007-on_PC}. For example, one study on the perishable goods supply chain integrates considerations of product spoilage and stochastic demand, proposing a network model that aims to optimise availability and net profit for all members of the supply chain \citep{Dagne2020-wy_PC}. Another paper explores the design of closed-loop supply chains for durable products, highlighting the complexities of managing return flows and the need for compliance with legislative recovery targets \citep{Jeihoonian2016-ln_PC}.

Similarly, the rapid growth of e-commerce has introduced distinct operational complexities, such as the need for efficient last-mile delivery and inventory management in a digital marketplace \citep{Janjevic2020-tn_PC}. In the context of e-commerce, \cite{Calzavara2023-um_PC} investigated different strategies for e-grocery logistics, comparing the costs and performances of various network designs to determine the most efficient approach in response to fluctuating online demand. With reference to the e-commerce sector, \cite{Liu2022-pi} further emphasised that an optimised network configuration may, in certain scenarios, influence the demand pattern itself, potentially leading to a situation where the initial optimisation no longer yields the best performance, thus highlighting the dynamic interplay between logistics design and consumer behaviour and demand. Another paper \citep{Agac2023-ya_PC} introduced a novel mathematical model using Mixed-Integer Nonlinear Programming for optimising blood supply chain networks, focusing on location, allocation, and routing. A variant of the Genetic Algorithm is applied, and the model's effectiveness is demonstrated through a case study in the Eastern Black Sea Region of the Turkish Red Crescent.

Several case studies illustrate these concepts in practice. Apple, for example, is known for its highly efficient global supply chain, which relies on a vast network of suppliers while maintaining strict control over key components. Zara, the fashion retailer, exemplifies agility and speed in supply chain management. Their approach involves nearshoring and vertical integration, enabling rapid response to changing fashion trends and quick turnaround from design to store shelves. Toyota's supply chain, particularly its Just-In-Time (JIT) methodology, has long been a model of efficiency. However, some disrupting events revealed the vulnerabilities of this approach, prompting the company to re-evaluate its strategy, demonstrating the necessity of balancing efficiency and resilience.

In conclusion, supply chain network design is a critical aspect of supply chain management that must navigate a range of complexities and challenges. Current issues such as disruptions, sustainability, digital transformation, and the tension between globalisation and rationalisation are driving significant changes in how supply chains are structured and managed. Insights from recent literature and practical case studies highlight the importance of adaptability, technological integration, and a balanced approach to designing supply chain networks. As the global business environment continues to evolve, supply chain managers must remain proactive in adapting their supply networks to meet both current challenges and future demands.

\subsection[Outsourcing (Byung-Gak Son)]{Outsourcing\protect\footnote{This subsection was written by Byung-Gak Son.}}
\label{sec:Outsourcing}
Outsourcing can be broadly defined as the practice of contracting out specific operations or services to external providers instead of performing them internally \citep{Simchi-Levi2021-gx_BGS}. Outsourcing has become a critical element in the modern businesses, especially in the fields of operations and supply chain management. Firms constantly need to make a boundary decision regarding how to carry out activities \citep{Fine2002-lm_BGS}. One way of doing this is internalisation that is the firm dedicates its own resources to carry out activities in-house for example, for the need have a better control of the processes \citep{Holcomb2007-rk_BGS}. Sometimes, many firms choose to organise their activities by outsourcing, meaning they rely on third-party providers to access the capabilities and resources they need \citep{Holcomb2007-rk_BGS}.

Historically, outsourcing was primarily driven by the desire to reduce costs. A notable example is the manufacturing outsourcing boom in the electronics industry during the 1990s, where companies outsourced to cut expenses and improve operational efficiency \citep{Harland2005-oh_BGS}. However, in recent years, outsourcing has evolved into a more sophisticated strategy. Modern companies now use it to enhance their competitiveness by acquiring specialised capabilities and cutting-edge technologies, helping them stay ahead in rapidly changing markets \citep{Holcomb2007-rk_BGS,Kremic2006-bo_BGS}.

There are many reasons why firms consider outsourcing. Firms' decisions to outsource are often closely aligned with their broader operational and corporate strategies, including goals related to growth, innovation, and maintaining a competitive edge \citep{Simchi-Levi2021-gx_BGS}. Outsourcing enables companies to focus on their strategic objectives while leveraging the capabilities of external partners to improve efficiency and flexibility. 

One of the primary reasons firms outsource is to achieve cost efficiency. By outsourcing in-house activities, firms can significantly reduce their capital expenditure and operational costs, freeing up their limited and valuable resources for other priorities \citep{Harland2005-oh_BGS,Hendry1995-gk_BGS}. This is particularly advantageous when outsourcing partners possess specialised capabilities or are located in regions with lower operational costs \citep{Simchi-Levi2021-gx_BGS}. 

Another key motivation for outsourcing is the ability to focus on core competencies. It is often impractical for companies to carry out all necessary activities internally. As a result, many firms choose to outsource non-core activities, allowing them to concentrate their limited resources on activities that are more strategically important \citep{Arnold2000-wd_BGS,Harland2005-oh_BGS,Prahalad2009-nh_BGS,Simchi-Levi2021-gx_BGS}.

Furthermore, outsourcing also grants firms access to specialised knowledge and advanced technologies that may not be available internally \citep{Harland2003-pn_BGS,Venkatesan1992-pa_BGS}. In industries where technology evolves rapidly and customer preferences change frequently, outsourcing offers a valuable means to stay updated without the heavy investment or risks of internal development \citep{Venkatesan1992-pa_BGS}. As a result, outsourcing knowledge creation to external partners has become increasingly common, particularly in high-tech sectors \citep{Ahuja2000-mc_BGS,Son2024-ah_BGS}. 

Lastly, outsourcing provides companies with scalability and flexibility, allowing them to adjust their operations in response to fluctuating demand \citep{Harland2005-oh_BGS,Holcomb2007-rk_BGS}. This is especially valuable in industries where demand is difficult to predict and products may become obsolete quickly \citep{Simchi-Levi2021-gx_BGS}. The ability to scale up or down through outsourcing helps firms remain agile and responsive to market changes.

Firms often have more than one motivation for outsourcing. A prime example is Apple, which strategically retains core activities such as research and development and marketing in-house—areas critical to its brand identity and product innovation, while outsourcing production to external specialist manufacturers like Hon Hai Precision and Pegatronics \citep{Mudambi2010-hn_BGS}. This approach enables Apple to achieve flexibility and cost efficiency without sacrificing the quality or premium image of its products. By heavily relying on outsourcing for manufacturing, Apple can concentrate on its core strengths: fostering innovation, enhancing its brand, and maintaining its leadership in the market \citep{Mudambi2010-hn_BGS}.

Despite the advantages discussed above, however, outsourcing presents significant challenges and risks. Many companies find their outsourcing experiences disappointing \citep{Harland2005-oh_BGS}, and a substantial number eventually choose to bring previously outsourced processes back in-house \citep{Handley2009-yi_BGS}. This means, firms must carefully weigh these risks against the benefits to ensure their outsourcing strategies align with long-term objectives. 

One of the primary risks of outsourcing is the potential loss of control, as companies hand over responsibility for certain functions to external providers \citep{Kremic2006-bo_BGS,Simchi-Levi2021-gx_BGS}. This can lead to various issues, particularly concerning quality, which becomes especially problematic when the outsourced function is customer-facing. In the early 2000s, many firms outsourced their customer service operations to low-cost locations for cost saving. However, this often resulted in significant challenges related to service quality, leading to widespread customer dissatisfaction \citep{Tate2008-je_BGS}. As a result, some companies were forced to bring these functions back in-house to regain control and improve service standards \citep{Treanor2007-ud_BGS}.

Another critical risk is the issue of hidden costs. While firms often view outsourcing as a means of achieving cost savings, there are frequently overlooked cost that arise throughout the process \citep{Kremic2006-bo_BGS}. These hidden costs can include contracting, coordination, and ongoing monitoring, which many firms fail to fully anticipate \citep{Kremic2006-bo_BGS,Simchi-Levi2021-gx_BGS}. For instance, managing relationships with external providers often requires additional resources, and discrepancies between expectations and actual service delivery can lead to further financial burdens eroding potential savings.

Another significant risk tied to outsourcing is the potential loss of competitive knowledge. When firms outsource essential functions, especially those related to core products or services, they run the risk of loss of competitive knowledge to competitors. This can happen when outsourcing providers work with multiple clients, inadvertently sharing insights across different companies \citep{Simchi-Levi2021-gx_BGS}. Additionally, over-reliance on external partners for innovation can stifle a company’s own capacity to innovate. As firms become dependent on the knowledge and capabilities of their outsourcing partners, they may lose their ability to generate unique ideas and solutions independently \citep{Kremic2006-bo_BGS,Simchi-Levi2021-gx_BGS}. 

In recent years, there has been a notable shift in outsourcing trends across industries, driven by the need for greater resilience in supply chains. This is because recent global events such as the COVID-19 pandemic, trade wars, and semiconductor shortages have exposed vulnerabilities in these extended supply chains, prompting firms to rethink their outsourcing strategies. The automotive industry, for instance, has begun moving away from outsourcing to regain control over critical components, such as electronic vehicle (EV) batteries and semiconductors \citep{The-Economist2022-ol_BGS}. Tesla's vertically integrated model, which includes in-house production of key materials and components, has inspired traditional automakers to shift towards a more integrated approach. Companies like Ford, BMW and VW are planning to make more EV motors in their factories \citep{The-Economist2022-ol_BGS}.

For more information on outsourcing, please refer to the book by \cite{Simchi-Levi2021-gx_BGS}, Chapter 9.

\subsection[Purchasing and procurement (Finn Wynstra)]{Purchasing and procurement\protect\footnote{This subsection was written by Finn Wynstra. Parts of this subsection build on some public video lectures developed by the author; see \url{http://www.youtube.com/@finnwynstrapsmrsm5732}}}
\label{sec:Purchasing_procurement}

\subsubsection*{Definition and organisational contribution}
\begin{figure}
\centering
{\includegraphics[width=0.65\linewidth]{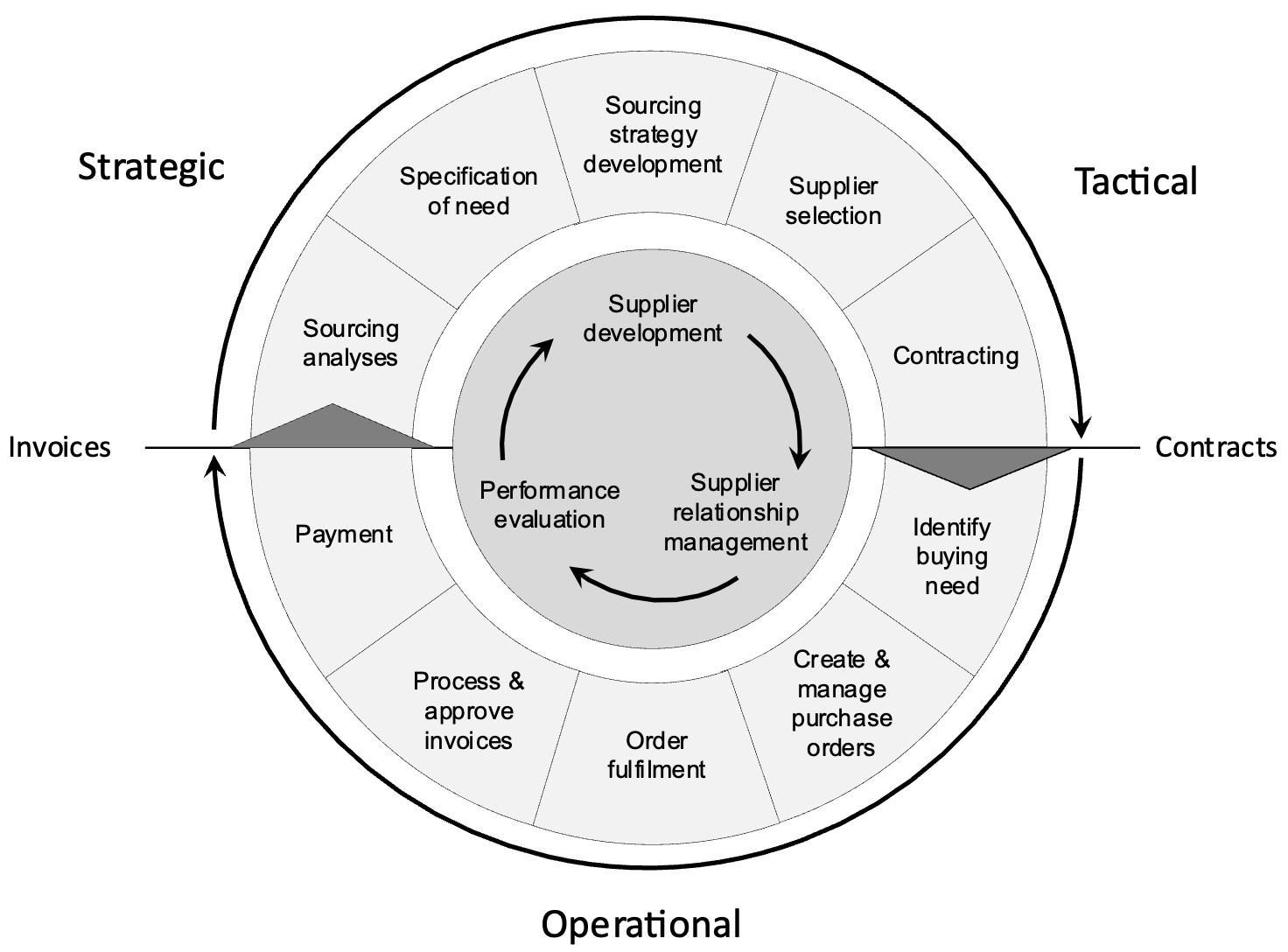}}
\caption{Purchasing and supply management process model. Source: \cite{Van-Raaij2016-zc_FW}, p. 14; published with the permission of the author.}
\label{fig:purchasing}
\end{figure}

Purchasing and supply management can contribute to organisational efficiency, risk management and innovation \citep{Axelsson1984-kx_FW}. In financial terms, this translates into cost reduction, assets optimisation and revenue growth. Growing emphasis is placed on the contribution to non-financial performance dimensions such as environmental and social sustainability, and governance in terms of the accountability, legitimacy and ethics of the organisation's activities \citep{Quarshie2016-se_FW}. The exact definition of purchasing and supply management -- or, in short: procurement -- has seen a shift in emphasis since the 1960s, when the field became subject of academic education and research \citep{Wynstra2019-am_FW}. The first textbooks focused on differences with consumer buying behaviour and on the tactical and operational purchasing process \citep[e.g.,][]{Webster1972-ls_FW}. By the 1990s, definitions started to explicate the variety of buying needs, such as the one by \cite{Van-Weele1994-nt_FW} that distinguished between direct (Bill of Material) and indirect spend. In the 2000s, definitions put the spotlight on the identification, development and maintenance of relations with suppliers \citep[e.g.,][]{Axelsson2005-qi_FW,Leenders2006-hn}. Consequently, purchasing and supply management is now seen to govern both tactical and operational, more short-term and internally oriented decisions and actions (purchasing) and strategic, more long term and externally oriented decisions and actions (supply), as depicted in figure \ref{fig:purchasing}.

The upper, strategic/tactical part of this process model is commonly referred to as ‘sourcing, and is the focus of this section. The bottom part of the model is referred to as ‘purchase to pay’. This particular process model thus captures parts of the procurement process that are often treated separately, dividing between tactical and operational processes \citep[e.g.,][p. 29]{Monczka2002-py} versus strategic processes \citep[e.g.,][]{Monczka1999-dt}.

\subsubsection*{Sourcing analysis}
Sourcing or spend analysis examines historic sourcing and business process data to improve or cut inefficient spend. It creates the foundation for spend visibility, compliance and control \citep{Pandit2008-ad_FW}. Spend analysis typically starts from four key elements in accounts payable data: invoice amount, supplier, spend category (see below), and cost centre (internal customer); see also \cite{Telgen2004-pp_FW}. After this data extraction, subsequent validation, classification, and analysis are required to derive actionable insights. 

\subsubsection*{Specification of need}
Specification of need pertains to both quantities and qualities. Especially for direct purchase items, need quantities are primarily derived from final demand forecasts (see \S\ref{sec:Forecasting}). Specification of needs in terms of qualities is typically a cross-functional process, where the procurement department gathers information and preferences from internal customers and technical specialists. This specification process is one of the typical areas where the mandate of the procurement function is being contested. In those organisations where there is no substantial mandate, procurement's role in the specification-setting process is usually highly limited, and they are relegated to processing orders based on the specifications provided by internal customers. This is very different in organisations where procurement does have a substantial mandate, for instance because of proven capabilities, achieved successes and general management support. In those situations, procurement is involved early on in the process, and has an explicit role in challenging possibly 'gold-plated' specifications. Also, current and potential suppliers can provide valuable inputs. Giving suppliers influence over the specifications of their offerings positively affects both the efficiency and effectiveness of new product development projects \citep{Suurmond2020-bo_AHER}.

\subsubsection*{Sourcing strategy development}
Every organisation buys different products and services, with varying technical, financial and market characteristics, and therefore would benefit from segmenting its spend into categories. A category strategy defines sourcing objectives, and how to select and interact with the supply base. One of the most common models to segment spend into categories is the portfolio approach of \cite{Kraljic1983-io_FW}. This model distinguishes four groups of sourcing categories (strategic, bottleneck, leverage and non-critical), based on profit impact (defined in terms of the volume purchased, impact on product quality or business growth, etc.) and supply risk (number of suppliers, storage risks and substitution possibilities, etc.). Each of the groups has a distinct recommended sourcing strategy: performance-based partnership for strategic items, secure and diversify for bottleneck items, competitive bidding for leverage items and systems contracting for non-critical or routine items \citep{van-Weele2022-pe_FW}. While this and other so-called portfolio models have been widely adopted, criticism has been raised that such models promote a static view of the product or service being bought and disregard the interdependencies between categories and between supplier relationships \citep{Dubois2002-ce_FW}.

\subsubsection*{Supplier selection}
Once a category strategy has been defined, is it possible to set up and execute an effective tendering and supplier selection process. This involves three choices: which tendering process to follow, which criteria to apply, and which selection method to use. 

\textbf{Tendering process}. To reduce the workload for suppliers and themselves, buying organisations can make a distinction between qualification and selection decisions in the tendering process. For the selection stage, four main process varieties exist. In a non-competitive purchase there are no competitive procedures or formal evaluations of supplier bids, while informal negotiations are somewhat more competitive. Closed tenders are on invitation only, whereas open tenders are open to everyone. The choice between these process varieties depends on the purchase situation. For instance, an open tender is typically chosen for large volume purchases, and where a buying firm is not fully knowledgeable about suitable suppliers \citep{de-Boer2001-yo_FW}. In the closed and open tender procedures, various requests may be sent out to suppliers: request for information (market sensing), request for proposal (when specifications are not fully set) and request for quotation (specifications are set). Together, these requests can be seen as a funnel process, although not all steps have to be executed each time \citep{Beil2010-uf_FW}. 

\textbf{Selection criteria}. One can distinguish two main levels of selection criteria: supplier (e.g, financial stability) and product/service offer (e.g., price). In the qualification stage, or when there is no actual product/service yet, buyers emphasise supplier features and only in the final selection consider product/service criteria. Regarding the latter, price, quality and delivery (reliability and flexibility) have traditionally been the most prevalent but other criteria such as sustainability (e.g., CO$_2$ emissions) and diversity (e.g., type of workers employed) are gaining importance. Another important choice is the relative importance of the different criteria, which is partly determined by the selection method. 

\textbf{Selection methods}. For supplier selection, various methods exist \citep{de-Boer2001-yo_FW}. Categorical methods are qualitative models mainly used for supplier qualification. Most common in final selection are linear weighting models, in which weights are allocated to criteria based on their importance – typically, either compensatory (no minimum score required for any of the criteria) or non-compensatory. A special group of ‘monetary’ models uses the concept of Total Cost of Ownership, by which all indirect costs before, during and after the transaction are calculated and added to the acquisition price to arrive at a total cost \citep{Ellram1993-ep_FW}. Recent ``Total Value Contribution'' approaches additionally include the revenue generating potential differences between sourcing alternatives \citep{Gray2020-io_FW,Wouters2005-iw_FW}. AI and Machine Learning offer potential benefits for supplier selection, through processing larger amounts of data, pattern recognition and dynamic scoring – creating additional opportunities for techniques such as Data Envelopment Analysis (DEA) and mathematical programming. When using (most of) these selection methods, buying organisations also need to choose a rule to assign scores to supplier bids for each criterion. Scores can be qualitative or quantitative, linear or non-linear, and absolute or relative \citep{SchotanusUnknown-is_FW}.

\textbf{Contracting}. Buying organisations can split their purchasing contract with a supplier into different levels of agreements: a purchase order (including quantities and delivery dates), which is covered by a general framework agreement (terms and conditions, assortment, pricing), which is in turn sometimes preceded by a letter of intent (strategic intentions, etc.). Obviously, just one single contract document could be applied too, as in the case of one-off buys.

With respect to contracts, there are two main groups of spend: physical products versus services and investment goods (capital expenditure). For physical products, contracts are traditionally based on fixed unit prices, possibly with volume discounts. Such contracts, however, are often subject to the risk of ‘double marginalisation’, which reduces supply chain profits because each actor makes its decisions considering only its own margin. Alternative contract types – such as quantity flexibility, buy back and revenue-sharing contracts - increase overall supply chain profits by making the supplier share some of the buyer’s demand uncertainty \citep{Chopra2021-lr_JGSD}.

For services and project-based capital expenditure, one can find a different range of contracts based on three main provisions or mechanisms: pricing, payment and activity allocation \citep{Van-Puil2014-se_FW}. Pricing has three main alternatives: fixed price, cost reimbursable and unit rates. Under fixed price pricing, the supplier performs an activity for a predetermined price. Under cost-reimbursable pricing (or: time and materials), supplier compensation is based on the actual resources used. Various intermediate forms have been developed, such as fixed price contracts with a price adjustment clause. Under unit rate pricing, buyers pay suppliers a predetermined rate per given amount of (highly standardised) output – the total volume of which is determined only afterwards. Payment mechanisms exist in two main alternatives: either the full amount is paid at once (lump-sum), or the payments are split up into instalments (milestone payments). The latter is used in case of a long project duration and when the supplier is incurring substantial expenses upfront.

Activity allocation provisions exist in five main varieties. In a construct contract, a supplier only needs to produce or construct the object, while in a design and construct contract, the supplier also designs the object. In an Engineering, Procurement and Construction (EPC) contract, the supplier additionally procures the materials and manage all subcontractors. The final two varieties allocate responsibilities to the supplier even after the delivery of the object. In a Design, Build, Finance and Maintain (DBFM) contract, a supplier arranges for the financing of the investments and the maintenance of the object and the buying firm pays a periodic service fee. In the case of a DBFMO contract, the supplier also operates the object. The choice for a specific activity allocation depends on how much risk a buying firm wants to allocate to suppliers, and how much expertise and capital a buyer has compared to suppliers. 

\subsubsection*{Outlook}
In a growing number of organisations, procurement will have the possibility if not the imperative to step up from just purchasing goods and services to becoming a function of real end-to-end value entrepreneurs. Whether the procurement function will be able to fulfill such an increasingly strategic role in a given organisation, depends on multiple factors including the capabilities and attitudes of procurement personnel and managers, its working relationships wither other functions, but surely also its effective application of digital technologies. While successful blockchain applications have become quite sparse by now, the uptake of Articial Intelligence (AI) applications in procurement is rapidly growing, particularly in areas such as negotiations with suppliers of non-critical items \citep{Van-Hoek2022-op} and other routine tasks such as tender preparations and (initial) bid evaluations \citep{Guida2023-cg}. As with the implementation of digital tools in the early 2000's, the rationale for these applications is largely that it will enable the procurement function to increase its efficiency in the operational and tactical processes, so that its critical resources can be dedicated to the more strategic processes. It is likely however, that AI will also become instrumental in these processes in the near future. To study these developments and associated challenges, procurement researchers can draw from a diverse and multidisciplinary set of both established and relatively new theories \citep{Tate2022-tz}.

\subsection[Relationship management (Paul Cousins \& Brian Squire)]{Relationship management\protect\footnote{This subsection was written by Paul Cousins and Brian Squire.}}
\label{sec:Relationship_management}
The origins of academic study on the management of buyer-supplier relationships can be briefly traced to the history of economics and specifically contracting through the theories of agency, finance, property rights and stakeholder theory \cite{Jensen1976-yb_BSPC}. Following the industrial revolution as firms moved towards `new' mass production techniques in the late 1920s and early 1930s scholars began to study the idea of `what is a firm' and think about where a firm begins and end.  

The seminal work of \citeauthor{Coase1937-ne_BSPC} which was first published in the journal \textit{Economica} in \citeyear{Coase1937-ne_BSPC} offered an economic explanation of why individuals should form businesses and the benefits and risks associated with these interactions.  This formed the basis for what became known at the `theory of the firm' and `transaction cost economics' (TCE).  Other authors such as \cite{Cyert1964-uy_BSPC} added behavioural aspects and the concept of `stakeholders' identifying interested internal and external influencers on the behaviour of the organisation and its markets. This was then formalised by \cite{Williamson1976-rk_BSPC,Williamson1985-wf_BSPC} in a series of seminal contributions in the 1970s and 80s which formalised the core tenets of TCE as we know them today. Subsequently, scholars of corporate strategy such as \cite{Dyer1996-zc_BSPC}, \cite{Barney1991-zi_BSPC} and moved away from analysis of the risk associated with individual transactions and towards the idea that competitive advantage and value creation can be obtained through inter-firm relationships processes and networks \citep{Porter2011-hz_BSPC}.  

This work was synthesised by the Industrial Marketing and Purchasing (IMP) Group in the early 1980s and 1990s under the leadership of \cite{Hakansson1982-rh_BSPC} to develop an empirical analysis of firm networks focussing on the relationship interdependencies and typologies of interactions ranging from short-term episodic to long-term ‘deep’ and embedded interactions \citep{Ellram2019-cv_BSPC}.  Collaborative business relationships are viewed as value adding process within the network of organisations or actors.  Building on this foundational work we suggest a working definition of buyer-supplier relationships as follows: 

\say{[Buyer-Supplier Relationships] ... refer to the interactions and collaborations between (and that are supported within) organisations that purchase goods or services and the various actors in the network that supply them. It involves managing these relationships effectively and efficiently [over time] to ensure a constant supply of resources that delivers ‘value’ for all actors involved in the transaction(s).}

\subsubsection*{Background}
The strategic importance of buyer-supplier relationships has been increasingly studied by scholars since the late 1970s and 1980s with the development of \citeauthor{Kraljic1983-io_FW}'s (\citeyear{Kraljic1983-io_FW}) seminal portfolio analysis tool, positioning spending based on its relative market availability (essentially a measure of risk exposure) and cost or value to suggest management strategies based on a four-quadrant model.  Moreover, development in buyer-supplier relationship thinking closely correlates to changes in the economic business environment and particularly in the paradigm changes in manufacturing which can be described as moving from craft, mass, lean and agile \citep{Lamming1993-dc_BSPC} and laterally towards mass-customisation.  Combining the portfolio analysis of Kraljic and his analysis of sourcing risk with the literature on contracts, firm boundaries and competitive advantage, led scholars to observe and develop a variety of buyer-supplier positioning and relationship measure concepts, tools and techniques \citep{Bensaou1999-tp,Cousins2003-hi_BSPC}. The practical implications of these portfolios were recognised by consulting firms and industry alike who used them to optimise spend, analyse risk, and manage their supplier relationships to better effect \citep{Cousins2003-hi_BSPC}.

During this time the business environment moved from local/regional sourcing to international and then global sourcing. The success of Japan following World War II as a major manufacturer employing innovative and novel manufacturing techniques began to change the way Western manufacturers built, designed and sourced their products -- moving from one-stop manufacturers to systems integrators, particularly true of the automotive and aerospace industries \citep{Hines1999-nb_BSPC}. Techniques such as Just-In-Time (JIT) and Total Quality Management (TQM) involved working and coordinating closely with suppliers \citep{Flynn1995-xb}. Practically, this meant manufacturers across many industries had to think much more strategically about how they would work with suppliers to compete. They restructured their supply bases to create tiered levels of suppliers, creating fewer but higher value suppliers, and started to work with their suppliers on new product development projects.

The next trend coincided with the rise of China (and since other parts of East Asia, India and currently Africa) as a low-cost manufacturing environment which encouraged large manufacturing organisations to outsource manufacturing and source products globally. This shift enabled low-cost production but also came with significant risks. More recently, organisations have slowed, or in some cases even reversed, these decisions as they seek to reshore production and diversify the locations of their supply bases in an effort to reduce a broad range of risks, including geographic concentration, geo-political challenges, and sustainability concerns. 

\subsubsection*{Theory lenses (firm, transaction, social interactions)}
Over the last fifty or so years there has been a substantial discourse on all aspects of buyer-supplier relationships creating a rich and varied body of knowledge.  It reasons that most problems and questions focused on buyer-supplier relationships are multi-faceted. For scholars in this area, before embarking on research it is important to consider the unit of analysis and then select an appropriate theoretical lens(es). Essentially the research in this area can be classified into three broad but inter-related theoretical areas, economic, strategic and behavioural. Firstly, the economic lens uses the theory of \cite{Coase1937-ne_BSPC} and specifically \cite{Williamson1985-wf_BSPC} which is Transaction Cost Economics (TCE). The unit of analysis here is the transaction at product or service level and focuses primarily on the risks of manufacturing within the boundaries of the firm or outside of it -- known as market vs hierarchy. The risks are known as transaction costs and contain a variety of characteristics that can potentially be modelled. That said, transaction costs are notoriously difficult to calculate.  The second major theoretical lens developed by \cite{Prahalad2009-nh_BGS} and extended by \cite{Barney1991-zi_BSPC} is the Resource Based View of the Firm (RBV) -- the focus here is on competitive advantage at the level of the firm, creating and managing value added relationships becomes a strategic competitive advantage to the focal firm \citep{Barney2012-yh}. The seminal work of \cite{Dyer1998-ei_BSPC} in this area builds on the previous ideas of the IMP of networks and dyads and theorises the concept of the `relational' view. This paper has been extended \citep{Dyer2018-sr_BSPC} to consider the value creation and capture benefits of building alliances and collaborations, viewing this as a dynamic capability. The third and final theoretical lens is `social capital'. Social capital can broadly be defined as the network of relationships among people who live and work in a particular society or organisation. Tools such as Social Network Theory (SNA) facilitate the mapping of these network relationships and to understand the strength of these ties and how information and knowledge flows between them \citep{Granovetter1983-yi_BSPC,Wu2005-ah,Kim2011-sz}. Leveraging these networks within buyer-supplier relationships can add significant value to the buyer and supplier organisations by facilitating innovations, learning, new product development and cost reduction and process improvement programmes \citep{Lawson2009-nx_BSPC}. These three theoretical lenses are not mutually exclusive however scholars tend to select one meta theory, this can be further supported by a secondary theory approach aimed at explaining some of the more nuanced issues within the research question \citep{McIvor2009-nj_BSPC}.   

\subsubsection*{Future impact: technology, sustainability, risk, supply network design}
The current political, economic, social and technological business environment is extremely difficult to navigate, fraught with supply chain disruptions, aggressive competition, sustainability pressures and move towards de-globalisation and protectionism. It is further compounded with profound changes in the way new technologies such as Artificial Intelligence (AI), Blockchain and the analysis of Big Data are impacting business and supply chain operations. These technologies are creating greater visibility and traceability allowing organisations to analyse and adjust the way they do business offering positive as well as negative consequences \citep{Gligor2021-rd_BSPC}. Facilitating relationships through the analysis of big data sets, help to analyse and resolve quality problems and maintain standards and improve innovation and problem solving \citep{Rahman2023-xn_BSPC}. However, it can also drastically reduce personal interactions leading to a lack of personal trust, socialisation and knowledge exchange \citep{Cousins2006-ui_BSPC,Lawson2009-nx_BSPC}. These relational bonds are essential for efficient and effective problem solving and innovation development and knowledge transfer \citep{Lawson2008-qo_BSPC}.. The impact of new technology while offering many advantages it can also negatively impact interfirm relationships. 

Inter- and intra-firm relationships are an essential process in how firms interact and manage their supply networks.  AI will inevitably impact this process as researchers and practitioners we must find ways to use it to enhance and deliver high value returns to mutual benefit \citep{Borah2022-js_BSPC}. 

\subsection[Stakeholder management (Stephen Brammer \& Stephen Pavelin)]{Stakeholder management\protect\footnote{This subsection was written by Stephen Brammer and Stephen Pavelin.}}
\label{sec:Stakeholder_management}
Stakeholder management is an increasingly critical aspect of operations and supply chain management (OCSM). Contemporary economic activity is increasingly complex, entailing coordination and collaboration within and between large networks of globally distributed actors that partner in activities through which a given product or service is brought into the market \citep{Co2009-jf_SBSP,Silvestre2023-cl_SBSP}. The efficiency and efficacy of the entire supply chain depend on successful coordination among diverse actors (including  customers, retailers, financial institutions, governments, logistics providers, distributors, manufacturers, and primary resource extractors and processors) within complex practices and processes \citep[including forecasting, planning, inventory management, scheduling, delivery, just-in-time sourcing, co-location, collaboration, and information sharing; see also][]{Najjar2023-gl_SBSP,Silvestre2023-cl_SBSP}. In this sense, the global economy is increasingly relational and the success of organizations and economies depends on the development and active management of relationships. These relational interdependencies have been shown to be especially significant in the recent geopolitical and pandemic threats to the global economy and society \citep{Devi2023-ew_SBSP,Montoya-Torres2023-uz_SBSP}.  

Stakeholder management research proposes that organisational success depends on considering the interests of multiple stakeholders in strategic decision-making \citep{Donaldson1995-sw_JRTB,Freeman1983-tg_SBSP,Jones1995-hg_SBSP,Laplume2008-xd_SBSP,Parmar2010-yc_SBSP}. Fundamental to stakeholder thinking is the suggestion that “if we adopt as a unit of analysis the relationships between a business and the groups and individuals who can affect or are affected by it, then we have a better chance to” create and distribute value to multiple actors globally \citep{Parmar2010-yc_SBSP}. Various stakeholder definitions have been proposed in previous research. \cite{Freeman1983-tg_SBSP} defines stakeholders as individuals or groups that could influence or be influenced by the activities of the firm while \citeauthor{Donaldson1995-sw_JRTB} (\citeyear{Donaldson1995-sw_JRTB}, p. 68) define stakeholders as “persons or groups with legitimate interests in procedural and/or substantive aspects of corporate activity”. In light of the breadth of actors encompassed by these definitions, considerable effort in stakeholder management research has sought to classify stakeholders to understand how they influence a focal firm \citep{Mitchell1997-su_SBSP}. \cite{Mitchell1997-su_SBSP} identified three attributes of firm-stakeholder dyads that raises their salience in management consideration: (1) Power refers to the ability of a stakeholder to influence, produce or effect behaviour, outcomes, processes, objectives, or direction; (2) Legitimacy relates to a claim’s consistency with expected behaviour, structures,  values, beliefs, norms, and rules; (3) Urgency refers to the stakeholder and claims that it is critical to the stakeholder and time-sensitive. Combinations of these relational attributes drive the level of strategic attention warranted within a firm’s consideration at a given point in time. 

Building on stakeholder classifications, research suggests several broad approaches to stakeholder management that an organisation can adopt, including proaction, accommodation, defence, and reaction \citep{Clarkson1995-zs_SBSP,Wartick1985-wo_SBSP}. Research has also classified the influence strategies of stakeholders, with  \cite{Frooman1999-qq_SBSP,Frooman2005-it_SBSP} proposing two basic influence strategies – coercion and compromise – adopted by stakeholders to influence a focal firm. Most research using stakeholder management differentiates between two fundamental motives for firms, including stakeholders, in their decisions, while recognizing that both are always at play to some degree. The first body of stakeholder thinking, what might be termed the instrumental approach, sees stakeholder management as important because managing stakeholder relationships generates significant expected returns for the organisation. A second body of research –- the normative approach –- emphasises the intrinsic value of taking stakeholders into account and sees stakeholder management as a moral responsibility of contemporary organisations. Studies on instrumental traditions have provided significant evidence that stakeholder management generates organisational benefits mainly through improved reputation, increased trust from stakeholders, an improved ‘license to operate’, and stakeholders acting reciprocally \citep{Berman1999-dx_SBSP,Choi2009-lc_SBSP,Clarkson1995-zs_SBSP,Hillman2001-zs,Jones1995-hg_SBSP}. Normative research in the stakeholder tradition has argued that moral imperatives should consider stakeholders, grounding arguments in social contract theory \citep{Donaldson1999-og}, property rights \citep{Donaldson1995-sw_JRTB}, and Kantianism \citep{Evan1988-sg_SBSP}. To some extent, research drawing on signalling theory has sought to bridge the instrumental and normative perspectives by exploring what attributions, inferences, and conclusions stakeholders attach to firms’ breadth and depth of stakeholder management activities \citep{Aronson2025-iw,Fu2022-mp}. 

To a significant extent, the major strands of stakeholder research in OSCM reflect the schism between normative and instrumental stakeholder thinking. To date, the most prominent strand of stakeholder OSCM research focuses on the roles played by stakeholders in supporting or inhibiting sustainable outcomes throughout firms’ operations and supply chains. Stakeholders are understood to exert pressure on firms for sustainable business conduct, thereby suggesting that firms pay attention to improving their environmental and social conditions \citep{Carter2011-qu_SBSP,Meixell2015-ur_SBSP}. This attention extends beyond individual firms’ operations, in that it also includes their direct \citep{Foerstl2010-vs_SBSP} and indirect \citep{Hartmann2014-em_SBSP} suppliers. Stakeholder pressures increase awareness of sustainability in the supply chain, push buying firms to adopt sustainability-related goals, and influence them in implementing sustainability in the supply chain \citep{Meixell2015-ur_SBSP}. Within this research firms’ CSR activities and reporting are understood to signal their commitments to stakeholders regarding social and environmental issues \citep{Friske2023-dx}. When stakeholders’ sustainability-related expectations are unfulfilled, irresponsible supplier behaviour may be projected onto buying firms, leading to adverse publicity, reputational loss, and costly legal obligations \citep{Bregman2015-cf_SBSP}. Thus, non-compliance with stakeholders’ requests for sustainability poses a risk to buying firms \citep{Hajmohammad2016-mi}, which this study refers to as supply chain sustainability risk (SCSR). Accordingly, SCSR is defined as “a condition or a potentially occurring event” residing “within a focal firm’s supply chain” which can “provoke harmful stakeholder reactions” \citep[p. 168]{Hofmann2014-jb_SBSP}. During times of global sourcing and ubiquitous information availability, SCSR poses a major challenge to buying firms \citep{Busse2017-ni_SBSP}.

Previous research suggests that, in general, a success factor for sustainable supply chain management (SSCM) is that buying firms should “reconceptualise who is in the supply chain. Rather than viewing NGOs and the like as adversaries, sustainable supply chains leverage the skills and abilities of these nontraditional chain members” \citep[p. 52]{Pagell2009-sw_SBSP}. Hence, an attentive and cooperative stance toward stakeholders is often  advisable for firms \citep{Meixell2015-ur_SBSP,Wong2015-bi_SBSP}. To identify, assess, and manage SCSR, firms must understand the differing perspectives, expectations, and values of stakeholders \citep{Wu2014-fc_SBSP}. Thus, when faced with the lack of visibility in the upstream supply chain, attention to stakeholders may be the strategic direction that firms should also pursue to identify SCSR, seeking to incorporate stakeholders’ SCSR knowledge. Stakeholders vary in numerous ways, including their interests and roles \citep{Wu2014-fc_SBSP}. Different groups of stakeholders can be interested in the economic, environmental, and social dimensions to different degrees \citep{Meixell2015-ur_SBSP}. Some stakeholders hope for the firm’s success (e.g., employees and customers), while others may not mind failure \citep[e.g., competitors and media;][]{Hofmann2014-jb_SBSP}. Therefore, firms should refrain from treating their stakeholders as homogenous aggregates; rather, they should  differentiate between them and dedicate specific attention to stakeholder groups \citep{Gualandris2015-ey_SBSP}.

Regarding stakeholder-OSCM research in the instrumental tradition, considerable attention has been given in OCSM research to the circumstances in which trust, collaboration, mutual investment etc. exist in firm-stakeholder relations, and how these support greater innovation, operational, and financial performance.  Empirical studies have often linked supply chain collaboration to performance improvement \citep{Co2009-jf_SBSP}. Managing the relationships among trading partners along the supply chain is an important question that must be addressed in supply chain collaboration \citep{Johnston2004-ga_SBSP}. A long-term perspective between the buyer and supplier increases the intensity of buyer – supplier coordination \citep{De-Toni1999-hq}, and supplier integration through strategic partnerships ‘‘will have a lasting effect on the competitiveness of the entire supply chain’’ \citep{Choi1996-be_SBSP}. \cite{Carr1999-ke_SBSP} also found that strategically managed long-term relationships with key suppliers have a positive impact on the firm’s financial performance.

Supply chain visibility entails trust and collaboration between trading partners. Higher levels of inter-organisational cooperative behaviours, such as shared planning and flexibility in coordinating activities, were found to be strongly linked to the supplier’s trust in the buyer firm \citep{Johnston2004-ga_SBSP}. Issues of trust and risk can be significantly more important in supply chain relationships because they often involve a higher degree of interdependency between competitors \citep{La-Londe2002-od}. The premise of trust and collaboration is that trading partners have comparable desires and commitments to collaborate. \cite{Morgan1994-wo_SBSP} argued that commitment and trust produce outcomes that promote efficiency, productivity and effectiveness. Otherwise, it would require an advocate with power and influence over its trading partners to mandate cooperation.

Moving beyond the reasons and stakeholder-OSCM research, considerable research has begun to examine the structural, attitudinal, and informational pre-cursors of successful stakeholder management. For example, research on supply chain visibility emphasises that relationships can be effectively managed only when information is visible across a supply chain and between supply chain partners. Supply chain visibility can be broadly defined as “traceability and transparency of supply chain process” \citep[p. 51]{Tse2012-pf_SBSP}. Buying firms often have low supply chain visibility because they possess little knowledge about indirect suppliers or cannot independently verify information about their components or practices  \citep{Pagell2009-sw_SBSP}. In particular, fast-moving industries such as the retail and fashion sectors often lack supply chain visibility beyond second-tier suppliers \citep{Roth2008-ng_SBSP}. \cite{Carter2015-tf_SBSP} argue that actors in supply chains often lack sufficient knowledge and visibility of their supply chain beyond first-tier suppliers (upstream) and direct customers (downstream), meaning that “what lies beyond the realm of [the] visible range simply emerges” \citep[][p. 90]{Carter2015-tf_SBSP}. This visibility boundary poses a severe management problem, since “beyond the visible range, the agent has no choice but to accept what happens there” \citep[][p. 90]{Carter2015-tf_SBSP}. Prior research has acknowledged that missing visibility in supply chains is a critical factor in effective supply chain risk management \citep{Durach2015-ym_SBSP}. Lack of visibility may cause knowledge  deficits, loss of control, and distrust, thereby enhancing sustainability and corporate social risks \citep{Busse2017-ni_SBSP}. The practical challenges of engaging inclusively with a wide range of geographically dispersed stakeholders have been emphasised in research \citep{Siems2021-yp}, with research suggesting that collective stakeholder orientations can help overcome barriers to effective stakeholder involvement \citep{Soundararajan2019-ax}. 

Research suggests that firms can adopt various stances toward their stakeholders, ranging from adversarial to welcoming \citep{Pagell2009-sw_SBSP}. \cite{Pagell2009-sw_SBSP} observed that leading firms reconceptualise who is in the supply chain, such that they regard not only their direct  buyers and suppliers as part of the supply chain, but also other stakeholders. Essentially, they suggest the opening up of firms to these stakeholders. Further studies have also found that some firms leverage the expertise and skills of stakeholders, resulting in better-informed managerial decision-making \citep{Wong2015-bi_SBSP}. \citeauthor{Wong2015-bi_SBSP} (\citeyear{Wong2015-bi_SBSP}, p. 56) argued that “feedback from [$\dots$] stakeholders represents key resources because (they) sometimes know more about the environmental  problems facing part of the supply chains than the focal firm”. Stakeholders can provide assistance, develop policies, engage in evaluation and monitoring, and identify improvement potential in a firm’s upstream supply chain with regard to sustainability \citep{Gualandris2015-ey_SBSP,Wong2015-bi_SBSP}. In the context of SCSR, leading firms have begun to proactively search for valuable information to help them identify their SCSR (or other objectives) by constantly scanning the external environment or conducting regular stakeholder consultations and round tables \citep{Foerstl2010-vs_SBSP}. 

Recent macro-environmental uncertainty and turbulence has encourage OSCM scholarship to explore the role of stakeholders in supporting organisational success in light of these conditions. Indeed, \cite{Freeman1983-tg_SBSP} himself drew attention to “the quantity and kinds of change that occur in the business environment” \citep[][, chapter 5]{Freeman1983-tg_SBSP}, necessitating a new paradigm of strategic management. 

The rapid growth and diversity of stakeholder-related research, both within and beyond OSCM, provide scholars with numerous valuable sources for further reading and inspiration. We have drawn the reader’s attention to some important review articles. For example, \cite{Aaltonen2016-lc_SBSP} review literature on stakeholder management in the project management field; \cite{Bhattacharya2021-rx_SBSP} review research on stakeholder theory in supply chain management; \cite{De_Gooyert2017-yt_SBSP} review stakeholder management research in the operations research field; \cite{Govindan2018-of_UR} examine stakeholder-related research in reverse logistics. Sustainable supply chain management has been reviewed extensively by \cite{Carter2011-qu_SBSP} and \cite{Sarkis2011-xf_SBSP}. Additionally, there are a number of excellent reviews on the development of stakeholder management outside OSCM, including \cite{Laplume2008-xd_SBSP}. Technology, especially AI and big data analytics, and government policy play a significant role in shaping how firm-stakeholder relationships evolve and are managed, and rapid changes in these in recent years will continue to show that stakeholder relationships play an increasingly salient role in OSCM. 

\subsection[Bullwhip effect (Stephen M. Disney)]{Bullwhip effect\protect\footnote{This subsection was written by Stephen M. Disney.}}
\label{sec:Bullwhip_effect}
\begin{figure}[!h]
\centering
{\includegraphics[width=0.95\linewidth]{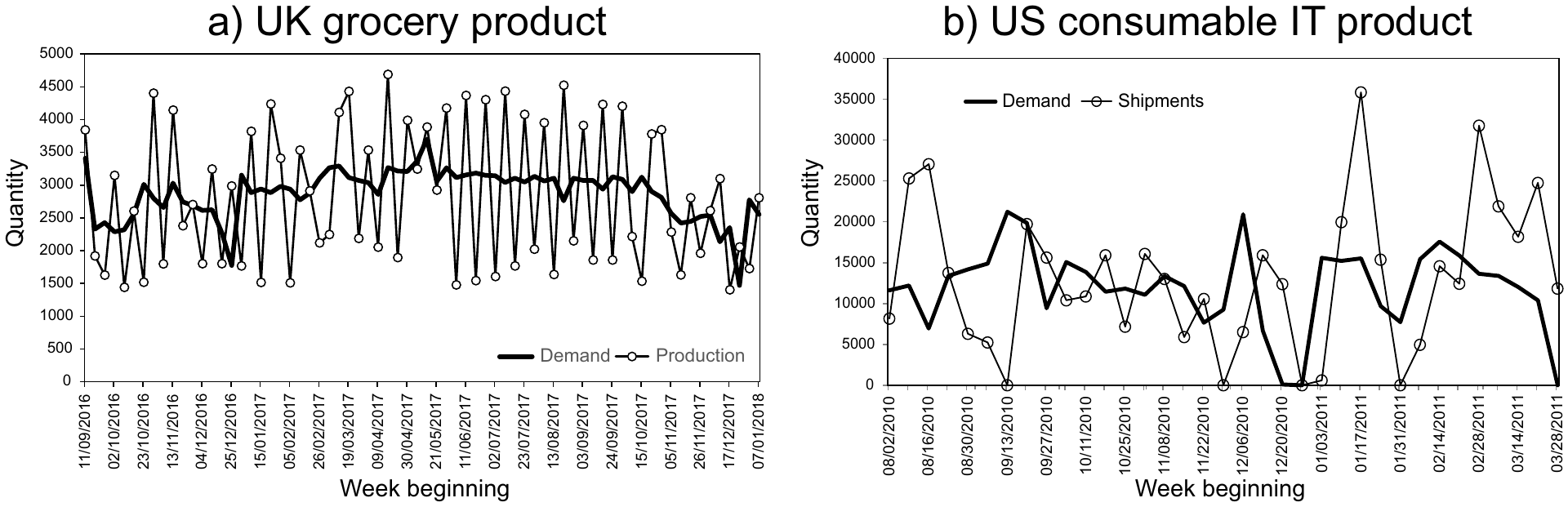}}
\caption{Examples of real-world bullwhip. Panel a) A UK grocery product, bullwhip ratio = 7.68. Panel b) A US IT consumable product, bullwhip ratio = 3.58.}
\label{fig:bullwhip_curves}
\end{figure}

The bullwhip effect refers to a supply chain phenomenon where the variance of customer demand is amplified in the production (or supplier) orders. Figure \ref{fig:bullwhip_curves}, panel (a) highlights the weekly demand and production for a grocery product. The production order variance is 7.68 times that of demand. This amplification is not uncommon in real supply chains. Panel (b) of figure \ref{fig:bullwhip_curves} illustrates the weekly demand and raw material order of a consumable IT product. The variance of the shipments to the assembly factory is 3.58 times that of the demand at the factory. Highly variable demand placed on the upstream echelons of a supply chain creates excess costs. For example, highly variable demand requires extra capacity to meet peak demands; extra inventory is needed to protect service levels; reduced transport efficiency; hiring, firing, and onboarding costs increase.

The bullwhip effect has been discussed in the economics literature for over 100 years \cite{Mitchell1924_SMD}. The dynamics of supply chains were well understood by \cite{Forrester1958_SMD}. Solutions to the bullwhip problem were proposed by \cite{deziel&eilon1967_SMD}, \cite{Towill1982_SMD} and \cite{John1994_SMD}. Academic interest in the bullwhip really gained momentum when \cite{Lee1997_SMD} coined the \emph{bullwhip effect} moniker and explored four causes of the bullwhip effect via the ratio in equation \eqref{eq:bullwhipRatio}. This paper has become one of the most highly cited papers in the operations and supply chain management field.
\begin{align}
\label{eq:bullwhipRatio}
    \text{Bullwhip ratio} = \frac{\text{Variance of the orders}}{\text{Variance of the demand}}.
\end{align}

\subsubsection*{Methodological approaches to studying bullwhip}
Control theory has long been one of the most powerful ways of solving \emph{linear} bullwhip problems. \cite{Simon1952_SMD} applied the Laplace transform to a production control problem described by differential equations. \cite{Towill1982_SMD} studied an inventory replenishment policy denoted IOBPCS (inventory and order-based production control system) in continuous time using the Laplace transform. \cite{John1994_SMD} added a work-in-progress feedback loop to the IOBPCS to create the APIOBPCS model (automatic pipeline, inventory, and order-based production control system). Continuous-time bullwhip measures are commonly determined by the system's \emph{noise bandwidth}. 

Most supply chains operate on discrete time, where one ordering decision per planning period is made. \cite{Vassian1955_SMD} studied a discrete-time replenishment system with \emph{z}-transforms, finding the inventory optimal \emph{order-up-to} (OUT) policy. \cite{Disney2001_SMD} used the \emph{z}-transform to study the dynamics of a vendor-managed inventory supply chain. The Fourier transform can be used to understand how supply chains create bullwhip under arbitrary (stationary and non-stationary) demands and (constant) lead times by investigating the system's frequency response. \cite{Dejonckheere2003_SMD} revealed the OUT policy with exponential smoothing and moving average forecasts always creates bullwhip, for all demand patterns and lead times. \cite{Li2014_SMD} extended these results, showing the OUT policy with Holt's forecasts always creates bullwhip, but the OUT policy with damped trend forecasts does not always create bullwhip for certain demand processes. Linear systems can also be studied directly in the time domain and using difference equations and expectation \citep{Hosoda2012_SMD}. Larger, more complex linear supply chain models can also be analyzed using state space techniques and matrix methods, \citep{Gaalman2006_SMD, Wang2017_SMD}.

For non-linear problems, simulation is a practical approach \citep{Ponte2022_SMD}. However, sometimes analytical progress can be made by studying time domain difference equations and probability density functions  \citep{Disney2021_SMD}. More qualitative approaches like system dynamics, \cite{Sterman2000_SMD}, influence diagrams \citep{Coyle1977_SMD}, and causal loop modelling \citep{Morecroft2020_SMD} can provide insight. The \emph{beer game} is a popular tabletop game for teaching the dynamics of supply chains \citep{Sterman1989_SMD}.

Surveys and secondary data can also be used to investigate bullwhip. At an individual product level, bullwhip appears almost inevitable. However, when measured at the company level, perhaps aggregated over a financial reporting period of many months, bullwhip may not exist \citep{Cachon2007_SMD}. Bullwhip has also been studied empirically: \cite{Fransoo2000_SMD} consider how to measure bullwhip in real supply chains (which may have convergent and/or divergent material flows). \cite{Disney2013_SMD} report on a project removing bullwhip from a global printer toner supply chain. \cite{Meng2018_SMD} improve the forecasting and production planning activities in a UK dairy product manufacturer to reduce bullwhip.

\subsubsection*{The replenishment policy}
Most bullwhip studies consider high-volume demand where the items are ordered and replenished every planning period. Planning periods range from several hours to a day in grocery supply chains, or weekly (most likely) to monthly in manufacturing environments. \cite{Hedenstierna2018_SMD} investigate the consequences of different planning cycle lengths; longer cycles have a natural pooling effect that reduces demand variance. With no ordering (capacity-related) costs, the OUT policy is the optimal replenishment policy for minimising inventory holding and backlog costs \citep{karlin1958_SMD}. When the objective function includes bullwhip-related (capacity-related) costs, the OUT policy is no longer optimal, \cite{Boute2022_SMD}; the dual base stock (DBS) policy is known to be the optimal policy \citep{Boute2015_SMD, Gijsbrechts2024_SMD}. However, the DBS policy is not a demand replacement policy. The DBS policy has a region of inaction where demand is not fully replaced and only $k$ units (where $k$ is the nominal capacity without overtime) are ordered. Generally, only numerical solutions for the two base stock levels in the DBS policy can be obtained \citep{Gijsbrechts2024_SMD}.

The high-performing linear proportional (POUT) policy, grounded in control theory, is amenable to analytical study. The POUT policy generalizes the OUT policy by incorporating two proportional feedback controllers into the policy, an established technique in the design of hardware systems. The feedback controllers in the POUT policy regulate the speed at which discrepancies in inventory and WIP levels are corrected \citep{Disney2003_SMD}. While the POUT policy requires careful set-up to avoid instability \citep{Disney2008_SMD}, it is the optimal linear quadratic regulator for minimising costs proportional to the variance of order and inventory \citep{Wang2016_SMD}. When piece-wise linear convex inventory costs and capacity costs associated with an installed capacity and overtime are present \cite{Boute2022_SMD} show the POUT performs well compared to the optimal non-linear DBS policy.

\subsubsection*{Lead times and the bullwhip effect} 
Increasing lead times will always increase the inventory variance, no matter the demand process. However, increasing the lead times does not always increase the OUT policy's order variance \citep{Gaalman2022_SMD}. Depending on the demand process, the bullwhip may increase or decrease in the lead time. For negatively correlated demand patterns, bullwhip exhibits an odd-even lead time effect, increasing or decreasing based on the parity of the lead time. It can even have more complex behaviour; a multi-period oscillating effect, or be increasing (decreasing) in the lead time when the lead time is short (long) and vice versa. 

Under stochastic lead times, the impact of the lead times is very sensitive to modelling assumptions.  If we assume past realisations of the lead time are representative of future lead times, the POUT policy bests the OUT policy when minimising inventory costs alone \citep{Disney2016_SMD}. Furthermore, setting the POUT policy to minimise inventory cost alone under stochastic lead times also removes bullwhip for free. However, dynamically forecasting lead times can create large amounts of bullwhip \citep{Michna2020_SMD}.

\subsubsection*{Demand processes} 
Most bullwhip studies assume demand is random. The simplest random demand is the independent and identically distributed (i.i.d.) demand. Forecasted with minimum mean squared error (MMSE) forecasts, i.i.d.~demand produces a bullwhip ratio of unity, as the OUT policy acts as a \emph{pass-on-orders} policy. 

More realistic demand patterns can be modelled using the auto-regressive, integrated, moving average (ARIMA) modelling framework of \cite{Box1976_SMD}. \cite{Ali2012_SMD} studied a data set of weekly demand for 1798 products from a European retailer. They identified the first order auto-regressive, AR(1), demand processes in 30.3\% of the time series, integrated moving average, IMA(0,1,1), processes accounted for 23.7\%, and i.i.d.~demand accounted for 16.3\%. The data set also contained higher-order ARIMA demands. ARIMA models represent many real demand patterns and have been studied extensively. \cite{Graves1999_SMD} considered IMA(0,1,1) demand, \cite{Lee1997_SMD} considered AR(1), \cite{Gaalman2006_SMD} considered ARMA(1,1) demand, \cite{Luong2007_SMD} studied AR(p) demand, and \cite{Gaalman2022_SMD} investigated ARMA(p,q) demand. \cite{Zhang2004_SMD} and \cite{Gilbert2005_SMD} have investigated how ARIMA demand is transformed into other ARIMA demand processes as it passes up the supply chain. 

\subsubsection*{Supply chain structure}
It is natural to consider how the bullwhip effect propagates in different supply chain structures. \cite{Chen2000_SMD} consider a multi-echelon supply chain. The sharing of demand information throughout the supply chain leads to a bullwhip effect that increases linearly, rather than geometrically, in a traditional supply chain \citep{Lee2000-gu_UR, Dejonckheere2004_SMD}. Whereas, a vendor managed inventory supply chain allows multiple supply chain echelons to act as one, effectively eliminating bullwhip transmission \citep{Disney2001_SMD}. 

Closed loop supply chains are popular areas for bullwhip studies \citep{Ponte2022_SMD}. Marketplace returns can have a positive effect on the production of new items. However, care must be taken as it is not always economical to remanufacture everything returned, even when the remanufacturing cost is less than the cost to produce new items \citep{Hosoda2021_SMD}. 

Issues with global dual sourcing have also been studied from a bullwhip perspective. The tailored-base-surge allocation scheme \citep{Allon2010_SMD}, places large constant orders to the offshore supplier, whereas the nearshore supplier satisfies peak demands with a shorter lead time. \cite{Boute2022_SMD} showed that local \emph{SpeedFactories} could maintain tighter inventory control than a single offshore supplier, facilitating the relocation of production back home before reaching price parity.

\subsubsection*{Emerging bullwhip research topics}
There are a number of critical areas related to the bullwhip effect that are relatively underexplored. For example, understanding the role of \emph{artificial intelligence/machine learning} (AI/ML) in forecasting demand and regulating production and distribution. There is a lot of industrial interest in using AI/ML to forecast demand. Presumably, the large and diverse datasets that can be used as input data will lead to more accurate forecasts. However, in cases where we know the optimal replenishment policy for a given cost function (for example, the OUT policy is the optimal linear inventory cost policy for a given set of forecasts), ML/AI may not be able to improve upon current best practice.

\emph{Geopolitical issues} such as tariffs, trade barriers, conflicts, and sanctions are altering supply chains and trading relationships. As companies restructure their supply chains, there is an opportunity to design out bullwhip problems by shortening lead times, removing supply chain echelons, re-shoring, and dual sourcing. The impact of bullwhip on the \emph{carbon emissions} and other \emph{climate consequences} generated by supply chains is also an important area for future bullwhip research. 

\subsection[Information sharing (Usha Ramanathan)]{Information sharing\protect\footnote{This subsection was written by Usha Ramanathan.}}
\label{sec:Information_sharing}
Supply chain (SC) information has become an integral part of global businesses. In the past few decades global SCs of commodities, including basic food, textile and medical supplies, have grown enormously. The critical importance of information and technology in supporting modern global SCs has been well recognised in the literature. A number of digital tools and techniques are enhancing operational efficiency of global SCs alongside of the SC information, including a transparent information exchange among SC players.

SC information facilitates several operations including warehouse management, inventory management, capacity planning, quality management, production control and very importantly accurate forecasting for planning production and demand management. Three decades ago, \cite{Aviv2007-nt_UR}, \cite{Lee2000-gu_UR} and \cite{Gavirneni1999-ii_UR} insisted on the importance of information sharing for forecasting and warehouse management. Later, \cite{Ramanathan2010-hq_UR,Ramanathan2011-bm_UR} demonstrated a strong link between the transparent information exchange and the forecast accuracy through empirical evidence from a case study of a soft drink manufacturing firm.

In the presence of collaborative networks, the global SCs have started sharing high volume of data, including traditional production data, warehouse and inventory data, alongside of transportation and customer data in real-time using Big Data (BD) platform \citep{Ramanathan2017-rh_UR,Singh2019-rw_UR,Balakrishnan2021-rl_UR}. Thanks to the advancement in technology, people and SCs remained connected through information without having physical contact or human intervention at the time of COVID. The latest connected technologies such as Internet of Things (IoT) technology, BD Analysis and Blockchain have brought several SC players together to exercise the power of information and automation, without needing to have human-intervention. 

\subsubsection*{Evolution of SC management with integration of information sharing}
Several prominent SC management practices have evolved in the past decades to nurture the global operations of SCs. Various strategic alliances of SCs decided the level of information sharing among partners \citep{Ramanathan2012-th_UR,Mentzer2001-qp_WPLD}. For example, vendor managed inventory (VMI) was introduced to support the business relationship between suppliers (or vendors) and retail stores in the 1970s. In the VMI, the information on inventory status, both in-store and warehouse, was shared by the retailers to the vendors. Based on the information shared, the vendor maintained the optimal level of inventory in line with the demand fluctuations \citep{Barratt2001-ly_CGTZ}. Speed and accuracy of the information supported the SC decision making and also helped avoiding bull-whip effect \citep{Ramanathan2010-hq_UR}. 

In late 80s continuous replenishment (CR) was adapted by the SC players who expected somewhat stable and consistent demand. Especially the functional products such as soap and soup followed this trend. However, in these SCs, information was not considered as the main element to create a perfect balance between the supply and the actual demand. On the other hand, the CR was not considered an effective approach for innovative or fast-moving products as it could not avoid the excess inventory due to lack of communication or lack of information sharing among the SCs \citep{Cachon2000-zz_UR,Fisher1997-lt_UR,Ramanathan2012-th_UR}. 

In 1990s, quick response (QR) and accurate response (AR) approaches were introduced in the SC management which needed a huge amount of real-time demand data to plan timely production, timely delivery, and timely replenishment to meet the customers’ ever increasing demand expectations \citep{Fernie1994-ic_UR,Fisher1997-lt_UR}. Both QR and AR have necessitated the importance of information exchange among these SC partners and developed electronic data interchange (EDI).

In early 2000s, another real-time information-based system called Collaborative Planning Forecasting and Replenishment (CPFR) framework was developed by Walmart and Warner-Lambert to enhance the response of businesses to the demand from the customers \citep{Seifert2003-db_UR}. This CPFR connected all partners through formal collaboration using information and smart technology \citep{Seifert2003-db_UR,Ramanathan2010-hq_UR}. Collaboration is considered a critical tool to achieve success in CPFR and emerged as a winning strategy for all SC partners to help avoid bullwhip effect with the help of transparent information \citep{Lee2000-gu_UR}. SC collaboration predominantly relies on network agreements and information sharing.

Transparent SC information on materials including raw materials, works-in-progress and finished products, order data, delivery information and local events impacting sales, can be shared seamlessly. This open information sharing approach connects all the SC partners from sourcing, production planning, inventory, and warehousing, until the final delivery which will help reducing the risk of stockout or excess inventory and hence it completely avoids bullwhip effect from SCs. The companies using the CPFR framework managed to achieve accurate forecasting of demand and hence the actual sales were meeting the forecasted demand. After the introduction of the CPFR, importance of data in SCs became very much prevalent. Following this, after 2010, a digital technology era has embraced the global SCs to gain resilience against SC risks. 

\subsubsection*{Role of information in modern SCs and reverse logistics}
In traditional SCs, materials and finished products moved from upstream to downstream SCs. On the other hand, the movement of goods from downstream to upstream was not usual, except for some special cases such as defective product-returns or excess stocks returns; the journey from downstream to upstream also carried a minimal level of information such as orders and payment details. In early 1990’s, information technology evolution has hit the forward SCs with the real-time information sharing among all collaborating partners around the globe to control SC operations such as production, store and deliver in line with the customers demand and preferences \citep{Wiengarten2010-ks_UR,Kembro2014-sf_UR}. At this point, forward SC was challenged by various newly introduced product-return policies and hence the introduction of reverse logistics (RL) was dominant feature of all retail SCs during this period \citep{Chen2011-rb_UR,Ramanathan2023-ld_UR}.

\cite{Carter1998-ys_UR} listed socio-economic and environmental benefits of RL through a detail literature review. Although recycling or remanufacturing are decided by the businesses, internal pressure from SCs and external pressure from various stakeholders such as government, policies, market, and competitors are considered as drivers of RL \citep{Govindan2018-of_UR,Ramanathan2023-ld_UR}. Here, SC information plays a crucial role to decide whether to use the RL or not. \cite{Ramanathan2023-ld_UR} expressed various causes of the reverse logistics, varies from simple product returns to end of product life cycle. Accordingly, every manufacturer and raw material supplier will need to know the information on reverse SCs and logistics to plan their remanufacturing or recycling activities as this will help them to decide the production level of new or existing products. Customer preference data including favourite products, changing trend, brand loyalty, location of purchase and interest in online or offline or omni channel purchase, is helping the SCs to react quickly to the demand of the market \citep{Zhang2023-mv_UR,Jain2022-xk_UR}.

Since the outbreak of COVID, majority of the global businesses have opted to have omni-channel strategy which required an increased level of SC services as the local delivery and pick-up points are ever increasing. This has made the SC information sharing mandatory rather than optional \citep{Ramanathan2024-ps_UR}.

\subsubsection*{Information sharing in the digital SC era }
In 21st century of digital era, global SCs use information from within the SCs and also use the market information effectively to integrate production, delivery and sales. Some of the SC information will not have any control by the SC players but this will be used for planning and combating the external risks.  SC resilience is gained with the power of information and BD Analysis. For example, simple GPS sensors attached in the transportation vehicles keep sending the data related to location of the products in the SC to alert suppliers while IoT sensors send data on quality of the perishable food items \citep{Ramanathan2023-ld_UR}. Some of the frequently used technologies in global SCs are radio frequency identification (RFID), blockchain technology, IoT, cloud computing, business data analytics, virtual and augmented reality \citep{Singh2018-fm_UR,Misra2022-hy_UR,Sharma2021-mw_UR}. Ikea, the leading furniture manufacturer, is using RFID for several decades to monitor inventories in-store. Barcode sales information from Ikea combines the inventory data and trigger the automated ordering system of suppliers to replenish on time. This helps avoiding any complexity in the SCs alongside of enhancing the traceability. Role of sensors and connected technologies were highly appreciated at the time of COVID when human intervention was at its record low \citep{Inman2024-cz_UR}.

After COVID, several global SCs started using the IoT technology comfortably in almost all SC operations staring from sourcing, production, delivery and recycling or remanufacturing, having real-time information and validation of the data through BD analysis.  Food SCs increasingly use IoT such as smart farming, sensors and tracking system to assure quality. The information from the sensors is analysed and integrated with the SC decision making \citep{Ramanathan2023-ld_UR}.

Other SCs namely fashion SCs and technology SCs are increasingly converging their business fully digitalised.  Modern world after the introduction of fast fashion is struggling with over production and less recycling. The SC information can help them to avoid risk and build resilience. \cite{Guo2023-bx_UR} explained the use of Blockchain in fashion SCs for information sharing to maintain sustainability. Similarly, retail omni-channel SCs need consumers preference data to utilise the warehouse space effectively and need consumers’ pickup location to plan logistics in a real-time basis. In a nutshell, SC information is indispensable for successful operation of all global businesses. 

\subsection[Transportation (Paul Buijs)]{Transportation\protect\footnote{This subsection was written by Paul Buijs.}}
\label{sec:Transportation}
Transportation is the backbone of operations and supply chain management, facilitating the movement of goods from raw material sources to end consumers. This process involves many segments, from global shipping lines that connect continents to the last-mile deliveries that complete the journey to the final destination. Each segment plays a distinct role, yet they work together to form a cohesive system that supports the seamless flow of products. Understanding these layers, their challenges, and the evolving trends within each is essential for achieving both efficiency and sustainability in supply chain operations.

Global production networks rely on complex, interrelated structures comprising multiple transportation modes -- maritime, air, pipeline, rail, and road -- connected by ports and terminals. Maritime transportation accounts for the majority of global freight volumes. Innovations in transport technology, particularly the introduction of the standardised container, have changed transportation by significantly increasing the efficiency and capacity of ports and terminals to handle large volumes of freight \citep{Fransoo2013-bb_PB,Notteboom2008-wo_PB}. As critical components of the modern transportation, containers have been instrumental in driving substantial growth in world trade since their introduction in 1956 \citep{Bernhofen2016-eb_PB}. Although responsible for a much smaller volume of freight, air transportation carries a significant share of world trade value due to its speed and suitability for high-value, time-sensitive goods. Pipeline transportation, though less commonly discussed, is the main mode for transporting energy resources like oil and gas over long distances. Rail and road freight play a minimal role in intercontinental transportation.

Research on global transportation networks is broad, covering topics such as network design (\S\ref{sec:Network_design}), routing, and scheduling \citep{Agarwal2008-vx_PB,Christiansen2004-rn_PB}, empty container movements \citep{Song2009-pi_PB}, and port and terminal operations \citep{Stahlbock2007-di_PB}. A significant portion of the academic literature addresses these topics through optimisation methods. Another line of research aims to describe current and predict future freight flows based on empirical data, often referred to as freight transportation modelling \citep{Holguin-Veras2013-cf_MJMB}. Freight transportation modelling can help identify the different roles and levels of criticality of ports and terminals in the global transportation network \citep{Bombelli2020-ad_PB,Verschuur2022-vi_PB} and can be used to design and evaluate transport policies. 

Recent transportation research focuses on the environmental sustainability (\S\ref{sec:Environmental_sustainability}) and resilience (\S\ref{sec:Risk_management_and_resilience}) of global transportation networks. Freight transportation is widely recognised as a major contributor to global carbon emissions \citep{International-Transport-Forum2023-zf_PB}, driving the current research focus on carbonising the sector. At the same time, freight transportation is slow to adapt and highly vulnerable to the impacts of climate change, while it will be crucial in enabling mitigation and adaptation efforts across other sectors \citep{McKinnon2024-bf_PB}. The considerable economic consequences of port shutdowns \citep{Rose2013-ee_PB} and recent blockages of maritime shipping routes have sparked increased scholarly attention on the robustness and (cyber)security of global transportation networks \citep{De-Liso2022-wd_PB}.

As goods transition from global routes to regional markets, the focus shifts to inland freight transportation, which plays an important role in connecting major ports to distribution centres and warehouses. Inland waterways, rail, and especially road are the main modes in this part of the transportation system. Road freight, often considered the most flexible mode, handles the majority of inland freight -- accounting for over 75\% of inland transportation in the European Union in 2022 \citep{Eurostat2024-ju_PB}. Efforts to encourage a shift from road to rail and inland waterways were initially aimed at improving the efficiency of the transportation system. More recently, these policies have expanded to also address sustainability concerns. However, despite significant investments and initiatives, particularly in Europe, the anticipated shift in transportation modes has largely failed to materialise \citep{Takman2023-cn_PB}.

A significant portion of inland transportation research focuses on the development and application of optimisation models to support decision-making in areas such as fleet management \citep{Crainic1998-cq_PB} and the planning and scheduling of transportation services \citep{Crainic2009-qp_PB}. Empirical and conceptual research in this area addresses topics such as mode choice and carrier selection \citep{Meixell2008-bl_PB}, examining the criteria shippers use to choose the most suitable mode of transportation by balancing cost, reliability, and service quality. Another critical area of research explores the decision-making process regarding whether to manage transportation in-house or outsource (\S\ref{sec:Outsourcing}) to specialised carriers, as well as how relationships between shippers and carriers evolve when outsourcing is chosen \citep{Wallenburg2009-fb_PB}.

Digitisation (\S\ref{sec:Digitalisation}) and information sharing (\S\ref{sec:Information_sharing}) has been a major focus of recent transportation research, with studies examining how digital platforms, real-time data, and market intelligence can enhance decision-making, improve efficiency, and foster more dynamic planning of transportation services \citep{Cichosz2020-qf_PB}. Another strand of recent research focuses on optimising transportation services across multiple supply chains. When applied to a single mode of transportation, such as road freight, this is often referred to as horizontal collaboration \citep{Buijs2014-ru_PB,Pan2019-jc_PB}. Synchromodality extends this concept by striving for dynamic transportation scheduling across multiple supply chains and multiple modes \citep{Acero2022-fo_PB}. Some researchers even envision a ``Physical Internet'', using the digital internet as a metaphor to achieve fully open and seamlessly interconnected private transportation networks \citep{Montreuil2011-sd_PB}. A key challenge with the ongoing transitions in road freight transportation -- such as those related to sustainability and digital transformation -- is the sector's high level of fragmentation. There are over 500,000 European road carriers, and 99\% of these companies have fewer than 50 employees \citep{Toelke2021-zm_PB}.

As goods progress towards their final destinations, the focus shifts to last-mile logistics, where products are ultimately handed over to end consumers -- whether through retail stores or e-commerce. The closer goods get to the final customer, the more challenging it becomes to achieve the economies of scale common upstream in the transportation system. Stringent service level targets, small order sizes, high delivery frequencies, and dispersed pickup and delivery locations hinder the efficient use of larger trucks, leading this part of the transportation system to rely on smaller trucks, vans, and even smaller vehicles to supply retail stores, restaurants, offices, and deliver parcels to consumer homes or pickup points. In the last mile, the negative externalities of transportation become more evident, as this stage occurs near where people live, work, and spend their leisure time \citep{Cardenas2017-lp_PB}. 

Much of the research attention in this area focuses on innovative solutions to mitigate these externalities, with particular emphasis on business-to-consumer e-commerce logistics \citep{Mangiaracina2019-tl_PB}. For last-mile logistics broadly, these include the adoption of low-carbon vehicle technology, e-commerce delivery through pickup points, off-hour delivery programs, dynamic loading and unloading bays, urban vehicle access restrictions, consolidation, and demand management \citep{Holguin-Veras2020-or_PB,Holguin-Veras2020-wm_PB}. Vehicle routing optimisation has long been a central focus in the academic literature on last-mile logistics \citep{Savelsbergh2016-tt_PB}, but recently, scholars have begun to recognise the significant impact of parking, loading, and unloading on last-mile efficiency \citep{Fransoo2022-ga_PB,Ghizzawi2024-uk_PB}. In e-commerce, there is ongoing debate regarding the sustainability of home delivery versus retail shopping \citep{Buldeo-Rai2021-bh_PB}. Consensus remains elusive, with some solutions -- such as pickup points -- often marketed as more sustainable but potentially increasing carbon emissions when consumer travel is considered \citep{Niemeijer2023-cb_PB}, highlighting the complexity of this area of study.

For those seeking a deeper understanding of the complexities across global, inland, and last-mile transportation, the textbooks by \cite{Rodrigue2024-ee_PB}, \cite{Tavasszy2013-xw_PB}, and \cite{Monios2023-nd_PB} provide a good starting point. \cite{Rodrigue2024-ee_PB} offers an accessible introduction to the geography of transportation. \cite{Tavasszy2013-xw_PB} present a comprehensive overview of freight transport modelling, systematically introducing the approaches used to support public transport policy analysis. \cite{Monios2023-nd_PB} deliver a state-of-the-art overview of urban logistics, addressing the unique challenges posed by modern demands such as e-commerce and sustainability.

\subsection[Emerging economies (Stefan Seuring \& Sharfah Ahmad Qazi)]{Emerging economies\protect\footnote{This subsection was written by Stefan Seuring and Sharfah Ahmad Qazi.}}
\label{sec:Emerging_economies}
The section briefly introduces institutional voids, forming a key challenge for managing supply chains in emerging economies. The next step argues for a more differentiated approach to SCM, considering differences among regions and countries. Looking at possible solutions, communication and coordination of suppliers and further supplier development related aspects are considered suitable approaches for addressing voids and strengthening the supply chain in emerging economies.

Emerging and developing economies span a wide range of countries. One persistent challenge for managing supply chains in such environments is typically that (governmental) institutions are less developed, leading to so-called institutional voids, which are typically seen as a core reason challenging the management of supply chain \citep[e.g.,][]{Zomorrodi2024-ej_SSSAQ}.

The related body of literature \citep{Parmigiani2015-kx_SSSAQ,Brenes2019-iz_SSSAQ} lists particularly product market voids and contracting voids on the dyadic, i.e. supplier-buyer level. Operational issues driving product market voids are, for example, lack of intermediaries or standards, in some cases also lack of communication infrastructure. Contracting voids lead to issues in enforcing written agreements, so even if they exist, there might not be a governmental body to turn to if the contract is not fulfilled. There are labour market voids on the supply network side due to a lack of education and, therefore, unavailability of a skilled workforce. Capital market voids are a particular issue as funding is unavailable or might be comparatively more expensive. Last, regulatory voids derive from corruption, unexpected regulatory conditions, or limited access to formal justice. 

These voids offer challenges for managing supply chains, where standardised (Western) approaches might come to limits. Recent contributions call for a more differentiated approach, as one-size-fits-all approaches might overlook the challenges and opportunities. In their Delphi Study covering six regions, \cite{Seuring2022-yc_SSSAQ} point to different needs for respective regions. Europe and North America serve as the developed world reference points against which other countries and regions are evaluated. For (selected countries in) Africa, the availability of SC finance options is a core challenge, particularly on the sourcing side, so materials and preproducts are made available \citep{El-Baz2019-lv_SSSAQ}. Digital technologies might enable some supply chain solutions not available so far, linking farmers and producers directly to markets and customers \citep[e.g.,][]{Schilling2023-hp_SSSAQ}. Brazil shows some similarities but also has challenges in having a qualified workforce available, causing additional challenges and risks in the supply chain \citep[e.g.,][]{Fritz2018-on_SSSAQ}. China seems comparatively closer to developed economies, so aspects of SC volatility and agility move more into the foreground. India and Pakistan show some similarities to the observations from Africa, while more emphasis is placed on having financial resources available for the manufacturing part of the SC and aiming for resilient supply chains. Last, Iran seems to be a very particular situation, which might be a consequence of its geopolitical isolation. This seems to lead to a high emphasis on circular SC solutions \citep{MahmoumGonbadi2021-he_SSSAQ}, as primary resources are in short supply.

These examples serve as a background for identifying suitable and adjusted solutions in related supply chains. It is evident that many of the mentioned aspects have their original cause in the mentioned institutional voids. In this respect, solutions could be derived by overcoming the related institutional voids, which are typically beyond the comprehension and achievability of supply chain management in a narrower sense, but link into the debate on, for example, Global Production Networks \citep{Yawar2024-gd_SSSAQ}. Within this body of literature, governance, power dynamics and stakeholder approaches are discussed, all find their equivalent in the (sustainable) supply chain management debate. 

On the level of the supply chain itself, there would be a wide range of suitable measures. Based on data from Kenya and Uganda, one kind of analysis has been provided by \cite{Seuring2019-qj_SSSAQ}, which builds on the conceptualisation of sustainable SCM as offered by \cite{Seuring2008-fb_SSSAQ}. There are arguments that both environmental \citep[e.g.,][]{Tate2019-kh_SSSAQ} as well as social aspects are particularly relevant in emerging economies \citep[e.g.,][]{Alexandre-de-Lima2021-oi_SSSAQ}, there are several strategies to manage related issues. 

The centrality of communication and coordination is much discussed and holds even more for emerging economy environments, where suppliers might not be aware of developed market demands \citep{Brix-Asala2020-dx_SSSAQ}. A central line of reasoning is that this should create win-win situations where environmental (and/or social) goals are achieved in line with economic ones. This is typically complemented by monitoring suppliers towards fulfilling minimum standards, recently advocated in respective supply chain due diligence legislation \citep{Buttke2024-ch_SSSAQ}. In many cases, this might require that stakeholders play a crucial role, moving beyond being divers of related developments \citep{Seuring2008-fb_SSSAQ}, thereby also acting as inspectors, such as in monitoring processes or even facilitators helping in implementing related process and indicators \citep[e.g.,][]{Liu2018-yn_SSSAQ}. Social norms are particularly relevant in emerging economy situations \citep{Silvestre2015-rt_SSSAQ} and are often challenging to monitor. However auditing and monitoring suppliers and, therefore avoiding trade-off situations are key demands for the long-term success of such ventures \citep{Seuring2019-qj_SSSAQ}. Following such a logic, often demand deep and continuous engagement with supplier and supply chain members along multiple tiers of the supply chain. In such settings, institutional voids challenge the inclusion of emerging economies \citep[Emerging economies in international SCs;][]{Silvestre2015-rt_SSSAQ}. To overcome this challenge, entrepreneurs -- in this case, emerging economy actors operating in weak institutional environments -- can acquire information, knowledge, and external resources via their relational networks.

\begin{figure*}[ht!]
    \centering
    \includegraphics[width=5.25in]{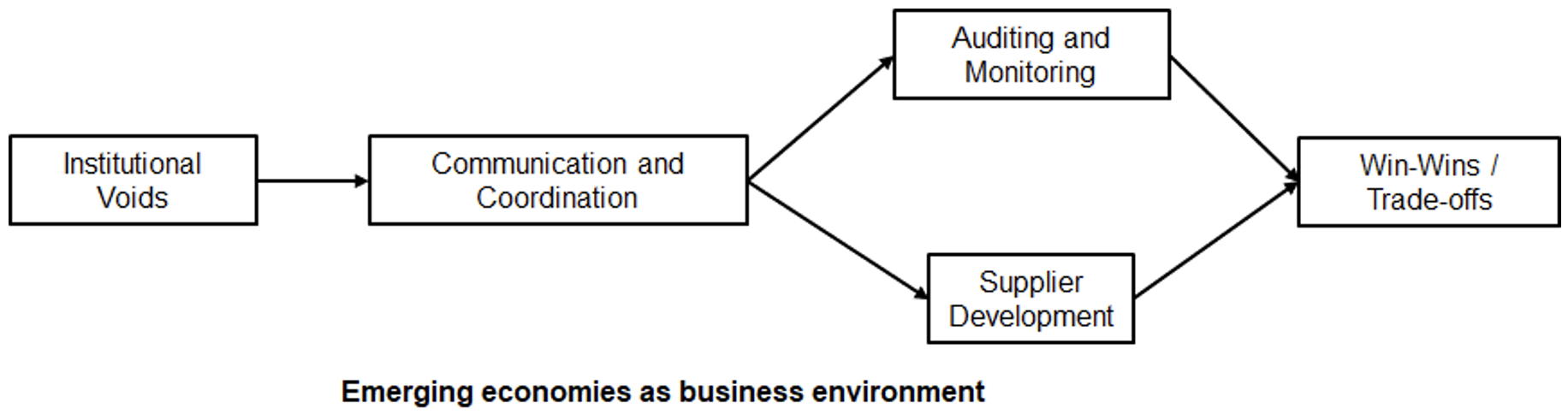}
    \caption{Core elements of managing supply chains in emerging economies.}
    \label{fig:emerging}
\end{figure*}

More particular solutions seem to be offered by supplier development measures, which can enable suppliers to take part in local and international supply chains \citep{Brix-Asala2020-dx_SSSAQ}. Case based research is a typical approach to getting access to such empirical fields, which might be required due to illiteracy or cultural differences, where researchers are seen as sceptical, so appropriate rigour-within-context has been argued for \citep[e.g.,][]{Halme2022-nt_SSSAQ}. In such settings, the applicability of indirect supplier development (e.g., evaluation and feedback, supplier rewards) and direct supplier development (e.g. training and evaluation, on-site consultation) have been assessed \citep{Brix-Asala2020-dx_SSSAQ}. Figure \ref{fig:emerging} sums up the line of reasoning.

Looking at future developments, there are a couple of issues which require further assessment. Among them is the role of social enterprises in supply chains \citep{Longoni2024-au_SSSAQ}, which play an important role in taking up and fulfilling the needs of large parts of the population in emerging economies. This links back into the already mentioned application of digital technologies (first of all smartphones) for managing businesses and the respective supply chains \citep{Schilling2023-hp_SSSAQ} and thereby contributing to the digital and sustainable transformation of supply chains \citep{Schilling2024-xs_SSSAQ}. It will be interesting to see how e.g. blockchain technology might contribute to such supply chains, for example by enabling trust \citep{Yavaprabhas2024-ze_SSSAQ} or what role artificial intelligence might have to play \citep{Fosso-Wamba2024-wa_SSSAQ}. Against such observations, research on supply chain management in emerging economies still seems to be in an early stage, opening up multiple opportunities for future research. It should be kept in mind to integrate researchers from such countries, thereby creating inclusive research.

\clearpage

\section{OM-SCM intersection}
\label{sec:OMSCM_intersection}

\subsection[Behavioural operations (Konstantinos V. Katsikopoulos \& Behnam Fahimnia)]{Behavioural operations\protect\footnote{This subsection was written by Konstantinos V. Katsikopoulos and Behnam Fahimnia.}}
\label{sec:Behavioural_operations}

Behavioural Operations Management (BOM) focuses on understanding and improving decision-making processes within supply chains and other operational settings by considering human behaviours and the individual, group, and organisational processes that lead to the behaviours. BOM emerged as a discipline after observing that human decision making often deviates from traditional theoretical models developed in neoclassical economics, ``hard'' operations research, and operations management. \citep{Croson2013-id_KKBF,Bendoly2015-fq_KKBF,Donohue2018-jk_KKBF}.

The foundations of BOM lie in disciplines such as behavioural economics and cognitive and social psychology \citep{Fahimnia2019-sd_KKBF}. BOM researchers employ various methodologies, including laboratory experiments, field studies, mathematical modelling, and computer simulation. Some of the broadly studied empirical phenomena include the bullwhip effect, coordination failures, the impact of incentives on supply chain performance \citep{Perera2020-gn_KKBF}, and the pull-to-centre effect in placing inventory orders under uncertain demand \citep{Schweitzer2000-qx_KKBF}. Key theoretical concepts include bounded rationality, heuristics and biases, and fast-and-frugal heuristics \citep{Katsikopoulos2013-ux_KKBF}. 

\subsubsection*{Theories}
Neoclassical economics and standard approaches to operations research and operations management have been assuming that individuals, groups, and organisations attempt and are able to maximise their subjective expected utility. However, since the 1950s, Herbert Simon (1955, 1956) (\citeyear{Simon1955-qb_KKBF,Simon1956-vu_KKBF}) -- Nobel prize winner in economics, and winner of the Turing award in computer science, occasionally wearing the hats of an operations researcher and organisation scientist -- and the many researchers influenced by him \citep{Viale2021-jf_KKBF,Gigerenzer2024-iv_KKBF} have been protesting that this assumption is a fantasy. Rather, the idea of bounded rationality suggests that agents must and do make decisions in the real world with limited information, time, and other resources such as computation. While Simon’s position is commonly framed as being that such practical decision making is always sub-optimal or clearly inferior to theory, it does not reflect Simon’s research program which sought to understand the conditions under which practice falls short of theory and under which it does not \citep{Gigerenzer2011-wk_KKBF,Katsikopoulos2013-ux_KKBF}.

Human bounded rationality has been captured by optimisation as well as heuristic mathematical models \citep{Katsikopoulos2023-sd_KKBF}. BOM applies both to tasks such as decisions under risk (i.e., choices between gambles), strategic interaction (i.e., games), operational decisions (e.g., inventory management), and more challenging decisions where uncertainty cannot be reduced to probabilistic risk \citep{King2020-oc_KKBF}. Examples of models include prospect theory \citep{Kahneman1979-qf_KKBF} and the priority heuristic \citep{Brandstatter2006-el_KKBF} for decisions under risk; inequity aversion theory \citep{Fehr1999-ew_KKBF} and the mirror tree \citep{Fischbacher2013-dw_KKBF} for strategic interaction; and prospect theory and the anchoring and adjustment heuristic \citep{Tversky1974-kf_KKBF} for inventory management. Optimisation models set up a utility or value function that reflects monetary and other considerations, such as perceptual distortion of probabilities (prospect theory) or aversion to monetary inequalities (inequity-aversion theory). To do so, free parameters might be needed, which enter relatively flexible mathematical forms. On the other hand, heuristic models may postulate the sequential processing of information without distorting values and probabilities (priority heuristic and mirror tree), or a simple additive form (anchoring and adjustment heuristic), typically employing just one or even zero free parameters. Empirical evidence suggests that both types of models have regions of superior performance on criteria such as predicting accurately out-of-sample or out-of-population, modelling cognitive processes, and being transparent to users \citep{Katsikopoulos2023-sd_KKBF}.

Can such models generate insightful explanations about supply chain and operational behaviours? Yes. For example, in the context of supply chains, prospect theory suggests that decision makers might overreact to demand fluctuations, amplifying the bullwhip effect. Behavioural agency theory, which examines how incentives influence decision making \citep{Pepper2015-fo_KKBF}, speaks to how aligning incentives among different stakeholders within a supply chain is crucial for coordination and performance. Misaligned incentives can optimise individual performance at the expense of overall efficiency. For example, if salespeople are rewarded based on short-term sales, they may push for larger orders, causing overstocking and increased holding costs. Behavioural game theories, such as inequity aversion theory, can help address how fairness, reciprocity, and altruism influence decisions \citep{Camerer2003-nc_KKBF,Fehr1999-ew_KKBF}. For instance, suppliers and retailers may engage in cooperative behaviour even when it is not the most profitable choice, driven by long-term relationship considerations. This theory helps explain why some supply chain partners maintain collaborative relationships despite short-term incentives to act opportunistically. Heuristics such as the mirror tree \citep{Fischbacher2013-dw_KKBF} can help understand why some decisions, such as a retailer rejecting a supplier’s offer, may take longer. Finally, the priority heuristic logically implies major empirical violations of the normative standard of expected utility theory, including the four-fold pattern of risk-attitude reversals, violations that cannot be all predicted at the same time by other theories \citep{Wakker1993-dr_KKBF,Katsikopoulos2008-az_KKBF}.

Finally, specific concepts can also be useful in BOM. One example is mental accounting which examines how people categorise and evaluate economic outcomes \citep{Thaler1999-fw_KKBF}. In supply chains, this can influence how costs and revenues are perceived and managed. For instance, managers might treat budget allocations for different departments as separate accounts, leading to suboptimal investment decisions that do not maximise overall supply chain performance.

\subsubsection*{Applications}
A notable BOM application is in inventory management \citep{Perera2020-gn_KKBF}. Research has shown that decision makers can improve performance by incorporating behavioural insights into demand forecasting and inventory policies. For example, retailers can use decision support systems that adjust for common biases such as the tendency to overreact to recent sales trends.

Another area is supply chain coordination \citep{Croson2014-bx_KKBF}. Behavioural studies have demonstrated that misalignment in goals and incentives between suppliers and buyers often creates issues. For instance, the beer game illustrates how lack of coordination and information sharing can exacerbate the bullwhip effect \citep{Sterman1989-oe_KKBF}. BOM interventions, such as aligning incentives and improving communication channels, have been effective in mitigating such issues. Additionally, collaborative planning, forecasting, and replenishment practices can help synchronise activities and reduce inefficiencies.

Behavioural approaches have also been applied to contract design \citep{Becker-Peth2013-yw_KKBF,Chen2019-ki_KKBF}. BOM highlights the importance of trust and fairness in negotiations. Studies have shown that contracts incorporating relational norms and fair profit-sharing clauses can enhance cooperation and performance \citep{Cao2015-mm_KKBF}. Firms can design contracts that include mechanisms for dispute resolution and performance monitoring, fostering long-term partnerships.

Case studies illustrate the impact of BOM interventions. In the retail sector, companies such as Zara have leveraged behavioural insights to optimise inventory and supply chain responsiveness. By understanding decision-making patterns, Zara has implemented agile supply chain practices that reduce lead times and align production with demand \citep{Sampath-M2024-wi_KKBF}. This approach not only improves inventory turnover but also enhances customer satisfaction by ensuring product availability.

In the healthcare sector, BOM has been instrumental in improving supply chain efficiency for medical supplies and pharmaceuticals. Behavioural interventions, such as enhancing communication and trust between suppliers and healthcare providers, have led to better coordination and reduced stockouts, improving patient care. For example, hospitals can implement inventory management systems that account for biases in ordering and usage, ensuring that critical supplies are always available.

The automotive industry also provides compelling examples of BOM ways of thinking. Toyota's production system incorporates principles of human behaviour, such as respect for people and continuous improvement (Kaizen). These principles have led to practices like just-in-time inventory management and lean manufacturing, which reduce waste and improve efficiency. The Toyota system practices have been interpreted as fast-and-frugal heuristics \citep{De_Treville2023-sp_KKBF}. 

The food and beverage industry has seen the implementation of BOM principles in managing perishable goods. Companies like Coca-Cola have optimised their supply chains by considering the urgency of replenishing stock and consumer purchasing patterns. By using analytics and behavioural insights, Coca-Cola can predict demand more accurately and adjust production schedules accordingly, minimising waste and ensuring product freshness.

\subsubsection*{Looking Ahead}
A criticism of the existing experiments is that participants are usually Western, educated, intellectual, rich, and democratic, commonly referred to as WEIRD \citep{Henrich2010-ic_KKBF,Ozer2014-zg_KKBF,Perera2024-em_KKBF}. This emphasises the need for more research incorporating diverse cultural contexts to ensure a holistic approach to BOM.

Researchers typically need ethics approval for conducting behavioural experiments. It is, however, argued that experiments should not mislead participants \citep{Katok2019-gm_KKBF}, although psychologists and economists sometimes disagree on this view \citep{Hertwig2001-wq_KKBF}. Emphasis should be placed on ethically acceptable supply chain contexts and avoid scenarios that might promote unethical behaviour. Crafting clear and unbiased cover stories for participants is essential to prevent misconceptions and ensure that data collection does not inadvertently bias participants' actions outside the experiment.

The behaviour of diverse actors within the supply chain should be understood, regardless of gender, culture, and background. For examples, studies can explore the behaviour of individuals and teams at various levels and echelons of the supply chain \citep{Ozer2014-zg_KKBF,Donohue2018-jk_KKBF}. 

As it should, sustainable decision-making is a key area which requires further investigation. Other emerging trends in BOM research include product development, retail supply chains, sourcing and procurement, healthcare supply chains, warehousing, and logistics.

BOM is an established, evolving field that addresses limitations of traditional models by incorporating human behaviour and asking its own research questions \citep{Fahimnia2019-sd_KKBF,Donohue2020-yr_KKBF,Katsikopoulos2023-sd_KKBF}. Research has generated invaluable insights on improving decision-making in supply chains and other operational settings, and there is every reason to expect that the field will help create more resilient, efficient, and ethical supply chains.

\subsection[Lean and agile (Elizabeth A. Cudney)]{Lean and agile\protect\footnote{This subsection was written by Elizabeth A. Cudney.}}
\label{sec:Lean_and_agile}
Supply chain management ensures the efficient flow of products, materials, services, and information from suppliers to customers, often through a complex network, directly impacting cost, quality, and customer satisfaction \citep{Le-Caous2021-bj_EC}. Organisations must balance efficiency with adaptability to stay competitive. Lean and Agile frameworks are widely used to address these challenges within supply chain management. Lean emphasises minimising waste and inefficiencies throughout the supply chain by focusing on continuous improvement efforts \citep{Holweg2007-ea_MKAC,Cudney2021-jy_EC}. Agile emphasises flexibility and responsiveness to change \citep{Amajuoyi2024-yl_EC}. Organisations use these frameworks together to provide a comprehensive approach to improving performance, customer satisfaction, and overall operational resilience \citep{Dahinine2024-wr_EC}, particularly in supply chain management.

Lean, rooted in the Toyota Production System (TPS), emphasises eliminating non-value-added activities or waste (muda) to deliver greater customer value. \cite{Womack1996-hj_EC} proposed a five-step thought process to lean transformation: identify value, map the value stream, create flow, establish pull, and seek perfection. These principles guide organisations to streamline processes, reduce costs, and improve lead times across production and supply chains \citep{Melton2005-gy_EC}.

Value stream mapping (VSM) is a key Lean tool that enables organisations to visualise the flow of materials and information through a process. Using a team-based approach, organisations use VSM to identify bottlenecks and inefficiencies, which are then systematically eliminated to improve operational performance \citep{Shah2007-lu_MKAC,Bhamu2014-md_EC}, which fosters a culture of continuous improvement \citep{Liker2004-gq_EC,Cudney2023-ml_EC}. This focus on operational excellence has proven effective in manufacturing and service industries, enabling organisations to reduce lead times, lower costs, and improve resource utilisation \citep{Cudney2011-ra_EC,Nunez-Merino2020-tj_TPIA}.

Lean principles are increasingly being applied beyond traditional manufacturing environments to service industries and healthcare, where waste reduction can significantly enhance service delivery and patient outcomes \citep{Santos2022-cj_EC}. A growing body of research has demonstrated Lean's impact on improving efficiency and customer satisfaction across sectors \citep{Mamoojee-Khatib2023-da_EC}.

Initially developed in the software development industry, Agile has been adapted for broader use in operations and supply chain management \citep{Rigby2016-rt_EC}. Agile frameworks, such as Scrum and Kanban, prioritise flexibility, iterative development, and rapid response to changing customer needs and market conditions \citep{Amajuoyi2024-yl_EC}. Agile's core values emphasise customer collaboration, individuals and interactions, and responding to change rather than following rigid processes or documentation \citep{Behrens2021-go_EC}.

In supply chain management, Agile provides a structure for organisations to respond quickly to demand variability, supply disruptions, or market shifts \citep{Palsodkar2022-us_EC}. For example, Kanban originated in Lean and is often used within Agile environments to manage workflow, track progress, and adapt tasks dynamically \citep{Zayat2020-np_EC}. Agile's iterative cycles, or sprints, facilitate continuous feedback, allowing teams to make quick adjustments and deliver incremental improvements \citep{Francisco2024-vz_EC}. This capability is particularly valuable in environments with high levels of uncertainty or fluctuating demand, such as during global supply chain disruptions like the COVID-19 pandemic \citep{van-Hoek2021-hh_EC,Ivanov2021-zn_TPIA,Kazancoglu2022-dr_EC}.

Recent research highlights the growing use of Agile practices in non-software industries, including manufacturing, healthcare, and retail, to improve operational flexibility \citep{Zuzek2020-cu_EC,Cano2021-rz_EC}. Agile methodologies are proving particularly effective in industries where rapid innovation and responsiveness are critical, such as technology and consumer goods \citep{Josyula2021-xb_EC}. As global supply chains become increasingly complex, Agile offers a valuable framework for managing complexity and ensuring resilience \citep{Kazancoglu2022-dr_EC}.

While Lean and Agile have distinct origins and focuses, their integration can significantly enhance efficiency and adaptability in operations and supply chain management. The Lean-Agile hybrid approach, commonly referred to as leagile, combines Lean's waste-reduction and process optimisation with Agile's adaptability and iterative development, enabling organisations to respond to changing market conditions while maintaining operational efficiency \citep{Piotrowicz2023-mt_EC,Silva2024-fs_EC}. By integrating Lean tools such as VSM with Agile's iterative cycles, organisations can identify bottlenecks, eliminate inefficiencies, and make continuous adjustments to meet customer needs \citep{Tripathi2021-pl_EC}.

For example, Lean's Just-in-Time (JIT) production system, which ensures that materials and products are delivered exactly when needed, aligns well with Agile's sprint-based development cycles. This synchronisation allows organisations to deliver high-quality products quickly without overproduction or excessive inventory \citep{Liker2004-gq_EC,Poppendieck2003-lv_EC,Gunasekaran2008-jk_EC}. Using cross-functional teams in both Lean and Agile further enhances collaboration and decision-making, enabling organisations to respond more effectively to changes in demand, production capacity, or supplier conditions \citep{Rigby2016-rt_EC,Naylor1999-cn_JGSD}.

Research also indicates that integrating Lean and Agile can significantly improve efficiency and customer satisfaction, as both frameworks emphasise customer value and continuous feedback \citep{Narasimhan2006-qt_EC}. This integrated approach is crucial in industries where speed, flexibility, and efficiency are critical to competitive success \citep{Power2001-qv_EC,Agarwal2006-oz_EC}.

Despite the benefits of integrating Lean and Agile, organisations may encounter challenges balancing Lean's focus on efficiency with Agile's emphasis on flexibility. Cultural differences between the structured, process-driven approach of Lean and the dynamic, adaptive nature of Agile can create friction within teams. Critical success factors for implementing Lean and Agile include collaborative relationships, trust, motivation, leadership capabilities, strategic management, human resource management, and organisational capabilities \citep{Virmani2018-xo_EC,Saini2019-np_EC}. Additionally, some organisations may struggle to implement both methodologies simultaneously due to differing operational requirements or resource constraints \citep{Power2001-qv_EC,Krishnamurthy2007-qq_EC}.

Organisations should adopt a phased approach to address these challenges, starting small and expanding as the integration matures. This approach involves initially implementing Lean and Agile in a few select areas or teams, allowing them to experiment, learn, and gradually align their practices. As the integration matures, more teams and areas can be brought into the Lean-Agile fold. Open communication and collaboration across cross-functional teams are essential to bridge gaps and ensure smooth integration \citep{Ding2021-jx_EC}. Tailoring the implementation to fit the organisation's and its customers' unique needs can also help mitigate potential issues \citep{Almeida2024-mr_EC}.

Lean and Agile frameworks are essential for modern operations and supply chain management. While Lean focuses on waste reduction and efficiency, Agile emphasises adaptability and responsiveness to change. Integrating these two methodologies enables organisations to streamline operations, reduce costs, and respond rapidly to market and customer needs. By combining the strengths of Lean's operational excellence with Agile's flexibility, organisations can achieve greater resilience, customer satisfaction, and overall performance.

\subsection[Circular economy (Kannan Govindan)]{Circular economy\protect\footnote{This subsection was written by Kannan Govindan.}}
\label{sec:Circular_economy}
A circular economy (CE) is a production and consumption model that aims to reduce waste and extend the lifetime of products and materials. The current 'take, make and dispose' economic model has serious negative impacts on the environment, human health, and social well-being \citep{Grossi2024-fc_KG,Pera2024-zi_KG}. These effects result in an unstable economy and the collapse of natural ecosystems, which calls into question human survival. As an alternative to this conventional economic model, circular economy has become a central focus for policy makers to promote cyclical thinking by closing the loop to minimise materials and energy consumption. With the closed loop concept, the value of the resources (including products, materials, and energy) can be preserved in the economy advantageously. The history and conceptual framework of CE was first introduced by \cite{Pearce1989-ux_KG} to study existing links between economic and environmental activities \citep{Prieto-Sandoval2018-sy_KG}. \cite{Boulding1966-zp_KG} discussed the closed loop concept by acknowledging the problems of limited availability of natural resources \citep{Merli2018-ue_KG}. Several other contributors were involved in creating the CE concept prior to \cite{Pearce1989-ux_KG}, as presented by \cite{Winans2017-du_KG}. With numerous benefits of implementing CE, scholars and practitioners became more interested in exploring its context, resulting in endless conceptualised interpretations. Such interpretations resulted in a wide range of definitions for CE; \cite{Kirchherr2023-ou_KG} revisited 221 definitions of CE, an expansion from their 114 definitions \citep{Kirchherr2017-gi_KG}. Although several definitions exist, a boilerplate definition was given by the Ellen MacArthur Foundation as ``an industrial economy that is restorative or regenerative by intention and design'' \citep{Ellen-MacArthur-Foundation2013-sj_KG}. Since there is no clear single definition of CE, concepts and theories of CE have been devised by various principles, including 3Rs (reduce, reuse, recycle), 6Rs (reuse, recycle, redesign, remanufacture, reduce, recover), industrial ecology, laws of ecology, regenerative design, cradle to cradle, blue economy, and biomimicry. In addition, the concept of CE has been developed with the political and social outlook of various global nations. China started adopting CE through a top-down approach, while other nations generated bottom-up policies. The difference between the approaches of different nations can be apparent even within their definitions. China defined CE as ``a general term for the activity of reducing, reusing and recycling in production, circulation and consumption'' \citep{Chinese-National-People-s-Congress2008-ap_KG}, whereas the EU chose a definition of ``where the value of products, materials and resources is retained in the economy for as long as possible and the production of waste is minimised'' \citep{European-Commission2014-co_KG,European-Commission2015-qy_KG}.

Despite these conceptual overlaps and wide range of interpretations, implementing CE in today's business environment is essential to balance the three pillars of sustainability (economy, environment, and society). Contrary to necessity, worldwide CE is still in the early stages of implementation, mainly focusing on `recycling' rather than other R's (such as reuse, reduction, and so on). To facilitate CE, researchers and practitioners are exploring various circular business models (CBMs) that decouple value creation from resource consumption. According to \cite{Linder2017-kv_KG}, CBM is defined as ``a business model in which the conceptual logic of value creation is based on exploiting the economic value retained in products after use in the production of new offers'' \citep{Lewandowski2016-tz_KG}. Better clarity about the design of such CBMs may result in better implementation of CE at both the macro and micro levels. The proposed CBMs should focus on three key areas: value creation, value transfer, and value capture \citep{Centobelli2020-nv_KG}. Although several CBM frameworks have been proposed in the literature \citep{Susur2023-np_KG,Abbate2023-vt_KG,Madanaguli2024-ma_KG}, a popular business model framework for CE was proposed by the \cite{Ellen-MacArthur-Foundation2015-zc_KG} as the `Regenerate, Share, Optimise, Loop, Virtualise, and Exchange' framework map (ReSOLVE). The CBM framework in the existing literature offers specific guidelines for different industry sectors, including supply chain, automotive, textile, mining, and medicine. Expansion of current CBMs is more focused on innovations among business models, whereas recent research is shifting from 3R and 6R strategies to focus on improved versions of CE, including 9R (Refuse, Rethink, Reduce, Reuse, Repair, Refurbish, Remanufacture, Repurpose, Recycle, and Recover) and 12R (above 9R’s and Research, Re-skill, Re-design, and Re-vision). Such improved CE concepts require innovation in business models, as discussed by \cite{Susur2023-np_KG}.  Business model innovations include themes of changing the existing linear business model with circular strategies, optimising existing CBMs through adding more CE strategies, and helping companies to produce the finest version of CBMs themselves from scratch.  

To create functioning CBMs and business model innovations, it is necessary to understand the basic metrics of CE as drivers/enablers/pressures, barriers/challenges, and practices. These metrics support practitioners to design appropriate CBM frameworks. Several studies focused on CE drivers \citep{Abdulai2024-vn_KG,Truant2024-gi_KG}, barriers \citep{Gonella2024-rt_KG,Nyffenegger2024-rd_KG} and practices \citep{Patra2024-xx_KG,Schoggl2024-qr_KG} with different application areas. Researchers group these metrics based on the context of the study. For example, \cite{Aloini2020-qo_KG} categorise the drivers of CE as institutional, economic, environmental, organisational, social, supply chain, and technological. \cite{Govindan2018-md_KG} categorise the drivers as politics and economy, health, environmental protection, society and product development, whereas the same study categories barriers as governmental issues, economic issues, technological issues, knowledge and skills, management issues, CE framework issues, cultural and social questions, and market challenges. In the CE literature, both the drivers and barriers are broadly categorised based on the sustainability pillars (economy, environment, and society) along with some management (stakeholders) and business elements (risk, supply chain, technologies). Based on the existing studies, it is difficult to conclude one potential driving force and barrier for CE implementation, because several connections exist among these drivers and barriers. Therefore, the potential drivers and barriers to CE implementation depend on the scope and nature of the case considered. Unlike CE drivers and barriers, practices are often associated with organisational performance \citep{Magnano2024-we_KG,Yin2023-kh_KG}. Several CE practices have been reported in the literature, and these practices are different for each industrial application. However, the overarching concepts of these practices follow the principles of design for 12R/9R, extended producer responsibility, industrial symbiosis, eco-industrial development, life cycle analysis, and nutrient recovery.

The performance related to CE practice is measured with indicators and different assessment models. Indicators have been considered one of the most important initiatives to communicate the effectiveness of CE implementation through the measurement and quantification of the process \citep{Saidani2019-kx_KG}. The existing literature proposes several circular indicators for measuring progress. For example, \cite{Saidani2019-kx_KG} proposed 55 circular indicators and \cite{De-Pascale2021-bu_KG} listed 61 circular indicators with three different levels of application (meso, micro, and macro). All indicators were set up based on ``R'' principles and dimensions of sustainability. Besides CE indicators, CE implementation performance is also assessed through various performance assessment methods. Mostly, these involve Life Cycle Analysis (LCA), Material Flow Analysis (MFA), Material Flow Cost Accounting (MFCA), Design for X (DfX), Design for Disassembly (DfD), Data Envelopment Analysis (DEA), Discrete Event Simulation (DES), Multi-Criteria Decision Making Methods (MCDM), and fuzzy approaches. Recently, these indicators and assessment methods were used to correlate CE practices with various organisational performances, including environmental performance, operational performance, and sustainability performance.

To improve performance through CE, recent studies \citep{Lu2024-dk_KG,Sahoo2024-de_KG,Liu2023-et_KG} began to integrate technologies within CE principles. Technologies identify useful methods to improve product lifecycles through real-life tracking and effective predictive analytics aligned with CE principles. Therefore, several studies argued that the effectiveness of CE implementation to improve organisational performance goes hand in hand with technology integration. Among various technologies, big data, Internet of Things (IoT), blockchain, and cloud computing are often used in the CE studies. Researchers began to explore various elements for the successful integration of Industry 4.0 technologies with CE, including studies of the challenges of digitisation-led CE \citep{Kannan2024-ap_KG,Costa2023-ox_KG} and digitisation-led business model innovations for CE \citep{Ranta2021-ok_KG}. With the abundant literature that has been published in the digital CE context, several researchers \citep{Sarc2019-kk_KG,Chauhan2022-tt_KG,Seyyedi2024-vk_KG} conducted a literature review on this digitisation-led CE, in which in-depth discussions have been made along with opportunities for future extensions.

Current research trends show that several studies are focused on business model innovation and technology integration with CE. However, the social dimension of sustainability is underexplored compared to economic and environmental aspects, which also receive limited attention \citep{Mies2021-ho_KG,Pitkanen2023-ut_KG}. The importance of consumer behaviour, a key CE stakeholder, is similarly undervalued. To achieve global sustainability goals like the SDGs, decolonisation, and net zero targets, integrating social, economic, and environmental dimensions, while considering stakeholder behaviour including leaders, regulators, customers, and media, is crucial. Digitisation is essential to enhance CE practices such as industrial symbiosis and accelerate progress towards these goals. To further strengthen the understanding of CE, it is worth reading the reports of \cite{Ellen-MacArthur-Foundation2013-sj_KG,Ellen-MacArthur-Foundation2015-zc_KG,Ellen-MacArthur-Foundation2015-lp_KG,Ellen-MacArthur-Foundation2016-rt_KG}. Other studies provide detailed discussions on definitions, taxonomy, business models, metrics, and digital revolutions happening with CE \citep[e.g.,][]{Geissdoerfer2017-oe_KG,Kirchherr2017-gi_KG,Ghisellini2016-nz_KG,Korhonen2018-om_KG,Murray2017-rl_KG,Lieder2016-wf_KG}.

\subsection[Environmental sustainability (Miriam Wilhelm \& Gerald Reiner)]{Environmental sustainability\protect\footnote{This subsection was written by Miriam Wilhelm and Gerald Reiner.}}
\label{sec:Environmental_sustainability}
Environmental sustainability in an operations and supply chain management context relates to the direct and indirect impact productive activities of a firm and its suppliers have on the natural environment. The World Commission on Environment and Development \citep{WCED1987-lv} defined sustainable development as ``development that meets the needs of the present without compromising the ability of future generations to meet their own needs''. This definition has been criticised for its all-encompassing scope, but it points to the evident inefficiency of our current products and production processes in their use of the planet's resources. In fact, the framework of planetary boundaries by the Stockholm Resilience Centre shows quite drastically that seven out of nine planetary boundaries are currently transgressed, making our planet increasingly uninhabitable for humans \citep{Richardson2023-qk_MWGR}. Because of these growing concerns, businesses are under strong pressure to minimise their adverse impacts on the environment by reducing the resources they use and the resulting footprint they leave behind. Primary activities contributing to their footprint are producing and transporting products \citep{Kleindorfer2005-au_MWGR}. 

The origins of environmental sustainability management in operations can be traced back to TQM (\S\ref{sec:Quality_management}) and lean production, as many of the tools and principles that apply to quality management are equally relevant for environmental improvements \citep{Corbett1993-no_MWGR}. The perceived synergy between quality management systems and environmental management was fuelled by the development of the international ISO 14000 standards that started in 1991 after the successful deployment of ISO 9000 standards \citep{Corbett2001-ea_MWGR,Pil2003-zu_MWGR}. In their seminal discussion of the parallels between TQM and environmental management, \cite{Klassen1993-gq_MWGR} point out that the ``cost of quality''\footnote{It should be noted that the concept of ``cost of quality'' is now considered outdated in the quality management literature.} includes both cost of defects and cost of prevention. Environmental costs similarly include costs related to pollution and pollution prevention.

Lean production (\S\ref{sec:Lean_and_agile}), and its core philosophy of eliminating waste in the process, has further contributed to the boundary expansion of TQM through the development of environmental management systems \citep{Corbett2006-tk_MWGR}. While waste originally focused on time, quality defects, and excess inventory, it is now being used effectively to reduce the use of natural resources and/or eliminate environmental waste \citep{Rothenberg2001-ah_MWGR,King2001-ls_MWGR}. \cite{Souza2012-kk_MWGR} lists seven types of waste that can be reduced through lean production, namely, overproduction, transportation, inventory, waiting time, motion waste, processing waste, defects.

The potentially synergistic perspective, as captured in the mantra ``lean is green'' \citep{Florida1996-ss_MWGR}, has formed the basis for almost all past research on lean production and sustainability. In one of the first studies on this relationship, \cite{King2001-ls_MWGR} find that lean production, as measured by ISO 9000 adoption and low chemical inventories, correlates with greater waste prevention and lower emissions. The other way around, improved environmental, health, and safety performance can aid plant-level productivity efforts \citep{Klassen2001-tp_MWGR}. However, subsequent research on the association between lean and green has produced inconsistent results \citep[e.g.,][]{Hajmohammad2013-zh_MWGR,Rothenberg2001-ah_MWGR}. For a review of the literature, see \cite{Garza-Reyes2015-rt_MWGR} and \cite{Dhingra2014-gb_MWGR}. \cite{Piercy2015-wo_MWGR} see the reasons behind the difficulties of exploiting the synergies between lean and green in a reductionist definition of lean as a mere JIT tool (\S\ref{sec:Lean_and_agile}). They argue that a full lean implementation, which includes changes in both internal operations and external supply chains, is more likely to reduce environmental impact. Thus, lean should be seen as a philosophy rather than a toolkit. Moreover, lean is not only limited to environmental benefits but can help to meet a wide range of sustainable outcomes, including supply monitoring, transparency, workforce treatment, and community engagement \citep{Piercy2015-wo_MWGR}. A later study by\cite{Distelhorst2017-bp_MWGR} confirmed that adopting lean production practices was associated with improved labour standard compliance in supplier factories, thereby contributing to ethical sustainability (\S\ref{sec:Ethical_sustainability}) as well. 

Beyond waste reduction, environmental sustainability also requires that companies track and measure their carbon footprint. The concept of corporate carbon footprint describes the total amount of direct and indirect Greenhouse gas (GHG) emissions that come from all company's activities \citep{Lee2012-tm_MWGR}. The Kyoto Protocol addresses six greenhouse gases: carbon dioxide, methane, nitrous oxide, sulfur hexafluoride, perfluorocarbons, and hydrofluorocarbons, all of which are measured as metric tons of their carbon dioxide-equivalent (CO$_2$) emissions. GHG emissions are further categorised into three groups or 'scopes' by the most widely used international accounting tool, the Greenhouse Gas Protocol. It distinguishes between emissions from one's operations (scope 1), emissions from purchased energy, heating, and cooling (scope 2), and emissions from any other supply chain process (scope 3).

Scope 3 emissions include indirect emissions from purchased products (upstream) and indirect emissions from sold products (downstream); see also \cite{Lee2012-tm_MWGR}. The latter includes the use and disposal of sold products at the end of life. According to the Carbon Disclosure Project, scope 3 emissions account for an average of three-quarters of a company’s emissions \citep{Carbon-Disclosure-Project2022-fx}. The importance of scope 3 emissions varies considerably by sector, and scope 3 emissions from energy-intensive industries are increasing faster than their scope 1 and 2 emissions \citep{World-Economic-Forum2021-id_MWGR}. Thus, it is not sufficient to consider direct emissions of a (manufacturing) company only, but it is important to consider the total emissions of the supply chain as well \citep{Jammernegg2018-sy_MWGR}.

However, only a few firms are currently accounting for scope 3 emissions. This will have to change as investors and regulators demand more disclosure. Starting in 2025, companies with European operations, including those headquartered elsewhere, will be required to report emissions across their supply chain, including scope 3 \citep{European-Commission2023-oi_MWGR}. One of the challenges is that there is currently no globally accepted standard for calculating carbon emissions, but different tools, frameworks, and methods co-exist in practice \citep{McKinnon2009-zr_MWGR}. For example, \cite{Wild2021-jt_MWGR} lists 18 different calculators commonly used in practice. 

A growing criticism regarding environmental sustainability in operations and supply chain management is, however, that companies typically focus on those areas of sustainability that are easier to measure and where metrics are quantifiable. This is best exemplified by Life Cycle Analysis (LCA), which is a prominent methodology to quantitatively assess the environmental impacts of goods and processes from ``cradle to grave'', that is, across the entire lifespan of a product and throughout the entire value chain (supply chain plus use and disposal phases); see  \cite{Hellweg2014-qy_MWGR}. LCA has traditionally focused on carbon emissions as a quantifiable metric but has often ignored impacts that are harder to measure or poorly understood, such as land use, impact on biodiversity, and human and ecotoxicological impact categories \citep{Finnveden2000-mm_MWGR}. A related criticism is that more emphasis has been on areas directly benefiting corporations, reflected in larger progress made in measuring and accounting for energy, water, waste, and packaging. Sustainability initiatives without direct cost savings have not advanced as far or as fast \citep[see also][]{Villena2021-au_MWGR}. Also, voices are getting louder that operations and supply chain management must go beyond a minimal harm approach. In other words, accounting for carbon along the supply chain, and reducing and preventing pollution through lean operations might not be sufficient to stop the transgression of planetary boundaries. Instead, operations and supply chain management must become regenerative by positively contributing to the social-ecological systems surrounding it \citep{Gualandris2024-ay_MWGR}. 

\subsection[Ethical sustainability (Devika Kannan \& Madan Shankar Kalidoss)]{Ethical sustainability\protect\footnote{This subsection was written by Devika Kannan and Madan Shankar Kalidoss.}}
\label{sec:Ethical_sustainability}
Ethical sustainability is a concept to identify an organisation's ability to understand and act on sustainable values and principles within the legal and moral norms of the given context. In contrast to other dimensions (economy, environment, and society) of sustainability, the ethical dimension has often been left out of the debate. Compliance with business ethics within the sustainability context improves brand value and reputation, resulting in good organisational performance and long-term financial gains \citep{Flores-Hernandez2020-pc_DKMSK}. If the taxonomy of ethical sustainability digs further, then 'ethics' can be defined as a practical philosophy that teaches us to practice certain habits and customs that lead to a better life. Within the context of sustainability, however, several researchers \citep{de-Bakker2019-mn_DKMSK,Hockerts2023-sz_DKMSK} define business ethics as the concept of evaluating moral ‘right’ and ‘wrong’. By understanding this morality, organisational leaders can make ethical decisions, and such ethical behaviour can improve customer satisfaction, shareholder loyalty, and employee satisfaction. Compared to sustainability, ethics in business is an older context. The earliest publication of business ethics can be seen from the early 1960s by \citep{Baumhart1961-gv_DK}, when the relationship between ethics and the values of business leaders was investigated. Several studies \citep{Donaldson1994-ee_DKMSK,Reidenbach1990-iy_DKMSK,Kolk2016-dv_DKMSK,Schneider2024-vc_DKMSK,Rydenfelt2024-dj_DKMSK} examine different scenarios of business ethics, and the role of sustainability began to gain momentum about thirty years after the Brundtland report was published in 1987 \citep{Khalili2015-fd_DKMSK,World_Commission_on_Environment_and_Development1987-zu_DPvD}.

Following this report, global nations began to explore different strategies to implement sustainability in their operations regardless of application area. Implementing sustainability or sustainable development involves multiple decision-making processes mostly involving human connection. To deal with such human connections and networks in the implementation process, issues of rights and interests must be addressed. The interpretation of ethical sustainability differs with contexts, applications, and cultures, which explains why no single definition is available in the literature. Some studies \citep{Vaupel2023-mv_DKMSK} classify ethical sustainability as a subset of social sustainability. This classification clearly demonstrates a need for a conceptual exploration of ethical sustainability with better definitions and different, more comprehensive thinking.

Despite the lack of clarity about definitions, organisations know that considering ethical sustainability in their operations can improve their reputation, extend their market reach, and make them more competitive \citep{Nicholson2019-fy_DK}. Therefore, practitioners and researchers have explored the ethical aspects of sustainability through empirical \citep{Baah2024-ci_DKMSK}, normative \citep{Becker2023-ct_DKMSK}, and conceptual \citep{Laasch2023-rr_DKMSK} methods. In the row, the business models \citep{Anshari2022-yg_DKMSK,Schaltegger2018-nt_DKMSK} and metrics for ethical sustainability have begun to be examined \citep{Raoult-Wack2002-el_DKMSK,Schaltegger2018-nt_DKMSK,Sharpe2019-fy_DKMSK}. A few studies have reported on drivers/metrics of ethical sustainability, but these works are compromised by limiting their focus to the single driver of `ethical climate'. According to \cite{Lee2017-co_DKMSK}, an ethical climate provides a collective ethical perception among employees, which influences each employee's attitudes and outcomes, along with the influence of internal ethical decision-making processes. Due to that fact, studies \citep{Lee2018-xu_DKMSK,Altahat2018-em_DKMSK} began to correlate the relationship between the impact of an organisation’s ethical climate and its sustainable performance.

Besides understanding the success factors/drivers/metrics, certain strategies were highlighted in the literature to implement ethical sustainability; corporate social responsibility (CSR) is one of the key strategies often associated with successful implementation of ethical sustainability. The definition of CSR can be found in \citeauthor{McWilliams2001-ss_DKMSK}'s (\citeyear{McWilliams2001-ss_DKMSK}, p. 117) work. \cite{Belas2022-jl_DKMSK} explore the relationship between CSR theoretical frameworks and the ethical factors on engineers' attitudes towards achieving sustainability. In business ethics, CSR offers more to ethical sustainability than other similar sustainable strategies, including circular economy (CE). 
 
Under ethical business development, several practices can be seen across the research, but for ethical sustainability, `ethical investment' is sought to be the most popular practice among researchers. Simply put, `ethical investment/responsible investment/socially responsible investment' is the investment made in ethical practices, such as the development of human welfare, regenerative capacities, community education, and investment in pro-environmental technologies and processes. Such investments are not constrained by profit motives. Several studies can be seen in the literature on different aspects of ethical investments; alternate terms such as responsible investments or socially responsible investments are also employed. Although several studies on ethical investments in general have been reported, only a few studies specific to ethical sustainability are available \citep{Richardson2009-zh_DKMSK,Becker2023-ct_DKMSK,Uddin2022-wh_DKMSK}. As discussed in the earlier parts of this chapter, ethical sustainability improves the organisation's reputation across value chain stakeholders. However, making these ethical sustainability practices visible to such potential stakeholders requires effective communication. There are several studies \citep{Parguel2011-eq_DKMSK,Boiral2019-yg_DKMSK,Hamrouni2023-cd_DKMSK} which provided specific guidelines for the preparation and publication of such an ethical communication report. In the literature, however, this reporting is often referred to as CSR reporting or sustainability reporting. Despite the reputation improvement from ethical sustainability reporting, organisations can occasionally experience a negative impact from activities commonly called ``greenwashing''. According to \cite{Bradford2007-ly_DKMSK} and \cite{Parguel2011-eq_DKMSK}, greenwashing is defined as ``tactics that mislead consumers about a company's environmental practices or the environmental benefits of a product or service''. Such greenwashing by companies makes the ethical reporting less significant and/or debatable. To understand more about the concept of greenwashing in ethical sustainability reporting, researchers \citep{Horobet2024-fi_DKMSK} began to look at various associated problems.

In addition to the discussions on ethical sustainability, one of the hottest topics that comes under ethical responsibility in recent years is ``digital responsibility''. After the intervention of Industry 4.0 technologies (such as blockchain, big data, IoT, AI) in a company's daily operations, there is an increased risk of misusing the collected data. Throughout the process of Industry 4.0 implementation in a company's activities, numerous data were collected from both people and materials involved. Sometimes, such data is too confidential to show publicly; therefore, the organisation or person responsible for handling such data should adhere to a strict ethical policy to protect this data from the public. Several studies \citep{Matarneh2024-cj_DKMSK,Cricelli2024-af_DKMSK,Santos2024-vh_DKMSK} admit that Industry 4.0 technologies are the main driver for the implementation of sustainability in any organisation, regardless of its application. This also includes the ethical dimension of sustainability, \cite{Hermann2021-by_DKMSK} listed several benefits of having AI-based equitable R\&D where injustice, inequalities, and discrimination were monitored and reported. However, such studies are  in the initial stages. A strong business model is needed to manage and optimise the integration of technologies in sustainability implementation by addressing all pillars of sustainability, including financial, environmental, legal, social, and ethical. Apart from the above-discussed topics, few attempts were made to explore the impact of ethical sustainability with the application context, including design \citep{Chan2018-st_DKMSK,van-Gorp2007-fw_DKMSK}, supply chain \citep{Agyabeng-Mensah2023-ub_DKMSK,Blowfield2000-qh_DKMSK}, and education \citep{Christensen2007-hj_DKMSK}.

As ethical sustainability is a relatively new and less explored topic of sustainability, there are several opportunities for future research. Among these, ethical leadership for sustainability is a topic worth exploring. Future studies can establish the relationship between the moral theory of business ethics with stakeholder theory and relational management theories. In addition to top leadership, it is necessary to understand employees’ perceptions on implementing and promoting ethical sustainability in companies. Legislation and policies remain the major obstacles for implementing ethical sustainability, so initiating new policies under circular economy and other sustainable supply chain approaches specific to ethical sustainability will help practitioners with their transition. Such studies will help researchers identify the impact of 'ethics of care' in sustainability implementation. In addition to leadership, there is room to analyse the behaviour of employees in an ethical climate or in a multidimensional organisational commitment. Further, this analysis can be expanded with an understanding of the connections between organisational sustainability performances. In terms of sustainability performance linked to ethical sustainability, no sustainability indicator currently focuses on ethical investments. There is a need for ethical sustainability indicators that can measure ethical investment footprints such as carbon footprints to measure environmental performance. Therefore, there is a need for regulatory and legal reforms to track and to improve ethical investment footprints. Communication of indicator outputs to different stakeholders through ethical sustainability reporting must be less biased. More studies are needed to check and identify ``greenwashing'' by the company during reporting. It is also important to study the relationship between sustainability ratings offered by external sources with ``greenwash'' activities. To identify such greenwash activities globally, Industry 4.0 technologies were integrated into the operations to increase transparency. To facilitate this process, a more open science approach is needed that will be limited to essential data but, at the same time, should make some potential data accessible to different value chain actors to improve collaboration opportunities and to move science forward. To balance these things, researchers need to develop more standards for digital responsibility such as certification as discussed in previous studies \citep{Anshari2022-yg_DKMSK,Lobschat2021-vk_DKMSK}. By addressing these existing gaps in ethical sustainability, it is possible to accelerate the process of achieving several global goals for sustainability, including SDGs, ESGs, net zero, and decarbonisation. To understand more about ethical sustainability, some key studies \citep{Parguel2011-eq_DKMSK,Schaltegger2018-nt_DKMSK,Richardson2009-zh_DKMSK,Consolandi2009-bl_DKMSK,Guerci2015-wn_DKMSK} highlight primary topologies such as ethics, sustainability, social sustainability, digital responsibility, and so on.

\subsection[Forecasting (Fotios Petropoulos \& Nada R. Sanders)]{Forecasting\protect\footnote{This subsection was written by Fotios Petropoulos and Nada R. Sanders.}}
\label{sec:Forecasting}
Forecasting is a critical aspect of operations and supply chain management, as forecasts are the foundation for all other business decisions. This includes decisions such as which markets to pursue, which products to produce, how much inventory to carry, and which sources of supply to use. Forecasts are the input to decisions in inventory management (\S \ref{sec:Inventory_management}), planning (\S \ref{sec:Planning}), new product and service development (\S \ref{sec:New_productservice_development}), sourcing and procurement (\S \ref{sec:Purchasing_procurement}), and project management (\S \ref{sec:Project_management}), among others. Whenever a decision is made it is always based on a forecast - whether a formal (systematic/statistical) or informal (judgmental) forecast. Forecasts aid operational, tactical, and strategic decisions within organisations and enable consideration of associated uncertainties. Poor forecasting results in incorrect business decisions and leaves the organisation unprepared to meet future demands. The consequences can be very costly in terms of lost sales and may even force a company out of business. 

It is important to differentiate between an \textit{actual forecast} -- an objective evaluation of how the future may unfold -- and a \textit{target} or \textit{goal} which is the desirable future position. In practice, it is advisable to start with an objective formal forecast and then plan towards achieving set objectives given that forecast. One of the most common approaches to forecasting involves the extrapolation of established patterns in sets of time-ordered observations that are recorded in fixed frequencies forming \textit{time series}.

A popular time series forecasting family of models is exponential smoothing (usually acronymised as ETS or ES). While originally developed to predict stationary time series (i.e., where there is no trend or seasonality), later developments allowed its extension to include additive or multiplicative forms of trend and seasonality \citep{Gardner2006-bv_FP,Hyndman2008-iu_FP}. Autoregressive Integrated Moving Average (ARIMA) is also popular for time series forecasting \citep{Box1976_SMD}. It involves regressing current values against past values and past errors once the time series has been rendered stationary through transformations and differencing. To select between the many available exponential smoothing models or ARIMA models, automatic selection criteria (such as information criteria that balance performance and complexity) have been proposed \citep{Hyndman2008_FP,Petropoulos2022REP_FP}. One challenge related to time series forecasting is the lack of observations needed to make such extrapolations. This problem is exacerbated by established Enterprise Resource Planning (ERP) systems which often limit the available history to only 3-4 years of data. 

Instead of simply using past values of the variable of interest to produce forecasts, regression-type approaches to forecasting are based on relationships across many variables. As an example, the forecast for the demand of a particular product (the variable of interest) could be a function of the types of promotion that the company is running in the next periods and the weather (the predictors or external regressors). Utilising variables that are ``internal'' to the company (such as promotions, marketing, price, availability, supply chain conditions) is straightfoward as accurate predictions for these predictors can be obtained; and such information can enhance the accuracy of the forecast of the variable of interest. Other ``external'' variables (such as macroeconomic indicators, competitors conditions) may be also used, but such predictors will be by definition harder to forecast. Popular regression approaches include ordinary least squares regression models, ordinal, logistic, Poisson, negative binomial regression models and generalised linear models. Extensions for ETS and ARIMA (namely ETSx and ARIMAx respectively) also allow for the consideration of such predictors.

Apart from formal/systematic approaches to forecasting, judgment can also be used in many facets of forecasting for operations and supply chain management. Reasons for using judgment include lack of data, expert information/knowledge not captured in hard data, and lack of expertise or systems to produce formal forecasts. However, when judgmental forecasting is practised, users need to be aware of associated biases, that include anchoring, availability, recency, hindsight, and confusion of noise with patterns \citep{Goodwin2017-eu_FP}. It is common that a hybrid approach is usually implemented: a formal forecast is initially produced (usually based on time-series forecasting approaches) and then judgmental adjustments are made on that formal forecast. Research has shown that such adjustments are common, usually upwards, and often have negative forecast-value added \citep{Fildes2009-tc_FP,Franses2009-ar_FP}. Finally, judgment may be also applied in other stages of the forecasting process, such as model selection \citep{Petropoulos2018-mt_FP} or model parameterisation. For overviews on the use of judgment in forecasting, see \cite{Lawrence2006-ey_FP} and \cite{fahimnia2020human_FP}.

Forecasting is also needed when deciding what new products/services to introduce, the need to build a new facility, expanding in new markets, and other long-term/strategic plans or technological aspects of forecasting. To this direction, suitable statistical tools include Bass and other diffusion models \citep{Meade2006-mc_FP}. Structured judgmental approaches that are based on (small) groups of experts are also popular and have shown to produce good results to this direction. Popular expert knowledge elicitation methods include the Delphi method \citep{Rowe1999-vs_FP}, structured analogies \citep{Green2007-ts_FP}, prediction markets \citep{Wolfers2004-mq_FP}, and interaction groups \citep{Van_de_Ven1971-vd_FP}.

One major lesson-learned through the research and practice of forecasting is that future patterns will never perfectly follow what happened in the past. In other words, identifying and applying a single “optimal” model may not be possible (or relevant) when dealing with the future. Instead, we can combine the forecasts from multiple models/experts. Forecasts combinations (even in their simplest forms) have shown to significantly improve forecasting performance and minimise uncertainty. \cite{Claeskens2016-hr_FP} offer a possible explanation on why the performance of forecast combinations is better than that of the individual forecasts. Forecast combinations can be applied across multiple formal forecasts or even to aggregate multiple judgments through concepts such as the ``wisdom of the crowds'' \citep{Surowiecki2005-ty_FP}. For an excellent review of forecast combinations, we direct the reader to the article by \cite{Wang2022comb_FP}.

Time series data are often organised in hierarchical structures that follow the operations and data structure of a particular organisation. For instance, in a retail setting, the bottom-level nodes of such an hierarchy could be the sales for each stock keeping unit (SKU) at each store; whereas aggregate levels could include total sales of each SKU across multiple stores, total sales of a single store, total/SKU sales within a region, sales for particular product categories, and others. Forecasts can be produced with the methods described for any/all levels of such a hierarchy. Implementing decision making using the forecasts of each level independently would lead to misalignments of decisions across the hierarchy as forecasts produced at different levels generally do not sum up. One way around this challenge would be for forecasts of a particular, single level to be propagated to other (upper or lower) levels via suitable summation or disaggregation. Alternatively (ideally), forecasts across many levels may be combined towards producing reconciled forecasts across the hierarchy. Such reconciled forecasts are usually more accurate and allow for aligned decision making. For an overview of hierarchical forecasting and forecast reconciliation, see \cite{Athanasopoulos2020-cx_FP} and \cite{Athanasopoulos2024-cn_FP}.

The products of the forecasting process could be point (mean) forecasts, prediction intervals (given a confidence level), probabilistic distributions, and/or scenarios. In any case, the forecasting process majorly benefits from ongoing monitoring of the forecasting performance on out-of-sample sets of observations. Towards this, performance measures can be used. Popular point-forecast measures include the mean absolute error (MAE), the root mean squared error (RMSE), the mean absolute percentage error (MAPE) which is very popular in practice, the mean absolute scaled error (MASE) which has better theoretical underpinnings \citep{HYNDMAN2006_FP,Koutsandreas2022_FP}. Popular measures of the estimation of uncertainty include the interval score (IS), the pinball score, the continuous ranked probability score (CRPS), and the energy score \citep{Gneiting2007-oj_FP}. Beyond performance measures, we also recommend the evaluation of forecasts on their utility in practice \citep{Yardley2021_FP}.

We direct the reader to the encyclopedic article by \cite{PETROPOULOS2022705_FP} for a comprehensive review of the forecasting field. The textbooks by \cite{HyndmanAthanasopoulos2021_FP} and \cite{FildesOrd2017_FP} offer detailed presentations of state-of-the-art methods and approaches to forecasting, while \cite{Vandeput2021-va_FP} presents a data-science perspective to supply chain forecasting. It is also worth-mentioning that the forecasting community is very active in the open-source software space; \texttt{forecast} \citep{Hyndman2022Rforecast_FP}, \texttt{smooth} \citep{Svetunkov2022smooth_FP}, and \texttt{hts} \citep{Hyndman2021hts_FP} are three notable packages for R statistical software. Finally, we cannot overlook the impact AI will have on revolutionising forecasting. AI will enable more accurate predictions through the analysis of vast datasets and real-time information, identifying patterns and trends that were previously undetectable. The predictive capabilities of AI will allow businesses to make more informed decisions, reduce risks, and optimise their strategies for future growth.

\subsection[Innovation (Juliana Hsuan)]{Innovation\protect\footnote{This subsection was written by Juliana Hsuan.}}
\label{sec:Innovation}
Innovation at the OM-SCM intersection entails management of technology and development of new products, services, and processes that link operational processes within the firm (e.g., R\&D, marketing, manufacturing, etc.) with the supply chain (e.g., suppliers and customers). Advancement of technologies, such as information and transportation technologies, has improved the speed, transparency and efficiency of transportation of goods globally. Emerging technologies, such as artificial intelligence, will inevitably prompt firms to innovate within and across the supply chain. Technological innovation within the firm can be push- or pull-, often realised from new product or service development (\S\ref{sec:Technology_management}). 

New product development (NPD) aims at the design of new products and/or refinement of existing products with respect to technological development of components. NPD decisions impact how suppliers are managed, be through strategic partnerships (e.g., early supplier involvement in design), or minimum involvement (e.g., outsourcing).  Outsourcing decisions in NPD lead to a certain degree of supplier-buyer interdependence, that can vary between arm’s-length relationship (e.g., off-the-shelf components) and strategic partnership (e.g., co-development of new components); see also \S\ref{sec:Outsourcing}. Process innovation typically relates to technology changes and adoption in production processes \citep{Carrillo2004-ko_JH,Lee2011-gb_JH}, administrative processes \citep{Kim2012-fy_JH}, and business processes \citep{Girotra2014-kk_JH}.

Anchored in the marketing literature, new service development (NSD) refers to the development of offering(s) not previously available to a firm’s customers, with focus on customer and market orientation, internal process organisation, and external networks \citep{Ordanini2009-ko_JH}. NSD includes topics, such as organising for NSD, NSD processes and stages, critical success factors and performance measurement, customer involvement, new service strategy, and new service design \citep{Papastathopoulou2012-cr_JH}. As new services are developed, the firm must continuously sustain its NSD competence \citep{Menor2008-na_JH}.

An innovation pursued by many firms is on product platforms, where a stream of derivative products can be efficiently developed, hence increasing commonality sharing among product variants \citep{Kim2017-tv_JH,Muffatto2000-ha_JH}. A notable example is Volkswagen Group that applies platform strategy to support its brand strategy for cost leadership (e.g., Skoda Oktavia, Seat Leon, and VW Golf) and differentiation (e.g., Audi TT and A3). The platform thinking is extended to VW Group’s manufacturing plants as well as service dealerships around the world. Adopting a platform strategy enables firms to develop customised products and services to gain from economies of scale and scope \citep{Magnusson2014-vx_JH}, often engaging in joint development of product and services \citep{Jagstedt2019-yl_JH}; see also \S\ref{sec:New_productservice_development}. Popular streams of literature on mass customisation \citep[e.g.,][]{Salvador2009-dc_JH,Fogliatto2012-pt_JH} and postponement \citep{Choi2012-io_JH} pose rich debate on the implications of NPD and NSD that are tailored to customer choices vis-à-vis optimal differentiation points in the supply chain \citep{Olhager2024-kr_DPvD}. 

Firms have to constantly innovate to stay competitive, and as such must consider how to revitalise their business models, by delineating how firms do business \citep{Zott2011-np_JH} and how new ideas and technologies are commercialised. Business model innovation paves the way for firms to acquire corporate transformation and renewal \citep{Demil2010-wt_JH}. Increasingly, many manufacturers are embarking on digital servitisation (DS) as a strategy to compete through value propositions that integrate products with development of services and software systems into the offerings \citep{Hsuan2021-mc_JH,Kohtamaki2019-we_JH,Vendrell-Herrero2017-pq_JH}. DS bridges a firm’s internal processes (e.g., NPD, NSD, manufacturing, IT) with the supply chain (e.g., value propositions and activities that go into the customer’s processes). Depending on the level of innovativeness, DS can be quite complex in terms of configuring products that are coupled with a wide range of services and software offerings. The question pounded by many manufacturers is on how to manage such complexity and how to strike the balance between innovation and standardisation, which is not trivial, as there are endless strategic options. While innovation prevents imitation from competitors, standardisation enables flexibility and gains from economies of scale and scope. As every manufacturer has its own strategic goals to stay competitive, there is no one-size-fits all strategy and there is no guarantee for success because balancing standardisation and innovation considers a wide range of tradeoffs that are specific to each manufacturer (\S\ref{sec:Servitisation}).

One approach to manage complexity is through modularisation strategies, where complex systems are decomposed into smaller pieces with well-defined interfaces so that each piece can be dealt with independently and yet operate together as a whole \citep{Baldwin2000-nb_JH}. Modularity enables customisation and increased variety of offerings \citep{Mikkola2003-sm_JH} through mixing-and-matching \citep{Sanchez1996-sx_JH}, commonality sharing \cite{Bask2011-ny_JH}, substitutability \citep{Mikkola2006-eg_JH}, upgrades \citep{Kamrad2013-rz_JH}, and more. Innovation can be achieved in various ways, such as in the form of modular innovation where it can be scaled up and substituted through upgrades of system families in platform-based designs. For example, by decomposing DS business model into product, service, and software architectures and respective resources, it is possible to determine the standard and specific offerings for creating the desired integrated configurations \citep{Hsuan2021-mc_JH}. Such process identifies the unique and generic resources required for innovation and standardisation respectively. 

Modularity can provide means for innovation and conceptualised as to how components can be put together for the optimal configurations. Through configuration, firms are able to get an overview of their capabilities and resources required (e.g., physical assets, operational, and human resources) vis-à-vis the competitive landscape and viability of supply chain resilience. Modular configurations typically are comprised of standard elements, hence low level of innovativeness. Conversely, integral configurations have high levels of innovativeness. As such, spectrums of modularity strategies exist along product (modular and integral), service (basic and advanced), and software (open and proprietary) dimensions, ranging from simplest (modular product, basic service, open software) to the most complex configuration (integral product, advanced service, proprietary software). 

Modularisation of physical products is conceptualised in terms of product architectures, ranging between modular and integral. Modular product architectures enable flexibility in platforms to create product variety, rapid introduction of technologically improved products as well as outsourcing decisions. Integral product architectures, on the other hand, aim at maximum performance and craftsmanship where knowledge sharing and interactive learning are enhanced \citep{Mikkola2006-eg_JH}. In modularisation of services, service architectures act as enablers of service agility, that is, to respond rapidly and effectively to changing market demands \citep{Voss2009-xz_JH}. Basic service systems typically offer generic services that can be offered to all customers cheaply (e.g., help desks, call centres, and service kiosks). Advanced service systems, by contrast, can be expensive as they are developed and tailored to specific customers’ needs (e.g., R\&D-oriented services, professional consulting, and high-tech remote monitoring services). Application of service modularity is increasing across industries, including, e.g., healthcare \citep{Vahatalo2015-np_JH,Peters2020-fz_JH,de-Blok2014-nj_JH}, tourism \citep{Avlonitis2017-hy_JH}, logistics services \citep{Bask2014-xy_JH,Yurt2023-lr_JH}, and education \citep{Sorkun2022-cr_JH}. In software systems, there are modular open digital architectures containing high data homogeneity and accessibility (e.g., blockchain platforms) as well as integral proprietary architectures characterised with heterogenous data and closed access (e.g., management of sensors systems). 

Innovation within a firm and across the supply chain creates research opportunities, especially those that are inter-disciplinary, both in methodologies and theoretical development \citep{Carrillo2015-fc_JH}. With increasing focus on sustainability, an increased effort is put on NPD and NSD on design-for-sustainability that has great focus on environmental and social responsibility, that inevitably fosters the emergence of sustainable supply chains. Such transitions entail the renewal or development of new business models, where innovation at OM-SCM interface provides the foundation for competitiveness. In NPD, design for sustainability has great focus on environmental and social responsibility, which links to sustainable supply chains (be linear or circular), where end-of-life products are recycled, remade, and reused \citep{ElMenshawy2024-ek_JH,Ulku2022-ec_JH}. It incorporates concepts such as design for environment, life cycle analysis, lean product designs, and design for disassembly (\S\ref{sec:Circular_economy}).

Innovation will continue to advance and shape our world beyond our imagination, and it is at the heart of all organisations. Its success (and failure) is dependent on how the supply chain is structured and how actors' relationships are managed (\S\ref{sec:Supply_Chain_Management}). Applying inter-organisational theories to NPD (transaction cost analysis, principal-agent theory, network theory, and resource-based view) adds rich insights into how to structure a supply chain collaborating organisations and how to manage a particular structure \citep[][\S\ref{sec:Relationship_management}]{Halldorsson2007-ci_JH,Halldorsson2015-pc_JH}. The book by \cite{Moreira2018-zp_JH} offers a collection of key topics on innovation and supply chain management. Refer to \cite{Baldwin2000-nb_JH} and \cite{Mertens2023-fc_JH} for the foundations of product modularity and a comprehensive review of this field, respectively. Finally, as interdisciplinary research continues to evolve in the field of servitisation, \cite{Kim2023-bf_JH} provides an overview of servitisation and product–service systems literature in Industry 4.0 and outlines a research framework for research on servitisation for innovation and circular economy.

\subsection{Technology management}
\label{sec:Technology_management}

\subsubsection[Adoption (Vaggelis Giannikas \& Ayse Begüm Kilic-Ararat)]{Adoption\protect\footnote{This subsection was written by Vaggelis Giannikas and Ayse Begüm Kilic-Ararat.}}
\label{sec:Adoption}
Technology adoption studies in the OM/SCM literature examine why and how technology is first adopted and -- often -- how it is subsequently diffused in an organisation (with the term `assimilation' used to describe both phases together). A simple, yet very useful way to categorise technology adoption research studies is by distinguishing whether they examine how an organisation (primarily companies but also non-business organisations) or an individual (employees in said organisations) adopt technologies. We will use this categorisation to structure the main part of this section.

At the organisation level, many OM/SCM researchers have used the innovation diffusion theory  \citep[IDT;][]{Rogers1962-zy_VGABKA}, a widely recognised theory which originates from sociology and seeks to explain how new advancements spread throughout a specific group. IDT has been applied in various operational contexts, such as in the adoption of supply chain service platforms \citep{Hong2021-bx_VGABKA}, and mobile supply chain management systems \citep{Chan2013-vr_VGABKA}, and the implementation of drone technologies \citep{Ali2024-kt_VGABKA}. IDT is very often used in conjunction with the TOE (technology-organisation-environment) framework, including in the examples listed above. TOE -- even though not technically a theory -- provides a lens to examine the different types of factors influencing an organisation's views towards adoption. In manufacturing, for example, it has been used to examine whether different determinants act as enablers or inhibitors to RFID adoption \citep{Wang2010-ni_VGABKA} or the adoption of AI-empowered industrial robots \citep{Pillai2022-dp_VGABKA}.

Another popular theory in relevant literature has been institutional theory; this is driven by the fact that organisations adopting technology operate in institutional environments that might influence their behaviour towards adoption \citep{Lai2006-lq_VGABKA}. \cite{Saldanha2015-ab_VGABKA} used institutional theory when studying the implementation of supply chain technologies to show that there is a need to reassess mental models developed in the West as they might not consider the institutional environments of emerging markets. Like IDT and TOE, institutional theory has also been used in conjunction with other theories, such as resource-based view, to demonstrate how firms can benefit from both external and internal technology adoption (and their interaction) to improve operational performance \citep{Zhang2009-xy_VGABKA} or with contingency theory to study how black swan external contingencies (like pandemics) impact technology adoption \citep{Tiwari2024-ps_VGABKA}. Contingency theory has also been used to investigate the moderating role of Industry 4.0 (I4.0) technologies on the relationship between lean production and operational performance improvement \citep{Tortorella2019-nu_VGABKA}. Finally, researchers have examined how learning and knowledge might influence adoption and how firms adapt and succeed in rapidly changing technological environments. \cite{Mishra2024-ym_VGABKA} conducted an in-depth case study, integrating dynamic capability and organisational learning theories, to explore how digital orientation supports the adoption of digital technologies for enhancing innovation.

Even though it is organisations that adopt the technologies, one must look into the behaviour of individual employees to understand better the technology adoption phenomenon \citep{Morris2010-fl_VGABKA}. The unified theory of acceptance and use of technology (UTAUT) proposed by \cite{Venkatesh2003-vc_VGABKA} is one of the most frequently used theories and researchers have used it -- among others -- to examine RFID adoption in the healthcare supply chain \citep{Yee-Loong-Chong2015-qv_VGABKA} and AI acceptance in SCM \citep{Hasija2022-le_VGABKA}. Recently, \cite{van-Dun2023-av_VGABKA} combined UTAUT and social exchange theory to explore the social enablers of employee acceptance of I4.0 technologies. Another established information systems theory, the technology acceptance model \citep[TAM;][]{Davis1989-kq_VGABKA} has also been used in operations and supply chain management; for example,\cite{Brandon-Jones2018-dz_VGABKA} used it to examine technology adoption in e-procurement. These popular theories have been used together with others to study technology adoption such as by \cite{Kamble2019-wu_VGABKA} who explored blockchain adoption in supply chains combining TAM and the theory of planned behaviour. Interestingly, there are cases where theories initially developed to study technology adoption at the individual level, are also used at the organisational one, such as \cite{Khan2024-fp_VGABKA}.

The studies listed so far in this section make use of established theories, that often come from elsewhere rather than the OM/SCM discipline. Due to the nature of the phenomenon under consideration and the strong focus on technology itself, many of the theories presented earlier were initially proposed or are extensively used by sociology and information systems scholars. However, OM/SCM scholars have occasionally examined technology adoption without the explicit use of a theory. In a recent example, \cite{Margherita2024-gq_VGABKA} studied the tensions that arise in Industry 4.0 adoption in lean production systems. \cite{Maghazei2022_CGS} investigated drone applications in warehousing, manufacturing and inventory management by looking at the relationship between adoption intention and the actual adoption. \cite{Leoni2022-eg_VGABKA} looked at how manufacturing firm performance is affected by AI, knowledge management processes and supply chain resilience as well as their mutual relationships. Another empirical example is on adopting data provision technologies to explore taxi drivers' routing decisions \citep{Lu2023-wx_VGABKA} showing that both operational efficiency and customer satisfaction can be improved using technology. Lastly, the impact of after-sales services on one's intentions to adopt a new technology has been explored empirically using models developed by \cite{Kundu2022-fk_VGABKA}. On the other side of the value chain, the OM literature has examined cases where technology is adopted for the design of new products rather than the management of relevant operations. For example, \cite{Singh2021-dm_VGABKA} studied the component selection and commonality problems to understand their impact on manufacturing and supply. 

At the same time, one can find recent studies which have chosen to navigate away from well-established theories, and apply less frequently used ones, to explore the role of technology adoption in managing operations and supply chains. For example, componential theory of individual creativity and valence theory -- relatively new theories to the OM/SCM area -- are used by \cite{Verma2022-ed_VGABKA} to explore employees' intentions to work in AI-based hybrid environments, concluding that human factors are as important as the benefits provided by the technologies for successful human-robot collaboration. In supply chains, where multiple actors are involved by nature, \cite{Sternberg2021-tt_VGABKA} used midrange theorising to understand why blockchain technology has seen very few successful implementations in interorganisational settings. Similarly, new supply chain specific theories, like the supply chain practice view, has been argued to be more suitable for the study of the adoption of digital procurement \citep{Kosmol2019-we_VGABKA}. \cite{Johansson2024-re_VGABKA} address contradictions between lean production systems and advanced digital technology adoption by using paradox theory. Finally, \cite{Liu2024-dw_VGABKA} developed a framework to investigate how supply chain resilience is affected by blockchain adoption, building on organisational information processing theory. They also highlight the role of transformational supply chain leadership in blockchain adoption.

Even though we focused on the `how' and `why' questions of technology adoption in our review, a noteworthy stream of literature is often concerned with the \textit{impact} of adoption, i.e. how adopting a technology will affect an organisation’s operations or supply chains. This can be done via ex-post impact studies (measuring the impact after solution deployment thus involving assessment of actual gains) or ex-ante impact studies (estimating the impact of technology adoption before the actual adoption thus providing insights into the expected gains). An example of an ex-post study is the work by \cite{Mcafee2002-hc_VGABKA} that aims to establish a link between technology adoption and subsequent operational performance improvement using a longitudinal case study. Similarly, an ex-ante approach is taken in \cite{Lee2007-so_VGABKA} where the value of RFID is measured via its impact on supply chain visibility. Since impact studies are –to their vast majority– context and technology dependent, the reader is referred to \S\ref{sec:Applications} for some examples.

We conclude by suggesting four research avenues technology adoption scholars might want to explore further in the OM/SCM discipline. Firstly, even though there exist theories that have proven to be suitable to study technology adoption phenomena over the years, the characteristics of some emerging technologies (see also \S\ref{sec:Emerging_technologies}) might make these theories less appropriate. As a result, new approaches and models to theorising might be required for a more detailed analysis of the subject matter. Similarly, researchers can explore the applicability of contemporary theories developed by academics in other social sciences and disciplines studying the impact of technology on organisations, especially those developed in sociology, psychology, economics and, of course, information systems. As technology advances, of particular importance is expected to be the role of people in the adoption and assimilation of different technologies, especially in those cases where human-technology collaboration is essential for the successful completion of a task. In OM settings in particular, where many activities and tasks take place in the physical world, the way people interact with technologies can be key to a successful adoption. Lastly, we have observed a lack of in-depth, longitudinal studies that cover both the pre-, during and post- adoption phases that would enhance our knowledge of the technology adoption phenomenon and provide a more holistic view of it.

\subsubsection[Emerging technologies (Omid Maghazei)]{Emerging technologies\protect\footnote{This subsection was written by Omid Maghazei.}}
\label{sec:Emerging_technologies}
Emerging technologies are often characterised as (radically) novel, fast-growing, and marked by high uncertainty and ambiguity \citep{Rotolo2015-ju_OM,Kapoor2021-dg_OM,Bailey2019-ey_OM}. Emerging technologies and their capabilities have long been at the forefront of attention for OSCM scholars\footnote{For instance, we refer to recent (2024–2025) calls for papers in leading OSCM and management journals on the impact of emerging technologies in sustainable humanitarian logistics (see \textit{International Journal of Production Research}) and healthcare (see \textit{Journal of Operations Management}). Special issues have also been published in the \textit{International Journal of Operations and Production Management} on emerging technologies and emergency situations \citep{Fosso-Wamba2021-dx_OM} and in \textit{Organization Science} on emerging technologies in organising \citep{Bailey2019-ey_OM}.}  \citep[e.g.,][]{Amoako-Gyampah1989-jy_OM}, as well as practitioners and consultants (the ``hype cycle'' framework developed by Gartner is a noteworthy example). Their strategic advantages, as well as their operational and economic impacts, have been extensively examined whenever a technology or a group of technologies emerges. For example, the advent of advanced manufacturing technologies (AMTs) in the early 1980s not only triggered a revolution in production but also offered OSCM scholars a strong basis for studying these then-emerging technologies in depth. Reviewing the existing literature on AMTs provides a reliable foundation for understanding their emergence and impact on industries. However, the unique characteristics of contemporary emerging technologies -- many of which are associated with ``Industry 4.0'' -- necessitate a reassessment of established concepts and theories.

Industry 4.0 technologies, such as additive manufacturing, advanced robotics, artificial intelligence, autonomous vehicles, blockchain, cloud computing, drones, and the Internet of Things enable companies to sense their environment, analyse information, collaborate within and across firms, and enhance task execution \citep{Mithas2022-zi_OM}. This section highlights some of the key differences between AMTs and Industry 4.0 technologies, focusing on strategic considerations, adoption barriers, and buyer-supplier relationships. It then examines the unique challenges of Industry 4.0 technologies and suggests reassessing methodological approaches to their study. Lastly, it briefly touches on what it means to classify a technology as ``emerging'', inviting reflection on the ontological and epistemological assumptions underlying these phenomena.

A review of publications on AMTs from 1986 to 1995 shows that strategic decision-making, planning, and economic evaluation were the most recurring themes in the early stages of research on AMTs, which were arguably emerging at the time \citep{Amoako-Gyampah1989-jy_OM,Maghazei2017-tq_OM}. One of the concepts that stands out in the early literature on AMTs emphasises adopting a ``strategic'' and ``systemic'' view\footnote{This concept is also reflected in the distinction between ERP ``systems'' and the ERP ``concept'', as discussed by \cite{Bendoly2005-jt_OM} and further detailed in \cite{Lewis2019-ox_OM}.} of planning and implementation \citep{Voss1986-xh_OM}. Similar strategic considerations are advocated for contemporary emerging technologies that can facilitate their integration and enhance their likely performance improvements \citep{Dohale2022-bv_OM,Venkatesh2024-ik_OM}. Given the technological interdependencies and ``convergence'' characteristic of the digital era \citep{Venkatesh2024-ik_OM}, such systematic approaches may be even more crucial for implementing Industry 4.0-type technologies in practice. Despite the importance of maintaining a clear strategic perspective on new technologies, observations from a specific category of emerging technologies, as documented by \cite{Maghazei2022_CGS}, reveal a ``shift'' from formal strategic and economic considerations to operational and supply chain concerns during early-stage assessments, with a focus on use case identification. 

A key challenge with emerging technologies, which have limited real-life implementations, is identifying or predicting adoption barriers before it becomes ``too late'' for practitioners. This challenge also applies to OSCM scholars, who, due to their ``process logic'' \citep[see][]{Lewis2019-ox_OM}, may overlook how an emerging technology unfolds beyond its initial phase. Although primarily ex-post analyses, past studies have well-documented barriers to adopting AMTs, such as short-term business strategies, poor strategic orientation, and incomplete risk assessment \citep{Lefley1994-ta_OM,Kotha2000-xb_OM}. Additional challenges include complex cost-benefit analyses, liquidity constraints, and the need to justify costs and identify economic rationales for the new technology \citep{Mithas2012-gf_OM,Mithas2012-ku_OM,Mithas2013-iu_OM,Sambasivarao1995-ix_OM,Spanos2009-bc_OM}. Barriers also arise when the selected technology misaligns with a firm's assets, including resources, capabilities, and competencies \citep{Meredith1987-nk_OM,Boyer1997-xc_OM,Sohal2006-hy_OM}. Other obstacles include inexperienced project managers or teams, insufficient support from top management, and a lack of technical expertise \citep{Bessant1985-sa_OM,Meredith1987-nk_OM,Dimnik1993-uj_OM,Sambasivarao1995-ix_OM,Small1997-mi_OM,Small1997-ug_OM,Co1998-fc_OM}.

While it is likely that many of these identified barriers remain applicable to contemporary emerging technologies, one challenge expected to intensify barriers to adopting emerging technologies and fully realizing their benefits involves uncertainty and the rapid pace at which these technologies are changing (cf. Moore’s law\footnote{Moore’s Law is the observation that the number of transistors on a microchip doubles approximately every two years, which enhances computing power and makes devices smaller and more affordable.}). \citeauthor{Bessant1985-sa_OM} (\citeyear{Bessant1985-sa_OM}, p. 98) argues that in situations where ``the rate of change in technology is increasingly fast'', the learning curve may lag behind, which can also exacerbate the ``problem of adaptation''. More broadly with a systemic view, ``the more profound and far-reaching the potential restructuring, the longer the time lag may be between the initial invention of the technology and its full impact on the economy and society'' \citep[][p. 91]{Lewis2019-ox_OM}. One way to overcome obstacles stemming from the fast-evolving nature of emerging technologies is exemplified by IKEA’s initiative to pilot and scale drone usage in its warehouses. They developed a new governance structure for experimentation, adopted a more dynamic approach to managing buyer-supplier relationships, and transitioned toward a use case-driven adoption model \citep{Netland2023-ty_OM}. This initiative has thus far resulted in 250 fully autonomous drones flying and counting inventories \citep{Kell2024-pa_OM}. 

One of the areas that is influencing the analysis of emerging innovations concerns with changes in buyer-supplier relationships. Traditional buyer-supplier relationship has been often characterised into two extreme of ``adversarial'' and ``collaborative'' modes \citep{Imrie1992-vs_OM}. The adversarial relationship is known as an exit, antagonistic, or arm's-length contractual relationship, whereas the collaborative relationship is referred to as a voice, cooperative, or obligational contractual relationship \citep{Gules1996-jn_OM}. Improving buyer-supplier relationships is identified as a success factor for implementing new technologies, as well as overall competitiveness of the company \citep{Lamming1986-cu_OM,Gules1996-jn_OM}. Evidence shows that new forms of buyer-supplier relationships are being exercised by companies to better accommodate and respond to emerging technologies’ characteristics and (re)structure new type of commitment, adaptability, flexibility, responsiveness, conflict management, appraisal mechanisms, and hardware-software integration. When the technological complexity and uncertainty increase, companies are more likely to increase collaboration with technology suppliers \citep{Abd-Rahman2009-nv_OM}. More recently, \cite{Kurpjuweit2020-fu_OM} advocate the implementation of new ``outside-in'' programs between startup suppliers and established firms and introduce a subset of such programs and name them ``startup supplier programs''. Startup supplier programs represent a new paradigm shift to see startups as potential suppliers as opposed to the traditional views that look at such collaborations through the lens of corporate venturing. 

In addition to the identified differences in established OSCM concepts, future research could further explore the ``dark side'' of emerging technologies, including their implications for behavioural operations, cybersecurity risks, sustainability impacts, user acceptance, and their effects on individuals and broader society \citep[see an exhaustive list of research questions for future studies in][]{Mithas2022-zi_OM}. For example, \citeauthor{Choi2022-qh_MALMA} (\citeyear{Choi2022-qh_MALMA}, p. 24) focus on human-machine conflicts in terms of worker welfare, privacy, security, health problems, and legal protection and (re)introduce sustainable social welfare ``to highlight that only by considering human welfare would the achieved social welfare be long-lasting''.

OSCM scholars may need to reassess their methodological approaches when grappling with uncertain, rapidly changing, and interrelated technologies \citep{Bailey2022-se_OM}. This need aligns with the growing interest among OSCM scholars in employing field experiments to analyse emerging technologies, such as \citeauthor{Kwasnitschka2024-qr_OM}'s (\citeyear{Kwasnitschka2024-qr_OM}) investigation of the effects of feedback using smartwatches, as well as field-testing—sometimes combined with prototyping—of technologies like live-streaming virtual reality \citep{Netland2023-ll_OM} and process mining \citep{Lorenz2021-tu_OM}. Furthermore, OSCM scholars could embrace multidisciplinary and multi-methodological approaches \cite[e.g., see][who combine empirical data with analytical modelling]{Bai2022-ys_OM} and reconsider new units of analysis \citep{Gaimon2017-il_OM,Venkatesh2024-ik_OM,Choi2022-qh_MALMA,Lewis2019-ox_OM}. 

After all, defining ``emergence'' presents ontological challenges due to the evolving nature, boundaries, and diverse applications of these technologies (e.g., drones in humanitarian logistics versus photography), their contextual variability (e.g., specific sectors or regions), and socio-technical aspects (e.g., technology in isolation or broader contexts). These characteristics may also create epistemological challenges in knowledge acquisition and validation, managing bias and uncertainty (often amplified by ``hype''), and balancing prescriptive versus descriptive knowledge. Furthermore, scholars' timing in engaging with such technologies and their perceptions of their emergence (e.g., perceived newness) can influence epistemological assumptions and inform different research designs. 

\subsubsection[Systems (Thanos Papadopoulos \& Imran Ali)]{Systems\protect\footnote{This subsection was written by Thanos Papadopoulos and Imran Ali.}}
\label{sec:Systems}
Information Systems (IS) refer to a set of technologies that span organisational boundaries and connect customers and supply chains \citep{Nunez-Merino2020-tj_TPIA}. They play a vital role in operations and supply chain management (OM/SCM), serving as the foundation for information, cash and physical flows to move within the supply chain processes \citep{Nunez-Merino2020-tj_TPIA}, as well as data-informed decision-making. In disruptions, IS could facilitate the continuity of these flows and create supply chain resilience. Therefore, the benefits of IS to OM/SCM could be summarised in coordination, enhancing production efficiency, responding to market fluctuations, adopting sustainable practices \citep{Kouhizadeh2021-wx_TPIA}, agility, resilience, and competitive advantage. IS applications used in OM/SCM include for instance, the early Electronic Data Interchange (EDI), Computer Aided Manufacturing (CAM), Enterprise Resource Planning (ERP), and more recently Internet of Things (IOT), Blockchain, Cloud Computing/ERP, and big-data analytics and artificial intelligence (AI); see also \cite{Gilchrist2016-gj_TPIA}, \cite{Toorajipour2021-xd_TPIA}, and \cite{Sharma2022-sn_TPIA}.

IS equip organisations with capabilities to manage diverse data types, ranging from structured information like inventory levels to unstructured data such as customer insights and real-time sensor inputs from IoT. The incorporation of advanced technologies -- including AI, Machine Learning (ML), Blockchain, and IoT -- into organisations and supply chains significantly improves decision-making, transparency, and operational efficiency \citep{Ivanov2021-zn_TPIA}. Such data influences critical decisions in areas such as production planning, inventory oversight, logistics, and customer service. To maximise the potential of data, organisations must utilise specific analytical techniques tailored to each dataset's characteristics. This might involve employing AI and ML for predictive analytics, using Blockchain for secure transactions, or leveraging IoT for immediate monitoring \citep{Papadopoulos2020-is_TPIA,Ali2023-qe_TPIA}. Effective systems management is essential for both operational and strategic choices; inadequate data analysis or mis-implementation of systems can result in inefficiencies, higher costs, and lost innovation opportunities. Therefore, organisations need to adopt appropriate advanced systems and associated data analysis methodologies to gain useful insights from the data and maintain competitiveness in an ever-changing business landscape.

Blockchain enhances security and transparency within supply chain management by facilitating decentralised systems crucial for tracking product authenticity. Analytical techniques such as hashing algorithms and consensus methods -- like Proof of Work (PoW) and Proof of Stake (PoS) -- are employed to uphold data integrity across blockchain networks \citep{Fahim2023-ed_TPIA}. This approach proves advantageous particularly in industries that require transparency and traceability since blockchain aids compliance efforts while verifying product authenticity.

The Internet of Things (IoT) generates massive amounts of real-time sensor data across various areas such as manufacturing and logistics. Connected devices facilitate the monitoring of machinery, shipments, and environmental factors. Stream processing solutions such as Apache Kafka and Flink allow for real-time analysis of this data while predictive analytics models help anticipate machine failures and optimise maintenance schedules -- thereby minimising operational disruptions \citep{Hernandez2020-uv_TPIA,Cakir2021-wc_TPIA}. In predictive maintenance scenarios, Markov models along with Hidden Markov Models (HMMs) analyse both historical trends and current sensor readings to enable proactive responses to equipment issues.

Time series data are essential for activities including demand forecasting, production scheduling, and inventory control. These datasets consist of sequential observations recorded at regular intervals typically managed through cloud-based Enterprise Resource Planning (ERP) systems where analytical models like Exponential Smoothing (ETS) and Autoregressive Integrated Moving Average (ARIMA) are commonly utilised \citep{Shmueli2016-yh_TPIA,Box1976_SMD}. For more complex forecasting requirements, hierarchical forecasting models can synthesise information from different levels, such as product categories or geographical areas, to maintain consistency throughout the supply chain \citep{Athanasopoulos2024-cn_FP}. Additionally, Vector Autoregression (VAR) models can analyse interdependent time series to enhance decision-making within interconnected systems \citep{Kaytez2020-bm_TPIA}.

Unstructured data, such as customer feedback or social media interactions necessitates, sophisticated analysis techniques like Natural Language Processing (NLP), which is often incorporated into Customer Relationship Management (CRM) systems. NLP methodologies such as Latent Dirichlet Allocation (LDA) alongside sentiment analysis reveal underlying patterns within extensive text datasets while evaluating customer sentiment; this information is valuable in refining product offerings and elevating service quality standards \citep{Zhou2023-cn_TPIA,Goodwin2017-eu_FP}. By capturing consumer preferences effectively through these models businesses can inform their marketing strategies accordingly.

In complex environments where multivariate data involving multiple simultaneous variables are prevalent -- a common occurrence in Product Lifecycle Management (PLM) and Supply Chain Management Systems (SCMS) -- analytical techniques such as multivariate regression or logistic regression are frequently used to predict outcomes based on relationships between independent variables. For example, multivariate regression could forecast product quality influenced by factors like temperature or material type. Additionally, Principal Component Analysis (PCA) -- a technique designed for dimensionality reduction -- simplifies large datasets by isolating key influencing factors which aids process optimisation \citep{Jolliffe2002-ig_TPIA}. Linear programming mixed-integer programming models also play a crucial role when optimising resource allocation transportation planning within supply chains ensuring effective management constraints surrounding capacity cost \citep{Bertsimas1997-wx_TPIA}.

Probabilistic modelling serves an important function when addressing inherent risks and uncertainties present in operation supply chains. Monte Carlo simulations are commonly employed to simulate risks associated with variable conditions, for instance in fluctuating demand supplier disruptions \citep{Rubinstein2016-xi_TPIA}. Bayesian Networks provide another probabilistic framework mapping risk factor relationships thereby assisting organisations develop robust contingency plans mitigating possible impacts on supply chains. These approaches prove invaluable in dynamic risk assessment-informed decisions reducing potential disruptions.

Each type of dataset -- whether it be time series, unstructured, sensor-generated, multivariate, or probabilistic -- demands specific analytical methods reinforced by suitable technologies and relevant modelling approaches. Time series datasets find efficient management through ERP systems utilising ARIMA hierarchical forecasting; unstructured datasets undergo processing via NLP sentiment analysis integrated CRM frameworks; IoT-generated sensor outputs benefit from real-time analyses employing predictive modelling stream-processing infrastructures; multivariate sets leverage regression PCA optimisation strategies; finally probabilistic risk assessments utilise Monte Carlo Bayesian Networks. By fusing these methodologies IS enterprises and SCs can enhance their decision-making capabilities, mitigate risks, and bolster their overall operational efficacy/efficiency, and performance.

We would refer the reader to the seminal articles by \cite{Gunasekaran2004-qs_TPIA} for an early but comprehensive review of the IS role in OM/SCM, and the later articles by \cite{Daneshvar-Kakhki2019-qf_TPIA}, \cite{Dutta2020-tf_TPIA}, and \cite{Sharma2022-sn_TPIA} for an overview of systems on OM/SCM. 

\subsection[Performance measurement and benchmarking (Pietro Micheli)]{Performance measurement and benchmarking\protect\footnote{This subsection was written by Pietro Micheli.}}
\label{sec:Performance_measurement_and_benchmarking}

Performance measurement and benchmarking are critical management processes, as they help organisations evaluate their efficiency, productivity, effectiveness, and overall success in achieving their strategic goals. In this chapter, we first review the main principles and tools, and then conclude by outlining the main challenges in their application. 

\subsubsection*{Performance measurement}
Performance measurement involves the systematic collection, analysis, and reporting of information related to attributes of individuals, teams, processes, and organisations \citep{Micheli2014-wq_PM}. For example, performance indicators can be deployed to assess the motivation of individual employees, the productivity of a team or function, the efficiency of a process, the financial results of an organisation, and the environmental impact of a network of firms. Measurement instruments range from highly standardised, quantitative performance indicators and targets to context-specific, qualitative case studies \citep{Beer2022-sb_PM}. Often, these tools are part of multi-dimensional performance measurement systems, such as the Balanced Scorecard \citep{Kaplan1996-sh_PM} and the Performance Prism \citep{Neely2002-rl_PM}. 

The first measurement instruments focused on financial and operational aspects of performance such as profitability, cost, efficiency and quality. Subsequently, performance measurement systems began to include also aspects related to resources (e.g., intellectual and social capital), and customers and stakeholders more widely \citep{Neely2002-rl_PM}, sometimes leading to the creation of measurement systems across firms. More recently, organisations have started to capture also aspects related to social and environmental outcomes \citep{Searcy2016-ae_PM}. This increase in scope has made performance measurement more prominent within and between organisations; this has also been facilitated by the much increased availability of data thanks to technological advances and the ability for organisations to collect data using a variety of sources. However, it has also led to challenges, as further discussed in the latter sections of this chapter.

Research has shown that the implementation of performance measurement systems can lead to several positive consequences. These include monitoring core processes, correcting deviations from intended trajectories and results, identifying strategic opportunities and new ways of operating \citep{de-Leeuw2011-gl_PM,Koufteros2014-vq_PM}, as well as improving alignment and coordination \citep{Mura2018-ef_PM} and linking strategy with value creation \citep{Melnyk2014-ei_PM}. In the context of inter-organisational networks, the introduction of a supply chain measurement system can lead to operational performance improvement, lower information overload, and higher supply chain integration \citep{Maestrini2018-ok_PM}, and using collaborative performance indicators to align partners in presence of performance-based contracts has a positive effect on performance \citep{Akkermans2019-mf_PM}. At the same time, measurement instruments have been found to negatively impact employee motivation \citep{Malina2001-nb_PM} and lead to gaming and an obsession with hitting targets \citep{Franco-Santos2018-jv_PM,Melnyk2010-dl_PM}. This is particularly pronounced when considering short-term targets such as quarterly or yearly ones, for example, in relation to sales (as discounts may be applied to hit quantity targets, although negatively impacting profitability). Similarly, gaming can be used to explain the lack of investment in longer-term projects, which would not provide positive results in a short time frame, such as the one to which financial rewards are tied \citep{Gray2014-nf_PM}.

These conflicting results have led researchers to explore the dynamics triggered by the design, implementation and use of performance measurement systems \citep{Pavlov2011-jn_PM}. For example, at the micro level, recent studies show how these systems can both reinforce and challenge perceptions of meaningful work by influencing whether individuals can successfully carry out their work tasks, identify the impact of their work on service users, and have a valued voice in their interactions with colleagues \citep{Beer2022-sb_PM}. At the macro level, authors have challenged the traditional performance measurement paradigm founded on control, objectivity and predictability \citep{Bititci2015-jt_PM,Pavlov2023-uz_PM}, and proposed alternative ones rooted in complexity theory and that explicitly take into account the increasing dynamism that characterises organisational environments. In relation to management practice, this means considering performance measurement more as a way to influence, support and provide direction, rather than to control, determine, and predict. 

\subsubsection*{Benchmarking}
Benchmarking is the process of comparing an organisation's performance, processes, products or services against those of competitors or industry leaders, as well as comparing the performance of units, teams and individuals within a single organisational context  \citep{Dattakumar2003-uq_PM}. As such, benchmarking is often supported by performance measurement tools, such as key performance indicators (KPIs), and results in the definition of performance targets and, more broadly, in the identification of particular areas of strength (that may act as the source of an organisation’s competitive advantage) or weakness.

Benchmarking can be classified into several types, each serving different purposes \citep{Anand2008-ve_PM}:

\begin{enumerate}[noitemsep,nolistsep]
\item Internal benchmarking: Comparing performance across different departments or units within the same organisation. This type helps in identifying internal better / best practices that can be replicated in other areas.
\item Competitive benchmarking: Comparing performance with direct competitors. It focuses on evaluating how well an organisation is performing against its rivals in the market.
\item Functional benchmarking: Comparing similar functions or processes with those of companies outside the direct competitive landscape, often in different industries. This type aims to find approaches and practices that are new to a specific industry context.
\item Generic benchmarking: Comparing operations and processes that are common across industries. It helps in identifying universal good/best practices that can be applied irrespective of the industry. Some are typically related to typical performance attributes (e.g., operational efficiency), whereas others are not (e.g., the design characteristics of supply chains affecting intrinsic resilience of systems).
\end{enumerate}

\noindent Whatever the type, the deployment of benchmarking processes typically has two main consequences:
\begin{itemize}[noitemsep,nolistsep]
\item Target setting: comparing with other organisations and identifying best performers (often indicated as ``the benchmark'') enables the definition of performance targets, intended as goals that are in line with market conditions and expectations.
\item Performance improvement: by comparing with competitors and industry leaders as well as with other functions, organisations can learn what works well and adopt new practices that can lead to significant improvements in performance.
\end{itemize}

The main intent of benchmarking is therefore to learn from others (internal and external to the organisation), improve and innovate processes, and enhance products and services, which often leads to improved customer loyalty and satisfaction, and profitability.

\subsubsection*{Challenges in performance measurement and benchmarking}
Despite their benefits, performance measurement and benchmarking come with several challenges:
\begin{enumerate}[noitemsep,nolistsep]
\item Identification of what to measure (so called measurand) and/or benchmark on: while some aspects may be widely established and standardised across organisations (e.g., how to calculate operational costs or greenhouse gas emissions), many measurands are defined differently depending on context or approach adopted (e.g., organisational culture, innovativeness). Importantly, measurands do not have to be related to tangible results, but also to structural (design of supply chains, as mentioned above) and perceptual aspects (e.g., organisational reputation). When measuring and comparing performance, it is essential that the definition of measurands is as unambiguous and consistently adopted as possible, although this may not be always possible or appropriate \citep{Gray2014-nf_PM}. 
\item Data quality and availability: Obtaining accurate, relevant, and timely data can be difficult. Poor data quality can lead to misleading conclusions and ineffective decision-making.
\item Cost and resource intensity: Implementing a comprehensive performance measurement and benchmarking system can be resource intensive, requiring significant time, effort, and financial investment.
\item Maintaining relevance: performance measurement tools and benchmarks must evolve with the organisation and industry. Indicators, targets and benchmarks that were relevant in the past may no longer be applicable and should therefore be regularly re-assessed and modified.
\item Unintended behavioural consequences: perhaps the main challenge in measuring and comparing performance is related to the negative effects that these processes may have on individual and collective behaviours. The literature on this subject is vast \citep[see, e.g.,][]{Bevan2006-rx_PM,Franco-Santos2018-jv_PM,Gray2014-nf_PM}, but the risks of gaming, tunnel vision, and measure fixation are still very high especially when there is pressure on the achievement and demonstration of results (vs. the adoption of a specific practice).
\end{enumerate}

\subsubsection*{Conclusions}
Performance measurement and benchmarking are essential tools for organisations striving to understand and improve their processes, products and services. While challenges exist, the benefits outweigh the difficulties, making performance measurement and benchmarking indispensable components of modern management practices. In this chapter, we have focused on extant research and practices: the sources cited are a good starting point to appreciate the richness of the work undertaken so far. In the future, these processes are likely to become even more relevant, as the success of new technologies – including business analytics and AI-based algorithms – will depend on the quality of information generated by measurement systems and on organisations' capacity to effectively compare with others and adopt new practices. Moreover, behavioural effects, both positive and negative, are likely to be more pronounced as measurement becomes more widespread and includes aspects increasingly related to human motivation, perceptions, and performance (see, e.g., so called ``Industry 5.0'').

\subsection[Risk management and resilience (Miranda Meuwissen \& Maximilian Koppenberg)]{Risk management and resilience\protect\footnote{This subsection was written by Miranda Meuwissen and Maximilian Koppenberg.}}
\label{sec:Risk_management_and_resilience}
Risk can be defined as the probability of an event multiplied with the impact of that event, where risks can have positive and negative implications \citep{Aven2016-wm}. We distinguish here between risks within an organisation, i.e., internal risks, and risks emerging in the supply chain and the wider environment of the organisation, i.e., external risks. Literature mostly assumes that risks are generally calculable. For instance, \cite{Chopra2021-lr_JGSD} illustrate implications of supply risks for inventory planning (see also \S\ref{sec:Inventory_management}, \S\ref{sec:Warehousing}, and \S\ref{sec:Planning}) based on data from the past. Recent advances are the combination of historical and real-time data to further optimise stochastic supply chain processes \citep{Wu2023-gf_MMMK}. Traditional risk management strategies often aim at preventing negative implications of adverse high-impact events, i.e., when negative risks materialise. An example are insurances which only pay out in case of an event, or contracts with suppliers that define repercussions in times of supplier failure. In the past decades, companies have built complex global supply chains to benefit from comparative advantages, e.g., due to lower labour costs in developing countries, which has, however, introduced new risks because the geographical scope of supply chains has tremendously widened and companies have outsourced substantial parts of their production processes \citep{Thun2011-im}. With increasing levels of uncertainty due to new types of risk and accumulating challenges, the nature and severity of disruptions are becoming more difficult to predict -- especially for external risks -- and standard risk management practices no longer suffice \citep{Finger2022-tt_MMMK}. This leads to a shift from risk management to resilience thinking. A resilient system is able to ensure the provision of its functions, even in the face of increasingly complex risks and uncertainties \citep{Meuwissen2019-gv_MMMK}. 

In a `makeable' world, which is harmonious and calculable, resilience is generally interpreted as robustness – aiming to keep systems in their status quo situation despite the occurrence of shocks. It is thus assumed that resilience can be `engineered' \citep{Ge2016-aa_MMMK}. However, in the current risk environment, the interpretation of resilience moves beyond this narrow view of robustness and includes capacities to adapt and transform operations and supply chains as a means to deal with the increasingly complex shocks and stresses \citep{Meuwissen2019-gv_MMMK}. Adaptation refers to relatively small changes, e.g., in inputs, production and marketing. In contrast, transformation involves fundamental shifts in structures, processes, relationships, behaviours and values \citep{Termeer2019-uy_MMMK}. 

Which adaptations and transformations should businesses and supply chains aim for in the current risk environment? Here, the resilience attributes give guidance \citep{Meuwissen2022-kt_MMMK,Reidsma2023-ru_MMMK}. They are rooted in the generic principles of resilience as proposed by the Resilience Alliance (2010). The premise is that systems which are grounded in resilience attributes, such as redundancy and diversity of resources, have the individual and collective competences to respond, adapt and transform to deal with a multitude of economic, social and environmental disruptions \citep{Meuwissen2019-gv_MMMK}. \cite{De-Steenhuijsen-Piters2022-vw_MMMK} summarised the resilience attributes as ``ABCD'' -- representing agency, buffers, connectivity, and diversity. Developing resilience attributes in businesses and supply chains proves difficult due to various factors, such as human mental models which tend to focus on maintaining status quo, experts being educated mostly in improving efficiency, and a series of vested interests creating lock-in situations \citep{Meuwissen2020-sc_MMMK}. Besides, there is additional complexity as incorporating resilience attributes also requires innovations in institutions, governance mechanisms and other systems of accountability, as well as changes in culture, individual behaviour and technology \citep{Ingram2023-wv_MMMK}.

Resilience at the operational (internal) and supply chain (external) level shows important interactions \citep{Dubey2021-yw_MMMK,Kamalahmadi2016-br_MMMK} where a company can increase its overall resilience either through making its operations and/or its supply chain more resilient. At the same time, resilience of operations can also be enhanced through improvements in the resilience of the remaining supply chain, thereby creating synergies between operational and supply chain resilience. Take the example of sourcing (see also \S\ref{sec:Purchasing_procurement}) and the resilience capacity of robustness. On the operational level, companies keep certain levels of inventories of raw materials as a buffer to ensure that operations will not be interrupted in case of a supplier failure \citep{Christopher2011-wv_MMMK,Ivanov2021-an_MMMK}. On the supply chain level, a high diversity of suppliers can have a similar effect where other suppliers of the same raw material may compensate the failure of one supplier \citep{Yang2012-eu_MMMK,Yu2009-jr_MMMK}. Higher diversity on the supply chain level will thus positively affect the robustness of operations as well as the resilience of the company overall. 

In addition to these synergies however, two important considerations need to be taken into account. First, improvements in resilience attributes (agency, buffers, connectivity and diversity) often involve trade-offs that must be considered by a company. These include costs and changes in risks. Coming back to the procurement example, higher inventories of raw materials imply higher costs \citep{Carvalho2012-vj_MMMK}. Diversifying the supply base will, however, also entail changes in the average price paid in procurement which impacts costs \citep{Burke2007-xj_MMMK,Heese2015-ne_MMMK,Yu2009-jr_MMMK}. Further, there is often a trade-off in the diversification of the supply base \citep{Yang2012-eu_MMMK,Yu2009-jr_MMMK}. In single sourcing, companies usually rely on the preferred supplier with the best prices, highest quality or highest reliability. When adding new suppliers to the supply base, the new supplier(s) will often perform worse in at least one of the three dimensions (price, quality and reliability) than the preferred supplier \citep{Yang2012-eu_MMMK,Yu2009-jr_MMMK}. Quantitatively evaluating the trade-offs regarding costs and changes in risks between different options to enhance resilience on the operational and the supply chain level is thus important in the optimisation of a company’s overall resilience (see also \S\ref{sec:Performance_measurement_and_benchmarking}). Second, for very high impact events at the supply chain level, operational resilience may not necessarily be able to compensate a lack of supply chain resilience \citep{Ivanov2021-an_MMMK}. For instance, automotive companies suffered a tremendous shortage in semiconductors in 2020 and beyond as the production for semiconductors was concentrated geographically to a few suppliers, which were affected by lockdowns \citep{Ramani2022-mk_MMMK}, and automotive companies had designed their supply chains in a lean and transactional way (see also \S\ref{sec:Lean_and_agile}) with usually only one or few suppliers of a given part \citep{Carvalho2022-yw_MMMK}. Even though certain levels of inventory generally need to be kept either at the automotive company or the input supplier \citep{Carvalho2022-yw_MMMK}, these could not compensate the disruptions which lasted for several months in the semiconductor industry \citep{Ramani2022-mk_MMMK}. The example illustrates that it is important to think beyond robustness and into adaptability and transformability of a business. As this requires commitments for the mid- to long-term future, sophisticated methods to identify the core areas for developing adaptability and transformability capacities are necessary to guarantee an efficient use of resources (see also \S\ref{sec:Forecasting}). 

With regard to the measurement of resilience, two approaches can be distinguished. The first considers the resilience of system \textit{outcomes}, i.e. whether system outcomes recover from shocks. Most of this work is quantitative. For instance, \cite{Lesk2016-qd_MMMK} quantify recovery likelihoods of global food supplies after extreme weather events. \cite{Slijper2022-ou_MMMK} estimate recovery rates of income after income shocks. Next to such empirical analyses, there are normative studies, such as the paper by \cite{Van-Voorn2020-gj_MMMK}. These authors develop an agent-based model to assess how shocks influence performance of a food supply chain. A shortcoming of the `outcome-based approach' is that shock-induced system adaptations and transformations, resulting in, e.g., the production of other outputs, remain unnoticed. Even more, it would be concluded that a system is not resilient as it does not recover to its prior outputs and output levels. However, the system might well be more resilient for future shocks.   

The second approach to measure resilience investigates the systems themselves, i.e. how do systems respond to shocks and challenges. This strand of literature is mostly based on participatory approaches. For instance, resilience capacities (robustness, adaptability, transformability) are studied through the scoring of statements \citep{Slijper2020-ox_MMMK}, biographical narratives \citep{Nicholas-Davies2021-xb_MMMK}, focus groups with multiple supply chain actors \citep{Soriano2023-my_MMMK}, bottom-up assessment of policy documents \citep{Buitenhuis2020-ms_MMMK}, and analysis of supply chain actors' actions during lockdowns \citep{Meuwissen2021-qv_MMMK}. Also assessments on the role of resilience attributes are part of this literature. \cite{Paas2021-lj_MMMK} measure the importance of resilience attributes in relation to systems' current capacities to deal with shocks. In a subsequent paper, \cite{Paas2021-at_MMMK} link resilience attributes, and the absence thereof, to critical system thresholds. \cite{Reidsma2023-ru_MMMK} identify resilience attributes for future systems. As all of these resilience attributes are context-specific, further research needs to address how resilience attributes materialise in different operations and systems. 

For further reading in the domain of risk management, \cite{Fan2018-qr} and \cite{Pournader2020-zx} review the literature on supply chain risk management. \cite{Munir2020-ch} analyse the importance of supply chain risk management for operational performance. \cite{Baryannis2019-vp} further discuss the potential of artificial intelligence as an emerging tool in supply chain risk management. With regards to the intersection of risk management and resilience, we recommend the book by \cite{Ivanov2021-an_MMMK}, as they explicitly consider risk and uncertainty as part of operational and supply chain decision making. In addition, the accompanying book \citep{Ivanov2024-vk_MMMK} presents the concept of stress tests. These tests are very useful in understanding possible consequences of disruptions – although one should prevent to restrict them only to risks for which data are available. Further reading is also recommended in the domain of standard risk management instruments, such as insurance. It is especially interesting to consider how these instruments can contribute to resilience. \cite{Vyas2021-jf_MMMK} postulate that agricultural insurance can be designed such that it stimulates adoption of more resilient practices. For further reading in resilience, the paper by \cite{Meuwissen2019-gv_MMMK} is recommended; it elaborates a resilience framework which can be applied to multiple types of systems, and lists a wide set of methods to measure various facets of resilience. For a reflexive perspective on the topic of resilience, we recommend reading \cite{Feindt2022-ub_MMMK} who conclude that systems face a resilience crisis. 

\subsection[Ripple effect (Dmitry Ivanov \& Alexandre Dolgui)]{Ripple effect\protect\footnote{This subsection was written by Dmitry Ivanov and Alexandre Dolgui.}}
\label{sec:Ripple_effect}

The ripple effect is one of the major phenomena studied in the field of supply chain resilience. Supply chain resilience is a firm’s capability to withstand, adapt to, and recover from disruptions to meet customer demand and ensure target performance \citep{Hosseini2019-uy_DIAD}. In short, supply chain resilience deals with disruptions in supply chain networks and develops capabilities to protect against and recover from these disruptions. One particular context in this field is disruption propagation, i.e., the effect when a disruption, rather than remaining localised or being contained within one part of the supply chain, cascades through the network and impacts the performance of the entire supply chain. And this is exactly what the ripple effect is about. Managing the ripple effect is crucial for decisions in risk management and resilience (\S\ref{sec:Risk_management_and_resilience}), inventory management (\S\ref{sec:Inventory_management}), network design (\S\ref{sec:Network_design}), global production networks, and bullwhip effect (\S\ref{sec:Bullwhip_effect}), among others.

The ripple effect in supply chains was first defined in academic literature by \cite{Ivanov2014-fr_DIAD}, building on disruption propagation studies by \cite{Liberatore2012-gp_DIAD}, \cite{Mizgier2013-es_DIAD}, and \cite{Ghadge2013-dt_DIAD}. \cite{Ivanov2014-fr_DIAD} stated that the ``ripple effect describes the impact of a disruption on supply chain performance and disruption-based scope of changes in the supply chain structures and parameters''. The ripple effect refers to multi-stage networks and the triggering of failures in the network elements similar to a domino or cascading effect. Its impact may include lost sales and delivery delays leading to adverse effects on the profitability of the supply chain \citep{Li2021-sg_DIAD}.

\begin{figure*}[ht!]
    \centering
    \includegraphics[width=4.5in]{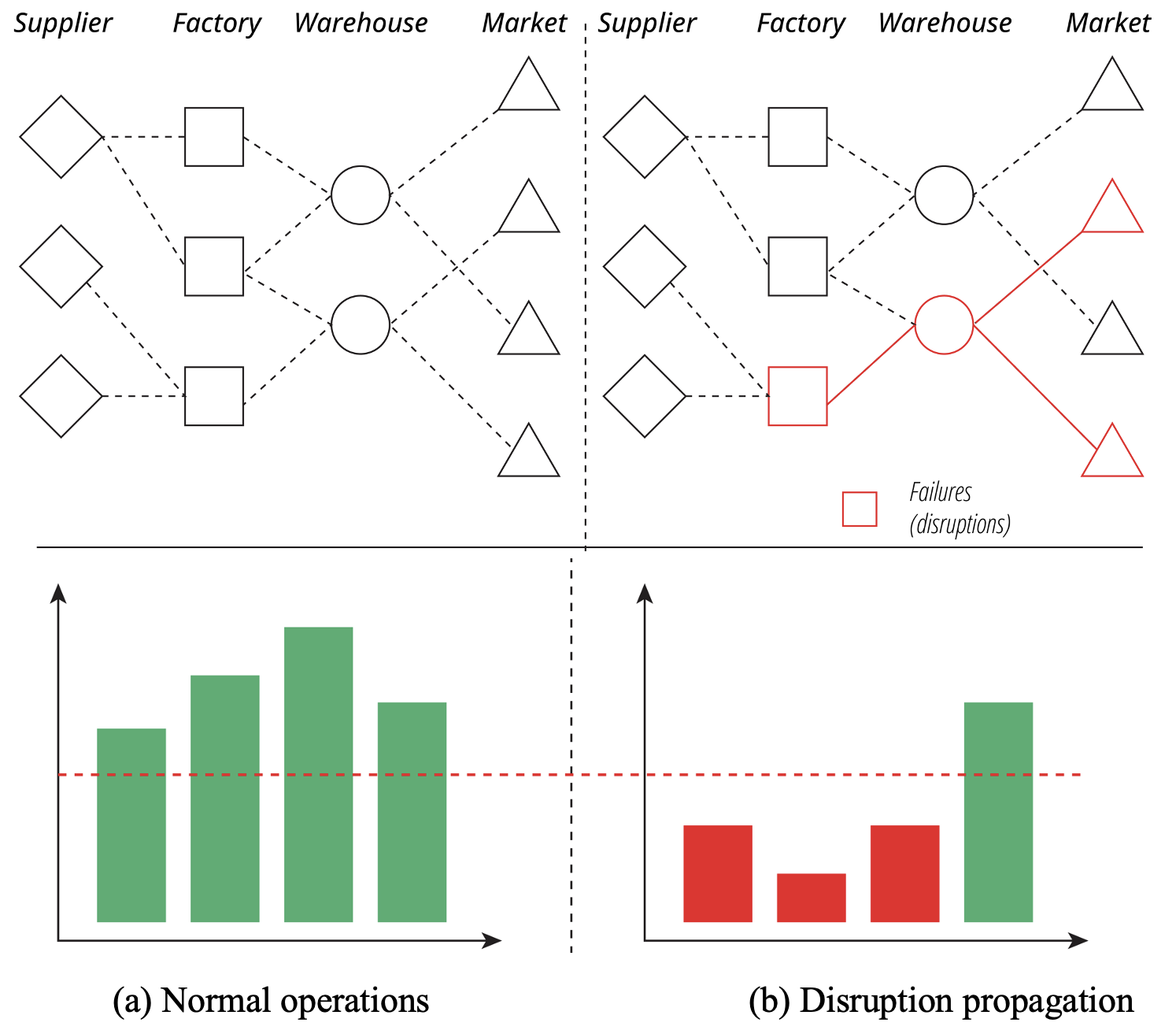}
    \caption{The ripple effect in the supply chain: changes in the network structure and associated performance degradation.}
    \label{fig:ripple}
\end{figure*}

Consider some practical examples the common features of which are summarised in figure \ref{fig:ripple} The earthquake and tsunami in Japan in 2011 disrupted multiple suppliers in the automotive industry and led to production breaks and material shortages worldwide. In May 2017, production at BMW was disrupted as a consequence of a supply shortage of steering gears. BMW’s 1st tier supplier, Bosch, was unable to deliver the steering gears since an Italian second tier supplier experienced production delays for certain steering parts due to internal machine breakdowns. In 2020, the COVID-19 outbreak caused production disruptions at many locations in China. Missing deliveries from/to China impacted the global supply chains and even the stock values of many multi-national companies.

The ripple effect is distinctively characterised by four major dimensions, i.e.:
\begin{itemize}[noitemsep,nolistsep]
\item Structural dynamics of the supply chain \citep{Pavlov2022-ff_DIAD}.
\item Low-frequency-high-impact effect \citep{Levner2018-vn_DIAD,Goldbeck2020-ee_DIAD}.
\item Disruption propagation through the network \citep{Brusset2023-tc_DIAD}.
\item Adverse impacts on resilience and performance \citep{Dolgui2020-fc_DIAD}.
\end{itemize}

Research on the ripple effect has essentially analysed how one or more disruptive events propagate through the supply chain, impacting material flows and performance. Management, modelling and assessment of the ripple effect became visible research avenues with a growing number and scope of contributions \citep{Dolgui2018-bk_DIAD}. The methodologies used include but are not limited to mathematical optimisation, simulation, game theory, control theory, data-driven analytics, network science, and reliability theory \citep{Dolgui2021-xj_DIAD}. Being different in scale and scope, the ripple effect studies share a common set of research questions, i.e.:
\begin{itemize}[noitemsep,nolistsep]
\item What are the reasons for the ripple effect? \citep{Ivanov2019-dr_DIAD,Chauhan2021-ej_DIAD}
\item What mitigation actions and recovery policies can be used to cope with the ripple effect? \citep{Dolgui2023-kc_DIAD}
\item How can supply chain stress tests be conducted to identify which network structures are especially prone to the ripple effect? \citep{Ivanov2017-ha_DIAD}
\item How can the severity of the ripple effect be measured? \citep{Li2020-hg_DIAD,Singh2021-db_DIAD}
\item How can digital technologies assist in managing the ripple effect? \citep{Ivanov2023-rs_DIAD}
\end{itemize}

The key findings to these questions are as follows. The common reasons for the ripple effect are lean operations reducing structural and process variety, the complexity of supply chains with low visibility, and insufficient coordination and collaboration between supply chain partners in managing disruptions \citep{Scheibe2018-tp_DIAD,Deng2019-lt_DIAD,Shi2022-to_DIAD,Ivanov2024-pb_DIAD}.

The scope and scale of the \textit{mitigation actions against the ripple effect} and recovery policies depend on redundancy (e.g., inventory pooling and multiple sourcing), flexibility (e.g., flexible manufacturing lines), agility (e.g., omnichannel distribution system), and visibility (e.g., digital supply chain twins); see also \cite{Ivanov2024-eg_DIAD} and \cite{Namdar2024-pa_DIAD}.

For running supply chain stress tests against the ripple effect, different methodologies can be used. Simulation, stochastic and robust optimisation, network science, machine learning and AI, and optimal control dominate the ripple effect stress-test domain \citep{Dolgui2021-xj_DIAD}. Simulation methodologies include discrete-event simulation \citep[DES;][]{Ivanov2020-ud_DIAD}, system dynamics \citep{Ghadge2022-hq_DIAD}, agent-based simulation \citep{Proselkov2024-kj_DIAD}, and Monte Carlo simulation \citep{Mizgier2013-es_DIAD}. 

DES has been the most popular method for the ripple effect analysis. DES is capable for a dynamic examination of disruption propagations and associated changes in supply chain structures, processes, and performance \citep{Llaguno2022-di_DIAD}. Network theory studies are centred around stress-testing of supply chain network structural connectivity and robustness \citep{Sokolov2016-ke_DIAD,Li2020-hg_DIAD}. In contrast to the simulation and optimisation methods, network-theoretical studies do not consider processes and material flow dynamics mostly restricting the scope of the analysis to the network (i.e., structure) level \citep{Li2021-sg_DIAD}. Looking beyond of merely structural assessments, Bayesian Networks became a popular method, particularly among researchers who are studying disruption propagation in complex networks due to its ability to connect disruption events, triggers, and consequences \citep{Garvey2015-oy_DIAD}.

The main \textit{application areas and insights} which can be obtained with the use of differ methodologies are summarised as follows. In the area of simulation and AI, the scope of analysis is comprised of the following areas:
\begin{itemize}[noitemsep,nolistsep]
\item Dynamic stress-testing of supply chain networks to the ripple effect \citep{Ivanov2023-rs_DIAD}.
\item Identification of critical suppliers and facilities triggering disruption propagation \citep{Mizgier2013-es_DIAD}.
\item Analysis of contingency/preparedness plans \citep{Liberatore2012-gp_DIAD}.
\item Recovery plan selection \citep{Pavlov2022-ff_DIAD}.
\item Prediction of performance impacts in supply chains for different disruption scenarios \citep{Ivanov2020-ud_DIAD}.
\item Dynamic analysis of disruption impacts on service level, inventory levels, delivery delays, and costs \citep{Ivanov2017-ha_DIAD}.
\end{itemize}

\noindent The network science studies have the following focus:
\begin{itemize}[noitemsep,nolistsep]
\item Identification of disruption propagation scenarios of different severity \citep{Pavlov2022-ff_DIAD}.
\item Stress-test of supply chain networks against node/link failures \citep{Li2020-hg_DIAD}.
\item Analysis of structural robustness to disruptions \citep{Sokolov2016-ke_DIAD}.
\item Identification of hidden suppliers and links \citep{Namdar2024-pa_DIAD}.
\end{itemize}

\textit{Measurement of the ripple effect} can be done in different ways. \cite{Kinra2020-oo_DIAD} proposed a method to measure the ripple effect based on maximum possible financial loss triggered by missing parts and components in the supply chain leading to lost sales at the customers. In other studies, service level indicators such as on-time delivery, fill rate, and demand fulfilment have been used \citep{Sawik2023-yv_DIAD}. The gaps between the planned and actual values of these indicators show the ripple effect severity. In addition, financial indicators such as annual revenues, costs, sales, and cash conversion cycle have been used in extant literature \citep{Ivanov2017-ha_DIAD,Badakhshan2023-vq_DIAD,Ivanov2024-ix_DIAD}. In network theory, such metrics like spatial complexity, network density, degree and betweenness centrality, connectivity, reachability, and structural robustness to analyse structural integrity in the presence of the ripple effect \citep{Li2020-hg_DIAD,Pavlov2022-ff_DIAD,Habibi2023-qr_DIAD}. Time-related metrics like time-to-recover and time-to-survive allow to quantify how long can the supply chain withstand the ripple effect and when it will recover \citep{Simchi-Levi2015-in_DIAD}. Finally, specific resilience and vulnerability indexes bounded between 0 and 1 can be considered to analyse the ripple effect impacts on the supply chain \citep{Ghadge2022-hq_DIAD,Ojha2018-ic_DIAD,Zobel2021-ep_DIAD,Hosseini2022-od_DIAD}.

We refer to the Handbook of the Ripple Effects in Supply Chains \citep{Ivanov2019-dr_DIAD}, Special Issue on the ripple effect \citep{Dolgui2021-xj_DIAD}, as well as literature review papers \citep{Dolgui2021-xj_DIAD,Ivanov2021-ue_DIAD,Llaguno2022-di_DIAD} for a comprehensive state-of-the-art analysis. \textit{Future trends} in the research on supply chain ripple effect can be seen in the following areas:
\begin{itemize}[noitemsep,nolistsep]
\item Digital twins, Metaverse and supply chain visibility \citep{Ivanov2023-rs_DIAD}.
\item Supply chain viability and Industry 5.0 \citep{Ivanov2022-bl_DIAD}.
\item Intertwined supply networks and cross-industry ripple effect \citep{Dolgui2023-kc_DIAD}.
\item Financial ripple effect \citep{Proselkov2024-kj_DIAD}.
\item Ripple effect in the presence of hyper-disruptions like COVID-19 pandemic and global trade system disruptions \citep{Ivanov2020-ud_DIAD,Brusset2023-tc_DIAD}.
\item Ripple effects in closed loop supply chains and circular economy \citep{Park2022-cr_DIAD}.
\end{itemize}

The simultaneous existence of different disruptive stressors (e.g., natural catastrophes, epidemics and pandemics, global geopolitical vulnerabilities, critical material shortages, climate change, and financial crises) along with a rapid development of digital technologies, industrial Metaverse and Industry 5.0 raises a number of novel research questions where novel and substantial contributions can be done.

\clearpage

\section{Applications}
\label{sec:Applications}

\subsection[Agriculture (Argyris Kanellopoulos)]{Agriculture\protect\footnote{This subsection was written by Argyris Kanellopoulos.}}
\label{sec:Agriculture}

Global problems related to climate change, depletion of natural resources, biodiversity loss, population's growth, urbanisation and dietary shifts, challenge current practices in agricultural and food production systems \citep{FAO20221_AK}. Moreover, consumers have become more aware about the impacts and the quality of the products they consume. As a result, businesses in agricultural supply chains (ASCs) try to improve their efficiency by redesigning their own operations but also the operations within the supply chains they control \citep{AHUMADA20091_AK}.

To (re)design efficient ASCs, novel technologies and alternative logistical concepts should be explored. The potential of sensor-based technologies, blockchains and decision support systems, to facilitate information sharing, transparency and decision-making, should be evaluated. In addition, alternative business models and supply chain structures that offer social benefits to farmers (e.g. short food supply chains) may also be considered \citep{KAMBLE2020101967_AK,LEZOCHE2020103187_AK}.

In operations and supply chain management field, quantitative methods and models have been used to support managerial decision making and to quantify the impacts of such alternative technologies and concepts \citep{zhu20181_AK}. These models provide guidelines to eliminate inefficiencies, and they unravel the relationships and trade-offs between different performance indicators. Applications of such models result in valuable insights for designing more efficient ASCs.

The objective of this section is to provide examples and to present key results of model-based applications in ASCs and operations management literature. To structure the section,, we use the theoretical framework of \cite{TSOLAKIS201447_AK} and we focus on strategic and operational decisions related to: establishing performance measuring systems; designing supply chain networks; selecting farming technologies; planning harvesting operations; planning transportation and inventory management operations; and ensuring food quality, food safety and traceability.

\subsubsection*{Establishing performance measuring systems}
An important first step towards redesigning ASCs is addressing the question of how to measure performance of alternative technologies and logistical concepts. A literature review study conducted by \cite{BANASIK20181_AK} showed that different sets of indicators have been used in planing, distribution and inventory management applications. \cite{IVODECARVALHO2022565_AK} proposed a universal framework to evaluate sustainability performance of companies operating in ASCs using qualitative comparative analysis and a content-based literature review. It was concluded that a large set of indicators is required to capture the complexity of measuring performance in ASCs. Literature reviews on sustainability assessment applications in agricultural and food supply chains revealed that life cycle assessment protocols have been accepted by researches and the industry as the performance standard in ASCs \citep{ARCESE2023100782_AK,VIDERGAR2021125506_AK}.

\subsubsection*{Designing supply chain networks}
Designing supply chain networks involves long term and investment-intensive decisions related to the optimal location, the number, and the capacity of facilities. Moreover, tactical and operational decisions related to supply and demand allocation, and material flows between network’s facilities are involved. In the literature, a wide variety of applications aim to improve efficiency and performance of supply chain networks \citep{JOSHI2022136_AK,YU2020121702_AK,ESKANDARPOUR201511_AK}. Specific features of agricultural and food supply chains can affect the network design process in ASCs. Such features include the seasonality of production, the uncertainty of yield and product quality, the product quality decay, the product-specific requirements for conditioned transportation and the specific traceability challenges and regulations of agricultural products \citep{Bloemhof20171_AK,bourlakis2008food_AK}. A variety of methods, e.g., Life Cycle Assessments, mathematical programming, multi-objective decision making and Stochastic programming have been used in applications that aimed to redesign networks in ASCs \citep{YADAV2022685_AK}. An informative literature review of model applications of network design studies in perishable food supply chains was conducted by \cite{JOUZDANI2021123060_AK}. According to the authors, in current applications for perishable products, the environmental impact of supply chain networks is measured mainly through CO2 emissions while other important environmental indicators that can affect network designing decisions are often ignored. Moreover, the social impacts are included only as salient aspects of the design process. The authors point to specific gaps in the literature of perishable network design studies and propose future research on modelling issues like traffic congestion, conditioned transportation, interest rates and uncertainty related to product perishability.

\subsubsection*{Selecting farming technologies}
Given the contribution of agriculture to environmental problems, selecting appropriate farming technologies has a substantial impact on supply chain efficiency and performance. In the agricultural economics literature, bio-economic models have been widely used for this purpose. These models optimise farmer's resource management decisions taking into account available production options and technologies \citep{BRITZ2021103133_AK}. Biophysical simulation models are used to quantify input-output parameters of environmental and ecological performance of the production options. Crop rotation, available resources and policy related constraints are used to determine feasible combinations of technologies. According to \cite{REIDSMA2018111_AK}, bio-economic farm models have been used to select farming technologies based on different indicators related to the use of agricultural inputs (like artificial fertilisers, agrochemicals, energy), groundwater quality, nutrient use efficiency, eutrophication, terrestrial and aquatic ecotoxicity, acidification, greenhouse gas emissions, crop diversification, biodiversity value, soil erosion, labour, working conditions, social benefits, eligibility for subsidies, food safety, animal welfare, animal health, landscape quality/values. In cases where the adoption of farming technologies require collaboration between farmers, or coordination between supply chain actors, agent based and muti-actor models have been used \citep{UTOMO2018794_AK,JONKMAN20191065_AK}. Applications of such models contribute in understanding the adoption mechanisms of sustainable technologies and provide managerial insights about the impact of a new technology to the decision making of individual actors in the chain.

\subsubsection*{Planning harvesting operations}
Harvesting is the final step and a crucial operation of agricultural production because it affects substantially the initial product quality, the supply of raw material and the economic performance of farming. Good yield predictions provide useful information for optimising planning and resource allocation during the harvest and post-harvest activities. In the literature, applications of machine learning and data mining techniques have been used to provide accurate yield predictions \citep{SHARMA2020104926_AK}. Such predictions are inputs to harvesting planning models which optimise decisions related to harvesting rates, routing/sequencing/scheduling of harvesters and resource allocation. Some applications also combine cultivation planning decisions e.g. determining optimal levels of cropping activities \citep{KUSUMASTUTI201676_AK}. Mathematical programming and simulations are commonly used methods in such applications. Features like harvesting time windows, labour and machinery capacity, perishability, storage, waste and uncertainty of yield and product quality are often included \citep{JARUMANEEROJ2021107694_AK, KUSUMASTUTI201676_AK}.

\subsubsection*{Planning transportation and inventory management operations}
Road transportation and logistics are directly related to the circulation of goods along the entire ASC and they play an important role in maintaining quality and compliance with performance criteria and regulations. Examples of transportation planning applications in ASCs are presented in \cite{TSOLAKIS201447_AK}. These applications provide case studies that focus on optimising transportation operations and the scheduling of road transportation in food and biomass supply chains. \cite{STELLINGWERF201880_AK} provided an overview of vehicle routing planning models that account for environmental impacts of transportation, while \cite{LIN2023103084_AK} optimised post-harvest operations with novel mobile precooling technologies for fruits and vegetables in short food supply chains. In ASCs, transportation planning problems occur also in field logistics, and agricultural routing planning. Representative applications of agricultural routing planning studies have been reviewed by \cite{UTAMIMA2022693_AK}. Applications sectors include orchards, vineyards, arable farming and biomass production. While objectives related to total finish time of machinery, task time, costs, and capacity utilisation are optimised. Applications of inventory management models in ASCs focus on optimising inventory, storage and material handling decisions \citep{BECERRA2021129544_AK}. An application article by \cite{PAAM2019104848_AK} focused on optimising inventory management decisions related to fruit in apple supply chains. A mathematical programming model was used to calculate trade-off between fruit loss and inventory costs. \cite{STELLINGWERF201880_AK} integrated inventory routing planning models in a case study of fresh food logistics between supermarket chains in the Netherlands. Economic and environmental benefits of collaboration were quantified.

\subsubsection*{Ensuring food quality, food safety and traceability}
The relationship between food quality, food safety and food waste has been thoroughly discussed in \cite{AKKERMAN20101_AK}. Often, in model-based applications in ASCs, food quality is treated as a continuous process of product degradation while food safety is treated as a binary state. It was demonstrated that preventing quality loss and ensuring food safety requires an increase in energy use due to temperature-controlled storage and distribution, which results in substantial environmental costs. Failing to preserve quality standards throughout the supply chain leads to wasted products which had environmental impacts without adding value \citep{AKKERMAN20101_AK}. \cite{BOSONA201332_AK} reviewed traceability systems in different agricultural sectors and identified benefits related to customer's satisfaction, food crises management, competence development, sustainability, and promotion of technological and scientific innovation. Moreover, \cite{zhu20181_AK} provided examples of applications in sustainable ASCs. RFID based traceability systems were evaluated and contributed in improving social dimension of sustainability either by mitigating food safety risks and allocating liability among supply chain actors or by ensuring animal welfare by incorporating restriction on livestock transport.

\subsection[Construction (Samudaya Nanayakkara \& Srinath Perera)]{Construction\protect\footnote{This subsection was written by Samudaya Nanayakkara and Srinath Perera.}}
\label{sec:Construction}
The construction industry makes a significant impact on the global economy. The estimated value of the global construction industry output in 2022 is over 15 trillion USD and approximately 13\% of the global gross domestic productions \citep{Wood2022-cc_SNSP}. The construction industry produces highly complex structures, necessitating collaboration across multiple supply chains \citep{Ashworth2018-kj_SNSP}. Modern Supply Chain Management Systems (SCMS) incorporate the latest Information and Communication Technologies \citep{Nanayakkara2021-sr_SNSP}.

This subsection offers a detailed exploration of key stages in the construction lifecycle phases and the uniqueness of construction supply chains and focuses on procurement aspects. Other parts address resource procurement in construction, challenges in supply chain management and examine advancements in e-procurement through its historical eras. By integrating these elements, the subsection provides a comprehensive overview of the complexities and advancements in operations and supply chain management in construction.

\subsubsection*{Key Stages in the Construction Lifecycle}
The construction lifecycle is not the only phase of construction activities that are carried out. The most common classification of lifecycle stages includes Pre-design, Design, Construction, Handover, Use/ Operation, and End of Life \citep{Cimen2023-if_SNSP}. 

\textbf{Pre-design stage:} This is the initial stage of a construction project and has a broader meaning than the planning stage. It focuses on defining project requirements, goals, and feasibility. The key activities are identifying key stakeholders, conducting feasibility studies, defining project scope and objectives, site selection and environmental assessments, budget estimation, and scheduling \citep{Cimen2023-if_SNSP}.

\textbf{Design stage:} creates detailed plans and specifications for the project. The key activities in design stage are creating architectural and engineering design, developing construction drawings and specifications, obtaining permits and regulatory approvals, cost estimation, and value engineering \citep{Cimen2023-if_SNSP}. There could be the development of sustainability designs, risk assessments and others \citep{Ceschin2016-ce_SNSP}. 

\textbf{Construction stage:} The construction phase is crucial, where the building is physically constructed according to design and plan. In this stage, construction supply chains support project objectives through activities like mobilisation, site preparation, and various works (foundation, structural, mechanical, electrical, etc.). This stage faces challenges such as weather, site conditions, supply chain issues, regulatory compliance, safety, quality assurance, and others \citep{Nafe-Assafi2024-hc_SNSP}. 

\textbf{Handover stage:} The completed project is transferred to the project owner in the handover stage. There are a couple of key activities, including final inspections and system testing, addressing punch/ snag list items, delivery of operation manuals and warranties, certification, and official handover to the owner \citep{Ceschin2016-ce_SNSP}.

\textbf{Operation stage:} This stage is sometimes defined as the operations and maintenance stage. During this stage, the completed construction object or built asset will be utilised by occupants or end users based on project objectives. The key activities in this stage are facility management and operations, regular maintenance and repairs, monitoring building performance and energy usage, and renovations and upgrades if needed \citep{Nafe-Assafi2024-hc_SNSP}. 

\textbf{End of life:} stage refers to where the asset reaches the end of its functional life. The key activities in this stage include decommissioning or demolition of the building, recycling or disposal of materials, site restoration or redevelopment, and possible repurposing \citep{Ceschin2016-ce_SNSP}.

\subsubsection*{The Uniqueness of Supply Chains in Construction}
A SC involves firms linked upstream (suppliers) and downstream (customers) to deliver products or services to consumers \citep{Mentzer2001-qp_WPLD}. Construction Supply Chains (CSC) differ from general SCs due to their unique characteristics, such as project-specific factors and construction processes. CSCs are often geographically dispersed with new suppliers and subcontractors. Therefore, each project represents a new supply chain \citep{Butkovic2016-vd_SNSP}. These factors create significant challenges in achieving supply chain maturity and performance, distinguishing CSCs from those in other industries.

The second distinctive feature of CSC compared to a common SC arises from the inherited construction procurements process flows. The highest decision-making power in CSC is held by the client, and main contractor, who are responsible for financing, designing, planning, and key activities. This zone can be identified as a construction procurement. The upstream of the CSC is lengthier and consists of specialist contractors, subcontractors, sub-subcontractors, suppliers, and manufacturers who bear some costs and heavily spend resources such as material, labour, and machinery. 

These unique characteristics make construction supply chains a distinct domain of knowledge and practices compared to supply chains in other sectors.

\subsubsection*{Construction Procurement}
\begin{figure}
\centering
{\includegraphics[width=0.95\linewidth]{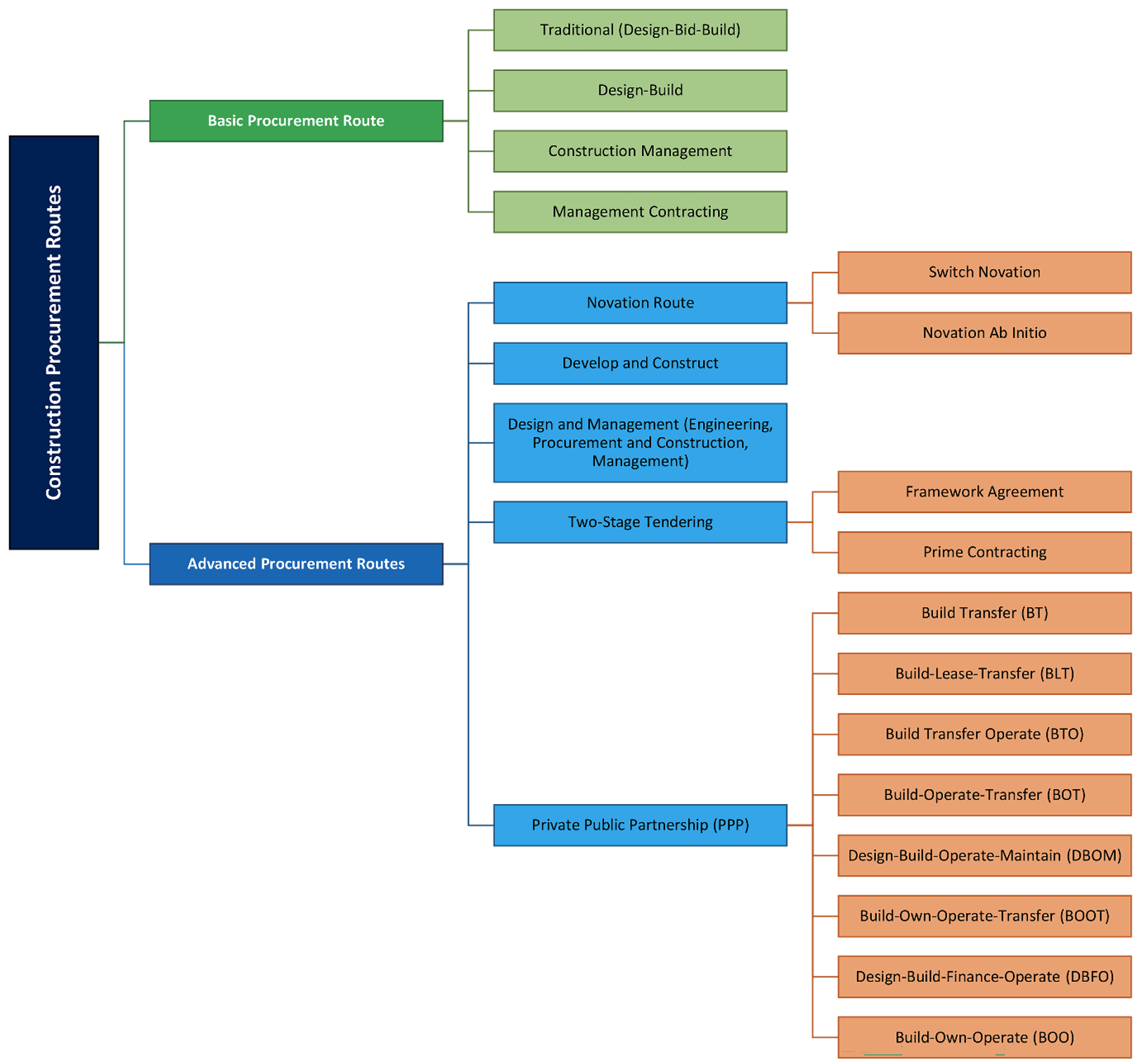}}
\caption{Construction procurement common routes.}
\label{fig:construction}
\end{figure}

Construction procurement can be recognised as one of the most complex procurement operations. During the pre-design stage, the client typically works with consultants to identify the most suitable procurement routes for achieving project goals and objectives in an optimum manner, ensuring value for money, and assessing the associated risks \citep{Aziz2007-uf_SNSP}. The construction procurement route can be divided into two main categories: basic and advanced procurement routes. Figure \ref{fig:construction} illustrates the most common routes within each category. The procurement route is the primary agreement type, especially between the client and the main contractor. Each construction procurement route has its own advantages and weaknesses. Therefore, it is essential to carefully evaluate the procurement route and the project's nature before moving forward with the construction procurement \citep{Oyegoke2009-tr_SNSP}.

\subsubsection*{Resources Procurement in Construction Projects}
\cite{Muya1999-cs_SNSP} has classified the previous research identifications into three types of construction supply chains: primary supply chain, support supply chain and human resource supply chain.

The primary supply chain in construction focuses on the procurement of materials (e.g., cement, sand, steel, etc.) essential for the project's execution commonly called as purchasing process. This process involves sourcing, purchasing, and delivering construction materials from suppliers or directly from the manufacturer \citep{Caldas2015-de_SNSP}. 

In construction, supply chains for plants and equipment (support supply chains) are vital for providing the necessary machinery and tools. This includes sourcing, acquiring, and maintaining equipment for tasks like excavation and lifting \citep{Perera2014-lg_SNSP}. Key activities involve selecting suppliers or rental companies, negotiating contracts, ensuring safety and performance standards, and coordinating logistics \citep{Edwards2009-tg_SNSP}.

In construction, human resource supply chains for procuring labour are essential for securing the skilled workforce needed to complete various project tasks. This process involves identifying, hiring, and managing both permanent employees and specialised subcontractors who perform specific functions such as electrical work, plumbing, or masonry \citep{Karim2006-wp_SNSP}.

\subsubsection*{Challenges and issues in construction supply chain management}
The construction industry faces numerous challenges across project phases, including design, construction, operation, maintenance, and decommissioning. Key issues identified include trust, knowledge management, integration, collaboration, and quality-related problems. SCM and SCMS often fail to adequately address these time, cost, quality, and trust issues within construction supply chains.

Low productivity, driven by a lack of skilled workers and outdated technology, leads to significant time-related issues and schedule overruns. Material delays, a frequent cause of project delays, impact multiple sequential activities \citep{Ala-Risku2006-bc_SNSP}. Quality and compliance issues are common in construction projects \citep{Zhang2011-gi_SNSP}. Financial challenges are notable due to the long payment cycles in construction. Subcontractors and suppliers often face delays and partial payments, exacerbated by inefficient supply chain processes \citep{Nanayakkara2021-sr_SNSP}. 

Trust is a critical factor in supply chains, yet it is often neglected in many CSC models. Short-term project nature negatively impacts essential success factors like long-term trust, collaboration, and risk-sharing. Researchers propose that improving supply chain maturity and transparency can enhance trust, though significant issues remain \citep{Koskela2020-yf_SNSP}.

\subsubsection*{Advancements in e-Procurement in the Construction Industry}
e-Procurement or electronic procurement is a vital function of any supply chain management system, and it was the primary function of forming the computer-based supplier interaction \citep{Puschmann2005-sx_SNSP}. 

The first e-procurement era began in the 1980s, initiated with centralised systems like Enterprise Resource Planning systems addressing internal information silos \citep{Nanayakkara2013-ca_SNSP,Nanayakkara2016-rr_SNSP}. Business-to-business communication happened through electronic media such as a floppy disk, compact discs (CDs) and emails \citep{Nanayakkara2015-vm_SNSP,Hewavitharana2019-wq_SNSP}. The second era of e-procurement was initiated in the 1990s, the enterprise computing era began with organisations adopting enterprise-wide infrastructures and the development of the Internet and suppliers could directly interact with e-procurement portals \citep{Afolabi2006-rv_SNSP}. The third era of e-procurement initiated in the 2000s, cloud computing revolutionised procurement and supply chains. System-to-system communication became common, enabling direct interaction between supplier and client systems directly \citep{Nanayakkara2014-jf_SNSP}. 

Blockchain technology was introduced in 2008 \citep{Nanayakkara2019-td_SNSP} and it could revolutionise procurement and supply chains by smart contracts and decentralised applications \citep[DApps][]{Perera2020-ld_SNSP}. Smart contracts automate and secure transactions with unprecedented accuracy and transparency, while DApps enable decentralised, peer-to-peer operations \citep{Nanayakkara2021-id_SNSP,Perera2021-cd_SNSP}. Blockchain’s advantages include immutable transaction records and enhanced trust in a trustless environment, reducing the need for intermediaries and lowering costs \citep{Perera2021-cd_SNSP}. Therefore, many researchers predict that blockchain and smart contracts will usher in a new era by addressing issues related to time, cost, quality, and trust in construction supply chains \citep{Nanayakkara2019-eg_SNSP}.

\subsubsection*{Conclusion}
Effective supply chain management is essential for construction projects due to the industry’s complexity and need for stakeholder coordination. Each project represents a new, dispersed supply chain, facing challenges like delays and financial inefficiencies. Advancements in e-procurement, especially blockchain and smart contracts, offer solutions by enhancing transparency, streamlining payments, and building trust. As the industry evolves, digital tools and innovative procurement will improve efficiency and sustainability.

\subsection[Public and nonprofit sectors (Claire Hannibal)]{Public and nonprofit sectors\protect\footnote{This subsection was written by Claire Hannibal.}}
\label{sec:Public_nonprofit_sector}
In what has been described as a `results-oriented management movement' \citep[][p. 325]{Poister1999-dc_CH}, the application of operations and supply chain management (SCM) principles to the public and nonprofit sectors has received increasing research attention over the last 30 years \citep{Ghobadian1994-ei_CH,Radnor2004-bx_CH,Taylor2014-lr_CH,Shevchenko2024-hs_CH}. Funded by the state, for most public services (e.g., registering to vote or applying for state benefit) there is no alternative choice of provider, therefore public services must be appropriate for all citizens. Frequently in the spotlight in relation to the measurement and management of performance \citep{Zheng2019-sm_CH,Garengo2021-yy_CH}, a key driver for the application of operations and SCM principles to the public sector has been governmental pressure to demonstrate improvement in the productivity and performance of public service provision \citep{Radnor2013-kg_CH}. The nonprofit sector is funded by grants, donations and service delivery contracts. The application of operations and SCM to the nonprofit sector has been identified as a mechanism to demonstrate the provision of continuously improving services that offer value for money to both funders and beneficiaries; accountability criteria that are essential to securing continued resources \citep{Moxham2007-hr_CH}. 

Against this backdrop of service improvement, the application of operations and SCM principles to the public and nonprofit sectors can be problematic, and the assumption that private sector operations improvement approaches can be directly applied to these sectors has been criticised \citep{Bush1992-ty_CH,Walley2013-wk_CH,Hazlett2013-wo_CH,Radnor2013-mx_CH}. For example, public sector organisations have to meet multiple, often conflicting, goals and are characterised by different ownership, funding, goal setting and control norms and expectations when compared with the private sector \citep{Goldstein2005-uc_CH}. These characteristics lead to complexities when attempting to `implant' improvement approaches taken directly from the private sector into the public sector context \citep[][p. 988]{Hazlett2013-wo_CH}. Similarly, the context in which nonprofits operate differs from that of the private sector. For example, nonprofits are required to be both cost-effective and have a positive impact on society, giving rise to vulnerabilities and tensions that may not be present in the private sector \citep{Shevchenko2024-hs_CH,Fagbemi2024-kg_CH}. The competitive landscape also differs, with humanitarian aid organisations, for example, having to rapidly collaborate with sectoral competitors to form consortia to deliver life-saving services \citep{Schiffling2020-dq_CH}; a requirement that would be unusual for most private sector companies. In response to these differing contexts, the application of operations and SCM principles to the public and nonprofit sectors often requires situational adjustment and operations management frameworks, including balanced scorecards for public and nonprofit sectors, have been developed \citep{Kaplan1999-vl_CH,Ronchetti2006-dl_CH}.

As noted, the continuous improvement of services, specifically in terms of quality, cost and experience, is often the focus of the application of operations and SCM principles to the public sector \citep{Dobrzykowski2019-nk_CH}. Comparison between different types of public sector organisations are common, with research focusing on emergency services, local and central government \citep{Walley2013-wk_CH,Radnor2013-zh_CH}. Studies situated in the public sector are wide ranging in terms of geography and context, examining for example, e-government in Slovenia \citep{Groznik2009-pw_CH}, procurement reform in South Africa \citep{Ambe2012-rm_CH}, public sector performance measurement in Germany \citep{Greiling2005-xu_CH} and supply chain innovation in public procurement in Finland \citep{Karttunen2022-fc_CH}. Much of the research on the application of operations and SCM principles to the public sector is situated in Europe and North America. For applications in developing countries the reader is directed to \cite{Mimba2013-hb_CH}, in South Africa to \cite{Ambe2012-rm_CH} and in the United Arab Emirates to \cite{Al-Dhaafri2020-ks_CH}.

The application of operations and SCM principles to the nonprofit sector emerged slightly later than that of the public sector. As with the public sector, performance management and measurement were initial themes, with published works emerging from 2007 \citep{Moxham2007-hr_CH}. The success of the application of operations and SCM principles to the nonprofit sector appears to be mixed, with earlier studies highlighting implementation challenges. For example, research highlights a disconnect between measurement criteria and performance goals \citep{Moxham2007-hr_CH,Moxham2009-hp_CH}, a performance measurement focus on the use of funds at the expense of continuous improvement \citep{Moxham2010-ar_CH} and variable responses to performance measurement from nonprofit stakeholders \citep{Beer2017-gm_CH}. More recently studies have showcased good practice as regards to operations and SCM in the nonprofit sector. Examples include collaborative service delivery \citep{Best2018-iu_CH} and cost effective operations \citep{Shevchenko2024-hs_CH}. Operations and SCM principles appear to be more embedded in the practices of nonprofits when compared with studies of 10 years ago and the benefits are being realised.  For example, recent research demonstrates how an awareness of SCM principles has a positive impact on organisational financial independence, thus allowing nonprofits to fulfil their mission \citep{Garcia-Dastugue2024-nf_CH}, and how the application of supply chain management to volunteer groups results in much needed rapid emergency response \citep{Shaw2024-fs_CH}.

As with the public sector, research situated in the nonprofit sector is wide ranging. Studies focus on, for example, performance management in a UK housing association \citep{Manville2013-jg_CH}, supply chain integration in food banks in the United States \citep{Ataseven2020-ea_CH}, reverse logistics in textile supply chains in Finland \citep{Zhuravleva2021-qv_CH} and the distribution of surplus food to individual beneficiaries in the United Arab Emirates \citep{Grassa2023-gg_CH}. The size and shape of nonprofits is also a fertile line of enquiry as the nonprofit sector is so diverse. Studies examine large nonprofits characterised by numerous paid staff, formal governance structures and an international reach \citep{Boateng2016-bl_CH,Moxham2009-hp_CH}, medium sized nonprofits with paid staff and a single country focus \citep{Best2018-iu_CH}, small nonprofits with a minimal number of paid staff and a large volunteer base \citep{Moxham2007-hr_CH}, short-term temporary collaborations between nonprofits \citep{Schiffling2020-dq_CH} and informal structures that rely solely on volunteers \citep{Shaw2024-fs_CH}. These studies show how the diversity of the nonprofit sector offers challenges and opportunities for the application of operations and SCM principles.

For a comprehensive discussion of the application of operations and SCM approaches to both the public and nonprofit sectors, the reader is directed to \cite{Poister2003-ck_CH}.  For a focused discussion on the application of operations and SCM to public services, the research handbook edited by \cite{Radnor2016-ox_CH} provides an excellent compilation of agenda setting articles. For nonprofits, \cite{Gazzola2022-oy_CH} provide a contemporary application of operations improvement tools and techniques that are relevant to the sector. Finally, whilst research on the application of operations and SCM to the public and nonprofit sectors is receiving increased research attention, we cannot overlook the continuing challenges facing both sectors. In a global context where the mantra for public and nonprofit organisations continues to be `how do to more with less?' the future of services provided by the public and nonprofit sectors is far from certain and applications of operations and SCM principles continue to be crucial.

\subsection[Healthcare (Wendy Phillips \& Linh Duong)]{Healthcare\protect\footnote{This subsection was written by Wendy Phillips and Linh Duong.}}
\label{sec:Healthcare}
In healthcare, the adoption of operations and supply chain thinking has played a pivotal role in enabling healthcare organisations engage with key partners to deliver quality healthcare products and services to patients through the efficient allocation of resources, effective delivery systems, and deployment of technology to support innovative healthcare provision \citep{Kc2020-lb_WPLD}. Healthcare supply chain management (HSCM) is defined as ``the management of people, processes, information, and finances to deliver medical products and services to consumers, in the pursuit of enhancing clinical outcomes and user experience, while controlling costs'' \citep[][p. 1334]{Betcheva2021-is_WPLD}. HSCM involves three core entities: healthcare product manufacturers, distributors or wholesalers, and service providers or procurement agencies, all of which directly contribute to the flow of products, services, finances, and information within the supply chain. These entities can span diverse organisational types, including private providers, public agencies, and non-profits \citep{Zhao2012-kl_WPLD}. Recent healthcare business models - such as specialist clinics, telehealth, and urgent care centres - particularly during the COVID-19 pandemic, have underscored the importance of operations and supply in improving quality of care, end-user experience, and cost-effectiveness \citep[e.g.,][]{Sampson2023-ze_WPLD,Thirumalai2023-xt_WPLD}.

Recent works \citep[e.g.,][]{Phillips2022-in_WPLD,Roscoe2022-hl_WPLD,Srai2020-ya_WPLD} have examined the role of global supply chains (GSCs) in healthcare, where globally connected organisations work together to bring products to market \citep{Kano2020-it_WPLD}.  GSCs allow healthcare organisations to access affordable products by leveraging economies of scale and scope, fostering economic value and technological advancements \citep{De-Marchi2020-jd_WPLD,Perez-Aleman2008-hw_WPLD}. However, healthcare organisations within GSCs often operate across multiple regions or countries, each with different regulations and policies \citep{Handfield2020-bm_WPLD}, and these organisations often have conflicting objectives \citep{Phillips2022-in_WPLD}. For instance, the Turkish government has mandated that foreign suppliers of certain healthcare products produce locally or through a local Turkish company \citep{European-Commission2024-jt}. Such challenges can cause inefficiencies, such as healthcare product shortages \citep{Muller2023-qi_WPLD}, as seen during the COVID-19 pandemic when nations struggled to procure essential supplies such as personal protective equipment (PPE), diagnostic tests, and oxygen.

The growing frequency and severity of geopolitical disruptions have heightened interest in supply chain resilience (SCR) to secure the continuous availability of critical healthcare products and services \citep{Bednarski2023-zm_WPLD,Finkenstadt2021-ko_WPLD,Holgado2023-vj_WPLD,Moradlou2021-fr_WPLD}. Reviews of the literature on SCR \citep{Sawyerr2019-bf_WPLD,Tukamuhabwa2015-hj_WPLD} have uncovered four key SCR strategies employed by healthcare organisations to mitigate disruptions: flexibility; redundancy; collaboration; and agility \citep{Scala2021-yt_WPLD}. Key themes emerging from the literature emphasise the need for faster delivery times, the movement of production closer to the point-of-care \citep{Betcheva2021-is_WPLD,Peters2023-ra_WPLD,Zavattaro2023-zn_WPLD}, and localised sourcing of components and production \citep{Gereffi2022-dn_WPLD,Kapletia2019-sb_WPLD}.

Studies on the transition from global to local healthcare supply chains (HSCs) suggest that local procurement is vital in preparing for future disruptions \citep{Harland2021-cb_WPLD}. For example, in response to the United Kingdom’s (UK) exit from the European Union (BREXIT), the majority of pharmaceutical companies planned to relocate production sites from the UK to EU states and distribution centres from the EU to the UK \citep{Moradlou2021-fr_WPLD}. Post-pandemic, similar strategies are likely to be adopted more frequently to shorten HSCs and enhance responsiveness and responsiveness \citep{Duong2024-pe,Finkenstadt2021-ko_WPLD,Xu2020-to_WPLD}.

However, a shift away from global HSCs presents two significant challenges that may render localisation impractical \citep{Panwar2022-dp_WPLD}. First, large-scale localisation requires substantial investments in facilities, labour and knowledge and can be particularly testing for developing countries which often have poor physical and logistical infrastructures as well as a lack of skilled labour \citep{Besiou2021-jr}. Second, dismantling GSCs could result in social and economic hardships for countries relying on international trade. Additionally, healthcare products and services are subject to strict regulations, requiring significant alignment between the regulatory frameworks of different countries within an HSC.  These regulatory challenges may further complicate localisation decisions \citep{Moradlou2021-fr_WPLD}. Consequently, many organisations, especially large ones, tend to adopt a cautious, ``wait-and-see'' approach, gathering information from partners to minimise uncertainty before making relocation decisions \citep{Roscoe2020-fp_WPLD}.

To achieve high-quality patient care while reducing costs and waste, HSCM faces significant obstacles \citep{Schneller2023-mv_WPLD}, underscoring the need for innovation and new technologies. With advancements in digital technology and the rise of the sharing economy, \cite{Wang2024-xf_WPLD}  review new service-sharing models in healthcare, offering fresh avenues for research. For example, many hospitals have implemented smart healthcare systems, enabling healthcare professionals to provide both online and offline diagnoses and treatment \citep{Lee2021-sf}, while wearable devices have been adopted for healthcare monitoring \citep{Huarng2022-nx}. 

It is worth noting a recent surge in the use of Artificial Intelligence (AI) applications within healthcare operations, driven by the increasing availability of data and advances in machine learning and data analytics \citep{Guha2018-gn_WPLD}. AI systems offer several potential benefits for HSCM, such as improving integration among supply chain partners, addressing demand forecasting challenges, enhancing risk and performance management, aiding inventory control, supporting greater resilience, and mitigating the ripple effects of disruptions \citep{Dubey2021-yw_WPLD,Ivanov2020-kd_WPLD}.

Despite clear evidence of the advantages of AI-enabled systems, their adoption is hindered by high implementation costs and a lack of confidence in their ability to deliver immediate results \citep{Adhikari2023-mf_WPLD}. Furthermore, the use of AI in healthcare raises issues related to data protection, patient safety, legal accountability, and ethical considerations \citep{Ghadge2023-zd_WPLD}. Lack of awareness and understanding of AI systems amongst healthcare professionals and patients further complicates their adoption \citep{Arfi2021-cg}. \cite{Kumar2022-ib_WPLD} have suggested several research directions to address these challenges, such as exploring how healthcare organisations can use information and communication technology (ICT) to securely share data with supply chain partners, developing blockchain solutions to reduce cybersecurity risks, and implementing regulatory policies to enhance supply chain cybersecurity.

In addition to SCR, localisation and AI, existing studies on healthcare applications over the last decade have covered several themes from the operations and supply chain management perspective. For example, business analytics \citep{Terekhov2024-ih_WPLD}, capacity strain \citep{Kim2024-or_WPLD}, decision making \citep{van-Rijn2024-lw_WPLD}, inventory management \citep{Yang2025-ta_WPLD}, quality of care \citep{Xu2025-gn_WPLD} and sustainability \citep{Berenguer2024-fh_WPLD}. Furthermore, studies on this topic are mostly empirical \citep{Kc2020-lb_WPLD} and modelling works \citep{Ali2022-nv_WPLD,Keskinocak2020-ec_WPLD}. For a comprehensive and up-to-date review of the role of operations and supply chains in healthcare, readers are encouraged to consult recent literature. \cite{Liu2024-hn_WPLD} present a framework for patient engagement in remote consultations; \cite{Moons2018-we_WPLD} analyse key performance indicators in hospital logistics; and for a broader strategic perspective, \cite{Schneller2023-mv_WPLD} provide a comprehensive overview of the issues facing HSCM and its role in healthcare strategy. These works are essential for understanding the modern requirements of HSCM, offering valuable case studies and insights into the complexities of contemporary healthcare supply chains. Moreover, they offer some suggestions for future studies. For example, future studies could adopt different methods or mixed methods to explore mechanisms that affect the effectiveness of healthcare operations management \citep{Kc2020-lb_WPLD}. Also, to increase the research impact in practice, closer collaboration between researchers and practitioners is vital for accessing actual and timely data \citep{Keskinocak2020-ec_WPLD}. Moreover, we call for more interdisciplinary studies, combining healthcare management with fields such as public sector, engineering and pharmaceutical manufacturing.

\subsection[Humanitarian operations (Marianne Jahre \& Maria Besiou)]{Humanitarian operations\protect\footnote{This subsection was written by Marianne Jahre and Maria Besiou.}}
\label{sec:Humanitarian_operations}
Three big disasters changed the way that we are thinking of humanitarian operations: the Hurricane Katrina in the USA in 2004, the Asian-Pacific Tsunami in 2004, and the Pakistan Earthquake in 2005. The disasters were followed closely by the media, making the general population aware of the big logistics challenges occurring in disaster response and the need for improvement. That was the trigger for the humanitarian operations research field to take off. 

\subsubsection*{Key concepts in humanitarian operations}
Humanitarian operations constitute a relatively new field of research in Operations Management (OM). With one exception \citep{Long1995-zy_MJMB}, the first OM journal papers date back to 2006 \citep{Van-Wassenhove2006-hd_MJMB,Altay2006-eu_MJMB}. A year later the first review was published \citep{Kovacs2007-bs_MJMB}. This has since been followed by more than 150 literature reviews, one of the latest being the one of \cite{Kembro2024-fa_MJMB}. The last two decades have not been busy only for the researchers, but even more for humanitarian organisations and other actors responding to mega-disasters like the 2010 Haiti earthquake, the Syrian refugee crisis, and the Russian-Ukrainian war. 

Similar with disaster management in general, humanitarian operations are not only about the response. The logistics involved in mitigation (attempting to reduce future disasters occurring), preparedness (minimising the consequences for affected populations by being prepared before disasters occur) and recovery (getting back to normal) are equally important parts of what is termed as the disaster management cycle and are interconnected \citep{Tomasini2009-ei_MJMB}. International humanitarian organisations (IHOs) including non-governmental organisations (NGOs), the UN and the International Red Cross \& Red Crescent Movement, play a central role in all four phases. For example, they operate programs like building wells or restoring medical centres to help the local population `building back better'. They also help local governments prepare better for future disasters, for example by prepositioning relief items or training local organisations. Traditionally, however, focus was on response. IHOs mostly operate in contexts where the local response teams (typically the fire brigade, police forces, medical emergency teams, local NGOs and military) cannot address the needs on their own. When a disaster strikes, the first to respond is always the local community and the beneficiaries themselves. International actors come in if asked by the government. 

A typical supply chain of activities starts with assessing and publishing needs and budgeting \citep{Keshvari-Fard2022-rq_MJMB}, followed by sourcing and procurement, transport, warehousing and distribution. Nowadays, humanitarian supply chains also include waste management. Some of these activities take place locally aiming to increase localisation \citep{Frennesson2022-ex_MJMB}, while others internationally. Which activities should take place locally and which internationally depends significantly on the available local resources: very often humanitarian organisations operate in countries where key infrastructures such as water, transport, and power are already struggling to meet in-country needs and placing further demand on these systems could give rise to tensions between humanitarian organisations and local communities \citep{Duong2024-pe}. Humanitarian organisations have an increasing understanding of such conditions, shown for example by their decision to implement cash-based assistance or not. The supply chain often uses a push system immediately after the disaster has occurred and changes to pull with more time and better information on the needs.

Items and services related to health, water and sanitation, food and shelter are provided by the HOs supported by (non-)earmarked funding from private and public donors and resources from private sector companies like storage, transport and relief items. Distributing cash to the beneficiaries to complement (in-kind) donated relief items has become a norm \citep{Juntunen2023-sx_MJMB} in an effort to also reduce failures of in-kind donations like equipment that quickly becomes redundant due to poor adaptation to local environmental conditions, and limited local resourcing to use, effectively staff or maintain it \citep{Tomasini2009-ei_MJMB}. Key resources also include information and communication systems, standards, supplier agreements \citep{Balcik2014-lh_MJMB} and last, but certainly not least, human resources. Experienced and skilled personnel is critical for humanitarian operations as is leadership \citep{Salem2022-fd_MJMB}. However, due to the limited funding available and the tough operating conditions, training and maintaining staff can be hard. The media’s main role is to increase awareness and hold the actors accountable for their actions, so that funding is used wisely and fair in accordance with humanitarian principles.

\subsubsection*{Key issues – literature reviews and empirical research}
Humanitarian operations often take place in resource austere settings with limited access to power, clean water, poor transport links, disrupted supply chains, limited cold chain as well as political volatility. Country access, the import and resupply of key equipment and medical consumables, security and communication issues and local environmental conditions all have an impact. The experiences in 2004/2005 clearly demonstrated the need for logistics preparedness \citep{Jahre2016-rt_MJMB,Stumpf2023-pt_MJMB} and coordination, collaboration and co-operation \citep{Balcik2010-qy_MJMB,Jahre2010-zr_MJMB,Kovacs2021-uq} to increase operational efficiency and effectiveness by closing gaps, reducing duplications, and managing resources better. Substantial efforts by the organisations followed. Until 2015 the most common strategy was strategic stock of relief items, but also other `stock' including vehicle fleets \citep{Pedraza-Martinez2011-is_MJMB}, emergency funds and rosters of trained human resources \citep{Jahre2017-zr_MJMB}. The use of postponement through non-earmarking of items, prepositioning of semi-finished goods, centralisation and outsourcing, collaboration \citep{Altay2014-rz_MJMB,Ergun2014-zo_MJMB} as well as use of flexible transportation and supply base have also been common. Providers must navigate a range of extremes, e.g. temperature, humidity, rainfall, etc., all of which can have a profound effect on the ability to effectively deliver humanitarian operations. Humanitarian organisations should always consider whether collaboration is appropriate given the need to maintain integrity and independence, and depending on aspects like interoperability of technologies and systems, the use and implementation of different regulations and standards, and information asymmetry \citep{Phillips2023-ka}. They should also consider more collaboration with the commercial sector: temporary supply chains dismantled once the disruption has been addressed are particularly important during humanitarian events, ensuring the delivery of critical supplies over a limited timeframe \citep{Muller2023-qi_WPLD,Duong2024-pe}.

During the past decade research and practice have expanded a lot and cover a range of themes. Table \ref{tab:humops} gives an overview of recent literature reviews within many of these themes, mainly from OSCM journals.

\begin{table}[h]
\centering
\caption{Recent literature reviews and their respective themes.}\vspace{0.25cm}
\begin{tabular}{ll}
\hline
Theme & Example of literature review \\
\hline
Cash distribution	& \cite{Juntunen2023-sx_MJMB} \\
Coordination	& \cite{Jahre2021-rq_MJMB} \\
Drones	& \cite{Rejeb2021-ye_MJMB} \\
Forecasting	& \cite{Altay2022-en_MJMB} \\
Health in disasters	& \cite{Dixit2024-fw_MJMB} \\
Human resources	& \cite{de-Camargo-Fiorini2022-du_MJMB} \\
Information sharing	& \cite{Negi2024-vh_MJMB} \\
Localisation	& \cite{Frennesson2021-xw_MJMB} \\
Modularity \& standards	& \cite{Paciarotti2021-nw_MJMB} \\
Optimisation models	& \cite{Kamyabniya2024-ba_MJMB} \\
Outsourcing	& \cite{Gossler2020-uo_MJMB} \\
Procurement	& \cite{Moshtari2021-vq_MJMB} \\
Refugees	& \cite{Dantas2023-ol_MJMB} \\
Role of media	& \cite{Abdulhamid2021-lc_MJMB} \\
Sustainability	& \cite{Anjomshoae2023-qd_MJMB} \\
\hline
\end{tabular}
\label{tab:humops}
\end{table}

The OSCM research community has come a long way since the mid-2000 with regards to the number of researchers involved, papers published and cooperation with practice. The focus has changed from logistics to supply chain/operations management \citep{Thomas2005-jq_MJMB}, securing a more holistic approach by increasingly taking the system into account \citep{Besiou2011-gi_MJMB}. However, there is an ongoing discussion of whether the published research match the needs in practice. \cite{Besiou2020-lf_MJMB} conclude that many papers tend to deal with traditional OM problems without much consideration for the specific disaster context. They provide advice on how to conduct relevant research. \cite{De-Vries2020-ag_MJMB} find that optimisation models for humanitarian operations need a paradigm shift towards cost-effectiveness maximisation and to provide more empirical evidence. Similarly, \cite{Kovacs2019-tq_MJMB} conclude that research must be contextualised and present a roadmap for higher research quality.

\subsubsection*{Closing remarks – the past, the present, and future}
\cite{Besiou2015-wf_MJMB} discuss the roles of different methodologies in ``solving'' problems and ``developing'' theory. For example, literature reviews, case studies and surveys help more in understanding how reality works, ``soft'' OR to study the behaviour of complex systems and ``hard'' OR when solutions need to be provided in specific subsystems. We agree and suggest our community to reflect on when to use which hammer, depending on the nail we want to hit. To conclude we briefly summarise the past and the present with suggestions for the future.

\textbf{Different methods}: In the beginning, research was characterised by conceptualisation based on literature reviews, case studies and OR in separate papers. During the past decade there has been more use of mixed methods, combining cases or surveys with OR and simulation. Mixed methods are becoming the norm in many high ranked journals, even if there are still many pure OR-papers, conceptual papers and case studies. What about the future? We expect to see more experiments, more use of big data and machine learning. 

\textbf{Different foci}: Research on humanitarian operations has traditionally focused much on response and preparedness. In the future we should see more on mitigation and recovery. Related to this we will see more focus on sustainability, i.e., going beyond efficiency and effectiveness in the traditional sense. Further, research has focused a lot on cost and response time, while recently more metrics like equity \citep{Breugem2024-db_MJMB} and pain \citep{Holguin-Veras2013-cf_MJMB,Wang2019-sp_MJMB} are considered important.

\textbf{Different disciplines}: Applications from OSCM have traditionally included facility location, supplier selection, network design \citep{Balcik2008-oa_MJMB,Dufour2018-za_MJMB}, inventory location-allocation, forecasting, transportation planning/routing supply chain risk management \citep{Gralla2024-ib_MJMB}. During the past few years, we have seen OSCM combined with other management disciplines like organisation theory, leadership, marketing, strategy, finance \citep{Ozer2024-ci_MJMB}. For the future, we expect to see more interdisciplinary research beyond management, combining OSCM for example with civil engineering, political science, and medicine.

The field started off relatively narrow, focusing on logistics in the response phase, demonstrating the importance of logistics to humanitarian practice and the humanitarian context to logistics research. During the past two decades the field has developed in depth and breadth, encompassing a multitude of themes. However, there is a danger of driving research too much toward areas and problems where data is readily available \citep{Gralla2024-ib_MJMB}, allowing use of fancy methods which are `in', and readily publishable rather than practically implementable. For example, we have seen a lot of focus on funding/financial flows lately. While getting funding is essential, we should not forget our field is operations, i.e., spending the available funding wisely and with the goal of saving lives.

\subsection[Manufacturing (Jiyin Liu)]{Manufacturing\protect\footnote{This subsection was written by Jiyin Liu.}}
\label{sec:Manufacturing}
Manufacturing is a central link in the supply chain, responsible for transforming raw materials into finished products to meet customer demands and create value. The efficiency and effectiveness of manufacturing operations directly impact lead times, cost structures, product quality, and customer satisfaction, all of which contribute to the overall performance of the supply chain. Many foundational concepts in operations management originated from manufacturing practices. Frederick Taylor's Scientific Management principles \citep{Taylor1911-gm_JL}, representing the studies on improving productivity in the late 19th and early 20th centuries as mass production arose, laid the groundwork for modern operations management. The need to make production systems more efficient also led to studies on plant layout \citep[see][]{Stephens2013-ec_JL} and material handling (see \S\ref{sec:Material_handling}).

With mass production and standardisation, quality control became vitally important in manufacturing, leading to the development of quality management practices focused on inspection, testing and process control \citep{Barrie-Wetherill1991-lo_JL}. Over time, quality management evolved to emphasise broader concepts such as customer satisfaction, total quality management (TQM), and continuous improvement. See \S\ref{sec:Quality_management} for more on quality management. While quality management principles are now widely applicable across all business sectors, they continue to play a crucial role in manufacturing.

Locating manufacturing facilities has been a critical decision in supply chain management, traditionally considering supply proximity, market access, and transportation costs. Globalisation and offshoring of production have profoundly reshaped the manufacturing landscape. In global supply chain networks, manufacturing facility location decisions must account for a broader set of factors \citep{MacCarthy2003-yi_JL}. These include the availability of skilled labour, the quality of infrastructure, political stability, the legal and regulatory environment, and the ease of doing business. Additionally, considerations such as cultural compatibility, environmental regulations, and proximity to innovation hubs have become increasingly important. These complexities make global location decisions critical to the success of the overall supply chain strategy.

A critical area of operations management for manufacturing is production planning and control (PPC). There used to be textbooks focusing specifically on PPC \citep[e.g.,][]{Bolton1994-it_JL}, but it is now usually included as an important part of Operations Management books \citep[e.g.,][]{Slack2019-hp_DPvD}. Within the PPC framework, the production planning level makes production and inventory plans for different products over multiple periods based on customer orders or demand forecasting to optimise the overall objective. The production plan for final products is translated to production plans of subassemblies and components at different levels according to the product structures. This is performed through a material requirement planning (MRP) system. The MRP evolved to manufacturing resources planning (MRP-II) considering processing capacities as well \citep{Sheikh2002-ng_JL}. Further expansion led to a company-wide system covering all business functions – Enterprise resources planning (ERP) that is widely used in different sectors not limited to manufacturing.

In manufacturing, except for raw materials, inventory is usually replenished gradually through production. Order quantity is thus optimised with an economic production quantity (EPQ) formula \citep{Taft1918-jy_JL} instead of economic order quantity (EOQ). While inventory acts as a safety buffer, it can also hide many problems and increase costs. The just-in-time (JIT) production system originated in Toyota \citep{Monden1998-uz_JL} aims to eliminate inventory and produce items only when they are needed. This has been extended to minimising all wastes in general, leading to lean manufacturing and lean operations (see also \S\ref{sec:Lean_and_agile}).

At a lower level of PPC, scheduling makes detailed plans to allocate specific resources over time to processing tasks. Since Johnson’s pioneering work \citep{Johnson1954-op_JL}, scheduling has quickly developed into a mature research field with books published \citep[e.g.,][]{French1982-eo_JL}. As the production environment can be very complex and involve more practical constraints, the gap between the scheduling theory and practice was noted and the need for more practical studies was pointed out \citep{Maccarthy1993-py_JL}. In recent decades increasingly more production scheduling studies have been addressing scheduling problems in practical manufacturing environments. This benefits manufacturing operations and enriches production planning and scheduling research. Meanwhile, scheduling problems in other sectors are widely studied as well. See \S\ref{sec:Scheduling} for more.

Different manufacturing industries have their specific characteristics which are reflected in their operations and supply chain management problems. We briefly discuss several manufacturing industries below. The automotive industry has highly complex and tiered supply chains involving thousands of suppliers globally. Safety and regulatory standards require stringent quality management. Just-in-time principle is commonly applied to minimise inventory costs and enhance efficiency. Besides the routine operations and supply chain management decisions, automotive manufacturers also make long-term plans for the allocation of products to production sites and investments in additional capacity in the global production network \citep{Fleischmann2006-sp_JL}. Increasingly consumers can choose the configurations for the vehicle they purchase. This makes it challenging to quickly fulfil the customised orders while keeping a low inventory of finished products. Reassignment of products in the production and distribution pipeline and inter-dealer trading have been demonstrated to be effective in reducing lead time and achieving quicker fulfilment \citep{Brabazon2006-xq_JL,Brabazon2010-sp_JL}.

The process industries include a wide range of businesses, such as the chemical and pharmaceutical industry, food industry, and steel industry. They often involve continuous or batch processes, different from discrete manufacturing as indicated by \cite{Van-Donk2006-nf_JL}. For continuous chemical processing, the cost of switching from one product to another is typically sequence dependent. Considering this, \cite{Cooke2006-si_JL} took a continuous economic lot scheduling problem (ELSP) approach to make production schedules. Pharmaceutical and food products are subject to high health and safety standards as well as variations in supply and demand. Therefore, quality and inspection must be considered carefully in designing a supply chain and selecting suppliers \citep{Pariazar2017-tc_JL}. Supply chain resilience is equally important. \cite{Lucker2017-eg_JL} studied the problem of enhancing resilience against disruption risks by considering mitigating inventory, double sourcing and agile capacity. In food supply chains, decision making needs to consider the fact that food products are perishable \citep[][also see \S\ref{sec:Agriculture}]{Huang2018-hr_JL}. While the food and pharmaceutical industries are directly related to human life, the iron and steel industry supports the economy by supplying materials to other industries such as automotive, construction and shipbuilding. There are various operations planning decisions in steel production. \cite{Tang2014-ld_JL} developed decision support systems for several such problems including batching problems, as well as a problem of allocating/reallocating slabs to customer orders which shares some similarities with the order fulfilment problem in the automotive supply pipeline described earlier. Product sequencing is also very important to minimise switching costs in finishing processes such as electro-galvanising lines \citep{Tang2010-eo_JL}.

The electronics and semiconductor manufacturing industry has been growing more and more important. Electronics assembly operations have been studied with various objectives. Throughput maximisation is a typical objective and is used in a telecommunication equipment assembly \citep{Kobza2002-fu_JL}. For a light-emitting diode (LED) array assembly, the objective was set to reduce the inventory level of the LED parts used as they may become obsolete with frequent change of product models \citep{Chang2013-li_JL}. Semiconductor chips are key components for electronics products. Semiconductor manufacturing includes wafer fabrication and chip assembly. Wafer fabrication involves hundreds of process steps with recirculating process flows. In such an extremely complex system, detailed operations scheduling usually adopts dispatching rules considering waiting time or job urgency. The mid-term production planning optimises product lot release times to the system considering process capacities, cycle times, lot priorities and work-in-process \citep{Kriett2017-nu_JL}, while the master production planning sets targets considering demands. The system of chip assembly and testing is also complex but less so than fabrication. While a mid-term planning model schedules product groups to the assembly and testing processes \citep{Zhang2017-ph_JL}, a more detailed scheduling problem can be solved to decide on machine setups \citep{Hur2019-cj_JL}, to maximise fulfilment of customer demands and maximise throughput. Allocation of wafer lots and chip lots to fulfil specific customer orders can be decided as lot-to-order matching problems in semiconductor supply chain settings to minimise unfulfillment and over-fulfilment \citep{Ng2010-rl_JL}.

Manufacturing processes and systems have been advanced with computer and digital technology, which in turn affect their operations and supply chain management problems. This applies to most industries while the machining industry is a typical example. Combining computer numerical control machines and automated material handling under central computer control resulted in the development of flexible manufacturing systems (FMS) capable of quickly responding to dynamic market demands of products in high variety and low volume, supporting mass customisation. See \cite{Maccarthy1993-nb_JL} for a classification of FMS. The flexibility also introduces additional complexity to the planning and control of the operation, e.g., extra loading decisions on setting up the machines and routing the parts, as first studies in \cite{Stecke1981-eb_JL}. Computer-integrated manufacturing (CIM) automates all processes and integrates data and information across the entire production system to achieve more efficient operations, improved accuracy, and faster time-to-market. See \cite{Hannam1997-rs} for more details.

Technology continues to advance rapidly in the 21st century. At the process level, additive manufacturing \citep{Abdulhameed2019-av_JL} has evolved from a prototyping tool into a potential method for efficiently producing individualised products. On a global scale, under frameworks such as Industry 4.0, new information and communication technologies, including the internet of things, cloud computing, big data analytics, and artificial intelligence, are being integrated into manufacturing systems to enable smart or intelligent manufacturing. This shift presents both challenges and opportunities for manufacturing operations management to leverage these technologies to fully realise their benefits. \cite{Zhou2022-ic_JL} conducted a systematic review of research on intelligent manufacturing and proposed future research directions under several themes.

\subsection[Professional services (O. Zeynep Aksin \& Elif Karul)]{Professional services\protect\footnote{This subsection was written by O. Zeynep Aksin and Elif Karul.}}
\label{sec:Professional_services}
Professional services make up a large and growing portion of the service sector. Management consulting, information technology (IT) services, architecture and engineering, legal services, and investment banking are prominent examples. Besides being an important sector in the economy, professional services act as a value-generation catalyst for many other private and public organisations.

The service management literature has attempted to characterise distinguishing features of professional services and proposed various schemes for their classification \citep{Schmenner_OZAEK,VonNordenflycht_OZAEK,Lewis_OZAEK}. Broadly speaking, the product of a professional service is a process. These service processes typically require knowledge work performed by professionals. So new product and service design (see \S\ref{sec:New_productservice_development}) implies process design (see \S\ref{sec:Process_design}) and job design (see \S\ref{sec:Job_design}). Professional services are tailored to customer needs. Their processes display high levels of customisation. Work in most professional service firms is project-based. Thus, project management (see \S\ref{sec:Project_management}) is an essential pillar of professional service management. Customer interaction is high and professional service firms rely on their customers as one of the key inputs in their production processes. Performance is driven by co-production between service professionals and clients. Many professional services are B2B and are active in service supply chains (see \S\ref{sec:Supply_Chain_Management}).  In summary, the operational performance of a professional service hinges on process design and management, the management of knowledge work, and systems that enable customer efficiency and effectiveness.

The process structure of a professional service drives operational performance outcomes including capacity, cost, and quality. Service processes can be characterised by their complexity, the number and elaborateness of the steps involved, and by their divergence, namely including a high level of executional latitude \citep{Shostack_OZAEK}. Professional services have a large share of high complexity-high divergence processes. A highly complex process requires longer throughput time and additional skills to perform the work. Taking a process view implies managing the entire customer experience in an integrated manner, however, complexity makes it challenging for one professional to handle the entire customer encounter. Expertise and task specialisation are inevitable but require handoffs that can create interruptions in a process. On the other hand, multi-skilled professionals will require higher compensation. Proper structuring of complex processes with these tradeoffs in mind leads to superior performance. Processes that combine multi-tasking by flexible cross-trained servers \citep{Iravani_OZAEK,Akşin2_OZAEK} are designed to balance capacity, quality, and labour costs. Process designs with multiple hierarchical stages are a common approach to managing these tradeoffs in settings with heterogeneous customer needs. In technical support, generalist gatekeepers refer certain cases to expert specialists \citep{Shumsky_OZAEK}. In health settings, triage plays a similar role \citep{Sun_OZAEK}. Collaboration and coordination among professionals \citep{Siemsen_OZAEK,Rahmani_OZAEK,Gurvich_OZAEK,Gurvich2_OZAEK,Roels_OZAEK} is critical to ensure seamless integration in a complex workflow. A process having high divergence, on the other hand, implies service workers who need to make judgments affecting the process flow. A standard operations procedure cannot be established, the process suffers from high variability, and quality is difficult to assess or manage. Processes involve discretionary tasks \citep{Hopp2_OZAEK,Debo_OZAEK}. Customer-intensive professional services have an inherent tradeoff between quality and capacity. Higher quality in outcomes requires spending more time with customers thus consuming capacity \citep{Anand_OZAEK}. Diagnostic services require tradeoffs between diagnostic accuracy and time taken for additional tests at the discretion of the professional \citep{Alizamir_OZAEK}. In expert service settings, like medical, legal, and consultancy services, clients may not be able to assess the appropriateness of the provided service giving these services a credence characteristic \citep{Debo2_OZAEK}.  

The main product of a professional service is knowledge work performed by employees with varying talent and experience levels. Professional service work is organised around projects completed by teams. Demand for these services strongly depends on quality and reputation, which is partially driven by the employees who perform the work. Team composition can act as a signal of quality for clients \citep{Talluri_OZAEK}. It is thus essential to manage the employee talent pool and understand the implications of team dynamics on performance. Managing this critical resource involves hiring and retaining the right employees and ensuring they continue to learn \citep{Gans_OZAEK,Arlotto_OZAEK,Musalem_OZAEK}. Designing operations and forming teams that enable the best learning outcomes is important \citep{Huckman2_OZAEK,Ramdas_OZAEK,Akşin_OZAEK}. Frequently teams in professional service work are fluid teams that are formed temporarily. Evidence from the team learning literature suggests that team familiarity leads to improved performance. However, diversity in experiences, tasks, and current and former team members are all shown to affect performance outcomes. Team formation that considers these tradeoffs is essential for a professional service. By scheduling or controlling the mix of tasks performed by professionals, operational outcomes like throughput, process time, and quality can also be optimised \citep{Pinker_OZAEK,Bray_OZAEK,Legros_OZAEK}.

Customers not only drive the demand for professional services but also play a role as an input in defining their outcome. Customer satisfaction and retention are critical for a sustainable professional service business. Past experiences determine contract renewals. Thus, effort drives retention in a service relationship \citep{Aflaki_OZAEK}. For a professional service, it is not just the effort of the service firm but also their clients’ effort that matters. Co-production refers to the joint production effort by service providers and customers. \cite{Xue_OZAEK} propose the concept of customer efficiency, arguing that customer work also needs to be managed just like employee productivity. The design of co-produced services requires determining how work is allocated between the parties, for instance, consultants and clients, and contracts need to be specified that manage the effort of both sides \citep{Xue2_OZAEK,Roels2_OZAEK}. \cite{Karmarkar_OZAEK} provide a more general analytical framework for value co-production in services. 

Technology is a major driver in the market for professional services and an enabler in their production and delivery. Developments in cloud computing have created a huge market for IT as a service. Different organisational structures that can exploit cost efficiencies, simultaneously creating value, have emerged in this domain. Given their technical nature, professional services in the technology arena are particularly affected by their customer capabilities leaving a space for intermediaries in their use \citep{Wu_OZAEK}. Differences in capabilities drive firms’ purchasing choices regarding outsourcing these services \citep{Chang_OZAEK} or keeping them in-house (see \S\ref{sec:Outsourcing}). For in-house professional services, the possibility of remote delivery has made the adoption of shared service structures that combine finance, HR, IT, and other shared services in one organisation more prevalent \citep{Akşin3_OZAEK}. Digital technologies have further enabled online outsourcing platforms for IT services \citep{Hong_OZAEK}. Online platforms have also led to changes in the labour market for professional services through the possibility of gig work. Web-based platforms that match service demands to service supply have enabled professionals to provide their expertise at their convenience while allowing firms to access capacity flexibly and timely \citep{Taylor_OZAEK}. Traditional professional service firms decide on which professionals to hire and retain, however in online service marketplaces this dynamic is two-sided and applies to workers' decisions regarding the platforms in which they participate \citep{Allon_OZAEK,Liu2_OZAEK}. More recently, artificial intelligence (AI) technologies present the potential to transform professional service work. As with most new technologies, despite their enormous prospects for value creation, they present new hurdles to overcome in business contexts \citep{Davenport_OZAEK}. AI is shifting how professionals work, possibly redefining the scope of their work \citep{Huang_OZAEK}. Decision support systems for knowledge work are incorporating AI features \citep{Gupta_OZAEK}. Integration of AI tools in decision-making by professionals requires a better understanding of when they are superior to human experts' judgment. Even when superiority is established, acceptance by stakeholders in processes, including the experts themselves, is not always ensured \citep{Dai_OZAEK,Vericourt_OZAEK}. Designing new processes that embrace the value-creation potential of these shifts in the technology landscape and that overcome major adoption barriers constitutes the future of professional services.

We direct the reader to the book by \cite{Maister_OZAEK} for more on managing professional service firms. The special issue on professional services \citep{Harvey_OZAEK} contains work focusing on operations management issues in these firms. \cite{Field_OZAEK} present exciting research directions under key themes in service operations, including a theme on the management of knowledge-based service contexts.

\subsection[Retail (Stanley Frederick W. T. Lim \& Bengü Nur Özdemir)]{Retail\protect\footnote{This subsection was written by Stanley Frederick W. T. Lim and Bengü Nur Özdemir.}}
\label{sec:Retail}
Retailing has been a fruitful application area for many operations and supply chain management topics, including inventory management (\S\ref{sec:Inventory_management}), pricing (\S\ref{sec:Pricing}), and purchasing (\S\ref{sec:Purchasing_procurement}), owing to its critical role in driving global economic growth. The retail sector generates millions of jobs and contributes 20\% to the United States' annual gross domestic product \citep{nrfretail_SFRTLBNO}. \cite{caro2020future_SFRTLBNO} define retailing as ``all the activities associated with the selling of goods to the final consumer''. To systematically examine these activities, we adopt a consumer-shopping journey framework, comprising \textit{before}, \textit{during}, and \textit{after purchase}. We then discuss key operational enablers underlying the efficiency of these consumer-facing activities, emerging topics of interest, methodologies employed, and further research opportunities.

\subsubsection*{Consumer-shopping journey framework}\label{framework_SFRTLBNO}
\textbf{Before Purchase:} This stage focuses on how consumers acquire product information, become aware of a brand, and decide to purchase. Traditionally, brick-and-mortar stores were the primary points of purchase. With omnichannel retail, these stores now also serve as catalogues where consumers can acquire information by seeing, touching, and comparing products. To capitalise on this, some retailers are opening showrooms that do not carry inventory. Showrooms are associated with higher demand \citep{bell2018offline_SFRTLBNO}, fewer returns \citep{bell2020customer_SFRTLBNO}, improved operational efficiency \citep{bell2018offline_SFRTLBNO}, and lower sensitivity to fulfilment lead times \citep{lim2024channel_SFRTLBNO}. In addition to brick-and-mortar stores and showrooms, consumers use online brands as an information acquisition channel \citep{gallino2014integration_SFRTLBNO}, leading to research on online retail design \citep{shi2013information_SFRTLBNO}.

Regardless of the channel, price is essential information to consumers, which is affected by retailers' distribution channel \citep{brynjolfsson2000frictionless_SFRTLBNO}, replenishment contracts \citep{lim2022scan_SFRTLBNO}, assortment strategy \citep{heese2018effects_SFRTLBNO}, and inventory constraints \citep{ferreira2018online_SFRTLBNO}. In addition to impacting prices, assortment (product variety) is central to the customer experience. \cite{wang2018prospect_SFRTLBNO} indicates that a wide assortment signals expertise, increasing purchase likelihood. Product availability is another key component; for example, in online retailing, low product availability generates pressure to buy quickly, leading to increased sales \citep{calvo2023disclosing_SFRTLBNO}. Having products available only online, or both online and in-store, is a strategic decision for omnichannel retailers \citep{ertekin2022online_SFRTLBNO}.

\textbf{During Purchase:} This stage consists of the actual buying and interactions with the retailer. In a physical store, interactions with sales associates shape the customer experience. \cite{ertekin2020assessing_SFRTLBNO} demonstrate that salesperson competence and friendliness are associated with return prevention. To improve sales associates' competence, retailers can invest in training. \cite{fisher2021does_SFRTLBNO} find that online training increases sales. The customer experience is also affected by the store environment, including ambience \citep{ertekin2020assessing_SFRTLBNO}, fixture types \citep{zhang2023effects_SFRTLBNO}, or fitting room designs \citep{lee2021managing_SFRTLBNO}.

The shift from the brick-and-mortar model to omnichannel retailing changes the way consumers and retailers handle fulfilment and transportation. In traditional in-store settings, consumers incur the effort to fulfil their orders and transport goods home. In omnichannel settings, retailers share part of the fulfilment and transportation effort. Various channel strategies offer customers different convenience levels. These include buy-online, pick-up-in-store \citep{gallino2014integration_SFRTLBNO}; reserve-online, pick-up-in-store \citep{jin2018buy_SFRTLBNO}; pick-up-from-locker/delivery trucks \citep{glaeser2019optimal_SFRTLBNO}; and buy-online, ship-from-store \citep{das2023order_SFRTLBNO}.

\textbf{After Purchase:} This stage focuses on the post-purchase customer experience, emphasising returns and customer feedback. Product returns are costly and harmful to the environment, but they are linked to increased sales and customer satisfaction. Consumers value return policies because of the potential mismatch between the product and their expectations. Also, some customers may be opportunistic by buying products with the intention of returning them after a brief use \citep{shang2017optimal_SFRTLBNO}. Therefore, retailers test different return policies involving return window \citep{ertekin2021does_SFRTLBNO}, reimbursement method \citep[e.g., store credit versus refund][]{abdulla2022consumers_SFRTLBNO}, or modes of return \citep[e.g., mail versus in-store][]{nageswaran2020consumer_SFRTLBNO}.

After making a purchase, customers often provide feedback. Star ratings and review sentiment influence future demand \citep{cho2022reading_SFRTLBNO}. Therefore, some retailers may create fake reviews for themselves and competitors \citep{luca2016fake_SFRTLBNO}, leading to customers questioning e-commerce platforms' credibility. To address these concerns, e-commerce platforms use strategies like purchase verification or utilising moderators to manually approve reviews \citep{kokkodis2022optional_SFRTLBNO}.

\textbf{Operational Enablers:} This stage focuses on backend operational elements, facilitating the entire consumer-shopping journey. Among these, inventory management has been a fruitful area. High inventory levels stimulate demand, which should be considered in inventory replenishment policies \citep{balakrishnan2004stack_SFRTLBNO}. Inventory models mostly assume that inventory data is accurate. Yet, substantial inventory records are inaccurate \citep{dehoratius2008inventory_SFRTLBNO}. ``Phantom stockout'' and ``phantom inventory'' are examples of inventory record inaccuracies \citep{chen2021fixing_SFRTLBNO}.

As indicated earlier, assortment is important for consumers, making its optimisation a key operational focus. Dynamic assortment models help retailers learn consumer preferences and optimise inventory replenishment \citep{ulu2012learning_SFRTLBNO}. Such models provide important managerial insights; for example, an optimal assortment is not always composed of the most dominated or proﬁtable products \citep{honhon2012optimal_SFRTLBNO}.

Fulfilment is another central topic, especially with the rise of online channels. Key areas include deciding the locations to fulfil orders \citep{acimovic2015making_SFRTLBNO}, the sequence in which orders are processed \citep{figueira2023impact_SFRTLBNO}, and whether or how to split a multiple-item order into separate shipments \citep{zhang2018package_SFRTLBNO}.

\textbf{Social Aspects and Emerging Issues:} Retailing and socio-environmental factors are mutually influential. First, consumers are more aware of their purchases' impact on the planet; hence, retailers must satisfy conscious shoppers. Sustainability initiatives, like providing fair-trade products, affect customers' store choices \citep{hampl2013sustainable_SFRTLBNO}. Second, retailers in developing economies experience unique challenges, like poor infrastructure or the lack of distribution channels, causing high inventory replenishment costs \citep{gui2019improving_SFRTLBNO}. Nanostores, a common retail model in such markets, operate under strict cash constraints \citep{boulaksil2018cash_SFRTLBNO}. Third, equity concerns relate to job design and workload. Responsible scheduling \citep{kesavan2022doing_SFRTLBNO}, spreading promotions to make store traffic manageable \citep{kalkanci2019role_SFRTLBNO}, or removing gender bias in job assignments \citep{corbett2024om_SFRTLBNO} show how retailers can incorporate equity. Research on responsible retailing is growing, as evidenced by dedicated special issues \citep[e.g.,][]{specialissuemsom_SFRTLBNO}.

\subsubsection*{Methods}\label{methods_SFRTLBNO}
Retail research employs a wide variety of methods, which we categorise as analytical, empirical, experimental, and numerical. Among analytical models, the traditional optimisation approach is common to provide a decision model to mathematically optimise retail variables, including assortment \citep{ulu2012learning_SFRTLBNO}, order replenishment \citep{balakrishnan2004stack_SFRTLBNO}, and phantom stockouts \citep{chen2021fixing_SFRTLBNO}. Theoretical models are also common. These stylised models are used to model pricing and consumer welfare \citep{gui2019improving_SFRTLBNO}, assortment \citep{heese2018effects_SFRTLBNO}, showroom network design \citep{lim2024channel_SFRTLBNO}, and fulfilment \citep{jin2018buy_SFRTLBNO}.

Empirical methods for causal inference are increasingly being used, in part due to the availability of real-world data. These methods use reduced-form regressions and structural models. Reduced-form models are used to generate insights related to showrooms \citep{bell2018offline_SFRTLBNO}, subscription models \citep{lim2021shopping_SFRTLBNO}, return policies \citep{ertekin2021does_SFRTLBNO}, omnichannel modes \citep{gallino2014integration_SFRTLBNO}, and pricing \citep{brynjolfsson2000frictionless_SFRTLBNO}. Structural demand models are popular for understanding consumer preferences and behaviours. These models have been used in research on pricing \citep{wang2018prospect_SFRTLBNO}, assortment \citep{honhon2012optimal_SFRTLBNO}, inventory replenishments \citep{lim2022scan_SFRTLBNO}, and fulfilment \citep{ertekin2022online_SFRTLBNO}.

Lab and online experiments are widespread given their ease of implementation and low costs. They are used in research on consumer preferences related to sustainability \citep{hampl2013sustainable_SFRTLBNO}, information precision \citep{lim2024channel_SFRTLBNO}, and store environment \citep{zhang2023effects_SFRTLBNO}. This experiment type is common in behavioural research, employing decision-making theories \citep{liang2024examining_SFRTLBNO}. Field studies are also conducted to analyse retail concepts such as store environment \citep{zhang2023effects_SFRTLBNO} and fitting rooms \citep{lee2021managing_SFRTLBNO}.

Among numerical methods, researchers utilise simulation to demonstrate performance of their optimisation or heuristic models \citep{acimovic2015making_SFRTLBNO}. This allows them to compare different policies and to test how the prescribed policy performs under various scenarios. Machine learning (ML) is another numerical method that is commonly used for prediction problems. Researchers utilise ML to examine online retailing design \citep{shi2013information_SFRTLBNO} and pricing \citep{ferreira2018online_SFRTLBNO}. A prediction method can be followed by an optimisation algorithm to provide prescriptions on variables like optimal location \citep{glaeser2019optimal_SFRTLBNO} or network expansion \citep{huang2019predictive_SFRTLBNO}.

\subsection*{Further reading and future research}\label{further_reading_SFRTLBNO}
\cite{rooderkerk2023advancing_SFRTLBNO} discuss key decisions, risks, and opportunities in omnichannel retail fulfilment models. \cite{caro2020future_SFRTLBNO} provide a review of retail research, categorising topics and methodologies and assessing their impact on both academia and industry. Meanwhile, \cite{mou2018retail_SFRTLBNO} offer a review focused on retail store operations decisions.

Future research can explore last-mile delivery innovations, which are essential to meet growing consumer demand in the retail sector \citep{lim2023hbr_SFRTLBNO}. This segment of the supply chain will be stress-tested in coming years. The integration of artificial intelligence into various retailing decisions, such as inventory or pricing, can also be investigated. Furthermore, the extant literature typically employs discrete choice models, such as the multinomial logit, where consumers are assumed to select a single item, or the multivariate logit, where consumers choose multiple items, each in a unit quantity. A more realistic approach is to consider basket-level choices, including varying quantities of multiple items \citep[see][for an application]{lim2024estimating_SFRTLBNO}. We hope these discussions inspire further studies and method contributions to the retail operations literature.

\subsection[Small and medium-sized enterprises (Kostas Selviaridis)]{Small and medium-sized enterprises\protect\footnote{This subsection was written by Kostas Selviaridis.}}
\label{sec:Small_and_medium-sized_enterprises}

Small and medium-sized enterprises (SMEs) matter – they make a sizeable contribution to gross value added and total employment as they represent the bulk of all firms internationally. In the United Kingdom, for instance, there were 5.6 million SMEs at the beginning of 2023, which corresponded to over 99\% of all companies registered in the country. These SMEs accounted for about half of total turnover in the UK private sector and three-fifths of employment \citep{Federation_of_Small_Businesses_2024-yq_KS}. Beyond their macroeconomic significance, SMEs play a key role in supply chains which, in many sectors, have become increasingly long and complex. In the context of global supply chains, SMEs supply raw materials, manufacture and export critical components, parts and finished goods, and provide core business services such as transport and logistics \citep{Kull2018-ub_KS}. They also possess specialised sets of knowledge and can be a source of technological innovations that help to address contemporary societal challenges including climate change and affordable and accessible healthcare \citep{Selviaridis2023-ul_KS}.

SMEs are defined mainly by the number of their employees, although there is no uniform definition internationally. In the European Union, SMEs are defined as companies with up to 250 employees and a maximum annual turnover of €50 million, or a balance sheet total of up to €43 million \citep{European_Commission_2020-ar_KS}. The UK is using the same defining criteria. In the United States, the SME definition is also based on workforce size and annual turnover, but it varies by sector – overall, an SME is a company who employs between 250 and 1,500 people and has a maximum turnover of \$47 million \citep{US_Small_Business_Administration2023-sg_KS}. Regardless of the definition used, the term ``SMEs'' incorporates a broad church of firms: from micro (less than 10 employees, according to the European definition) and small (between 10 and 50 employees) companies to medium-sized enterprises. These categories of firms can differ significantly with respect to the challenges they face and their ability to function as knowledgeable suppliers, service providers, or customers in the supply chain. SME firms can also be distinguished based on their technological knowledge intensity: high-technology SMEs  have considerable growth potential and drive innovation and productivity improvements more than low-technology SMEs do \citep{Sato_2023-ly_KS}. 

The operations and supply chain management (OSCM) literature has identified some distinctive characteristics of SMEs vis-à-vis large firms, and discussed the advantages and limitations of small businesses that shape their contributions in supply chains. By virtue of their less formal, flatter governance structures and agile decision-making processes, SMEs are often responsive to changing circumstances and flexible in handling crises \citep{Ketchen2021-kw_KS}. They are also fast in exploiting market opportunities and introducing new products and services at pace, relying on founding entrepreneurs as their driving force \citep{Kull2018-ub_KS}. On the other hand, SMEs face finance and human resource restrictions; have limited production and distribution capabilities; and lack connections and market reputation \citep{Selviaridis2021-pn_KS,Zaremba2016-mn_KS}. These limitations are particularly germane to micro and small businesses, compared to medium-sized enterprises which tend to have more resources available. 

OSCM scholars have examined why and how these SME peculiarities influence the ability of small businesses to collaborate and integrate with their larger supply chain counterparts to achieve a variety of performance outcomes including, among others, quality improvement and innovation \citep[e.g.,][]{Rezaei2015-fr_KS}. For instance, research has shown that innovative small firms have difficulty in collaborating with large buying organisations for innovation purposes. This difficulty stems from SME limitations in terms of resources, capacity and social capital. It also relates to institutional failures and weak abilities of large buying firms to engage with small supplier firms with specific needs including fast-paced decisions and cash flow \citep{Knight2022-iw_KS}. Public policies targeting technology-intensive SMEs seek to address these issues and foster collaborative innovation in supply chains by, for example, incentivising SMEs and buying organisations to invest in collaborative R\&D projects, connecting SMEs to buying organisations and their incumbent suppliers, and supporting the development of SMEs as (possible) suppliers \citep{Selviaridis2022-yb_KS,Selviaridis2024-le}. 

Supplier development is an important element of a firm's sourcing strategy (\S\ref{sec:Outsourcing}) and while such support programmes are not restricted to SMEs, they are especially relevant in the context of efforts to improve the capabilities and productivity of small suppliers. Supplier development programmes originally focused on quality improvement (Wagner, 2010), but their scope has been extended to digitalisation and sustainability goals, among others \citep[e.g.,][]{Jia2021-zq_KS}. In the public sector context (\S\ref{sec:Public_nonprofit_sector}), many governments seek to increase the involvement of SMEs in public contracting, either as direct suppliers or subcontractors in the case of high-value contracts \citep{Harland2019-es_KS}. SMEs are also important partners in efforts to promote sustainability and resilience goals in supply chains: increasing spend with SME suppliers helps to create jobs and wealth in local communities, reduce carbon emissions, and diversify the supply network of a buying organisation \citep{Mills2022-rb_KS,Selviaridis2023-ul_KS}. Supplier diversity programmes seeking to increase engagement with ``minority'' suppliers (e.g., based on gender, ethnicity or disabilities) are directly relevant to the SME population, as most of these minority suppliers are simultaneously micro or small businesses \citep{Bateman2020-zb_KS}.

In addition to SME support initiatives, OSCM research has examined supply chain finance solutions \citep{Caniato2019-hj_KS}. In brief, these seek to ensure the financial stability of suppliers – especially SME suppliers – against a backdrop of major crises such as the crunch of 2008. Effective cash-flow management is imperative for SMEs given their finance limitations – to this end, supply chain finance aims to optimise financial flows within the supply chain \citep{Hofmann2011-en_KS}. The solutions, typically led by financial institutions or technology providers, involve collaboration among supply chain counterparts to ensure effective and fair outcomes such as paying small suppliers in a timely fashion.
	
Research has also emphasised SME capabilities in procurement, supply chain management (SCM) and logistics \citep{Arend2005-xh_KS,Gelinas2004-ic_KS,Maloni2017-lb_KS}, among other areas. Regarding procurement, for instance, SMEs often have limited resources to pursue structured supplier selection, contracting, and supplier relationship management practices \citep{Ellegaard2006-nq_KS}. On the other hand, they can leverage the boundary-spanning capacity of their founder-managers to effectively manage their supplier relationships \citep{Son2019-dm_KS}. Overall, however, there has been considerably less attention to the roles and activities of SMEs as customer firms who need to manager their supply chains, relative to studies focusing on SMEs as supplier firms.

Researching the contributions of SMEs in supply chains presents methodological and theoretical issues. Methodologically, the preference for multiple-respondent survey designs has meant that the perspective of SMEs is largely neglected because micro and small firms often lack multiple informants. \cite{Kull2018-ub_KS} have highlighted this challenge and argued for inclusive research designs using single-respondent approaches and expanded units of analysis. Furthermore, data triangulation strategies typically used in qualitative research are more challenging to pursue in that SMEs are not obliged to disclose their performance and business conduct to the extent that large firms do, meaning that valuable sources of secondary data may be in short supply. Compensating for such limitations might involve using participant observation techniques, diaries, and data collection through SME representative organisations such as national associations.

Theoretically, studying SMEs can help us understand better supply chain phenomena which are inherently inter-organisational. Consider incentive alignment for example, a key tenet of SCM \citep{Lee2002-lr_KS,Selviaridis2018-dl_KS} – it has been studied mainly from the perspective of large firms, with the unit of analysis being a focal buying firm or the relationship between a buyer and its first-tier supplier. However, incentive alignment along multi-tier supply chains requires understanding the motivations, goals and constraints facing small firms upstream the supply network. In the context of supply chain cyber security, for instance, SME suppliers of products, services and software are often regarded as ``weak links''. And yet, these firms are reluctant or unable to invest in cyber defences \citep{Melnyk2022-qb_KS}. Unless we examine the perspective of these SME suppliers farther upstream, we cannot fully grasp why that is the case, and how incentives can be aligned across the supply chain to ensure cyber security.

In conclusion, OSCM research has only relatively recently began to pay attention to SMEs. For a thoughtful review of SME research in SCM, readers are directed to \cite{Kull2018-ub_KS}. Selviaridis and colleagues offer insights regarding how innovative SMEs can be more effectively integrated in supply chains, and the role of targeted public policies in this respect \citep{Selviaridis2021-pn_KS,Selviaridis2022-yb_KS}. Moving forward, as a discipline we need to understand better how SMEs can function as knowledgeable buyers and customers in supply chains, and the capabilities they need to effectively manage their own supply chains. We must also unpack what SCM entails for micro firms, including young, technology-based businesses who are good at developing novel products but may not necessarily have an established supply chain to source materials, and manufacture and distribute their innovations. More generally, pushing the frontiers of SME-oriented OSCM research requires synthesising operations and supply chain knowledge with insights from the longstanding research on small business, family business, and entrepreneurship \citep[e.g.,][]{Arend2005-xh_KS,De-Massis2018-dt_KS}.

\subsection[Utilities (Henk Akkermans \& Wendy van der Valk)]{Utilities\protect\footnote{This subsection was written by Henk Akkermans and Wendy van der Valk.}}
\label{sec:Utilities}
Utilities are a backbone of modern society, providing essential services such as electricity, water, gas, and telecommunications. As such, the management of utility assets, i.e., the often large-scale, expensive, and technically complex physical assets underpinning these services, has a direct impact on societal well-being, economic growth, and environmental sustainability. Operations and supply chain management principles play a critical role in ensuring these assets are effectively managed across their entire life cycle, from planning (\S\ref{sec:Planning}) and acquisition, to operations (\S\ref{sec:Operations_Management}) and maintenance (\S\ref{sec:Maintenance}), and eventual disposal or renewal (\S\ref{sec:Circular_economy} and \S\ref{sec:Material_handling}). 

Utility assets, such as power plants, water treatment facilities, and energy grids, provide the infrastructure from which essential services can be rendered, and are hence of critical importance to society at large. They are also often long-lived, with life spans stretching decades, making decisions regarding these assets highly important, as their effects may be substantial and last for a long time. As such, effective management of utilities is, either directly or indirectly, imperative for meeting the United Nations' Sustainable Development Goals (SDGs), such as access to clean water (SDG 6) and energy (SDG 7), or sustainable cities and communities (SDG 11). Finally, they are quite often state-owned or heavily regulated, partly due to their natural monopoly characteristics. The large capital requirements, network-based operations, and economies of scale make parallel systems inefficient, further emphasising the need for careful operations and supply chain management strategies to optimise performance while adhering to regulatory, safety, and sustainability requirements. Without such strategies, the societal and economic consequences of infrastructure failure or underperformance can be catastrophic \citep[see e.g.,][]{Frangopol2007-oo_HAWV}, as evidenced by events such as the collapses of Morandi bridge in Italy in 2018, and the one in Baltimore in 2024. The risks have been signalled by frequent reports of aging bridges, outdated power grids, and vulnerable water systems. Nevertheless, research on operational excellence for utilities is limited. For overviews of utilities-focused empirical research studies and of  industry-specific studies see \cite{Akkermans2024-et_HAWV} and \cite{Joglekar2016-il_HAWV}. 

For these reasons, the management of utility assets must adopt a long-term, systemic view that takes into account the various and deeply interconnected phases of their existence, known as the asset life cycle: planning, acquisition, operations, maintenance, and end-of-life (EOL); see \cite{Browning2008-zm_HAWV}. With decisions made in one phase affecting performance in subsequent phases, effective utility asset life cycle management is essential for ensuring their longevity, efficiency, and sustainability.

The \textit{planning phase} of the asset life cycle is critical because it sets the stage for all future decisions. In this phase, forecasting (\S\ref{sec:Forecasting}) plays a central role in determining demand, budgeting, and setting performance expectations for the assets to be acquired or constructed. For instance, demand forecasting in the energy sector can guide decisions about how many new power plants to build or how to expand grid capacity to meet future energy demands. Once planning is complete, the \textit{acquisition phase} involves sourcing and procurement (\S\ref{sec:Purchasing_procurement}) strategies to secure the necessary assets. Here, managers of utilities (often public) engage with a limited number of (often private) contractors through tendering processes that are subject to European and national procurement laws. This results in a lock-in for managers, that need to activate relationships for specific acquisition projects with a long yet temporary nature, while these relationships also need to be reproduced and sustained beyond individual projects \citep{Manning2017-il_HAWV}. Hence, management of long-term relations between partnering organisations needs to be balanced alongside contracting for specific projects \citep{Chakkol2018-ar_JRTB,Roehrich2024-sn_JRTB}. An interesting development in response is that some organisations are moving away from a project-by-project approach and towards contracting strategies that address the need for efficiency, speed, and standardisation when facing large-scale challenges \citep{Frederiksen2021-fu_HAWV}. 

Research furthermore highlights the challenges associated with risk allocation between asset owners and contractors: shifting too much risk to contractors can discourage participation, leading to fewer bids and higher costs, as seen in cases like bridge renovation projects in the Netherlands \citep{Vestergaard2024-vc_HAWV}. Performance-based contracting, which aligns contractor payments with the achievement of predefined goals, has emerged as a potential solution. However, such contracts require collaboration and deep understanding between service providers and utilities to ensure smooth operations throughout the asset’s life cycle \citep{Akkermans2019-mf_PM}. See \cite{Van-der-Valk2021-uh_HAWV} and \cite{Selviaridis2015-pr_HAWV} for reviews on (performance-based) contracting; and \cite{Fang2024-bb_HAWV} for a recent empirical example. 

Once the assets are \textit{operational}, the focus shifts to maximising their performance, reliability, and efficiency. This phase of the asset life cycle involves routine and preventive maintenance to ensure the asset functions as expected, while minimising downtimes and extending the useful life of the asset. Maintenance strategies have evolved significantly in recent years, particularly with the rise of condition-based maintenance, which relies on real-time monitoring and data analytics to predict when maintenance is needed based on the asset's condition \citep{Akkermans2024-vu_HAWV,uit-het-Broek2020-jp_HAWV}. This approach, enabled by digital technologies such as IoT, big data, and AI, improves asset reliability while reducing unnecessary maintenance costs. Also promising is the use of digital twins, which allow compiling a digital shadow of past behaviour \citep{Ladj2021-wx_HAWV} based on historical data of operations, conditions, and maintenance activities. However, there are challenges associated with the digitalisation of asset management. Many asset owners struggle when adopting digital solutions for managing utilities due to organisational silos, data-sharing barriers, and misaligned incentives between manufacturers, asset owners, and service providers \citep{Holmstrom2019-if_HAWV}. Overcoming these challenges requires enhanced supply chain coordination and collaboration. The concept of supply chain control towers -- centralised platforms for real-time data integration and decision-making -- has been proposed as a way to align operations and maintenance activities across stakeholders \citep{Gerrits2022-hp_HAWV,Topan2020-zk_HAWV}.

The final phase of the asset life cycle -- \textit{disposal} or \textit{end-of-life} (EOL) -- is increasingly being viewed through the lens of sustainability and circularity. Traditionally, utilities have not focused much on asset disposal, often resulting in significant environmental impacts. For example, old oil wells in Pennsylvania, many of which were abandoned without proper decommissioning, continue to leak methane into the atmosphere, contributing to climate change \citep{Aflaki2024-rd_HAWV}. The same risks apply to newer energy technologies such as wind turbines and solar panels, where proper disposal and recycling strategies are not always considered at the design stage \citep{Atasu2021-ea_HAWV}. As a consequence, sustainable business models for utility asset management that incorporate circular economy principles need to be developed, such as designing assets for easier disassembly and recycling, as well as developing closed-loop supply chains for recovering valuable materials from decommissioned assets, thereby reducing reliance on scarce natural resources. For instance, in road and waterways infrastructure, identification and separation of waste and recyclable material streams pose serious challenges because of the extended lifetime these materials have been in service. At the same time, the benefits of improved reuse and recycling would be substantial: Concrete and its main ingredient cement not only constitute an important resource for public infrastructures, but also contribute substantially to CO$_2$ emissions.

As utilities face increasing pressure to modernise ageing infrastructure, improve performance, and reduce environmental impacts, the application of these operations and supply chain management principles will become even more critical. One key area for further investigation is the further integration of end-of-life considerations into the planning and acquisition phases. Given the long life cycles of utility assets, the environmental and social impacts of disposal are often underestimated or ignored. Incorporating life cycle assessment tools into early-stage decision-making can help address these issues and promote more sustainable asset management practices \citep{Akkermans2024-vu_HAWV}. Another promising area for research is the use of advanced analytics and AI to improve forecasting, operations, and maintenance decisions. AI can analyse vast amounts of data from sensors and digital twins to predict asset failures and optimise maintenance schedules. However, these technologies also raise new challenges related to data governance, privacy, and security, which need to be carefully managed to avoid unintended consequences \citep{Angelopoulos2023-xv_HAWV}. Finally, the development of innovative contracting models that foster greater collaboration and alignment between utilities and their contractors is an area in need of increased attention. Programmatic tendering, for example, offers a way to bundle multiple projects into a single contract, creating economies of scale and encouraging innovation in project delivery \citep{Frederiksen2021-fu_HAWV}. Empirical research is needed to understand the operational conditions that lead to successful outcomes under such models.

From planning and acquisition to operations, maintenance, and EOL, each phase requires careful consideration of technical, financial, and sustainability factors. By embracing digitalisation, fostering collaboration across the supply chain, and integrating sustainability into decision-making processes, utilities can ensure that their assets continue to provide reliable and efficient services for generations to come. 

\clearpage

\section[Conclusions (Ana Barbosa-Povoa)]{Conclusions\protect\footnote{This subsection was written by Ana Barbosa-Povoa.}}
\label{sec:Conclusions}

This paper aims to provide an extensive review of the core principles and practices of operations and supply chain management (OSCM), tracing the field's evolution from its historical roots to its current challenges and innovations. It showcased how OSCM enables organisations to build sustainable competitive advantages. The paper also emphasised the growing importance of integrating broader societal, environmental, and technological trends into OSCM.

The structure of the paper was organised into four major sections. First, it explored key topics in operations management (OM), including diverse important topics that span from more strategic ones, such as operations strategy and capacity management to more operational ones, such as scheduling and maintenance. Next, it transitioned into supply chain management (SCM), addressing topics like supply chain strategy, stakeholder management, and information management, among others, as well it looks into the impacts of recent trends such as digitalisation and emerging economies. The paper also discussed the growing intersection between OM and SCM, particularly in areas such as sustainability, ethics, risk management, and innovation, which are becoming increasingly relevant in today’s global economy. Finally, real-world applications of OSCM principles were examined across various industries, namely agriculture, construction, public and nonprofit sectors, healthcare, humanitarian operations, manufacturing, professional services, retail, small and medium-sized enterprises, showcasing the practical impact of OCSM methodologies and solutions.

\subsubsection*{Future Directions and Research Opportunities}
While the paper has covered a wide range of topics, it has also revealed several areas where future research and practice could further enrich the field of OSCM. These areas of research represent both challenges and opportunities for scholars and practitioners:

\textbf{Enhanced Focus on Sustainability:} As environmental and social sustainability become critical imperatives for organisations, future research must deepen its exploration into how sustainability can be embedded into every aspect of OSCM. The paper highlighted discussions on the circular economy, environmental sustainability, and ethical supply chains. Future studies could investigate how organisations can integrate sustainability goals, such as decarbonisation, into their supply chain strategies without compromising operational efficiency. Collaboration among supply chains towards the common share of sustainable goals is another area of great potential where symbiotic supply chains are an example that should be pursued.  Further, exploring the role of sustainable innovations -- such as renewable energy sources, sustainable materials, and waste reduction technologies -- within supply chain frameworks could provide more actionable insights. Another key area could be to examine the impact of governmental regulations and how policy changes drive shifts towards greener supply chains.

\textbf{Technological Advancements and Digital Transformation:} Digitalisation, artificial intelligence (AI), machine learning (ML), and the Internet of Things (IoT) are already revolutionising operations and supply chain management (OSCM). However, the paper highlights that many organisations remain in the early stages of adopting these transformative technologies. Future research could focus on how AI and ML can improve predictive analytics for demand forecasting, automate supply chain decisions, and create smarter, more resilient networks. There is also potential in advancing decision-support tools that combine these technologies with optimisation techniques, providing more strategic insights for OSCM decision-making. Additionally, blockchain technology offers promising opportunities for enhancing transparency and traceability within supply chains, particularly in industries like pharmaceuticals and food, where ethical sourcing and safety are paramount. Further studies could investigate how real-time data sharing through IoT can improve coordination across supply chain networks, increasing visibility and mitigating risks related to demand fluctuations and supply disruptions.

\textbf{Behavioural Operations and the Human Factor:} Human decision-making plays a crucial role in OSCM, and behavioural biases can significantly impact operational outcomes. The paper introduced the concept of behavioural operations, which highlights how psychological factors influence managerial decisions. Future research could focus on the development of decision-support systems that account for human biases, such as overconfidence or the tendency to anchor decisions on past experiences. Understanding how to better manage the interplay between human judgment and technological systems (like AI or automated tools) will be key as more companies adopt hybrid decision-making models. Furthermore, exploring the dynamics of workforce management in an increasingly automated world, including how human-robot collaborations can be optimised, could also provide valuable insights.

\textbf{Risk Management and Supply Chain Resilience:} The disruptions caused by global events such as the COVID-19 pandemic have underscored the need for resilient supply chains. The paper addressed risk management and the ripple effect, but more work is needed to develop frameworks that help companies build resilience. Future studies could examine how supply chain managers can proactively design systems that anticipate and adapt to sudden disruptions, whether from pandemics, geopolitical events, or climate-related risks. Research could also explore how companies can balance the need for lean, cost-efficient operations with the need for redundancy and buffer capacity in critical areas of their supply chains. There is also a need for new models that combine traditional risk management practices with innovative concepts like supply chain decentralisation, onshoring, and the use of local suppliers to mitigate global risks.

\textbf{Collaboration and Ethical Supply Chains:} As global supply chains become more complex, the need for collaboration among diverse stakeholders is increasingly important. The paper’s exploration of ethical supply chains suggested that there is growing pressure on organisations to ensure their operations adhere to ethical standards, such as fair labour practices and responsible sourcing. Future research could investigate how companies can develop stronger relationships with their suppliers, customers, and communities to foster collaborative networks that prioritise ethical outcomes. This could include exploring how digital platforms facilitate transparency and accountability across the supply chain. Additionally, studying the role of cross-sector partnerships (e.g., between NGOs and corporations) in promoting ethical standards could offer new insights into managing supply chain challenges in a socially responsible manner.

\textbf{Broader Industry Applications and Contextual Adaptation:} While the paper provided applications of OSCM principles in industries such as agriculture, construction, and healthcare, there is significant room for research on how these principles can be adapted to other underexplored sectors. For instance, humanitarian logistics, which deals with the challenges of delivering aid in crisis situations, presents unique supply chain issues such as resource scarcity, high uncertainty, and the need for rapid response. Future work could investigate how supply chain strategies used in commercial sectors can be tailored to fit the unique constraints of humanitarian operations. Similarly, studying OSCM in emerging economies, where infrastructure and technological adoption levels vary widely, could reveal innovative approaches to overcoming challenges in resource-limited environments.

In conclusion, this paper has provided a thorough examination of the current state of OSCM, outlining the challenges and opportunities that lie ahead for both researchers and practitioners. By focusing on key areas such as sustainability, technological advancements, behavioural operations, risk management, and ethical supply chains, future work can push the boundaries of OSCM, helping organisations navigate the complexities of global supply chains while making meaningful contributions to societal goals. The ongoing evolution of OSCM will require adaptability, innovation, and a continued emphasis on collaboration across industries and academia where multidisciplinary approaches should be explored. Through such efforts, OSCM will continue to be a powerful driver of organisational success and societal well-being.

We hope you have found this collection on the OM-SCM field insightful. While we believe it serves as a valuable reference, we acknowledge it captures only a moment in time—reflecting our past, present, and the path forward.

\clearpage

\section*{Disclaimer}
The views expressed in this paper are those of the authors and do not necessarily reflect the views of their affiliated institutions and organisations.

\section*{Data availability statement}
No data have been used in this article.

\section*{Acknowledgements}

Fotios Petropoulos would like to thank Alexandre Dolgui and Mohamed Zied Babai for inviting this article for publication to the \textit{International Journal of Production Research}, as well as all the other co-authors who accepted his invitation to be part of this encyclopedic article.

Henk Akkermans and Wendy van der Valk would like to thank Finn Wynstra (RSM Erasmus University) and Luk van Wassenhove (INSEAD) who co-authored the \textit{Journal of Operations Management}'s Special Issue editorial that their subsection builds on.

Ana Barbosa-Povoa thanks FCT – Foundation for Science and Technology, I.P., under the project UIDB/00097/2020.

Alistair Brandon-Jones thanks Maneesh Kumar (University of Cardiff) and Andrea Chiarini (University of Verona) for their support in developing this content.

Salar Ghamat's work is supported by Canada Research Chairs Program (CRC, Tier 2, in Business Analytics in Supply Chain, CRC-2022-00130).

Maximilian Koppenberg's work was supported by 4TU.Federation under the project 4TU.Redesign.

\clearpage

\appendix
\section{List of acronyms}\label{sec:acronyms}

\noindent
AGV: Automated Guided Vehicle \\
AI: Artificial Intelligence \\
AIV: Autonomous Intelligent Vehicle \\
ALFP: Assembly Line Feeding Problem \\
AMR: Autonomous Mobile Robot \\
AMT: Advanced Manufacturing Technology \\
APIOBPCS: Automatic Pipeline, Inventory, and Order-Based Production Control System \\
APM: Association for Project Management \\
APP: Aggregate Production Planning \\
APSs: Advanced Planning Systems \\
AR: Accurate Response \\
ARIMA: AutoRegressive Integrated Moving Average \\
ARIMAx: AutoRegressive Integrated Moving Average with eXternal regressors \\
ARP: Age Replacement Policy \\
ASC: Agricultural Supply Chain \\
ASRS: Automated Storage and Retrieval System \\
ATO: Assemble To Order \\
AVS/RS: Autonomous Vehicle Storage and Retrieval System \\
B2B: Business-To-Business \\
BAP: Buffer Allocation Problem \\
BD: Big Data \\
BoP: Base of Pyramid \\
BRP: Block Replacement Policy \\
CAM: Computer Aided Manufacturing \\
CBAM: Carbon Border Adjustment Mechanism \\
CBM: Circular Business Models \\
CD: Compact Disc \\
CE: Circular Economy \\
CI: Customer Involvement \\
CIM: Computer-Integrated Manufacturing \\
CM: Corrective Maintenance \\
CO$_2$: Carbon dioxide \\
CODP: Customer Order Decoupling Point \\
CONWIP: Constant Work-In-Progress \\
CPFR: Collaborative Planning, Forecasting, and Replenishment \\
CPM: Critical Path Method \\
CR: Continuous Replenishment \\
CRAFT: Computerised Relative Allocation of Facilities Technique \\
CRM: Customer Relationship Management \\
CRPS: Continuous Ranked Probability Score \\
CSC: Construction Supply Chain \\
CSR: Corporate Social Responsibility \\
DApps: Decentralised Applications \\
DBFM: Design, Build, Finance, and Maintain \\
DBFMO: Design, Build, Finance, Maintain, and Operate\\
DBS: Dual Base Stock \\
DCS: Distributed Control System \\
DEA: Data Envelopment Analysis \\
DES: Discrete-Event Simulation \\
DfD: Design for Disassembly \\
DfX: Design for X \\
DP: Demand Planning \\
EDI: Electronic Data Interchange \\
EFQM: European Foundation for Quality Management \\
ELSP: Economic Lot Scheduling Problem \\
EOL: End-Of-Life \\
EOQ: Economic Order Quantity \\
EPC: Engineering, Procurement and Construction \\
EPQ: Economic Production Quantity \\
ERP: Enterprise Resource Planning \\
ES: Exponential Smoothing \\
ESG: Environmental, Social and Governance \\
ETO: Engineer To Order \\
ETS: ExponenTial Smoothing \\
ETSx: ExponenTial Smoothing with eXternal regressors \\
EU: European Union \\
EV: Electronic Vehicle \\
FCFS: First-Come-First-Serve \\
FLP: Facility Layout Problem \\
FMS: Flexible Manufacturing Systems \\
GHG: GreenHouse Gas \\
GSC: Global Supply Chain \\
HMM: Hidden Markov Model \\
HO: Humanitarian Operations \\
HR: Human Resources \\
HSC: Healthcare Supply Chain \\
HSCM: Healthcare Supply Chain Management \\
I4.0: Industry 4.0 \\
IBM: Inspection-Based Maintenance \\
ICT: Information and Communication Technology \\
IDT: Innovation Diffusion Theory \\
IHO: International Humanitarian Organisation \\
IIoT: Industrial Internet of Things \\
IMP: Industrial Marketing and Purchasing \\
IOBPCS: Inventory, and Order-Based Production Control System \\
IoT: Internet of Things \\
IS: Information Systems \\
IS: Interval Score \\
IT: Information Technology \\
JIT: Just-In-Time \\
JRP: Joint Replenishment Problem \\
KPI: Key Performance Indicator \\
LCA: Life Cycle Analysis \\
LDA: Latent Dirichlet Allocation \\
LED: Light-Emitting Diode \\
LIP: Large Interorganisational Project \\
LP: Linear Programming \\
LSS: Lean Six Sigma \\
MAE: Mean Absolute Error \\
MAPE: Mean Absolute Percentage Error \\
MASE: Mean Absolute Scaled Error \\
MBNQA: Malcolm Baldrige National Quality Award \\
MCDM: Multi-Criteria Decision Making \\
MES: Manufacturing Execution System \\
METRIC: Multi-Echelon Technique for Recoverable Item Control \\
MFA: Material Flow Analysis \\
MFC: Material Flow Control \\
MFCA: Material Flow Cost Accounting \\
MFFLP: Multi-Floor Facility Layout Problem \\
MHE: Material Handling Equipment \\
MIP: Mixed-Integer Programming \\
MILP: Mixed-Integer Linear Program \\
ML: Machine Learning \\
MP: Master Planning \\
MRP: Material Requirements Planning \\
MRP-II: Manufacturing Resource Planning \\
MTO: Make To Order \\
MTS: Make To Stock \\
MvB: Make-versus-Buy \\
NGO: Non-Governmental Organisation \\
NLP: Natural Language Processing \\
NPD: New Product Development \\
NSD: New Service Development \\
OEE: Overall Equipment Effectiveness \\
OEM: Original Equipment Manufacturer \\
OM: Operations Management \\
OPEX: OPerational EXcellence \\
OR: Operational Research \\
OS: Operations Strategy \\
OSCM: Operations and Supply Chain Management \\
OT: Operational Technology \\
OUT: Order-Up-To \\
PCA: Principal Component Analysis \\
PDM: Predictive Maintenance \\
PLC: Programmable Logic Controllers \\
PLM: Product Lifecycle Management \\
PM: Project Management \\
PMBOK: Project Management Body of Knowledge \\
PMI: Project Management Institute \\
PMTO: Purchase and Make To Order \\
PoS: Proof of Stake \\
POUT: Proportional Order-Up-To \\
PoW: Proof of Work \\
PPC: Production Planning and Control \\
PPE: Personal Protective Equipment \\
PP\&S: Production Planning and Scheduling \\
QR: Quick Response \\
R\&D: Research and Development \\
RBV: Resource Based View \\
ReSOLVE: Regenerate, Share, Optimise, Loop, Virtualise, and Exchange \\
RFID: Radio Frequency IDentification \\
RL: Reverse Logistics \\
RMSE: Root Mean Squared Error \\
SALBP: Simple Assembly Line Balancing Problem \\
SC: Supply Chain \\
SCADA: Supervisory Control And Data Acquisition \\
SCM: Supply Chain Management \\
SCMS: Supply Chain Management System \\
SCOR: Supply Chain Operations Reference \\
SCR: Supply Chain Resilience \\
SDG: Sustainable Development Goal \\
SDP: SemiDefinite Program \\
SI: Supplier Involvement \\
SKU: Stock Keeping Unit \\
SME: Small- and Medium-sized Enterprises \\
SNA: Social Network Theory \\
SPC: Statistical Process Control \\
TAM: Technology Acceptance Model \\
TCE: Transaction Cost Economics \\
TOE: Technology-Organisation-Environment \\
TPS: Toyota Production System \\
TQM: Total Quality Management \\
UAFLP: Unequal-Area Facility Layout Problem \\
UK: United Kingdom \\
UTAUT: Unified Theory of Acceptance and Use of Technology \\
VAR: Vector AutoRegression \\
VMI: Vendor Managed Inventory \\
VSM: Value Stream Mapping \\
WIP: Work-In-Progress \\

\clearpage

%% References
\bibliographystyle{elsarticle-harv}
\bibliography{refs}

\end{document}